\documentclass[11pt]{article}

\RequirePackage[OT1]{fontenc}
\RequirePackage{amsthm,amsmath}
\RequirePackage{natbib}
\usepackage{multirow}
\usepackage{graphicx}
\usepackage{enumitem}
\usepackage{amsfonts}
\usepackage{gensymb}
\usepackage{bigints}
\usepackage{multicol}
\usepackage{xr}
\usepackage{authblk}
\usepackage{fancyhdr}
\usepackage{subfig}
\usepackage{hyperref}
\begin{document}

\title{\bf Fire seasonality identification with multimodality tests}

\author{Jose Ameijeiras--Alonso$^{1,2,*}$,
 Akli Benali$^{3,\dagger}$, 
Rosa M. Crujeiras$^{1,\ddagger}$, 
Alberto Rodr\'iguez--Casal$^{1,\ddagger}$ 
and Jos\'e M.C. Pereira$^{3,\dagger}$.
}
\date{%
    $^*$KU Leuven, 
    $^\dagger$University of Lisbon and 
		$^\ddagger$Universidade de Santiago de Compostela
}
\footnotetext[1]{Supported by the Project MTM2016--76969--P from the Spanish State Research Agency (AEI) co--funded by the European Regional Development Fund (ERDF), the Competitive Reference Groups 2017--2020 (ED431C 2017/38) from the Xunta de Galicia through the ERDF.}
\footnotetext[2]{Supported by the FWO research project G.0826.15N (Flemish Science Foundation), GOA/12/014 project (Research Fund KU Leuven) and was partially supported by the grant BES--2014--071006 from the Spanish Ministry of Science, Innovation and Universities.}
\footnotetext[3]{Supported by Centro de Estudos Florestais (CEF) Strategic Project (UID/AGR/00239/2013).}
\maketitle

\begin{abstract}
Understanding the role of vegetation fires in the Earth system has become an important environmental problem. Although fires time occurrence is mainly influenced by climate, human activity related with land use and management has altered fire patterns in several regions of the world. Hence, for a better insight in fires regimes, it is of special interest to analyze where human activity has influenced the fire seasonality. For doing so, multimodality tests are a useful tool for determining the number of fire peaks along the year. The periodicity of climatological and human--altered fires and their complex distributional features motivate the use of the nonparametric circular statistics. The unsatisfactory performance of previous nonparametric proposals for testing multimodality, in the circular case, justifies the introduction of a new approach, accompanied by a correction of the False Discovery Rate with spatial dependence for a systematic application of the tests in a large area between Russia and Kazakhstan. 
\end{abstract}

\noindent%

\section{Introduction and motivation}\label{introfires}

Vegetation fires are caused by several factors, being their occurrence strongly influenced by climate. In general, in the different areas of the world, the climatological conditions favor the occurrence of fires concentrated around one specific annual season. For instance, attending to climatological reasons, in most of the temperate regions over the Tropic of Cancer (approximately $23.5\degree$ north latitude), fires occur from May to September, when dry conditions prevail \citep{LePageetal10}. But apart from climate, human activity also influences the fire regimes in some cases, altering the seasonality of fires occurrence. Fires are employed for many purposes related to land use practices, with preferential timings. For example, fire is used by humans for hunting, pasture management, clearing fields for agriculture, eliminating crop and forest harvest residues, manage fuels and reduce wildfire risk or clear brush and drive away wild animals. Analyzing in which regions the seasonality due to human burning activities can be separated from fires occurring during climatological seasons helps to identify where fire is used as a land management tool \citep{Magi12}. 

In this paper, a new tool for analyzing how many fire seasons can be identified in a certain region is presented. The issue of determining the number of fire seasons can be translated into the statistical problem of testing the number of modes, defined as local maxima of the density function. When studying this random variable, fires periodicity (jointly with the possibility of having a single season of fires both in December and January) must be accounted for, motivating the use of circular statistics to analyze this kind of data \citep{XuSchoenberg11,Benali17}. From a parametric approach, \cite{Benali17} tackle this problem using a mixture of two von Mises (circular) distributions. However, the human--altered seasons of fires can be caused by several factors and as the shapes counting the number of human fires show \citep[][Figure~5]{Korontzi06}, the underlying structure of the wildfires can be very complex (presenting, e.g., asymmetry). In such context, simple parametric models may not capture appropriately the data characteristics. This fact motivates the considertion of nonparametric techniques for determining the number of fires seasons with a testing approach. For scalar (real--valued) data, different alternatives have been presented in the statistical literature, some of them based on the idea of the \textit{critical bandwidth} defined by \cite{Silverman81} and others using as a test statistic the \textit{excess mass} introduced by \cite{MulSaw91}. 
In the circular case, just \cite{FisMar01} provide an approach for testing multimodality, using the $U^2$ of Watson as a test statistic, but computational results show the poor calibration in practice even for ``large'' values of sample size (see Section~\ref{simstud}). The proposal for solving the multimodality testing problem presented in this paper considers an adapted version of the excess mass statistic for circular data. A correct calibration is guaranteed using a bootstrap procedure, where resampling is based on a nonparametric estimator (a modified kernel density estimator) of the circular density function.

{The study area used herein, straddling the border between Russia and Kazakhstan, is one of the main agricultural regions in the world. Fire is widely used here \citep{LePageetal10}, both before planting and after harvest, resulting in a multimodal fire season pattern \citep[][Figure 2]{Hall16,Benali17}.} The method proposed in this paper will be applied in each grid \textit{cell} of size $0.5\degree$ in the study area (represented in Figure \ref{fignfires}). The analysis of the number of seasons in the different cells can be used as an indicator of human activity, studying in which of these cells more fire seasons appear than those expected under regional precipitation seasonality patterns. Previous studies \citep[see, for example,][]{LePageetal10,Benali17} have shown that, in the study region, the dry season, when fire weather severity is higher, lasts from June to September. The Summer fire season coincides with this period, but it seems that there is also another annual fire season that occurs earlier, in early Spring (March and April), when climate conditions are not conducive of wildfires. It is this temporal mismatch between an observed fire season and the meteorological conditions most suitable for fire that, generally, is indicative of anthropogenic vegetation burning, as opposed to fires of natural origin.

Another issue that needs to be taken into account for the practical application is related with the spatial area division. The study area is separated in cells of size $0.5\degree$ (see Figure \ref{fignfires}) and the nonparametric test must be applied systematically in each cell. In this context, a \textit{False Discovered Rate} (FDR) procedure is required in order to control the incorrect rejections of the null hypothesis, that is, the identification of unimodal fire regimes as multimodal. Note also that the temporal pattern of fires can be spatially correlated with the neighboring cells. To identify in which cells fire occurrence is expected to be similar, the land cover information can be employed to somehow reflect the spatial relation between cells. Then, an adaptation of the \cite{BenjaminiHeller07} proposal will be applied in order to correct the FDR accounting for the spatial dependence of the data.

Summarizing: a nonparametric testing procedure for determining the number of modes in a circular density will be presented. This procedure is designed with the goal of determining the number of fire seasons in the Russia--Kazakhstan area, treated in a lattice division in such a way that the test is applied systematically in each cell of the lattice, requiring therefore a FDR correction. 

The organization of the paper is the following: Section~\ref{method} details the circular excess mass approach for testing the null hypothesis that the data underlying distribution has $k$ modes. The method is validated in Section~\ref{simstud}, presenting a complete simulation study and comparing the new proposal with the one by \cite{FisMar01}, in terms of empirical size and power. The FDR correction, accounting for the spatial dependence of the data, and the analysis of the number of fire seasons in the study area is done in Section~\ref{datafires}. Some final comments and discussion are given in Section~\ref{conclusions}. Details on the models employed in the simulation study; a complete description of the calibration function used to generate the resamples in the bootstrap procedure with some theoretical background; and the construction of the fire patches cells where a similar fire behavior is expected (with resemblance to land use), are provided in the appendix.

\section{Statistical tools: a nonparametric test for circular multimodality}\label{method}

Directional data, observations on directions, arise quite frequently in many natural sciences and in particular in wildfires modeling \citep[several examples are provided in][]{Ameijeiras18}.  
The need of circular statistics appears when the periodicity must be taken into account and the sample can be represented on the circumference. As mentioned before, this is the case of the wildfires dataset where a strong seasonal pattern appears.

Given a circular random variable $\Theta$, with probability density function $f$, the goal is to test if the number of modes of $f$ (season of fires), namely $j$, is equal to a given value $k\in\mathbb Z^+$ (climatological season of fires), against if it is greater than $k$. In general, rejecting $H_0$ will represent evidence of human influence on fire seasonality in the region. The statistical testing problem can be formulated as assessing
\begin{equation}\label{testh0} 
H_0: j= k \quad \mbox{vs} \quad H_a: j>k.
\end{equation}

\begin{figure}[t]
 \centering
\begin{tabular}{cc}
\subfloat{
    \includegraphics[width=0.49\textwidth]{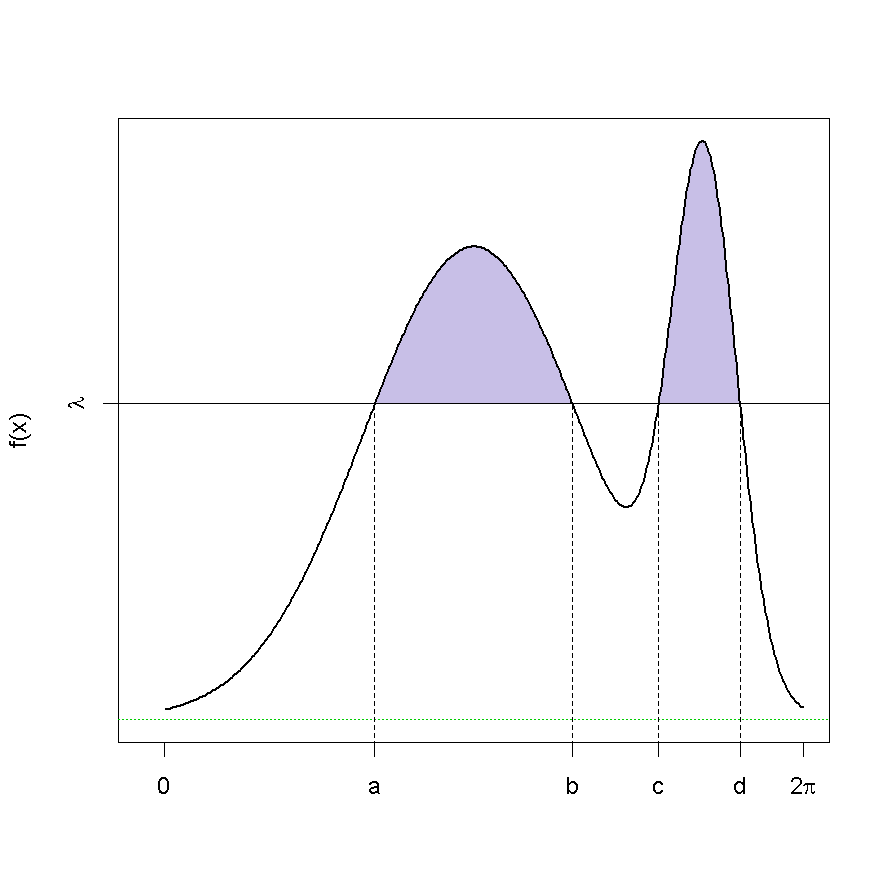}}
\subfloat{
    \includegraphics[width=0.49\textwidth]{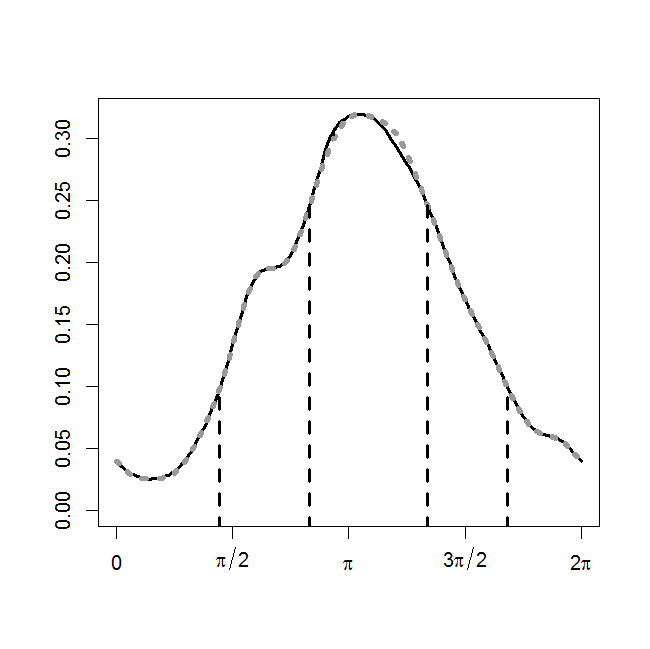}}\\
\end{tabular}
 \caption{Left: theoretical excess mass function for two modes (in gray), i.e., largest probability of mass exceeding the level $\lambda$ (horizontal line) when taking two arcs. Right: kernel density estimation with critical concentration parameter for one mode, $\hat{f}_{\nu_1}$ (dotted gray line), and calibration function, $g$ (solid line). The sample ($n=200$) was obtained from the model MC7. Dashed lines: neighborhoods where the $J$ functions are defined for $\widehat{\theta}_1$ and $\widehat{\theta}_2$.}
 \label{figpaper}
\end{figure}

For $H_0$ with a general $k$, the excess mass, introduced by \cite{MulSaw91}, can be used as a test statistic, which requires the construction of an empirical excess mass function for $k$ modes. Given a sample $\boldsymbol{\mathit{\Theta}}=(\Theta_1,\ldots,\Theta_n)$ from $\Theta$, the empirical excess mass is defined as
\begin{equation*}
E_{n,k}(\mathbb{P}_n,\lambda)=\underset{{C}_1(\lambda),...,{C}_k(\lambda)}{\mbox{sup}} \left\{ \overset {k} {\underset {m=1} \sum} \mathbb{P}_n({C}_m(\lambda))-\lambda ||{C}_m(\lambda)||\right\},
\end{equation*}
where the supremum is taken over all families $\{C_m(\lambda) : m = 1, \cdots, k\}$ of disjoint closed arcs, $||{C}_m(\lambda)||$ denotes the set measure, $\mathbb{P}_n(C_m(\lambda))=(1/n) \sum_{i=1}^n \mathcal{I}(\Theta_i \in C_m(\lambda))$ and $\mathcal{I}$ is the indicator function. An example of the theoretical excess mass is provided in Figure~\ref{figpaper} (left) for illustration purposes. A way of determining the plausibility of the null hypothesis ($f$ has $k$ modes) is by observing if the difference $D_{n,k+1}(\lambda)=E_{n,k+1}(\mathbb{P}_n,\lambda) - E_{n,k}(\mathbb{P}_n,\lambda)$ is ``large''. Using these differences for different thresholds ($\lambda$ values), the test statistic for (\ref{testh0}) is
\begin{equation}\label{emstat} 
\Delta_{n,k+1}=  \underset{\lambda}{\max } \{D_{n,k+1}(\lambda) \},
\end{equation}
and the null hypothesis that $f$ has $k$ modes is rejected for large values of $\Delta_{n,k+1}$.

\subsection{Results for scalar random variables}\label{linch}
As it has been previously mentioned, the test statistic in~(\ref{emstat}) was originally proposed for scalar random variables. For the linear case, the asymptotic behavior of the excess mass was provided by \cite{ChengHall98}. Under some assumptions, which include $f$ being class--one and the existence of a finite number of stationary points, which are the modes and antimodes (denoted as $x_1,\ldots,x_{2k-1}$), jointly with some regularity conditions on $f$ in a neighborhood of these points (see Appendix~\ref{calibrationfun}); the distribution of $\Delta_{n,k+1}$ is independent of unknowns except for the factor
\begin{equation}\label{dich}
d_i=\frac{|f''(x_i)|}{f(x_i)^3} \mbox{, with } i=1\ldots,2k-1.
\end{equation}
If $f$ has $k$ modes, \cite{ChengHall98} also proved that the distribution of $\Delta_{n,k+1}$ can be approximated by $\Delta_{n,k+1}^*$ calculated from bootstrap resamples generated from a `calibration distribution' with $k$ modes. For guaranteeing an asymptotic correct behavior, this calibration function, namely $g$, must satisfy that its associated values of $\hat{d_i}=|g''(\hat{x_i})|/g(\hat{x_i})^3$ in its modes and antimodes, denoted as $\hat{x_i}$, converge in probability to the value of $d_i$ in (\ref{dich}), as $n \rightarrow \infty$, for $i=1,\ldots,2k-1$ (see Appendix~\ref{calibrationfun}). These ideas can be adapted to the circular case, with the complexity of defining an adequate calibration function to generate bootstrap samples. In what follows the construction of this calibration function is illustrated.

\subsection{The calibration function}\label{calfunc}

The construction of an adequate calibration function $g$ must be done in the following way: \hypertarget{obji}{(i)} preserving the structure of the data under the assumption that $f$ has $k$ modes and antimodes; \hypertarget{objii}{(ii)} verifying that $\hat{d}_i=|g''(\widehat{\theta_i})|/g(\widehat{\theta_i})^3$ converges in probability to $d_i=|f''(\theta_i)|/f(\theta_i)^3$, for $i=1,\ldots,2k$, as $n \rightarrow \infty$, where $\theta_i$ and $\widehat{\theta_i}$ are, respectively, the modes and antimodes of $f$ and $g$; \hypertarget{objiii}{(iii)} satisfying some regularity conditions (see Appendix~\ref{calibrationfun}). Note that an abuse of notation was made in this part as now $d_i$ and $\hat{d}_i$ refer to the circular case. 

The calibration function $g$ will be obtained as follows. First, for estimating the unknown circular density $f$ in \hyperlink{obji}{(i)}, given the random sample of angles $\boldsymbol{\mathit{\Theta}}$, the kernel density estimator is employed. This estimator is defined as

\begin{equation*}
\hat{f}_{\nu}(\theta)=\frac{1}{n} \overset{n}{\underset {i=1} \sum} K\left(\theta;\Theta_i,\nu \right) \mbox{, with } \theta\in[0,2\pi), 
\end{equation*}
where $K(\cdot;\Theta_i,\nu)$ is a kernel function, centered in $\Theta_i$ and concentration parameter $\nu$ \citep[see, e.g.,][]{Oliveiraetal12}. The chosen kernel function is the wrapped normal density with mean direction $\Theta_i$ and concentration parameter $\nu\in(0,1)$. This specific kernel leads to the following representation for the kernel density estimator:
\begin{equation}\label{kernel}
\hat{f}_{\nu}(\theta)= \frac{1}{2\pi n} \overset{n}{\underset {i=1} \sum}   \left( 1 + 2 \overset{\infty}{\underset {p=1}\sum} \nu^{p^2} \cos (p (\theta-\Theta_i)) \right) \mbox{, with } \theta\in[0,2\pi).
\end{equation}

The kernel density estimator given in (\ref{kernel}) preserves the structure of the sample, depending the number of modes on the concentration parameter $\nu$. With this particular kernel, the number of modes of $\hat{f}_{\nu}$ is always a nondecreasing function of $\nu$ \citep{Huckemann14}. Hence, with the aim of preserving the structure of the data under the $k$--modality hypothesis, i.e., objective \hyperlink{obji}{(i)}, an analogue of the critical bandwidth of \cite{Silverman81}, namely the critical concentration, can be employed as the concentration parameter for (\ref{kernel}). The critical concentration is defined as
\begin{equation}\label{nucrit}
\nu_k=\max\{\nu :\hat{f}_\nu \mbox{ has at most } k \mbox{ modes}\}.
\end{equation}
A representation of the kernel density estimation, employing $\nu_1$ as the concentration parameter, can be observed in Figure~\ref{figpaper} (right, dotted gray line). A unimodal estimation can be observed and a second mode will appear between $\pi/2$ and $\pi$ if a larger value of $\nu$ is taken. With this concentration parameter, $\hat{f}_{\nu_k}$ should provide a correct estimation of the density function and also of the modes and antimodes locations. Nonetheless, a good estimator of $f''(\theta_i)$ is also needed in order to ensure \hyperlink{objii}{(ii)}, i.e., $\hat{d}_i$ converges in probability to $d_i$, as $n \rightarrow \infty$. In that case, for correctly estimating $f$ and $f''$ different concentration parameters are required. $f''$ can be properly estimated taking the value of $\nu$ which minimizes the asymptotic mean integrated squared error expression of $\hat{f}''_{\nu}$, replacing $f$ by a mixture of $M$ von Mises \citep[a similar procedure for estimating $f$ was proposed by][]{Oliveiraetal12}. If $\nu_{\tiny{\mbox{PI}}}$ denotes this parameter, then $d_i$ can be estimated from the sample with

\begin{equation}\label{dihat}
\hat{d_i}=\frac{|\hat{f}''_{\nu_{\tiny{\mbox{PI}}}}(\widehat{\theta_i})|}{\hat{f}_{\nu_k}(\widehat{\theta_i})^3}  \mbox{, with } i=1\ldots,2k.
\end{equation}

To construct the calibration function $g$ from $\hat{f}_{\nu}$, satisfying \hyperlink{objii}{(ii)} and \hyperlink{objiii}{(iii)}, two modifications are needed. First, it is necessary to remove the $t$ saddle points denoted as $\zeta_p$, for $p\in\{1,\ldots,t\}$, this will be done below with the $L$ function. Secondly, the density estimator will be modified, with the function $J$, in a neighborhood of the estimated turning points in order to obtain that $|g''(\widehat{\theta_i})|/(g(\widehat{\theta_i}))^3$ is equal to the value of $\hat{d_i}$ in (\ref{dihat}), for $i\in\{1,\ldots,2k\}$. Then, the calibration function will be obtained by modifying the kernel density estimator with the critical concentration in the stationary points, using a similar procedure as in \cite{Ameijeiras17}. In particular, the employed calibration function is

\begin{equation}\label{gfunc}
g(\theta;\nu_k,\nu_{\tiny{\mbox{PI}}},\boldsymbol{\varsigma})= 
  \begin{cases}
J(\theta;\theta_i,\nu_k,\nu_{\tiny{\mbox{PI}}},\varsigma_i) & \mbox{if } \theta \mbox{ is in a neighbourhood of } \widehat{\theta_i}, \\
L\left(\theta;z_{(2p-1)},z_{(2p-1)}\right) & \mbox{if } \theta \mbox{ is in a neighbourhood of } \widehat{\zeta_p},\\
\hat{f}_{\nu_k}(\theta) & \mbox{in other case.}
	\end{cases}
\end{equation}
Functions $J$ and $L$ are respectively applied in each $i \in \{1,\ldots,2k\}$ and $p\in\{1,\ldots,t\}$. The complete characterization of $g$ is provided in Appendix~\ref{calibrationfun} and an example of its representation is given in Figure~\ref{figpaper} (right, continuous line) where the effect of the $J$ function can be observed.  

Using the calibration function defined in (\ref{gfunc}), the proposal in this paper for testing the null hypothesis (\ref{testh0}) is to consider a bootstrap procedure in order to calibrate the excess mass statistic defined in (\ref{emstat}). Given the sample $\boldsymbol{\mathit{\Theta}}$, $B$ resamples $\boldsymbol{\mathit{\Theta}}^{*b}$ ($b=1,\ldots,B$) of size $n$ are generated from $g(\cdot;\nu_k,\nu_{\tiny{\mbox{PI}}},\boldsymbol{\varsigma})$. If $\Delta_{n,k+1}^*$ is the excess mass statistic obtained from the generated resamples, for a significance level $\alpha$, the null hypothesis will be rejected if $\mathbb{P}(\Delta_{n,k+1}^*\leq \Delta_{n,k+1}|\boldsymbol{\mathit{\Theta}})\geq 1- \alpha$.

\section{Simulation study}\label{simstud}

The aim of the following simulation study is to analyze the performance of the testing method, using the bootstrap procedure proposed in Section~\ref{method}. The empirical size and power of the new method is also compared with the other existing proposal for testing multimodality for circular data, the one introduced by \cite{FisMar01}. For testing $k$--modality, they suggest using the $U^2$ statistic of \cite{Watson61} as a test statistic, that is
$$
U^2=n \int_{0}^{2\pi} {\left[ {F_n}(x)-F_0(x)- \int_{0}^{2\pi}{( {F_n} (y)-F_0(y))dF_0(y)} \right]^2dF_0 (x)},
$$
estimating $F_0$ (circular distribution function) employing a kernel distribution estimation with $k$ modes. In this simulation study, the distribution function associated to $\hat{f}_{\nu_k}$ is employed to estimate $F_0$ and its associated distribution is used for generating the bootstrap resamples to calibrate the test statistic.

Samples of size $n = 50$, $n = 200$ and $n = 1000$ ($n = 100$ instead of $n = 1000$ in power studies) were drawn from 25 different distribution, ten of them unimodal (M1--M10), ten bimodal (M11--M20) and five trimodal (M21--M25), including unimodal symmetric models, mixtures of them and reflective asymmetric models. These distributions models are described in Appendix~\ref{app_sim_models}. For each model (M1--M25) and sample size, 500 sample realizations were generated. Conditionally on each sample, for each test, 500 resamples of size $n$ were drawn using the calibration function of each test ($g$ for the new proposal and $\hat{f}_{\nu_k}$ for the $U^2$ statistic). Results are reported for significance levels $\alpha=0.01$, $\alpha=0.05$ and $\alpha=0.10$.

Results are organized as follows: Table~\ref{estsimcirc1} shows empirical sizes \hyperref[estsimcirc1]{(a)} and power \hyperref[estsimcirc1]{(b)} for testing $H_0:j =1$. Table \hyperref[estsimcirc2]{\ref*{estsimcirc2}(a)} and \hyperref[estsimcirc2]{(b)} show the same results (empirical sizes and power, respectively) for $H_0:j =2$.

From Tables \hyperref[estsimcirc1]{\ref*{estsimcirc1}(a)} and \hyperref[estsimcirc2]{\ref*{estsimcirc2}(a)}, the poor calibration of the \cite{FisMar01} proposal can be observed. Even for sample size equal to 1000, sometimes, the percentage of rejections is under the significance level, as in the distributions where unimodality is tested: M1, M2, M4, M8, M9 or M10; or the models where bimodality is assessed: M12, M14, M16 or M17. For other scenarios, as in models M3 and M5 (unimodality) and M11 and M20 (bimodality), the percentage of rejections is above $\alpha$. 

For the new proposal, as shown in Tables \hyperref[estsimcirc1]{\ref*{estsimcirc1}(a)} and \hyperref[estsimcirc2]{\ref*{estsimcirc2}(a)}, a good level accuracy is obtained in general, with the exception of model M3. Even for small sample sizes ($n=50$), when the null hypothesis of unimodality is tested, the percentage of rejections is close to the significance level $\alpha$, except in the commented case M3, and also on models: M1 ($n=50$), M4 ($n=200$), M8 ($n=50$), M9 ($n=200$) and M10 ($n=50$), where the percentage of rejections is slightly below the significance level. For testing bimodality, when the sample size is equal or larger than $n=200$, our proposal seems to calibrate correctly, except for model M11 where, the percentage of rejections is slightly below $\alpha$. When the sample size is not large enough, our new proposal presents a conservative performance in the leptokurtic models, such as model M3. In this last model, this behavior is corrected when considering a larger sample size ($n=2000$).

Power results in Tables \hyperref[estsimcirc1]{\ref*{estsimcirc1}(b)} and \hyperref[estsimcirc2]{\ref*{estsimcirc2}(b)} show that the new proposal, which seems to be the only one which is well calibrated, appears to have also good power, in terms that the percentage of rejections increases with the sample size. The method detects the clearly rejection of the null hypothesis on the bimodal model M11 and on the trimodal models M21 and M22. This new proposal also detects the small blips, for example on models M14, M15 (bimodal) and M25 (trimodal), although, in other distributions, it has some troubles, when these small blips represent a low percentage of the data and also the sample size is small, like, for example, in model M10. In the difficult cases, with almost overlapping peaks, such as models M13 (bimodal), M23 (trimodal) and M24 (trimodal), when the sample size is small, the new method has some difficulties to detect the rejection of unimodality (with $n=50$ in M13) and the rejection of bimodality (with $n=100$ in M23 and with $n=50$ in M24), but, as expected, the percentage of rejections increases with $n$.

\begin{table}
\scalebox{0.58}{
\begin{tabular}{|c |c|c|c c c |c c c |}
\hline
(a) & Unimodal &  $\alpha$ & 0.01 & 0.05& 0.10 & 0.01 & 0.05& 0.10 \\ \hline
\multirow{4}{*}{\includegraphics[width=15mm]{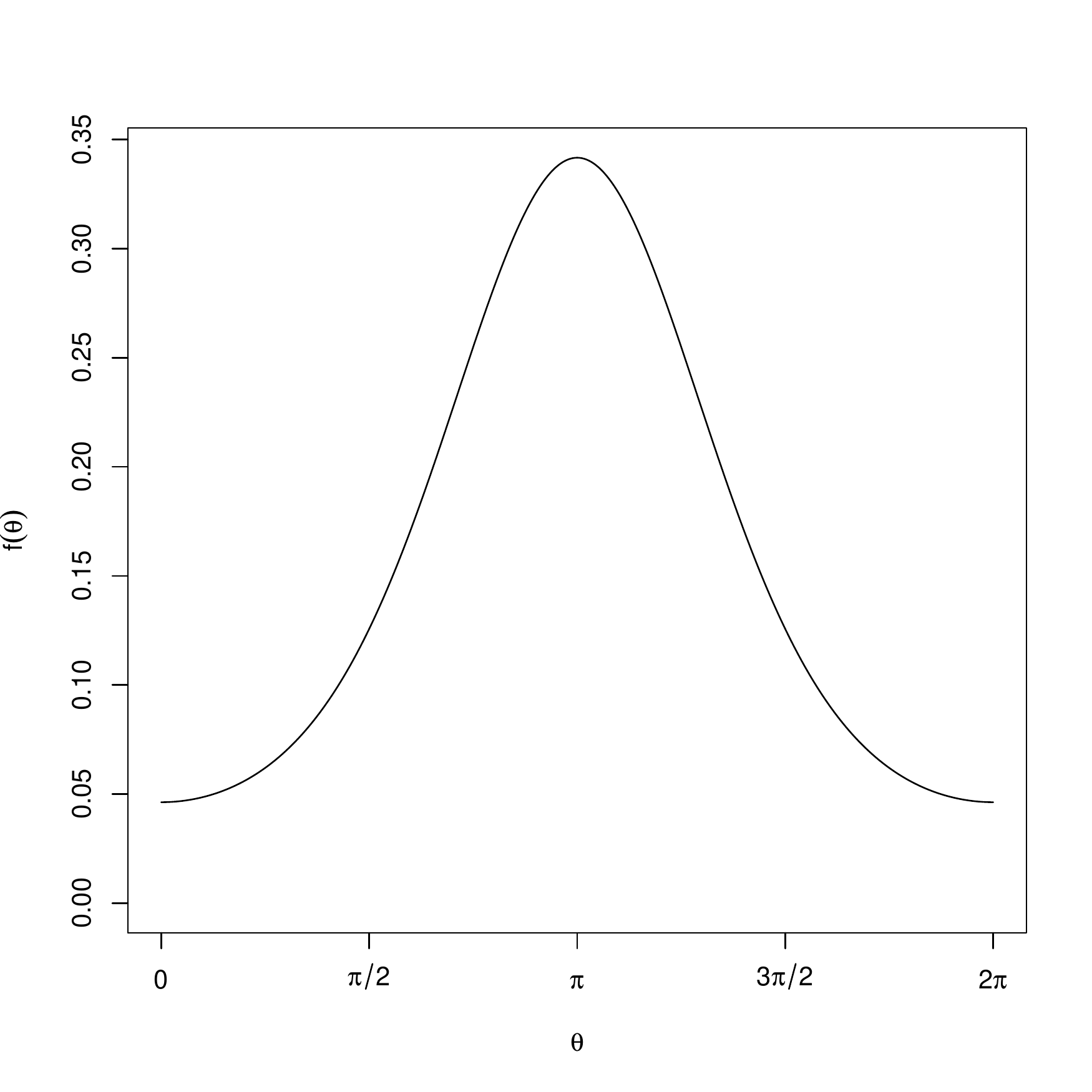}}& \multirow{4}{*}{\includegraphics[width=15mm]{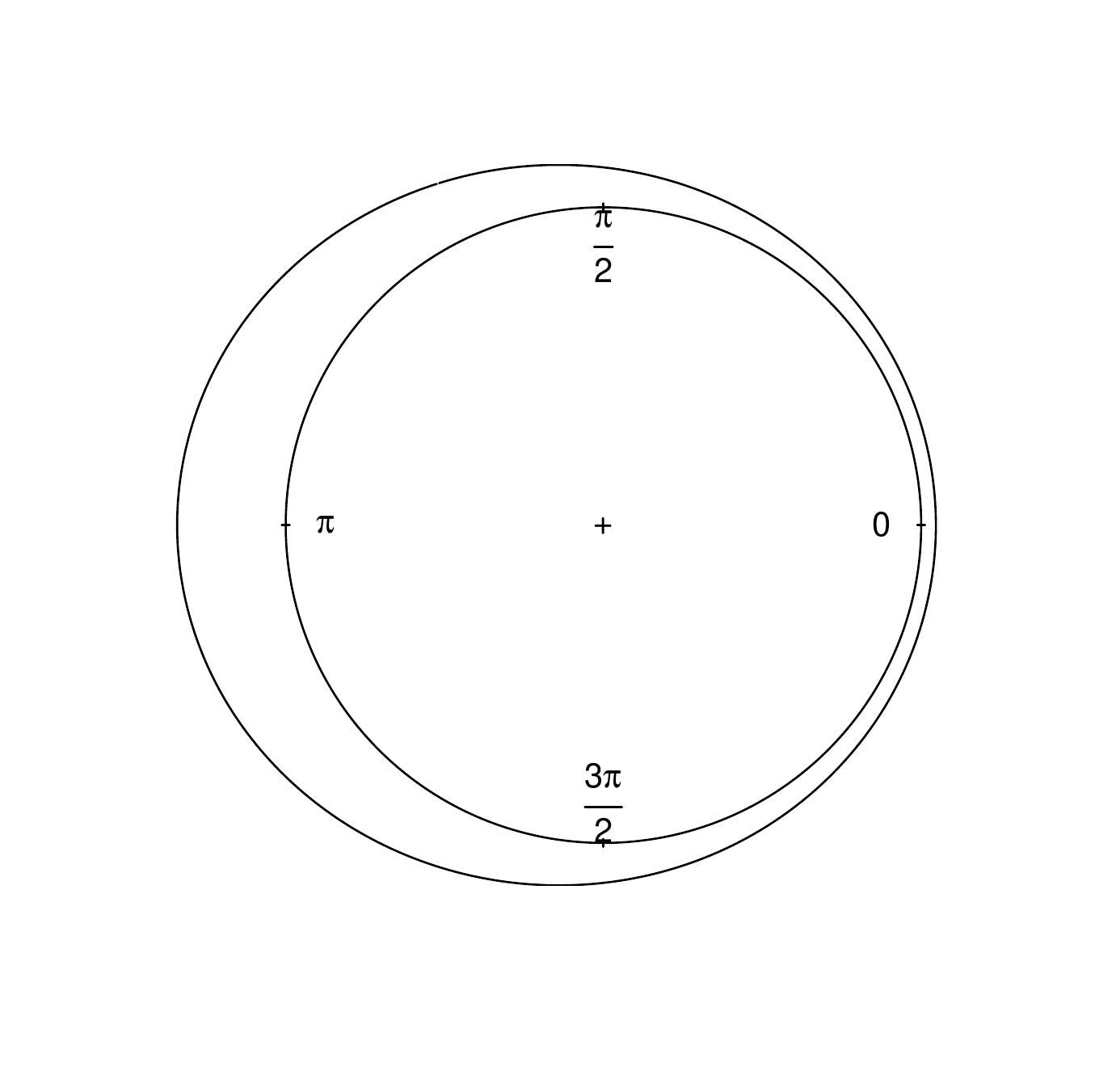}}& M1 &\multicolumn{3}{c|}{$U^2$ of Watson} &\multicolumn{3}{c|}{Excess mass}  \\ \cline{3-9}
 &  &  $n=50$ & 0.008(0.008) & 0.028(0.014) & 0.064(0.021) & 0.002(0.004) & 0.022(0.013) & 0.070(0.022)  \\ 
   &  &   $n=200$ & 0.006(0.007) & 0.030(0.015) & 0.064(0.021) & 0.004(0.006) & 0.034(0.016) & 0.074(0.023) \\ 
   &  &  $n=1000$ & 0.004(0.006) & 0.032(0.015) & 0.058(0.020) & 0.006(0.007) & 0.052(0.019) & 0.088(0.025) \\  
\hline
\multirow{4}{*}{\includegraphics[width=15mm]{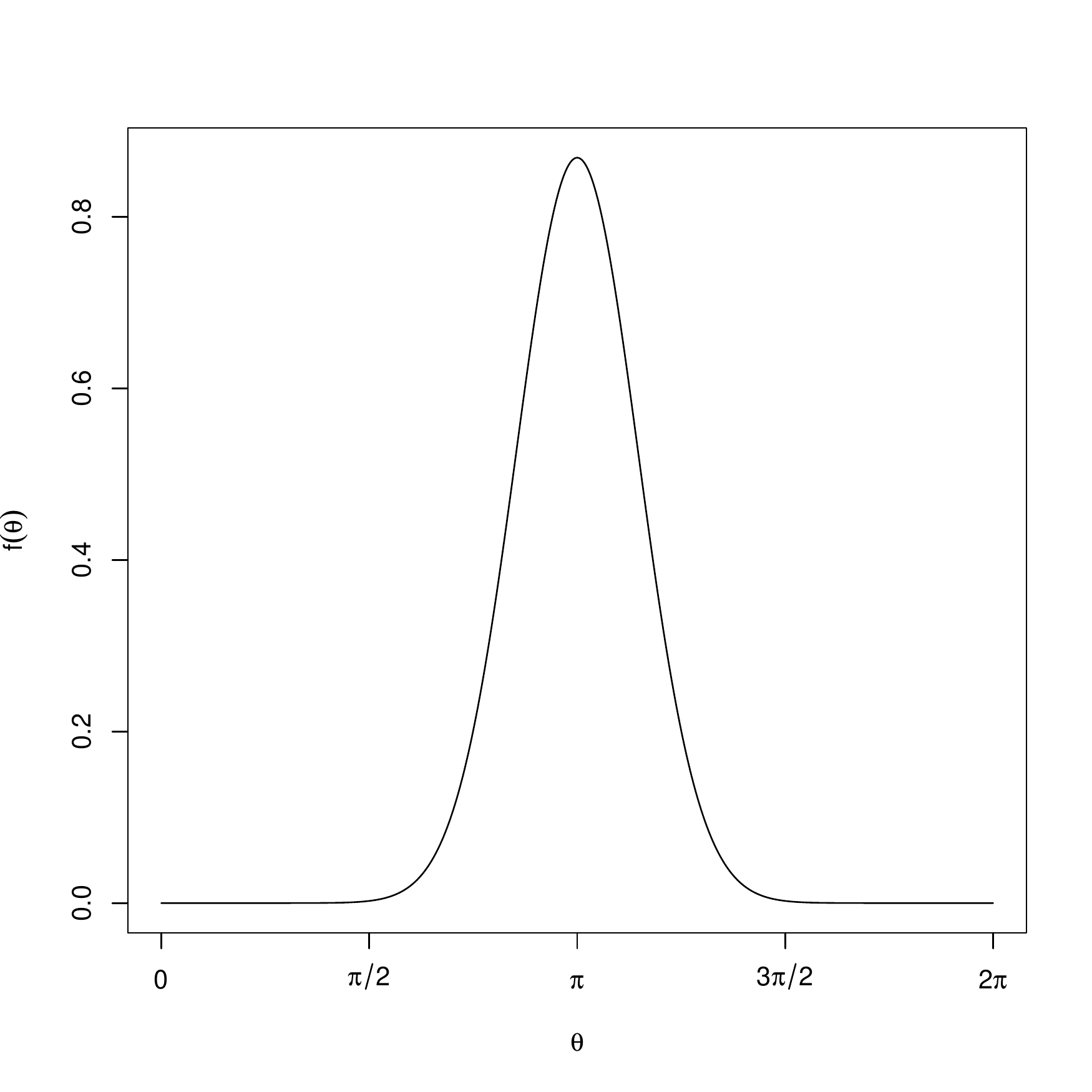}}& \multirow{4}{*}{\includegraphics[width=15mm]{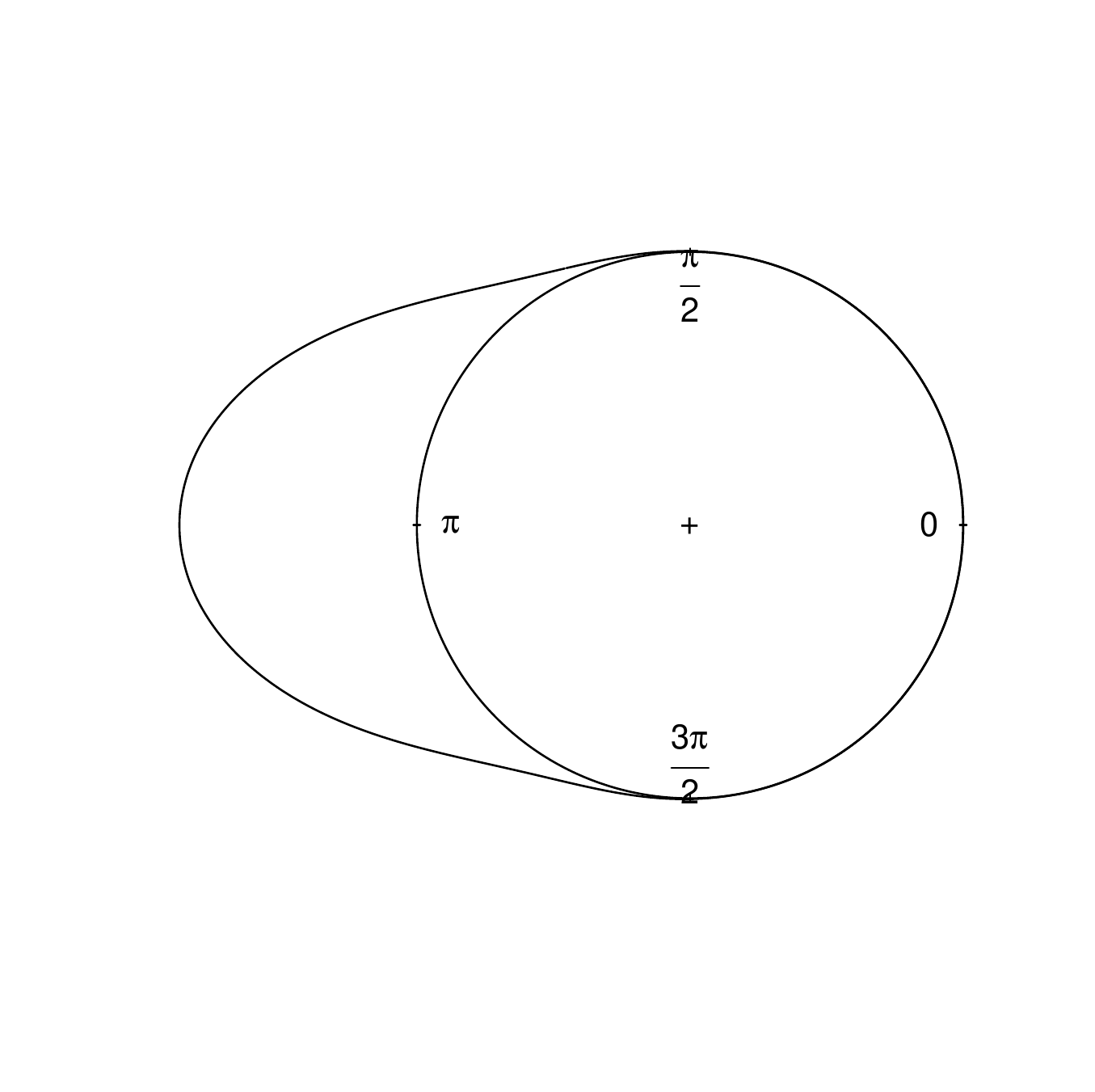}}& M2 &\multicolumn{3}{c|}{$U^2$ of Watson} &\multicolumn{3}{c|}{Excess mass}  \\ \cline{3-9}
 &  & $n=50$ & 0(0) & 0.022(0.013) & 0.068(0.022) & 0.012(0.010) & 0.054(0.020) & 0.094(0.026) \\ 
   &  & $n=200$ & 0(0) & 0.012(0.010) & 0.058(0.020) & 0.008(0.008) & 0.038(0.017) & 0.092(0.025) \\ 
   &  & $n=1000$ & 0.004(0.006) & 0.008(0.008) & 0.032(0.015) & 0.006(0.007) & 0.040(0.017) & 0.086(0.025) \\  
\hline
\multirow{4}{*}{\includegraphics[width=15mm]{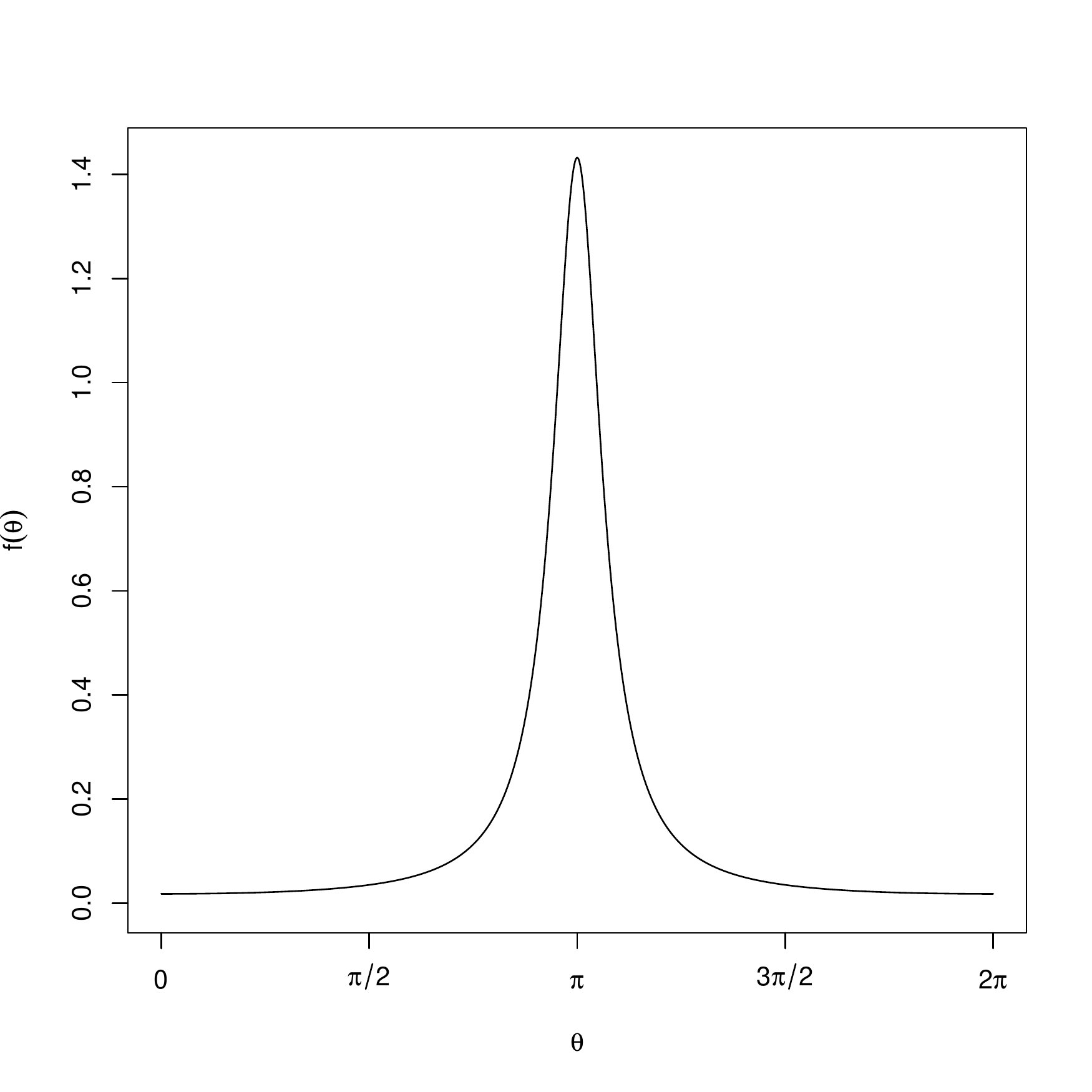}}& \multirow{4}{*}{\includegraphics[width=15mm]{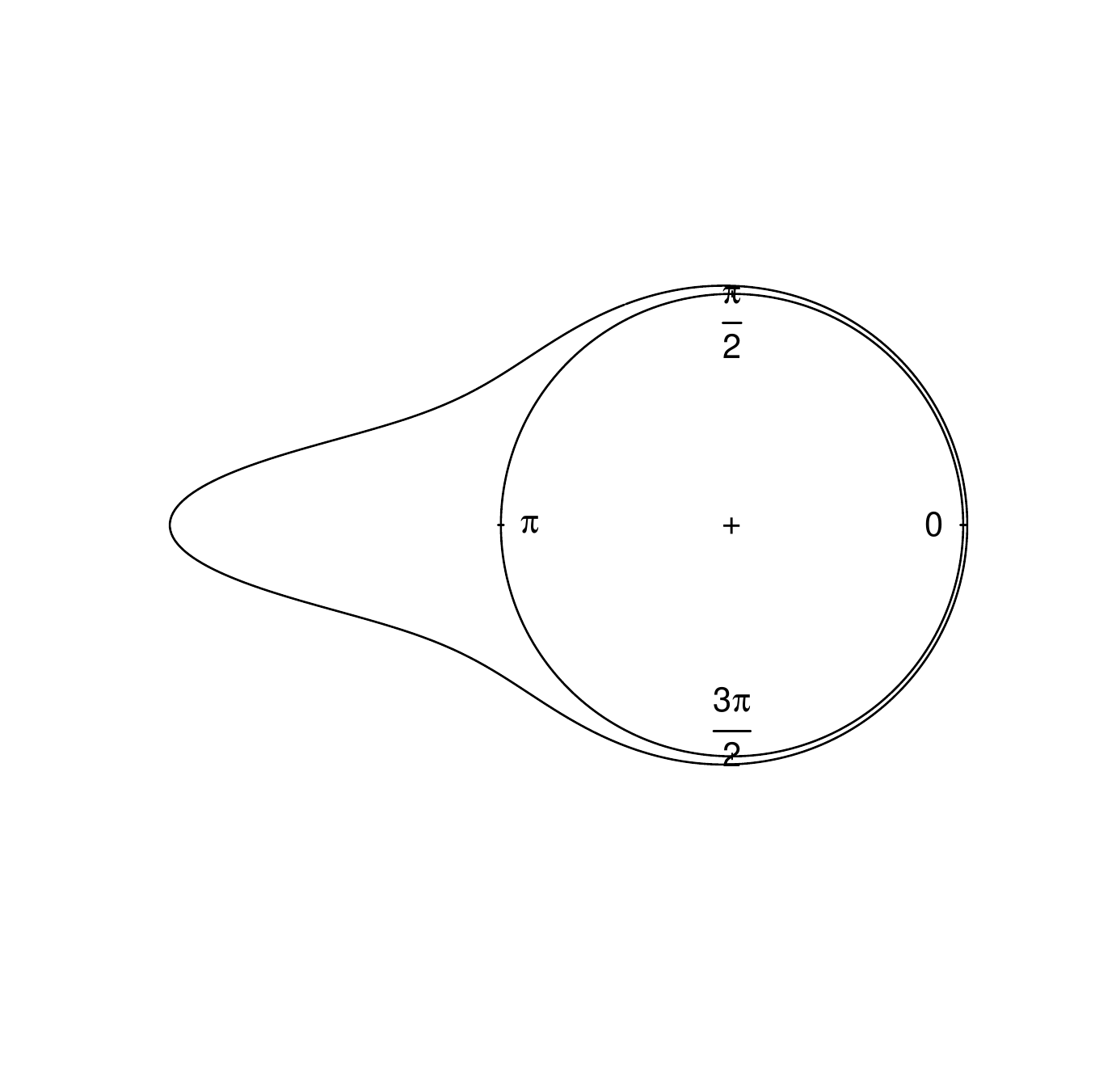}}& M3 &\multicolumn{3}{c|}{$U^2$ of Watson} &\multicolumn{3}{c|}{Excess mass}  \\ \cline{3-9}
 &  & $n=50$ & 0.654(0.042) & 0.828(0.033) & 0.886(0.028) & 0(0) & 0.018(0.012) & 0.038(0.017) \\ 
   &  & $n=200$ & 0.708(0.040) & 0.822(0.034) & 0.884(0.028) & 0.002(0.004) & 0.012(0.010) & 0.024(0.013) \\ 
   &  & $n=1000$ & 0.574(0.043) & 0.718(0.039) & 0.770(0.037) & 0(0) & 0.014(0.010) & 0.022(0.013) \\ 
\hline
\multirow{4}{*}{\includegraphics[width=15mm]{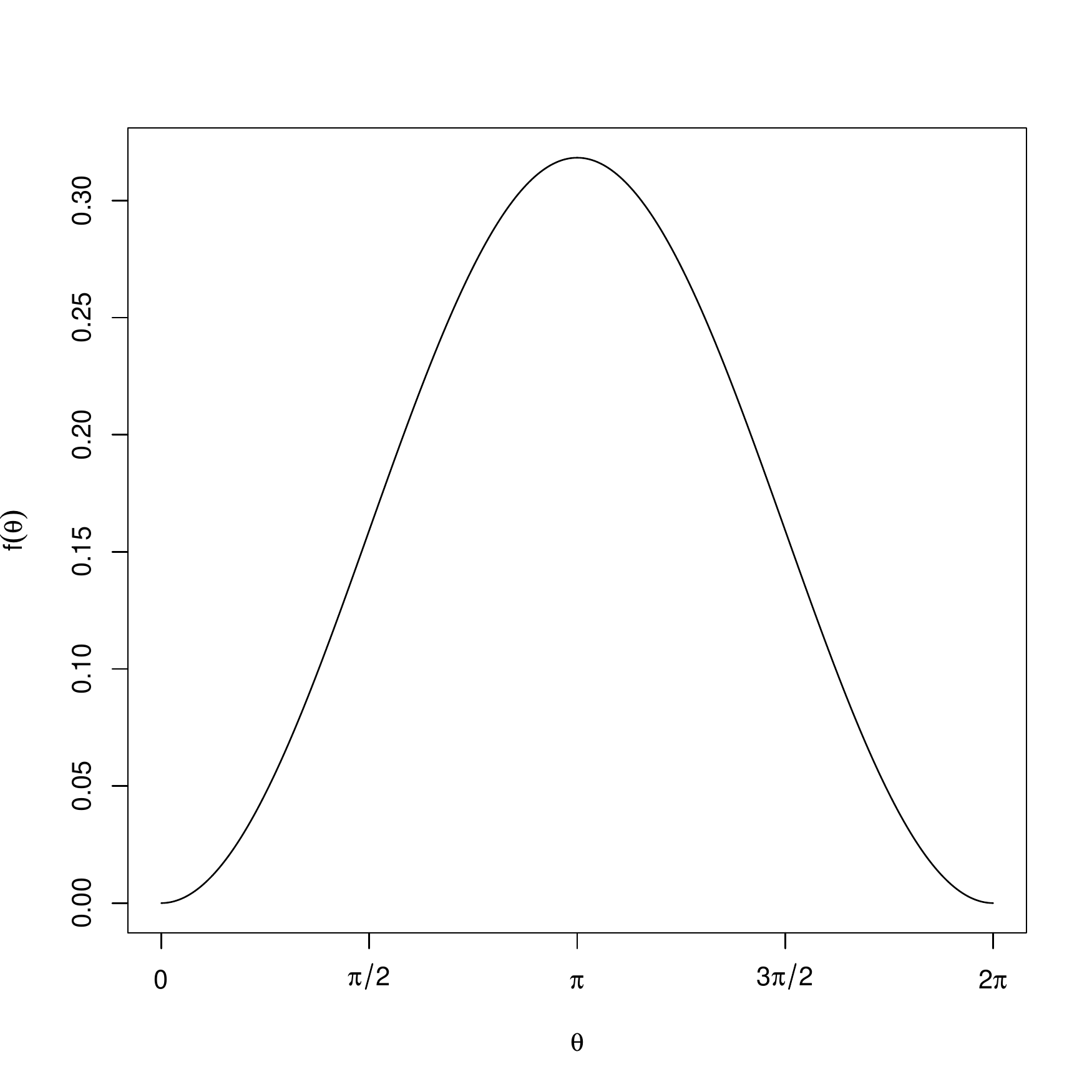}}& \multirow{4}{*}{\includegraphics[width=15mm]{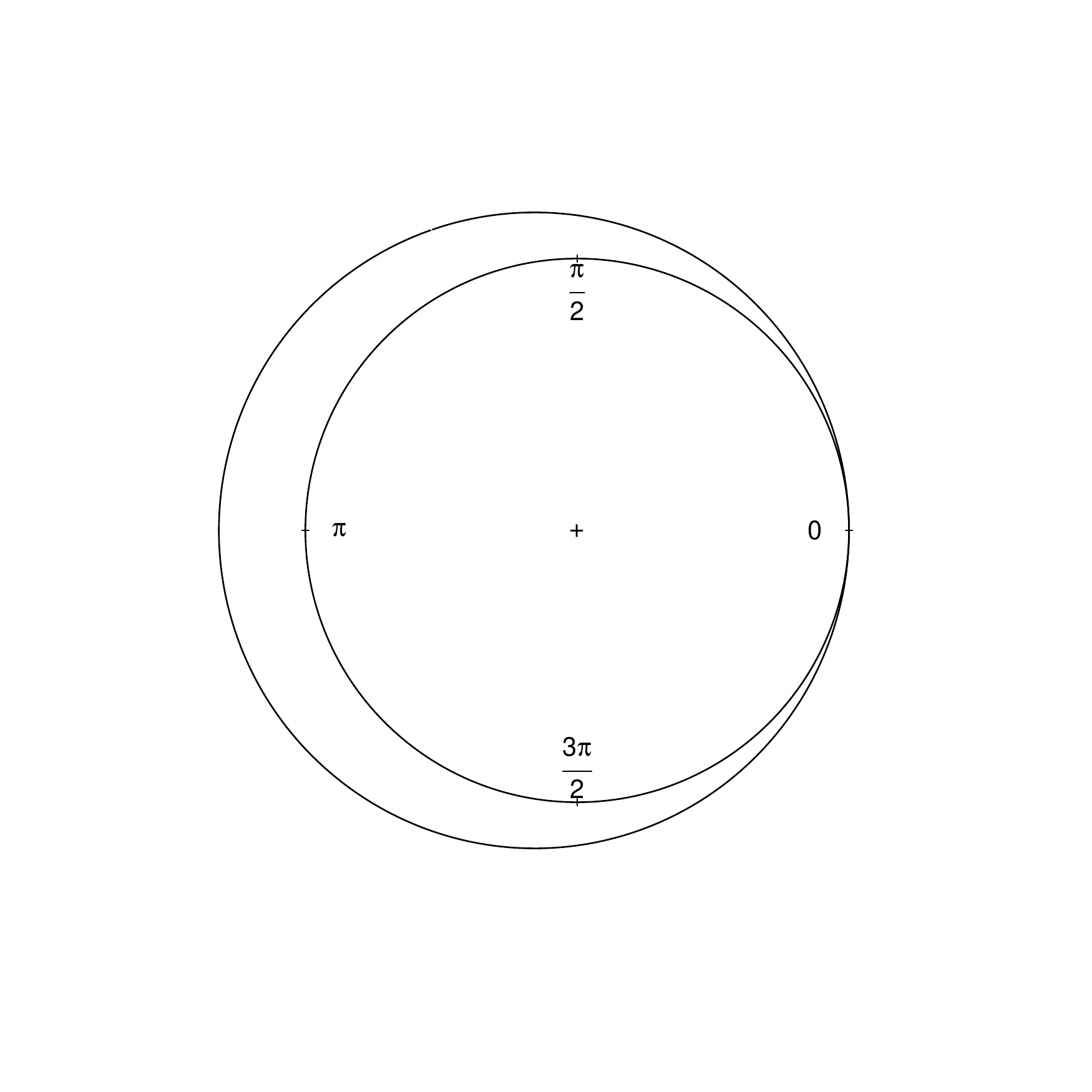}}& M4 &\multicolumn{3}{c|}{$U^2$ of Watson} &\multicolumn{3}{c|}{Excess mass}  \\ \cline{3-9}
 &  & $n=50$ & 0(0) & 0.024(0.013) & 0.066(0.022) & 0.006(0.007) & 0.044(0.018) & 0.114(0.028)  \\ 
   &  & $n=200$ & 0(0) & 0.020(0.012) & 0.042(0.018) & 0.008(0.008) & 0.026(0.014) & 0.068(0.022) \\ 
   &  & $n=1000$ & 0.004(0.006) & 0.024(0.013) & 0.058(0.020) & 0.01(0.008) & 0.046(0.018) & 0.082(0.024) \\ 
\hline
\multirow{4}{*}{\includegraphics[width=15mm]{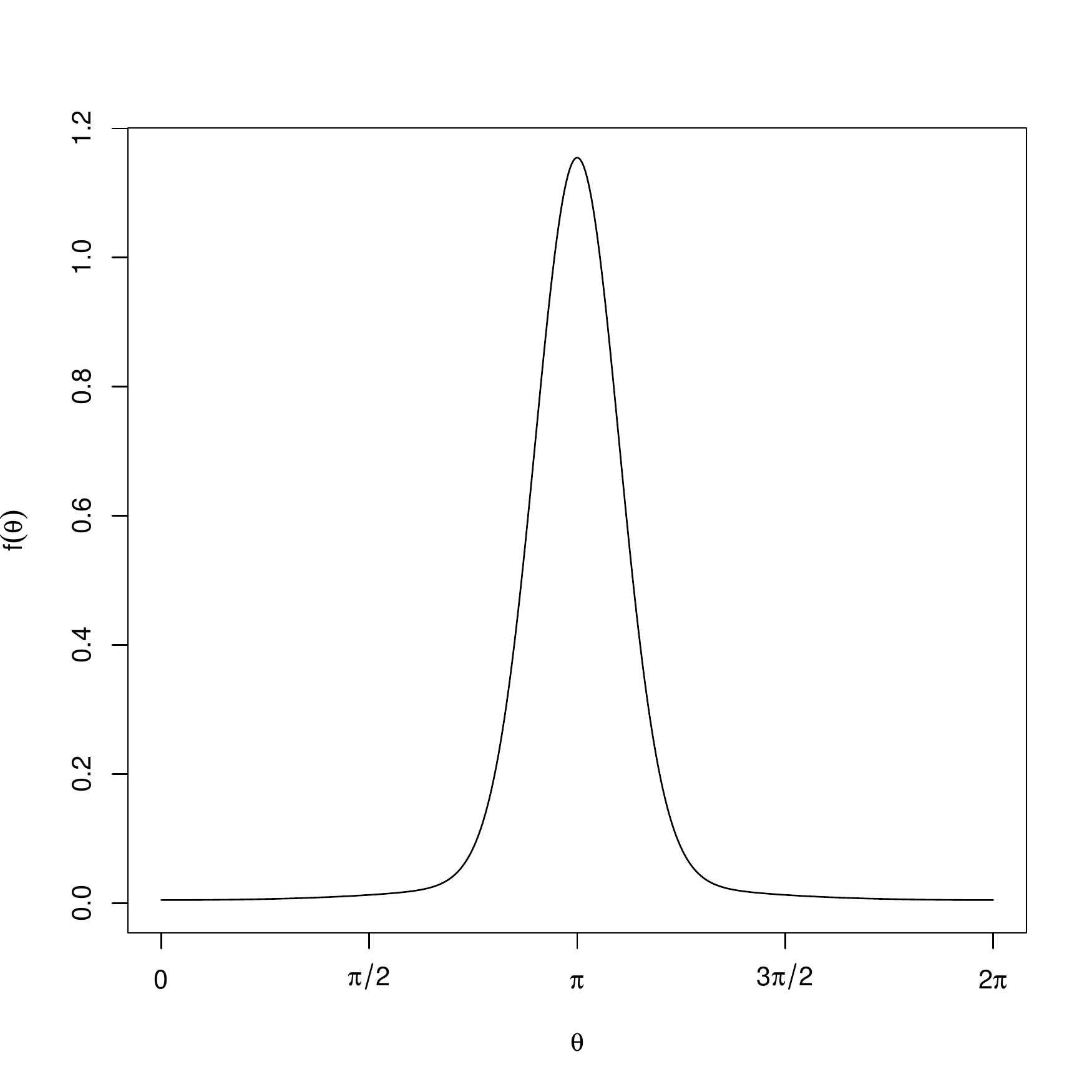}}& \multirow{4}{*}{\includegraphics[width=15mm]{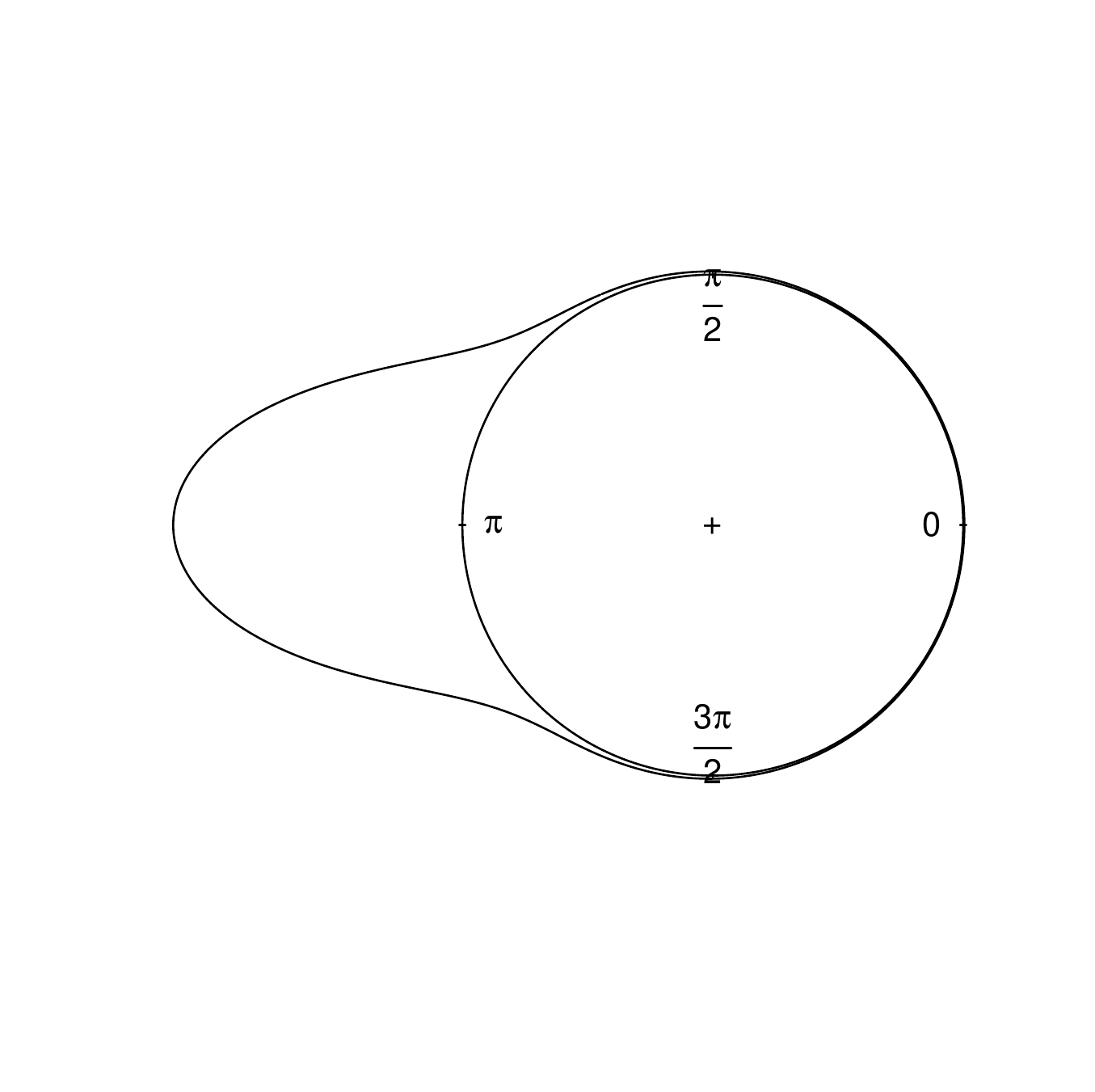}}& M5 &\multicolumn{3}{c|}{$U^2$ of Watson} &\multicolumn{3}{c|}{Excess mass}  \\ \cline{3-9}
 &  & $n=50$ & 0.370(0.042) & 0.520(0.044) & 0.598(0.043) & 0.008(0.008) & 0.044(0.018) & 0.104(0.027) \\ 
   &  & $n=200$ & 0.648(0.042) & 0.776(0.037) & 0.848(0.031) & 0.014(0.010) & 0.036(0.016) & 0.076(0.023) \\ 
   &  & $n=1000$ & 0.498(0.044) & 0.648(0.042) & 0.728(0.039) & 0.012(0.010) & 0.040(0.017) & 0.074(0.023) \\  
\hline
\multirow{4}{*}{\includegraphics[width=15mm]{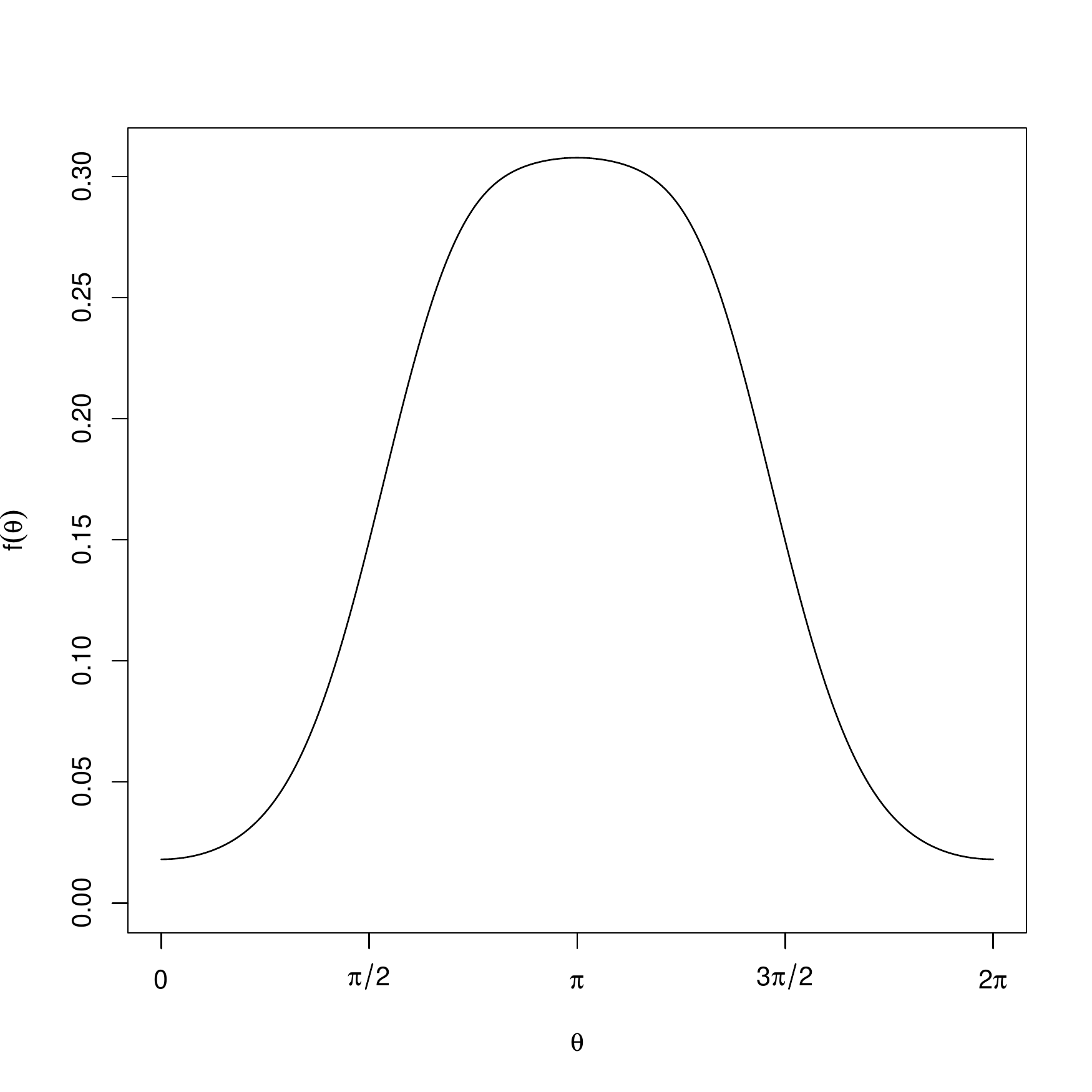}}& \multirow{4}{*}{\includegraphics[width=15mm]{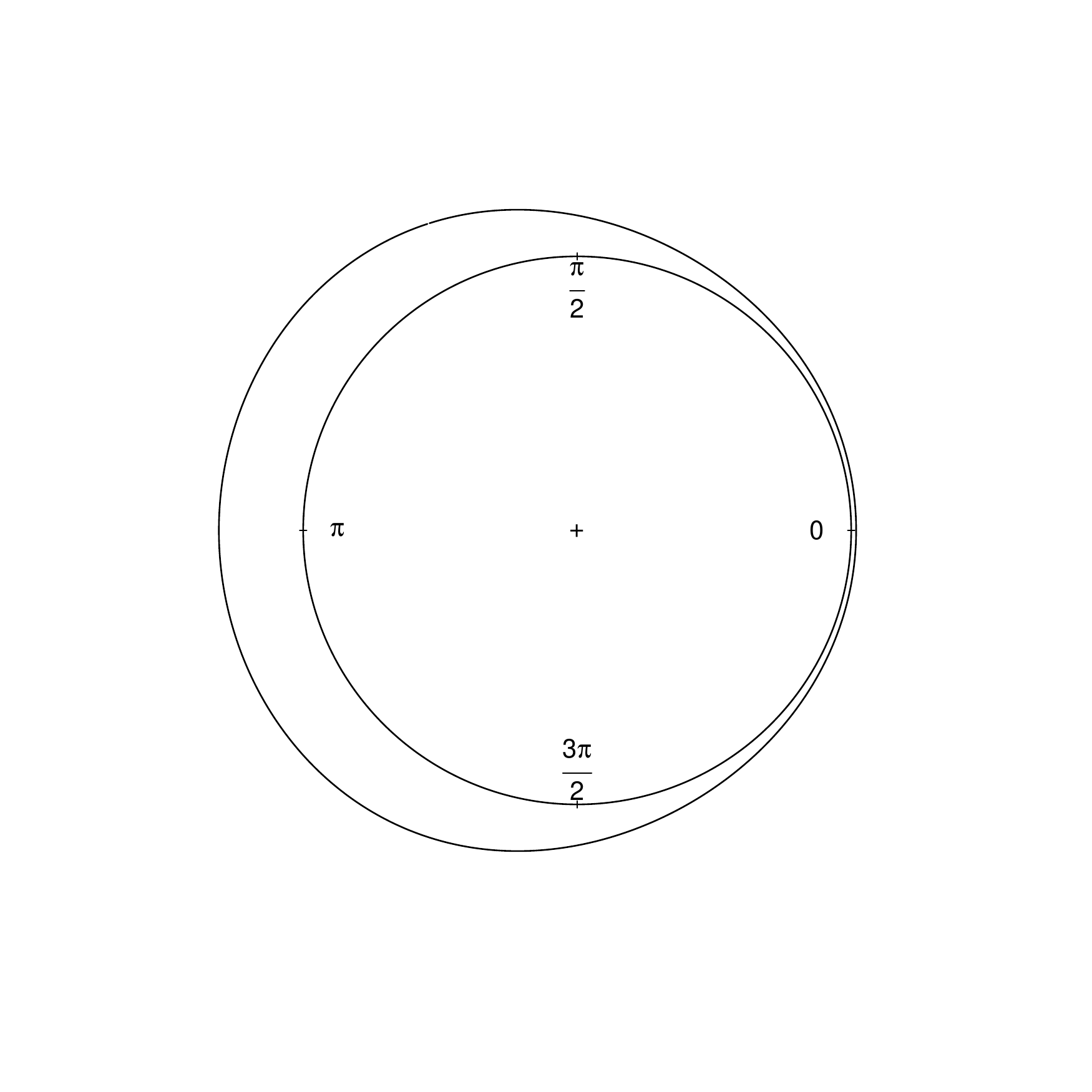}}& M6 &\multicolumn{3}{c|}{$U^2$ of Watson} &\multicolumn{3}{c|}{Excess mass}  \\ \cline{3-9}
 &  & $n=50$ & 0.006(0.007) & 0.040(0.017) & 0.064(0.021) & 0.002(0.004) & 0.046(0.018) & 0.094(0.026) \\ 
   &  & $n=200$ & 0.004(0.006) & 0.044(0.018) & 0.108(0.027) & 0.020(0.012) & 0.058(0.020) & 0.118(0.028) \\ 
   &  & $n=1000$ & 0.016(0.011) & 0.070(0.022) & 0.132(0.030) & 0.006(0.007) & 0.048(0.019) & 0.098(0.026) \\   
\hline
\multirow{4}{*}{\includegraphics[width=15mm]{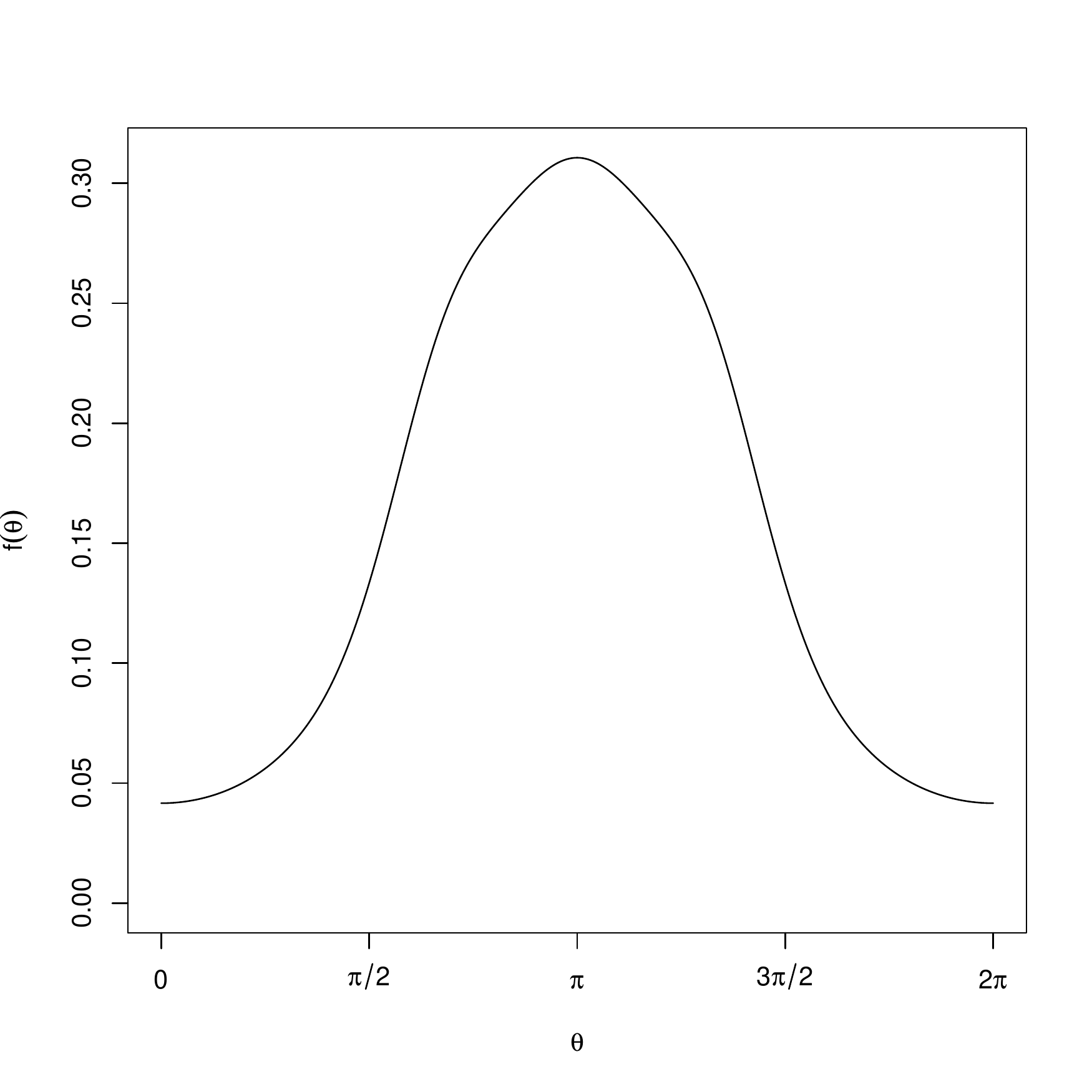}}& \multirow{4}{*}{\includegraphics[width=15mm]{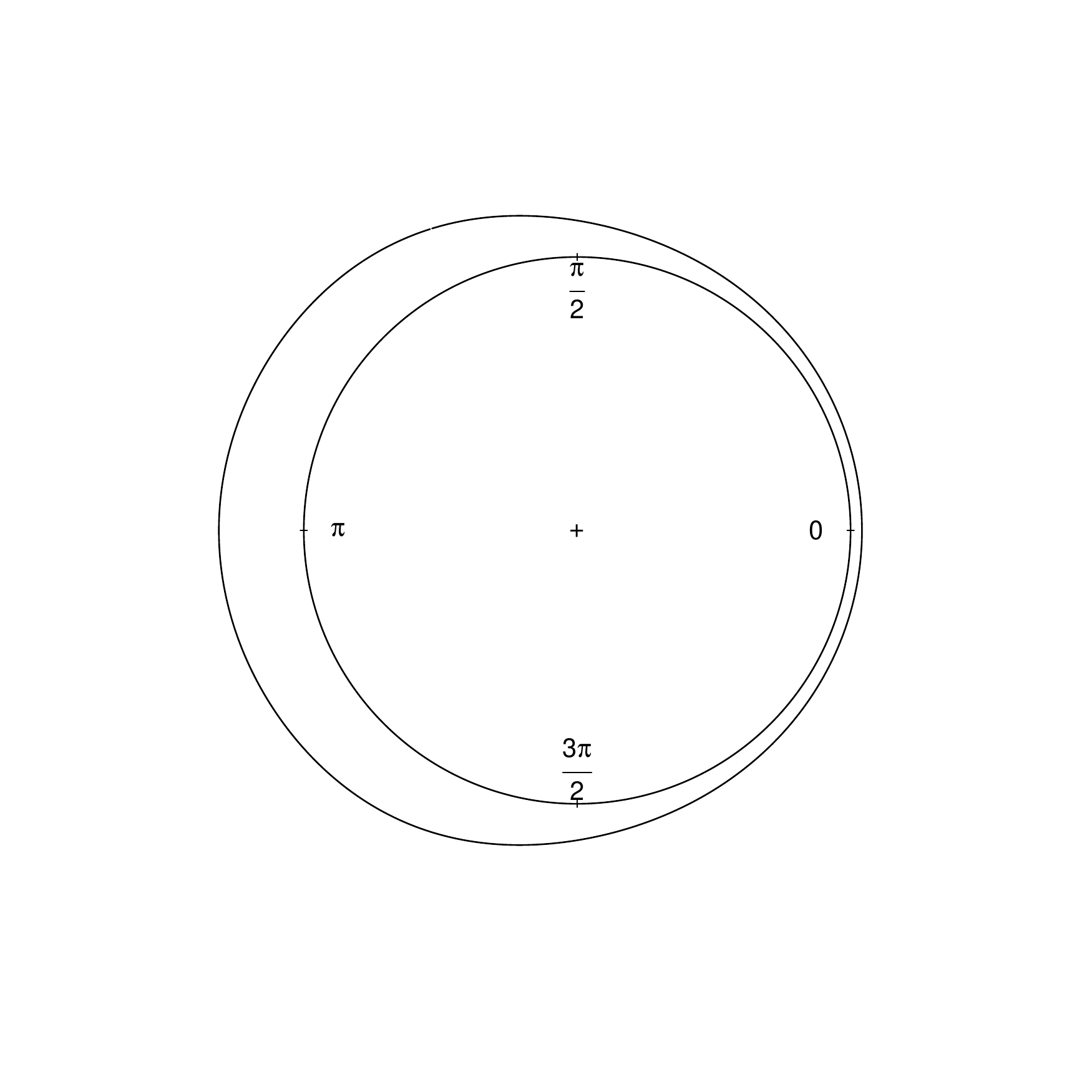}}& M7 &\multicolumn{3}{c|}{$U^2$ of Watson} &\multicolumn{3}{c|}{Excess mass}  \\ \cline{3-9}
 &  & $n=50$ & 0.002(0.004) & 0.020(0.012) & 0.052(0.019) & 0.008(0.008) & 0.034(0.016) & 0.086(0.025) \\ 
   &  & $n=200$ & 0.012(0.010) & 0.048(0.019) & 0.096(0.026) & 0.012(0.010) & 0.064(0.021) & 0.114(0.028) \\ 
   &  & $n=1000$ & 0.010(0.009) & 0.050(0.019) & 0.092(0.025) & 0.004(0.006) & 0.040(0.017) & 0.094(0.026) \\   
\hline
\multirow{4}{*}{\includegraphics[width=15mm]{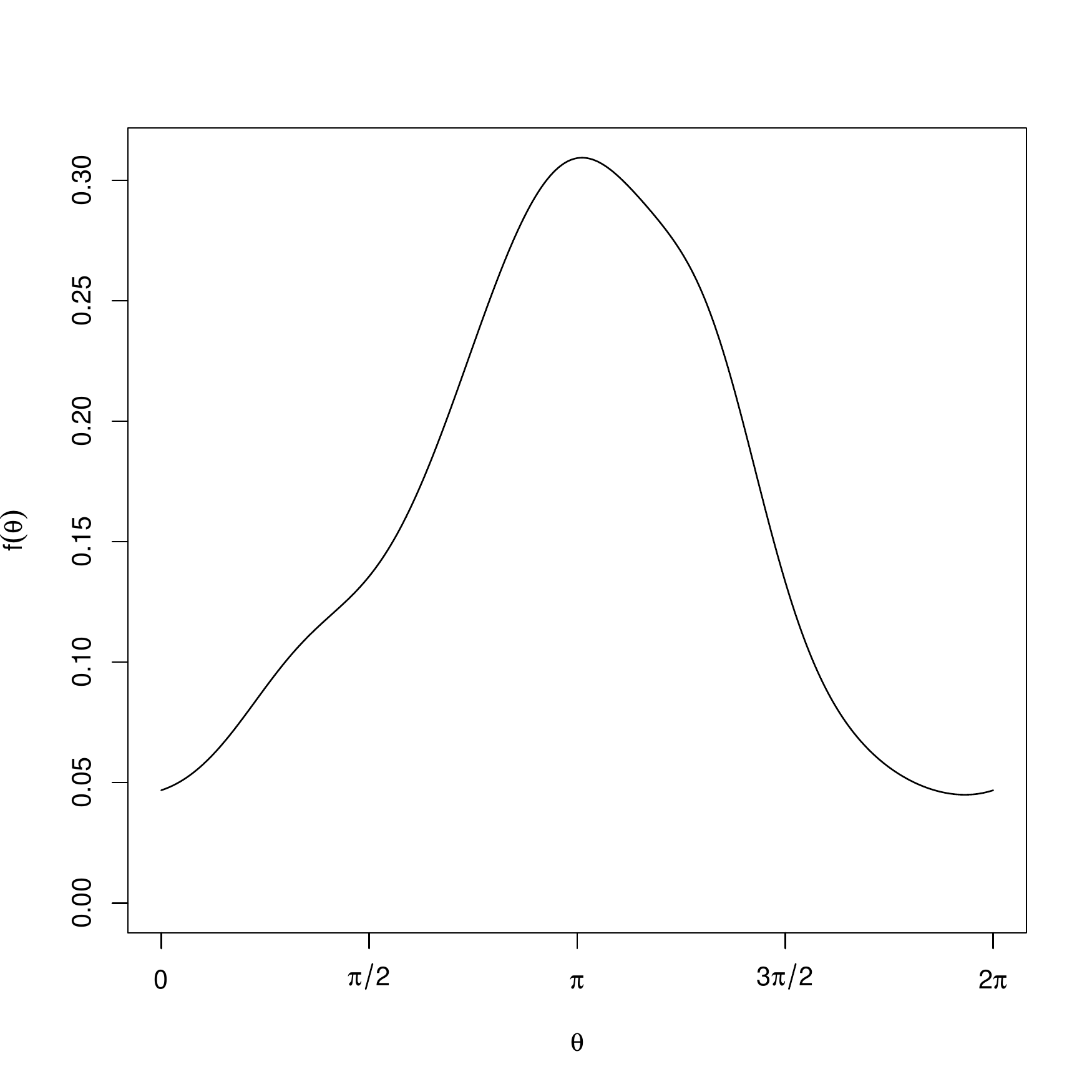}}& \multirow{4}{*}{\includegraphics[width=15mm]{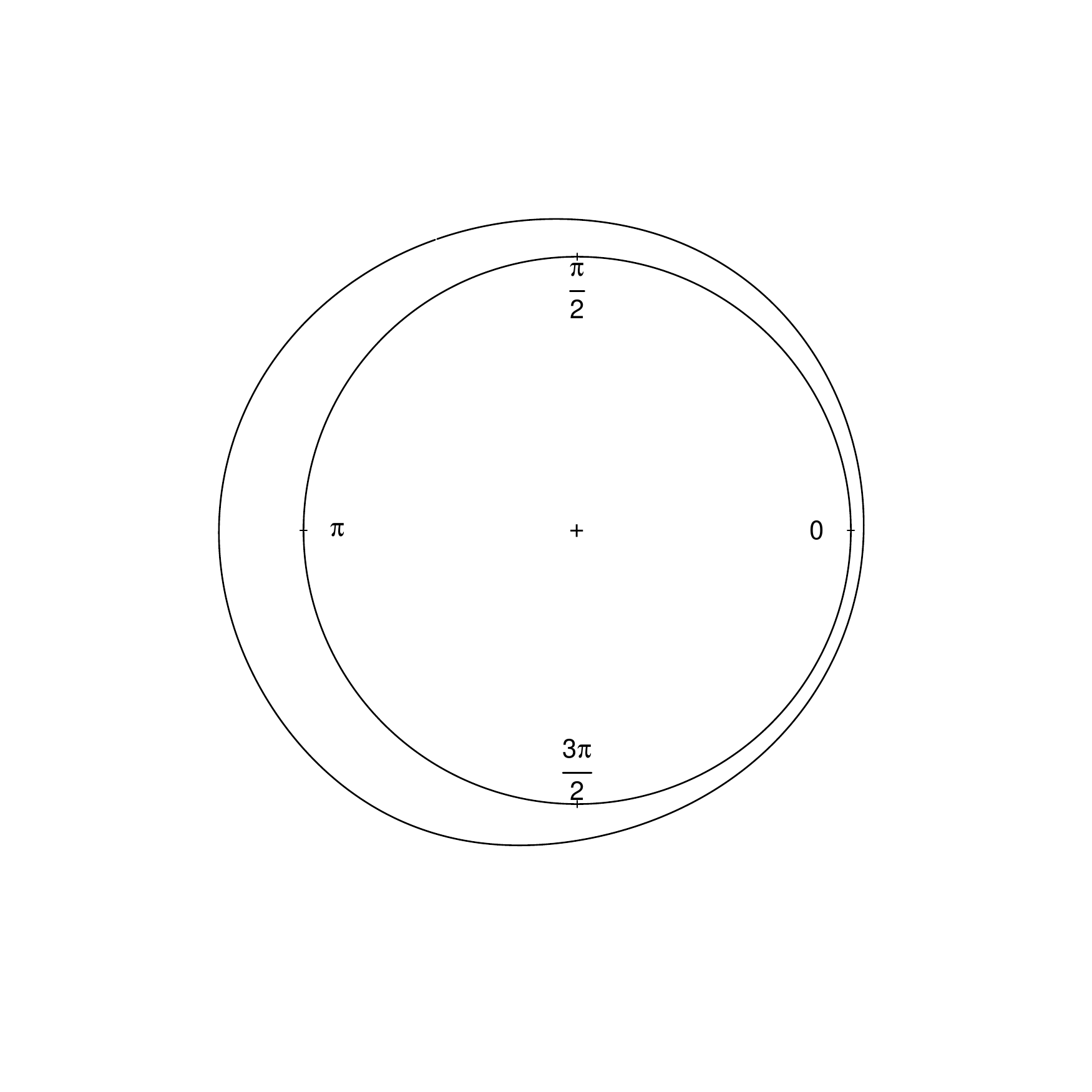}}& M8 &\multicolumn{3}{c|}{$U^2$ of Watson} &\multicolumn{3}{c|}{Excess mass}  \\ \cline{3-9}
 &  & $n=50$ & 0.004(0.006) & 0.030(0.015) & 0.062(0.021) & 0(0) & 0.024(0.013) & 0.044(0.018) \\ 
   &  & $n=200$ & 0.004(0.006) & 0.032(0.015) & 0.052(0.019) & 0.006(0.007) & 0.044(0.018) & 0.100(0.026)  \\ 
   &  & $n=1000$ & 0.002(0.004) & 0.018(0.012) & 0.048(0.019) & 0.010(0.009) & 0.048(0.019) & 0.100(0.026) \\  
\hline
\multirow{4}{*}{\includegraphics[width=15mm]{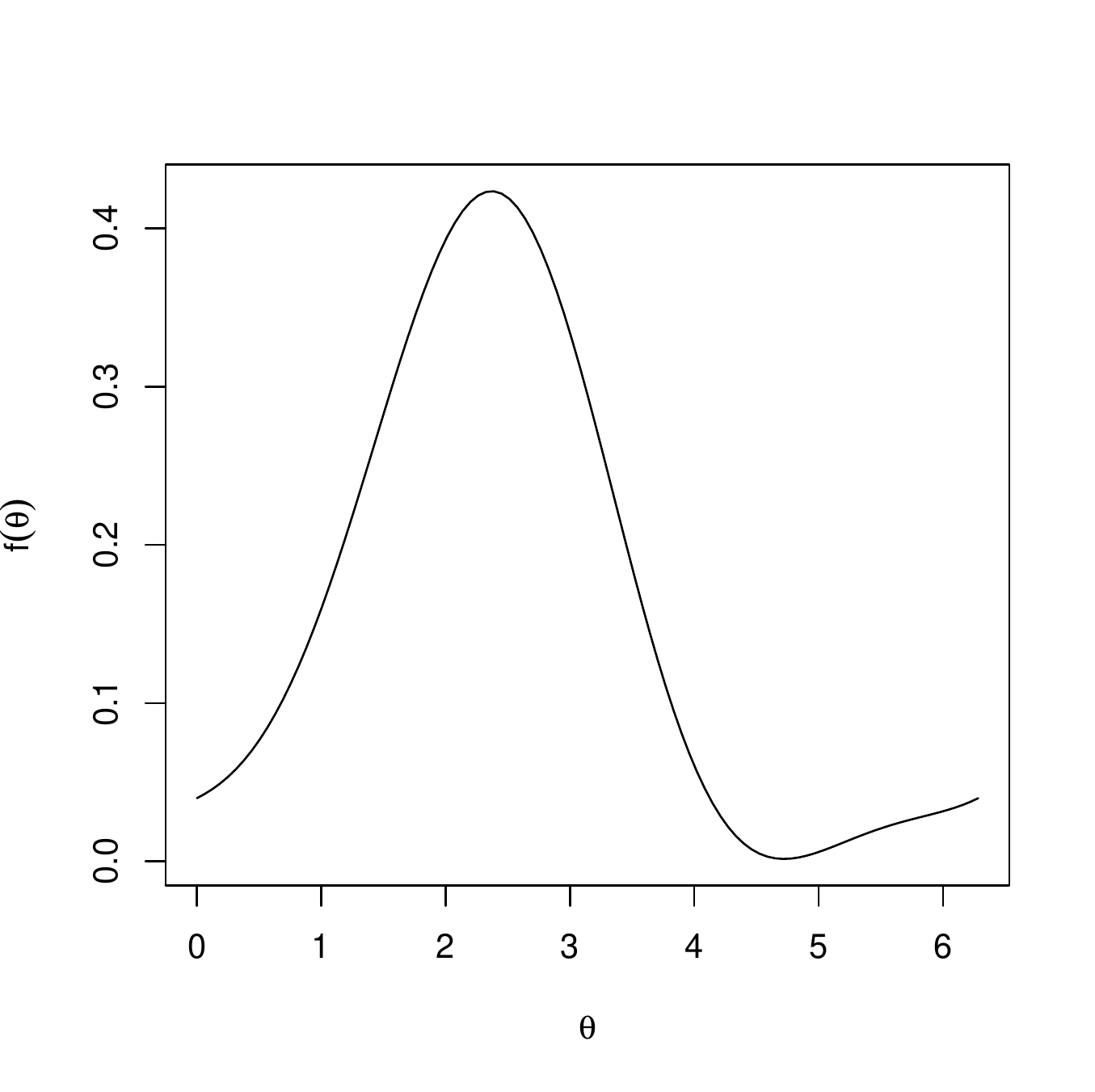}}& \multirow{4}{*}{\includegraphics[width=15mm]{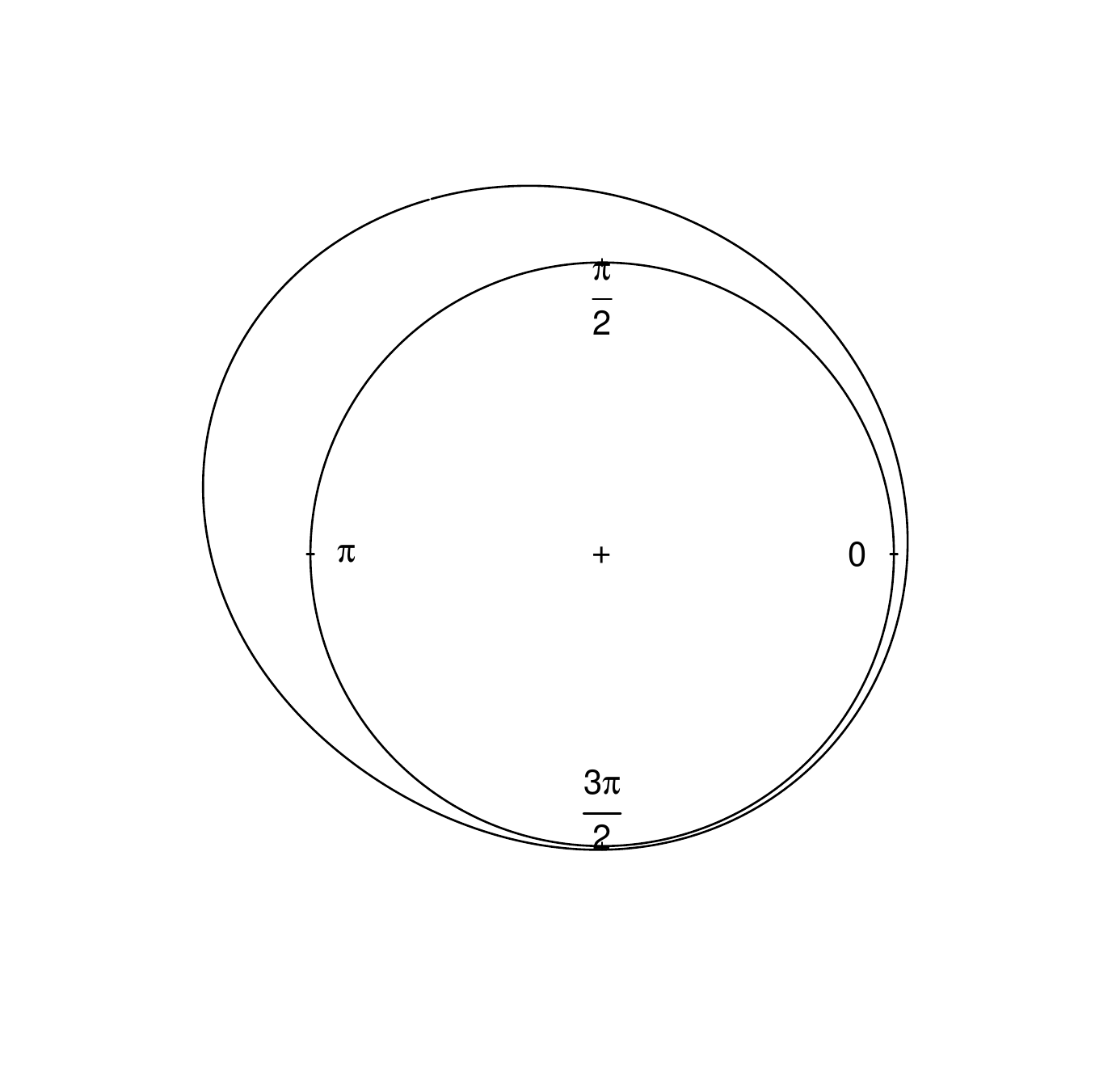}}& M9 &\multicolumn{3}{c|}{$U^2$ of Watson} &\multicolumn{3}{c|}{Excess mass}  \\ \cline{3-9}
 &  & $n=50$ & 0.010(0.009) & 0.054(0.020) & 0.114(0.028) & 0.004(0.006) & 0.038(0.017) & 0.064(0.021) \\ 
   &  & $n=200$ & 0.008(0.008) & 0.020(0.012) & 0.076(0.023) & 0.008(0.008) & 0.030(0.015) & 0.062(0.021) \\ 
   &  & $n=1000$ & 0(0) & 0.010(0.009) & 0.044(0.018) & 0.004(0.006) & 0.036(0.016) & 0.084(0.024) \\  
\hline
\multirow{4}{*}{\includegraphics[width=15mm]{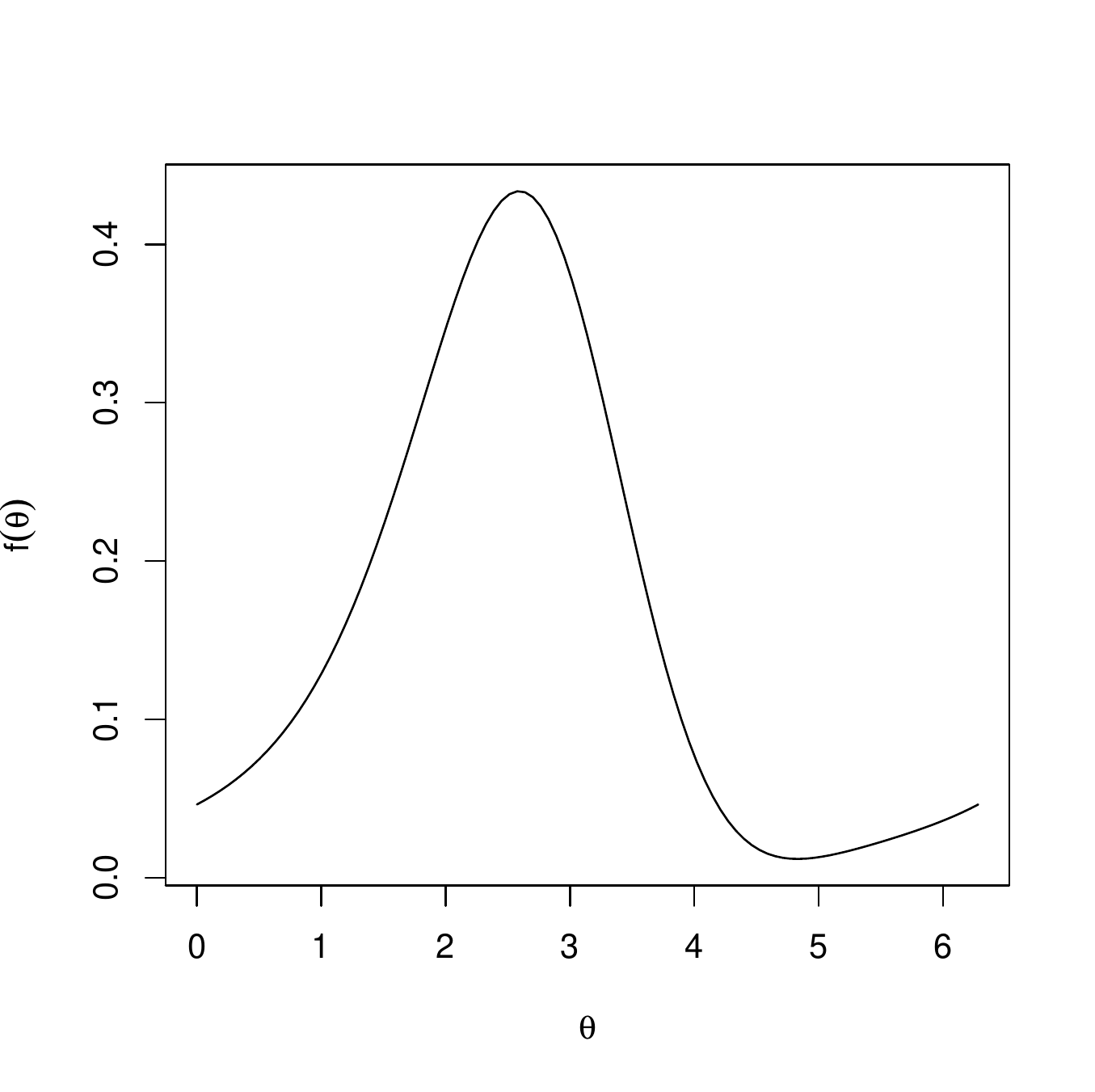}}& \multirow{4}{*}{\includegraphics[width=15mm]{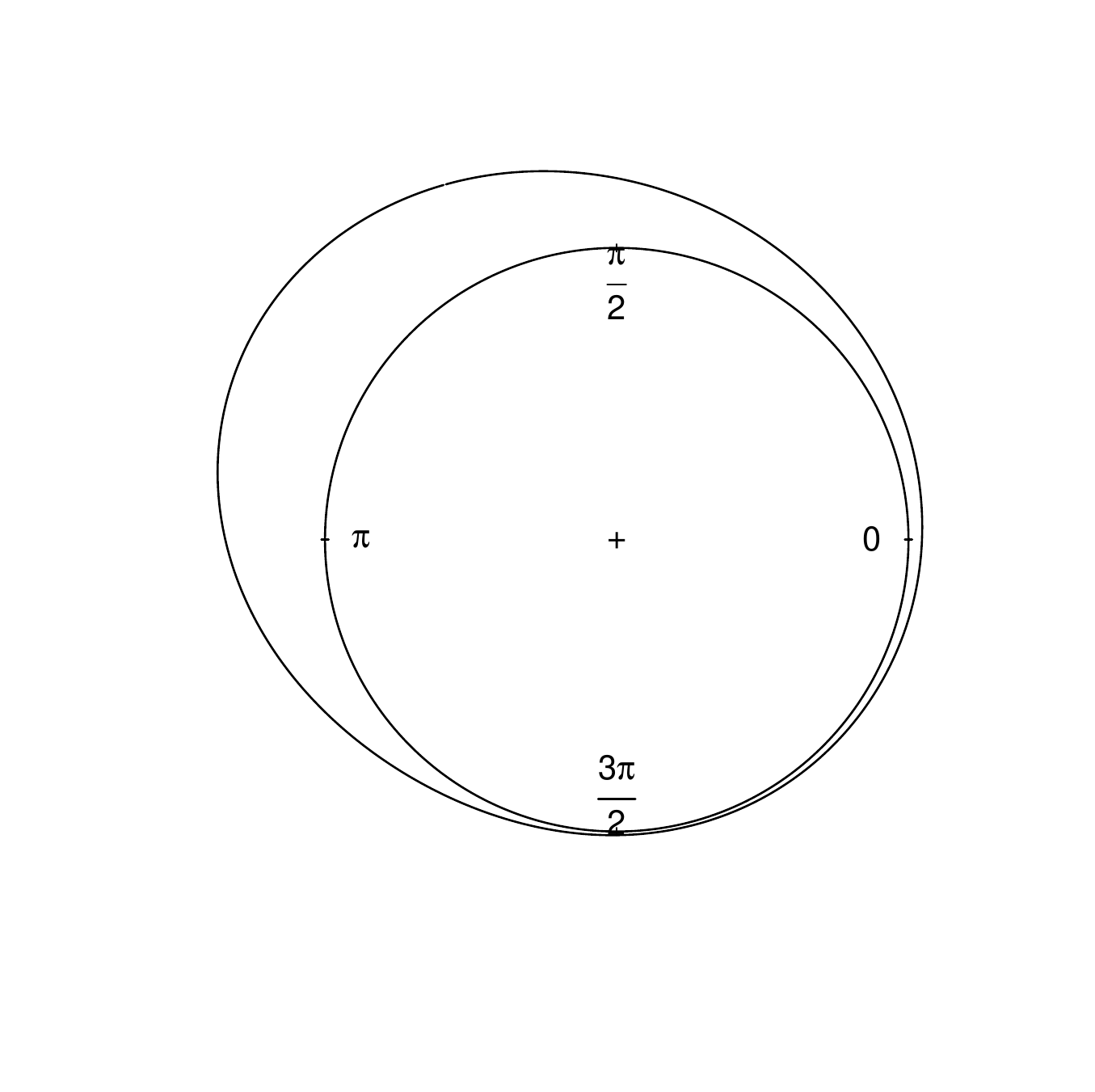}}& M10 &\multicolumn{3}{c|}{$U^2$ of Watson} &\multicolumn{3}{c|}{Excess mass}  \\ \cline{3-9}
 &  & $n=50$ & 0.014(0.010) & 0.056(0.020) & 0.106(0.027) & 0(0) & 0.030(0.015) & 0.070(0.022) \\ 
   &  & $n=200$ & 0.012(0.010) & 0.064(0.021) & 0.122(0.029) & 0.012(0.010) & 0.038(0.017) & 0.082(0.024) \\ 
   &  & $n=1000$ & 0.004(0.006) & 0.026(0.014) & 0.064(0.021) & 0.002(0.004) & 0.038(0.017) & 0.086(0.024) \\ 
\hline \hline
(b) & Bimodal &  $\alpha$ & 0.01 & 0.05& 0.10 & 0.01 & 0.05& 0.10 \\ \hline
\multirow{4}{*}{\includegraphics[width=15mm]{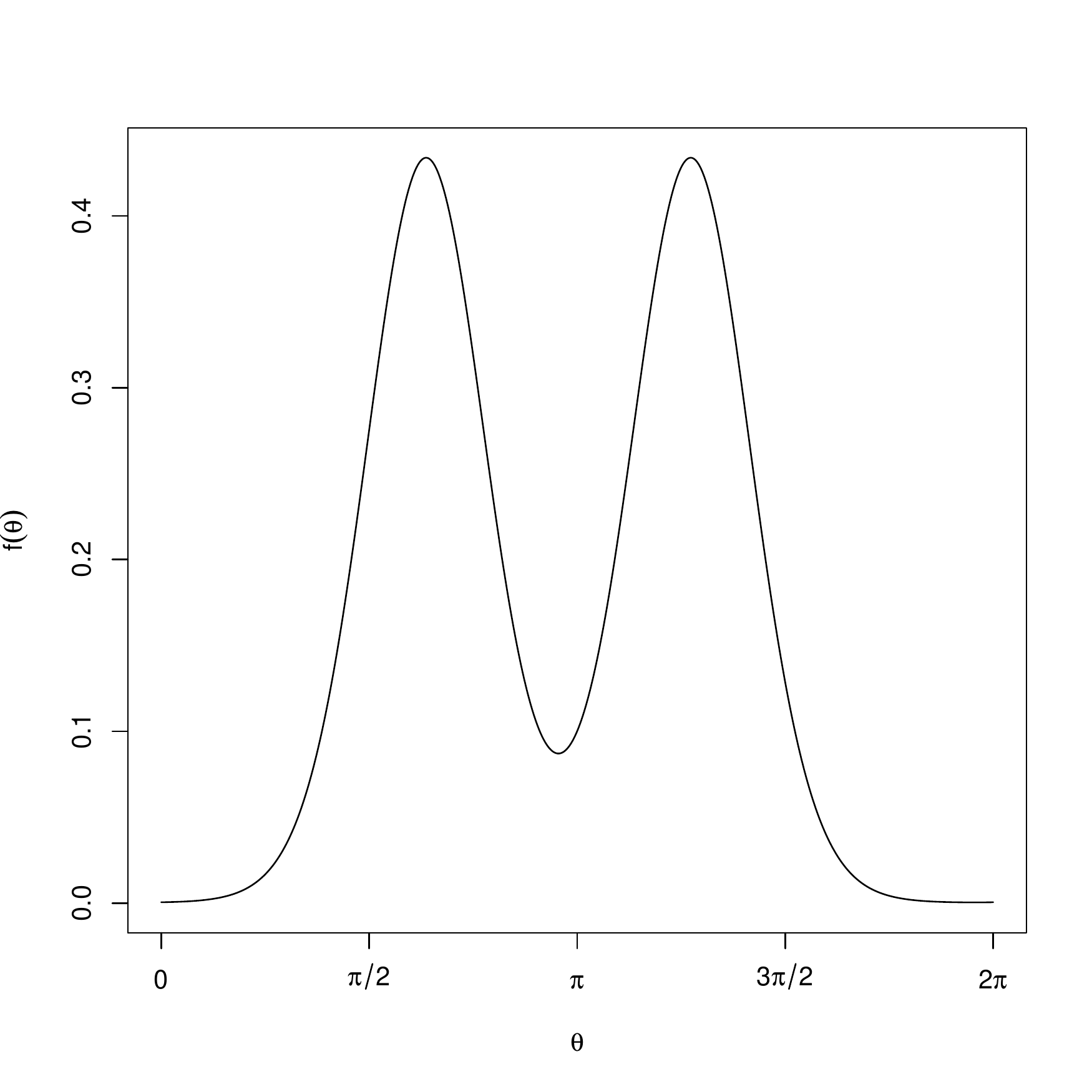}}& \multirow{4}{*}{\includegraphics[width=15mm]{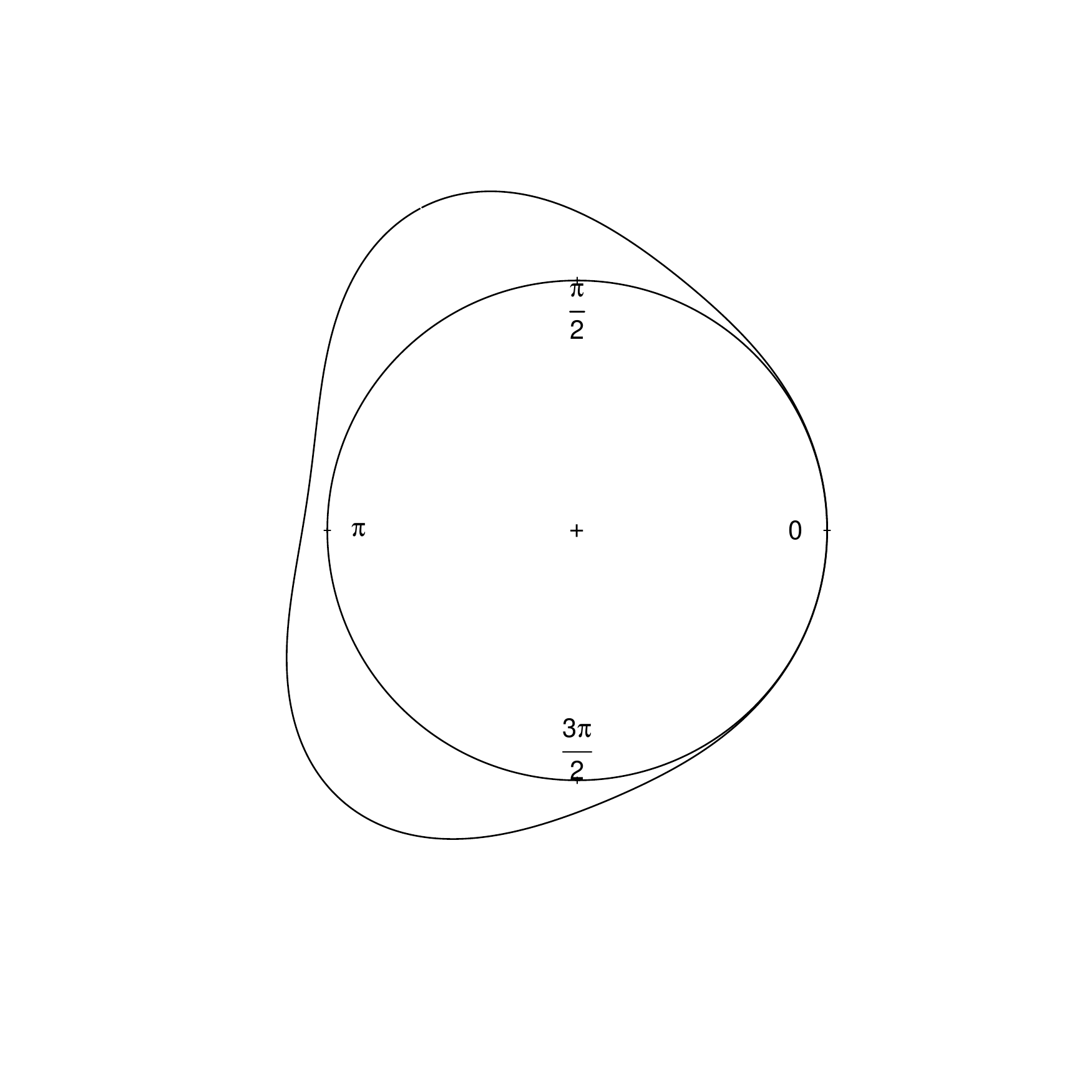}}& M11 &\multicolumn{3}{c|}{$U^2$ of Watson} &\multicolumn{3}{c|}{Excess mass}  \\ \cline{3-9}
 &  & $n=50$ & 0.782(0.036) & 0.920(0.024) & 0.958(0.018) & 0.534(0.044) & 0.758(0.038) & 0.854(0.031) \\ 
   &  & $n=100$ & 0.978(0.013) & 0.996(0.006) & 0.998(0.004) & 0.914(0.025) & 0.968(0.015) & 0.984(0.011) \\ 
   &  & $n=200$ & 1(0) & 1(0) & 1(0) & 0.996(0.006) & 1(0) & 1(0) \\   \cline{4-7}
\hline
\multirow{4}{*}{\includegraphics[width=15mm]{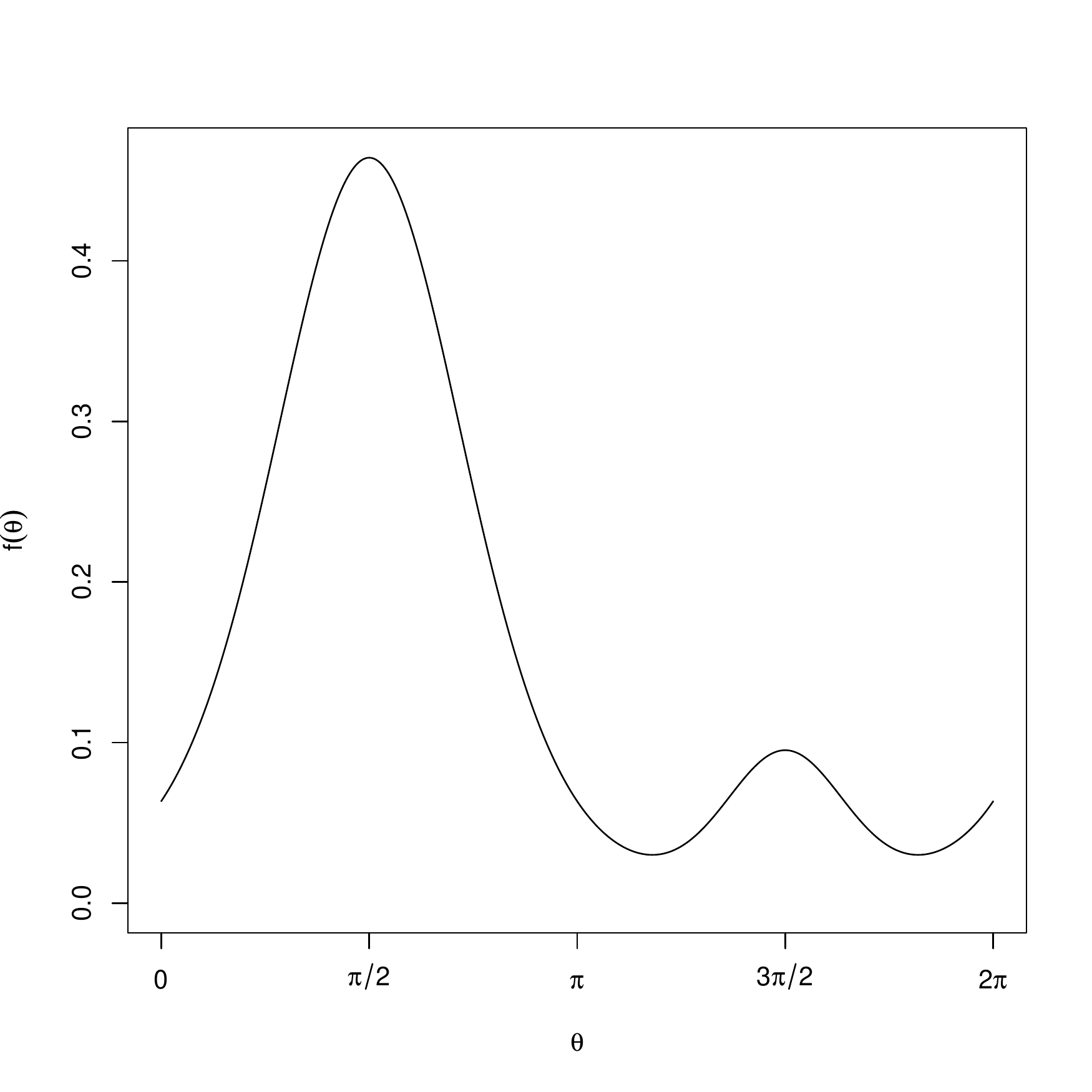}}& \multirow{4}{*}{\includegraphics[width=15mm]{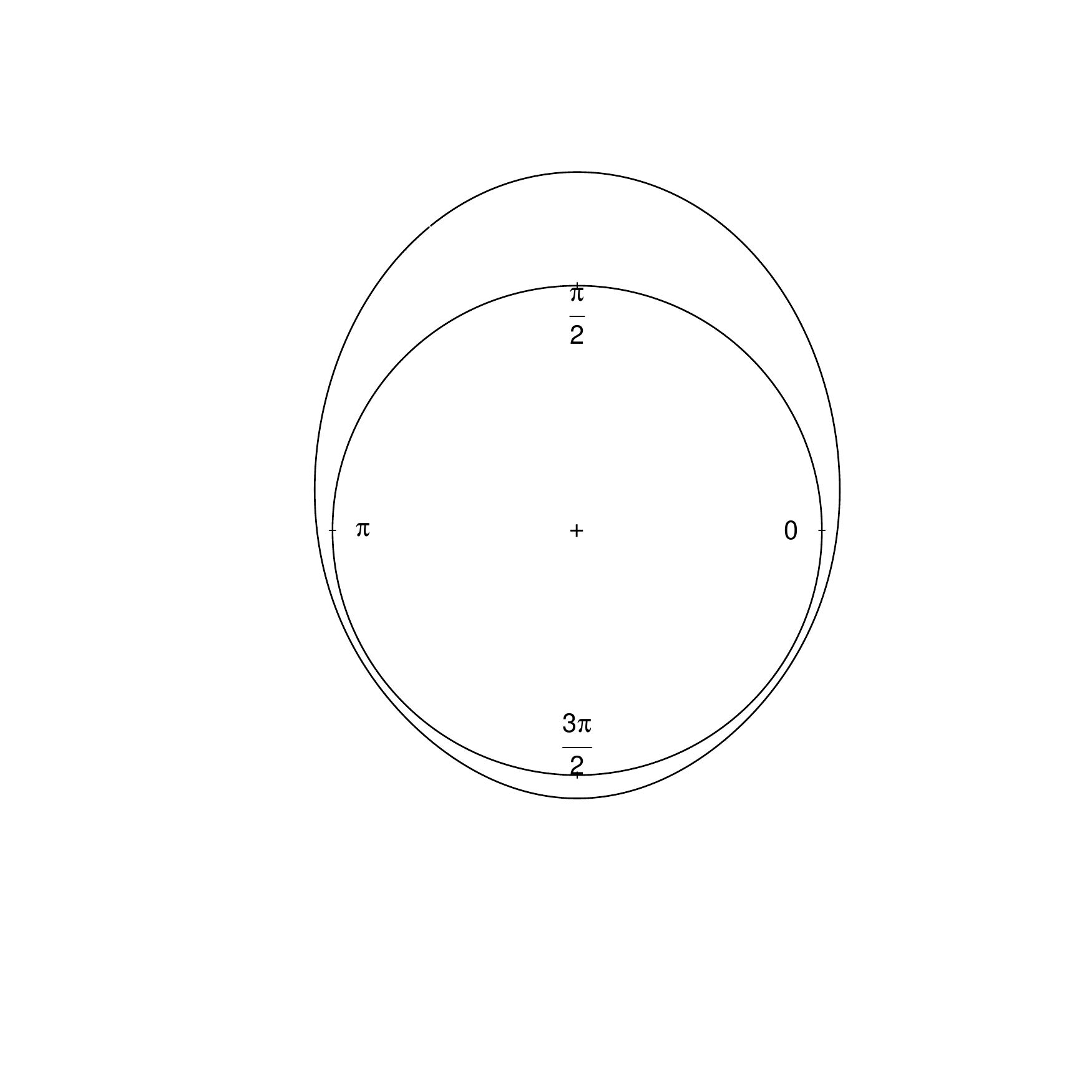}}& M12 &\multicolumn{3}{c|}{$U^2$ of Watson} &\multicolumn{3}{c|}{Excess mass}  \\ \cline{3-9}
  &  & $n=50$ & 0.338(0.041) & 0.548(0.044) & 0.654(0.042) & 0.002(0.004) & 0.040(0.017) & 0.076(0.023) \\ 
    &  & $n=100$ & 0.594(0.043) & 0.758(0.038) & 0.830(0.033) & 0.010(0.009) & 0.058(0.020) & 0.112(0.028) \\ 
    &  & $n=200$ & 0.880(0.028) & 0.940(0.021) & 0.956(0.018) & 0.040(0.017) & 0.116(0.028) & 0.238(0.037) \\   
\hline
\multirow{4}{*}{\includegraphics[width=15mm]{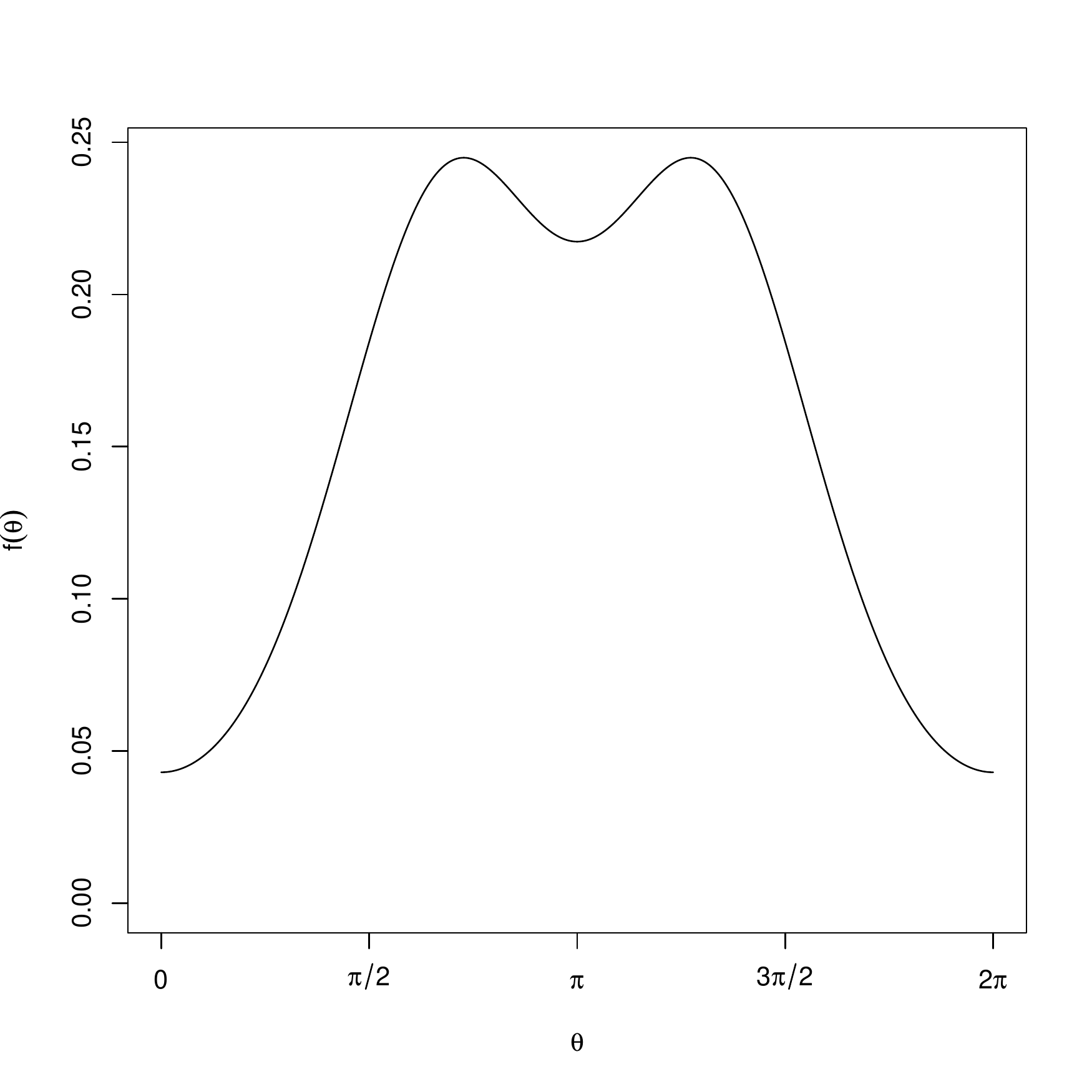}}& \multirow{4}{*}{\includegraphics[width=15mm]{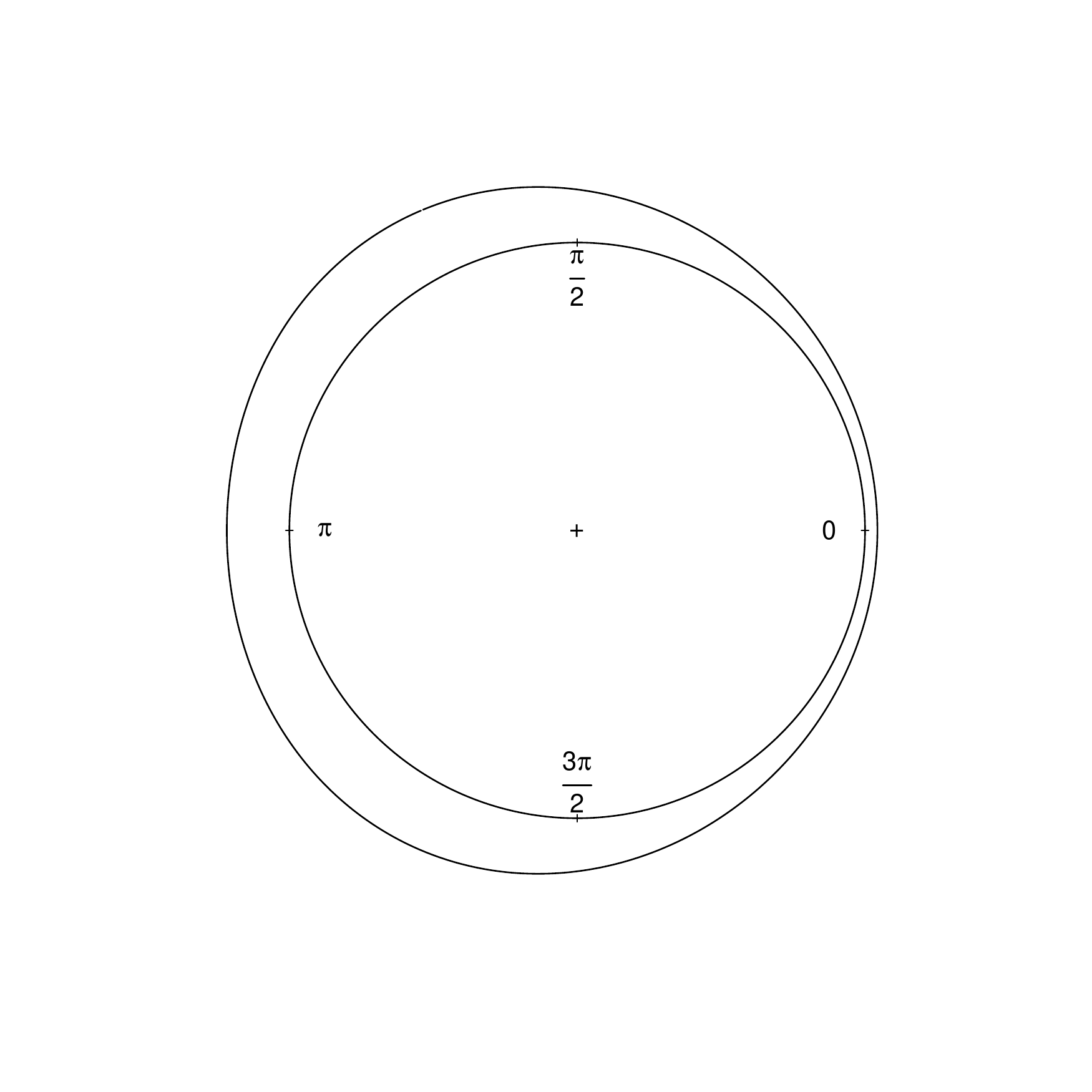}}& M13 &\multicolumn{3}{c|}{$U^2$ of Watson} &\multicolumn{3}{c|}{Excess mass}  \\ \cline{3-9}
  &  & $n=50$ & 0.014(0.010) & 0.070(0.022) & 0.130(0.029) & 0.022(0.013) & 0.072(0.023) & 0.140(0.030) \\ 
    &  & $n=100$ & 0.038(0.017) & 0.124(0.029) & 0.196(0.035) & 0.032(0.015) & 0.084(0.024) & 0.156(0.032) \\ 
    &  & $n=200$ & 0.066(0.022) & 0.170(0.033) & 0.246(0.038) & 0.044(0.018) & 0.102(0.027) & 0.204(0.035) \\
\hline
\multirow{4}{*}{\includegraphics[width=15mm]{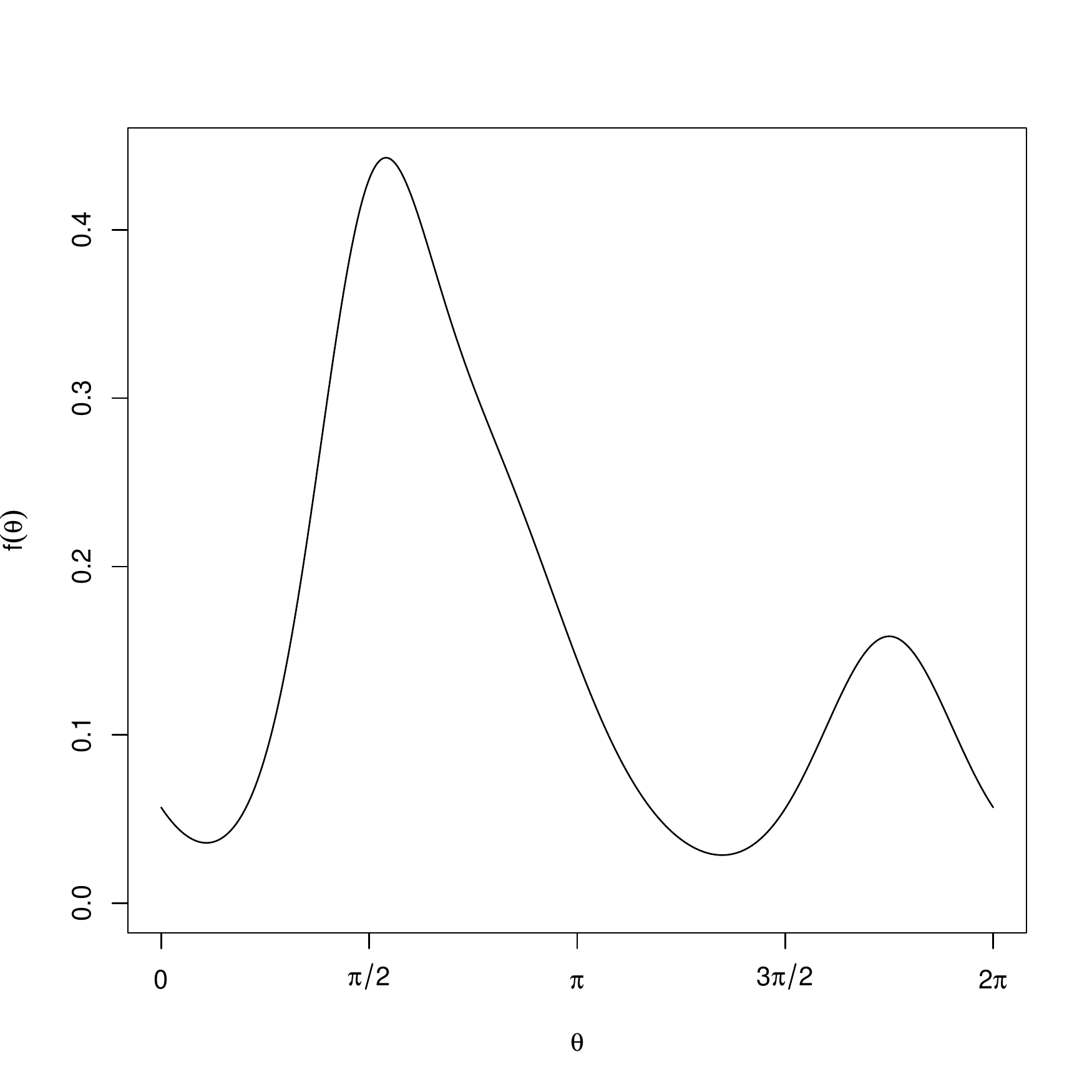}}& \multirow{4}{*}{\includegraphics[width=15mm]{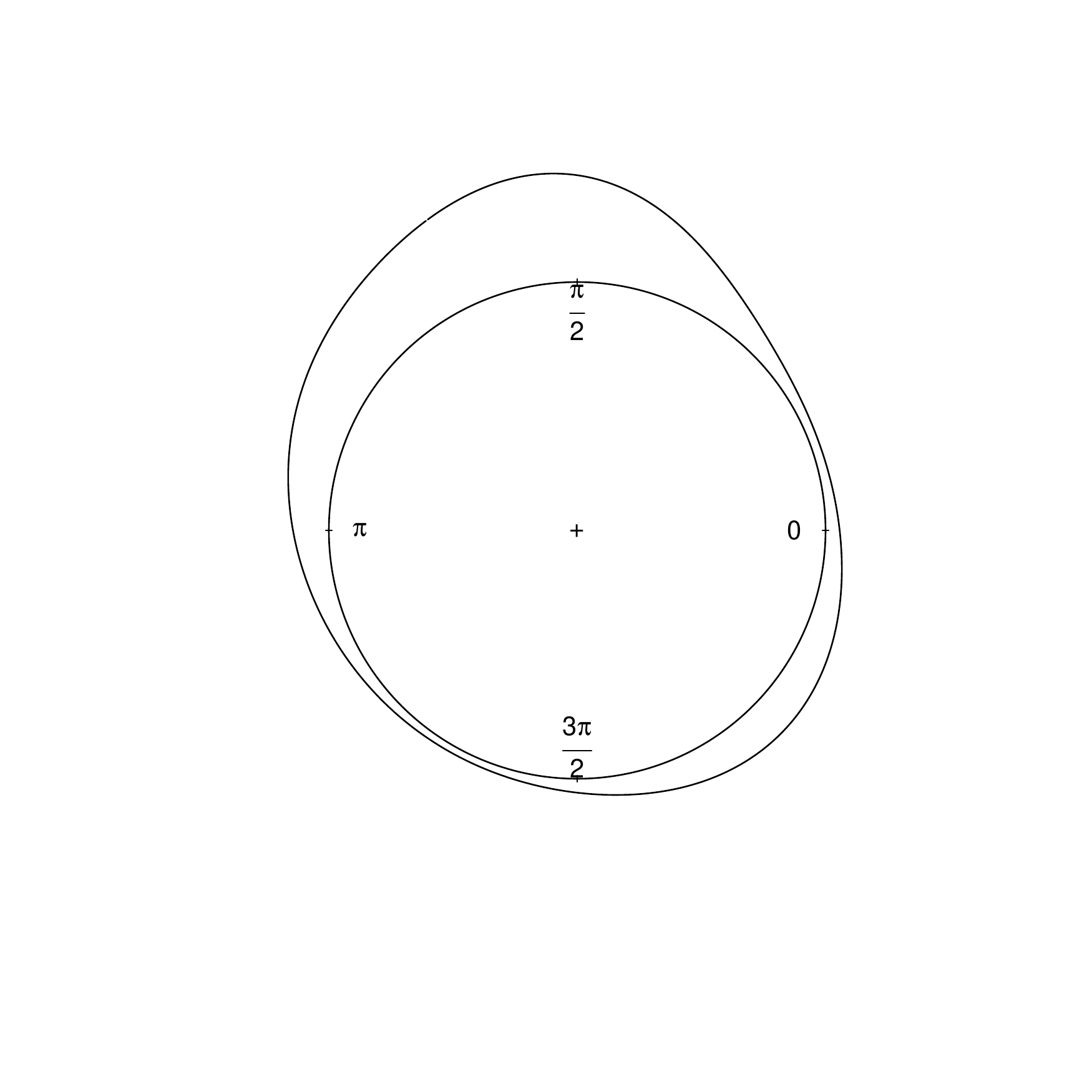}}& M14 &\multicolumn{3}{c|}{$U^2$ of Watson} &\multicolumn{3}{c|}{Excess mass}  \\ \cline{3-9}
  &  & $n=50$ & 0.476(0.044) & 0.730(0.039) & 0.840(0.032) & 0.044(0.018) & 0.164(0.032) & 0.284(0.040) \\ 
    &  & $n=100$ & 0.828(0.033) & 0.952(0.019) & 0.976(0.013) & 0.208(0.036) & 0.438(0.043) & 0.554(0.044) \\ 
    &  & $n=200$ & 0.990(0.009) & 0.998(0.004) & 1(0) & 0.578(0.043) & 0.796(0.035) & 0.870(0.029) \\   \cline{4-7}
\hline
\multirow{4}{*}{\includegraphics[width=15mm]{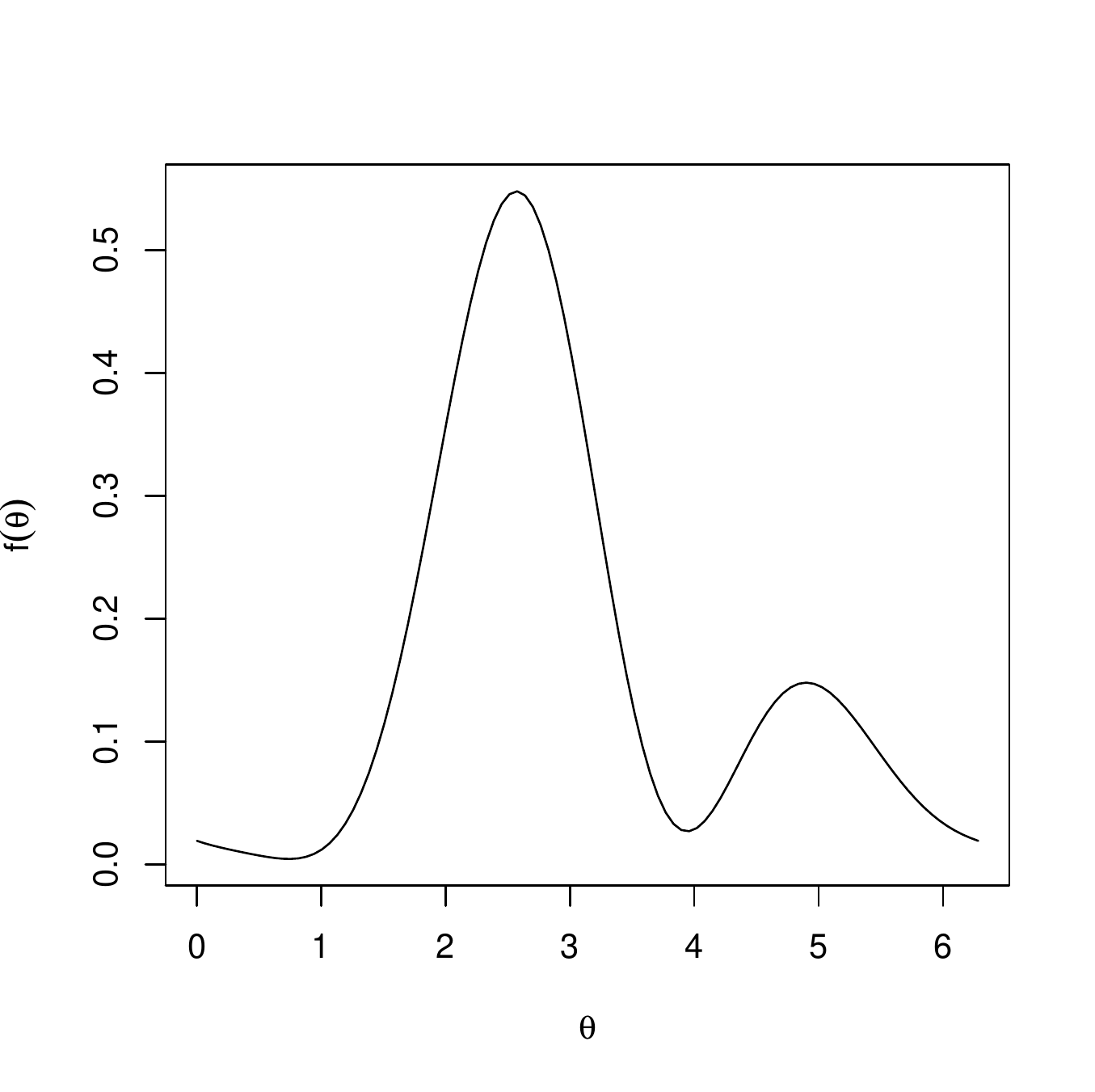}}& \multirow{4}{*}{\includegraphics[width=15mm]{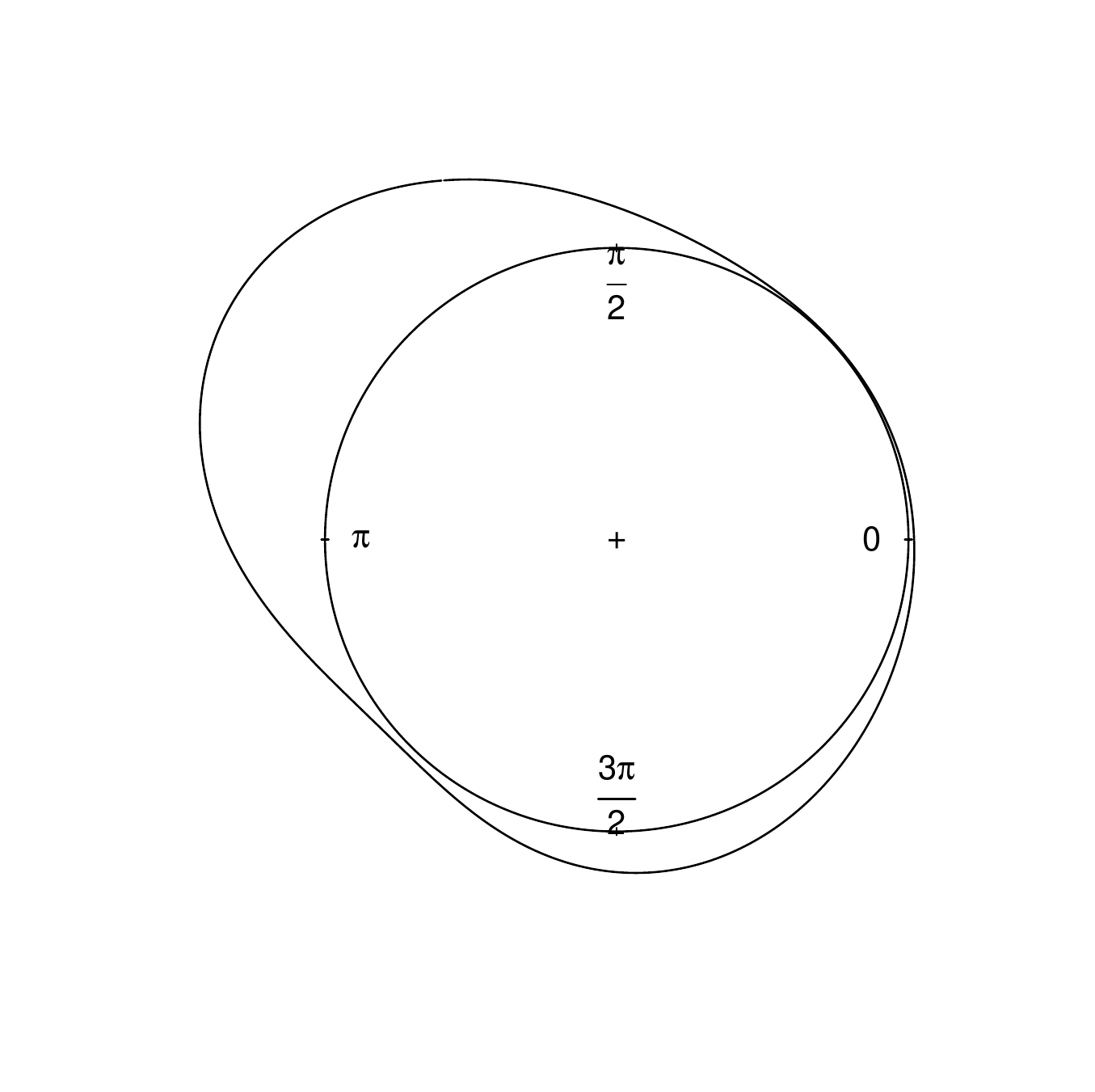}}& M15 &\multicolumn{3}{c|}{$U^2$ of Watson} &\multicolumn{3}{c|}{Excess mass}  \\ \cline{3-9}
 &  & $n=50$ & 0.678(0.041) & 0.852(0.031) & 0.908(0.025) & 0.026(0.014) & 0.118(0.028) & 0.212(0.036) \\ 
   &  & $n=100$ & 0.894(0.027) & 0.956(0.018) & 0.968(0.015) & 0.128(0.029) & 0.318(0.041) & 0.452(0.044) \\ 
   &  & $n=200$ & 0.986(0.010) & 0.996(0.006) & 0.998(0.004) & 0.406(0.043) & 0.644(0.042) & 0.752(0.038) \\ 
\hline
\end{tabular}
}
\caption{Percentage of rejections for testing $H_0:j=1$ vs. $H_a:j>1$, with 500 simulations (1.96 times their estimated standard deviation in parenthesis) and $B=500$ bootstrap samples. For models under the null (a): M1--M10, and under the alternative hypothesis (b): M11--M15. First and second column: linear and circular representations.}
\label{estsimcirc1}
\end{table}

\begin{table}
\scalebox{0.58}{
\begin{tabular}{|c |c|c|c c c |c c c |}
\hline
(a) & Bimodal &  $\alpha$ & 0.01 & 0.05& 0.10 & 0.01 & 0.05& 0.10 \\ \hline
\multirow{4}{*}{\includegraphics[width=15mm]{mc9}}& \multirow{4}{*}{\includegraphics[width=15mm]{mc9c}}& M11 &\multicolumn{3}{c|}{$U^2$ of Watson} &\multicolumn{3}{c|}{Excess mass}  \\ \cline{3-9}
 &  & $n=50$ & 0.006(0.007) & 0.024(0.013) & 0.066(0.022) & 0.010(0.009) & 0.034(0.016) & 0.068(0.022) \\ 
   &  & $n=200$ & 0.024(0.013) & 0.064(0.021) & 0.090(0.025) & 0.002(0.004) & 0.024(0.013) & 0.058(0.020) \\ 
   &  & $n=1000$ & 0.038(0.017) & 0.094(0.026) & 0.132(0.030) & 0(0) & 0.038(0.016) & 0.074(0.023) \\ 
\hline
\multirow{4}{*}{\includegraphics[width=15mm]{mc10}}& \multirow{4}{*}{\includegraphics[width=15mm]{mc9c}}& M12 &\multicolumn{3}{c|}{$U^2$ of Watson} &\multicolumn{3}{c|}{Excess mass}  \\ \cline{3-9}
 &  & $n=50$ & 0.012(0.010) & 0.048(0.019) & 0.096(0.026) & 0.002(0.004) & 0.028(0.014) & 0.060(0.021) \\ 
   &  & $n=200$ & 0.010(0.009) & 0.032(0.015) & 0.066(0.022) & 0.006(0.007) & 0.030(0.015) & 0.082(0.024) \\ 
   &  & $n=1000$ & 0(0) & 0.004(0.006) & 0.022(0.013) & 0.004(0.006) & 0.040(0.017) & 0.088(0.025) \\  
\hline
\multirow{4}{*}{\includegraphics[width=15mm]{mc11}}& \multirow{4}{*}{\includegraphics[width=15mm]{mc9c}}& M13 &\multicolumn{3}{c|}{$U^2$ of Watson} &\multicolumn{3}{c|}{Excess mass}  \\ \cline{3-9}
 &  & $n=50$ & 0.006(0.007) & 0.026(0.014) & 0.058(0.020) & 0.004(0.006) & 0.042(0.018) & 0.100(0.026) \\ 
   &  & $n=200$ & 0.002(0.004) & 0.020(0.012) & 0.048(0.019) & 0.008(0.008) & 0.046(0.018) & 0.104(0.027) \\ 
   &  & $n=1000$ & 0.014(0.010) & 0.044(0.018) & 0.098(0.026) & 0.006(0.007) & 0.056(0.020) & 0.112(0.028) \\  
\hline
\multirow{4}{*}{\includegraphics[width=15mm]{mc12}}& \multirow{4}{*}{\includegraphics[width=15mm]{mc9c}}& M14 &\multicolumn{3}{c|}{$U^2$ of Watson} &\multicolumn{3}{c|}{Excess mass}  \\ \cline{3-9}
 &  & $n=50$ & 0.014(0.010) & 0.060(0.021) & 0.100(0.026) & 0.004(0.006) & 0.034(0.016) & 0.054(0.020) \\ 
   &  & $n=200$ & 0.002(0.004) & 0.020(0.012) & 0.046(0.018) & 0.004(0.006) & 0.036(0.016) & 0.086(0.025) \\ 
   &  & $n=1000$ & 0(0) & 0.012(0.010) & 0.030(0.015) & 0.002(0.004) & 0.034(0.016) & 0.082(0.024) \\  
\hline
\multirow{4}{*}{\includegraphics[width=15mm]{mca3}}& \multirow{4}{*}{\includegraphics[width=15mm]{mca3c}}& M15 &\multicolumn{3}{c|}{$U^2$ of Watson} &\multicolumn{3}{c|}{Excess mass}  \\ \cline{3-9}
	 &  & $n=50$ & 0.048(0.019) & 0.152(0.031) & 0.238(0.037) & 0.008(0.008) & 0.028(0.014) & 0.066(0.022) \\ 
   &  & $n=200$ & 0.040(0.017) & 0.100(0.026) & 0.170(0.033) & 0.012(0.010) & 0.034(0.016) & 0.074(0.023) \\ 
   &  & $n=1000$ & 0.002(0.004) & 0.036(0.016) & 0.076(0.023) & 0.010(0.009) & 0.040(0.017) & 0.092(0.025) \\  
\hline
\multirow{4}{*}{\includegraphics[width=15mm]{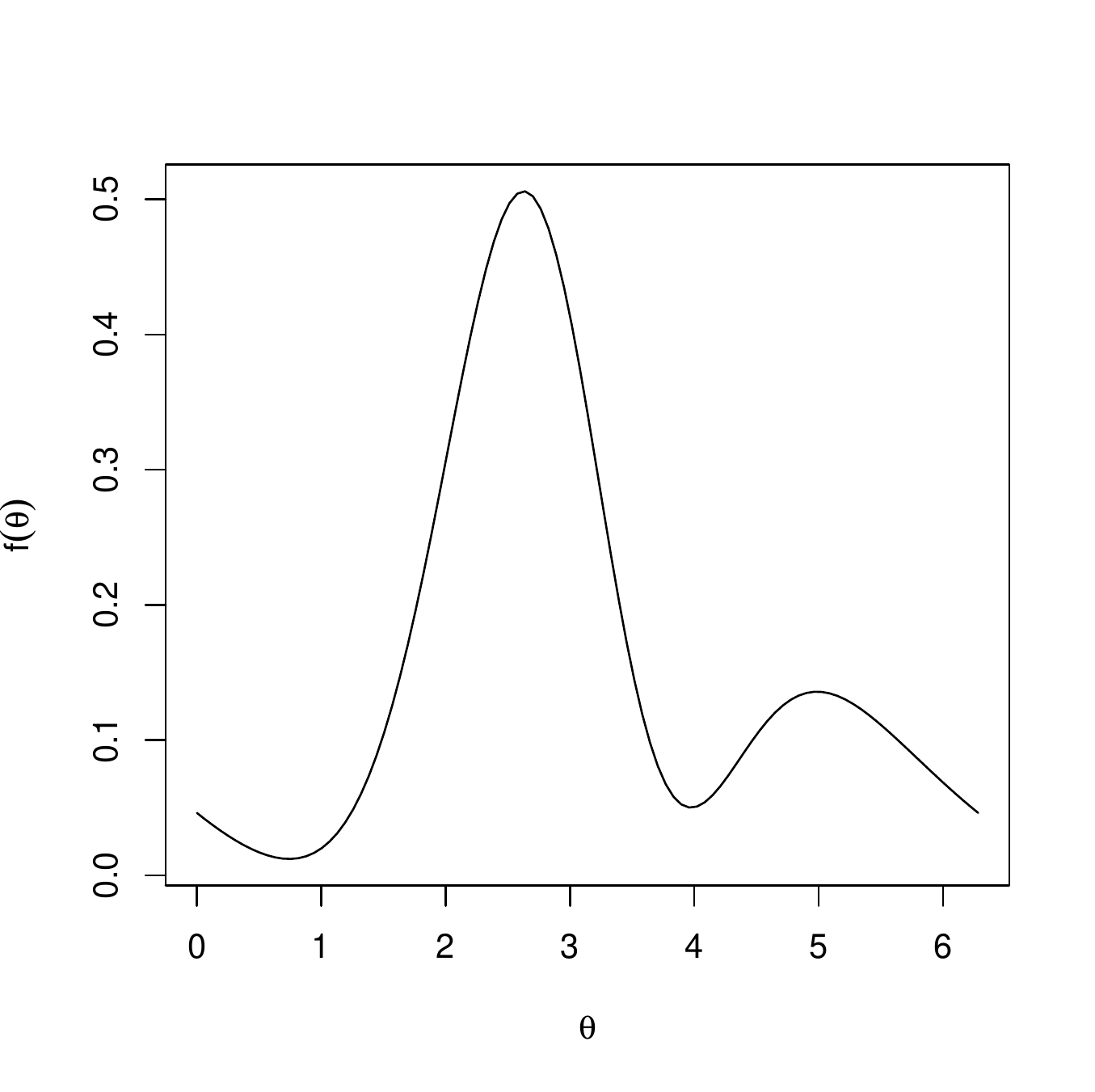}}& \multirow{4}{*}{\includegraphics[width=15mm]{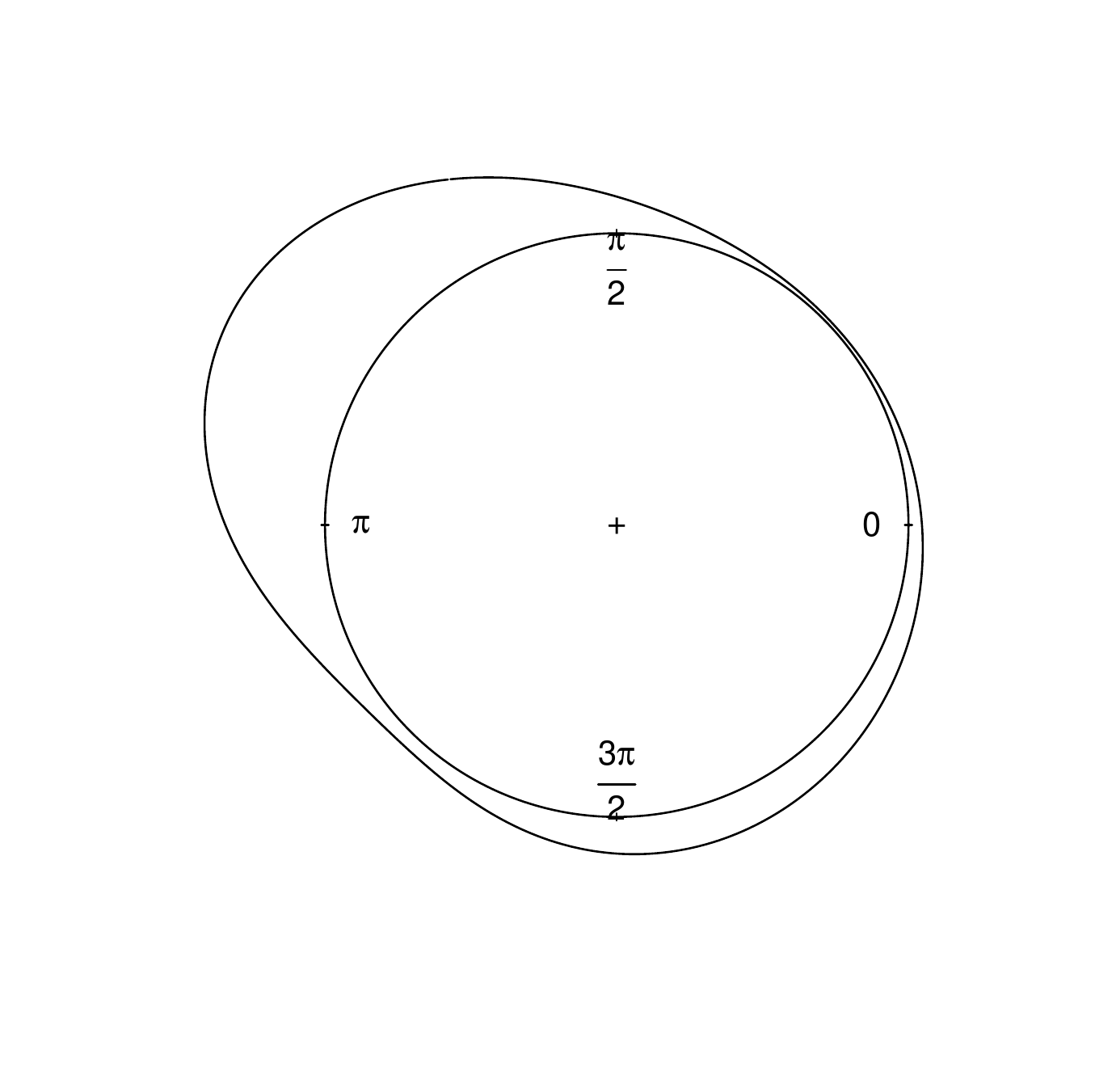}}& M16 &\multicolumn{3}{c|}{$U^2$ of Watson} &\multicolumn{3}{c|}{Excess mass}  \\ \cline{3-9}
   &  & $n=50$ & 0.030(0.015) & 0.084(0.024) & 0.156(0.032) & 0.006(0.007) & 0.024(0.013) & 0.060(0.021) \\ 
   &  & $n=200$ & 0.026(0.014) & 0.068(0.022) & 0.118(0.028) & 0.002(0.004) & 0.040(0.017) & 0.078(0.024) \\ 
   &  & $n=1000$ & 0.002(0.004) & 0.010(0.009) & 0.034(0.016) & 0.010(0.009) & 0.048(0.019) & 0.100(0.026) \\ 
\hline
\multirow{4}{*}{\includegraphics[width=15mm]{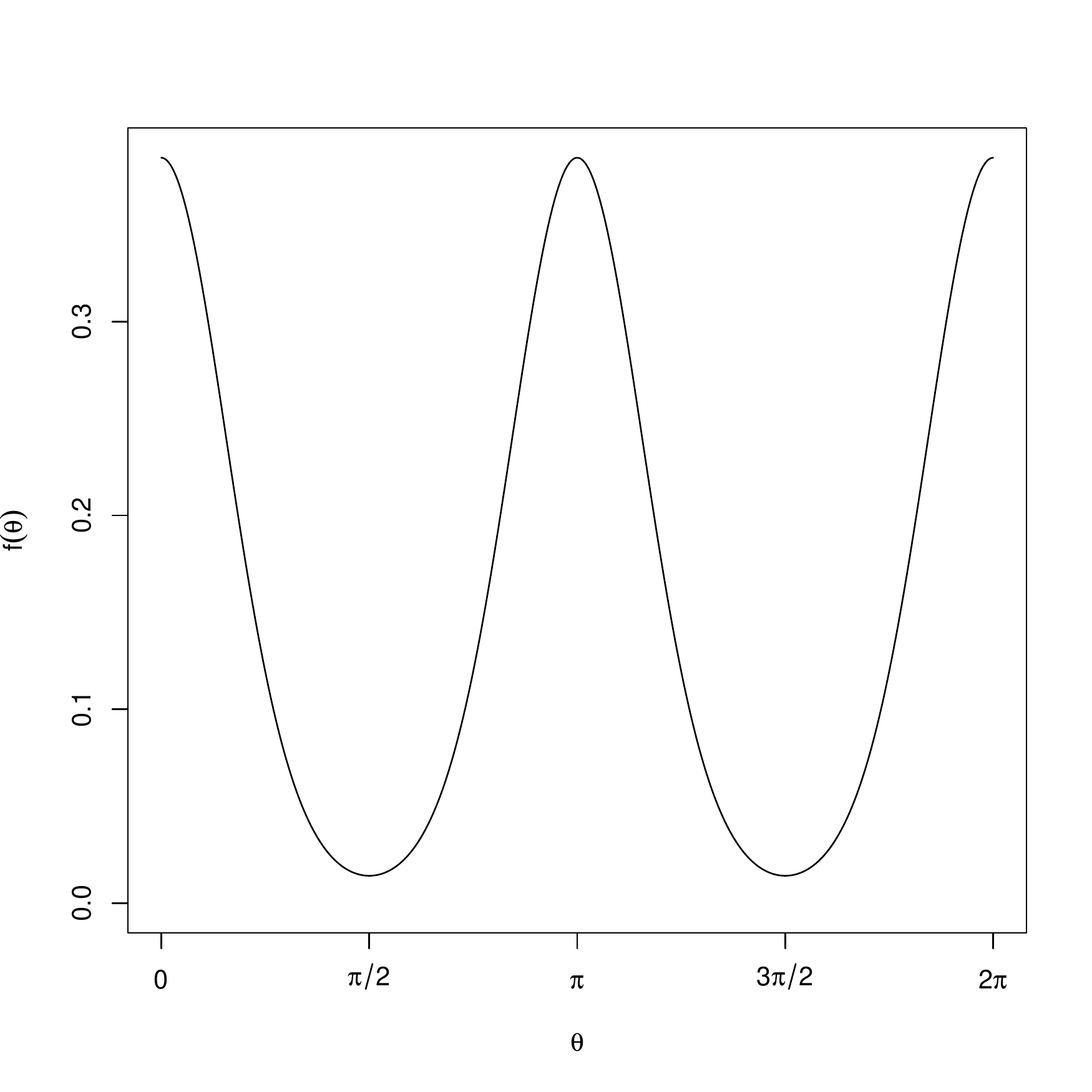}}& \multirow{4}{*}{\includegraphics[width=15mm]{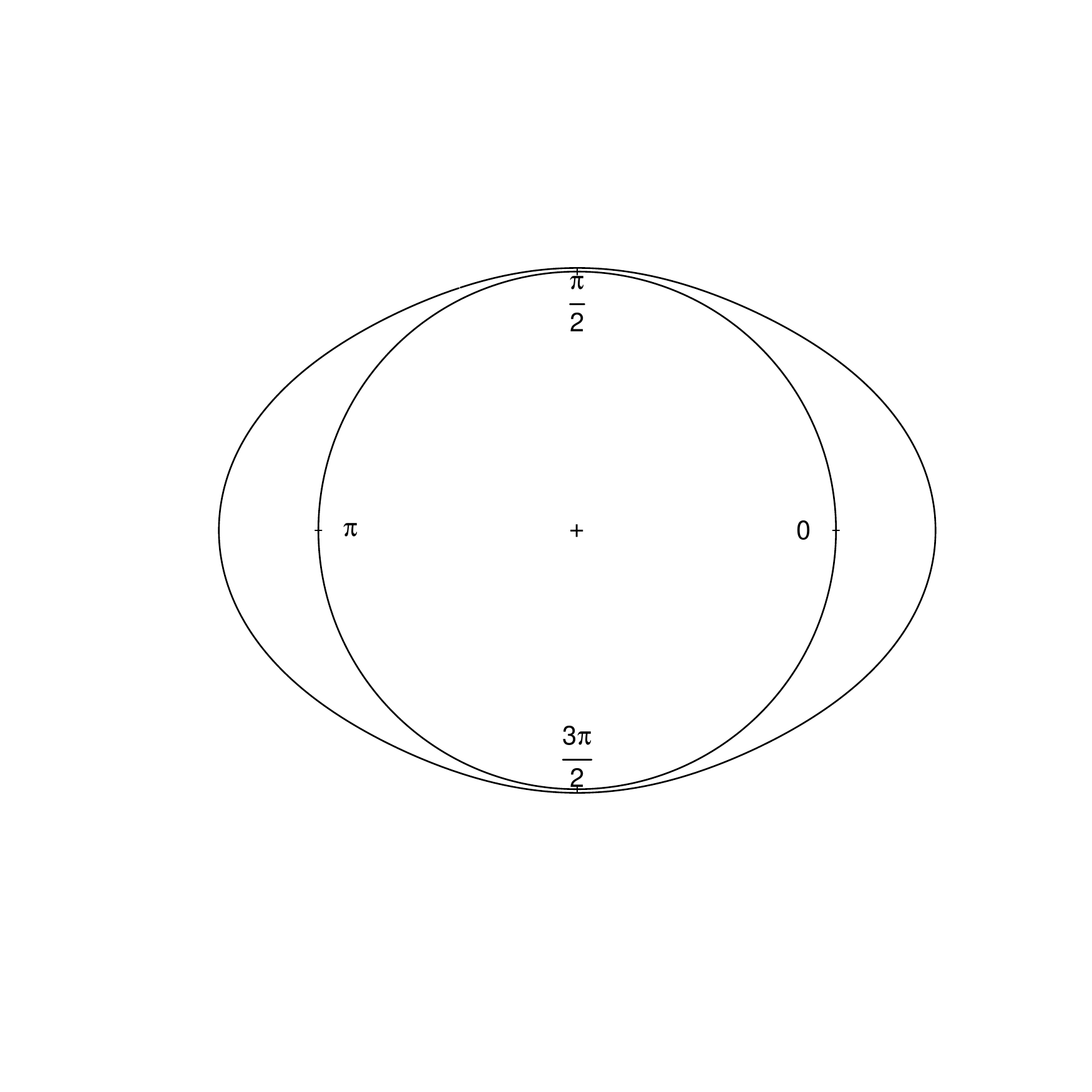}}& M17 &\multicolumn{3}{c|}{$U^2$ of Watson} &\multicolumn{3}{c|}{Excess mass}  \\ \cline{3-9}
 &  & $n=50$ & 0.004(0.006) & 0.012(0.010) & 0.044(0.018) & 0.004(0.006) & 0.030(0.015) & 0.080(0.024) \\ 
   &  & $n=200$ & 0(0) & 0.012(0.010) & 0.030(0.015) & 0.004(0.006) & 0.036(0.016) & 0.066(0.022)  \\ 
   &  & $n=1000$ & 0.002(0.004) & 0.022(0.013) & 0.052(0.019) & 0.002(0.004) & 0.032(0.015) & 0.072(0.023) \\  
\hline
\multirow{4}{*}{\includegraphics[width=15mm]{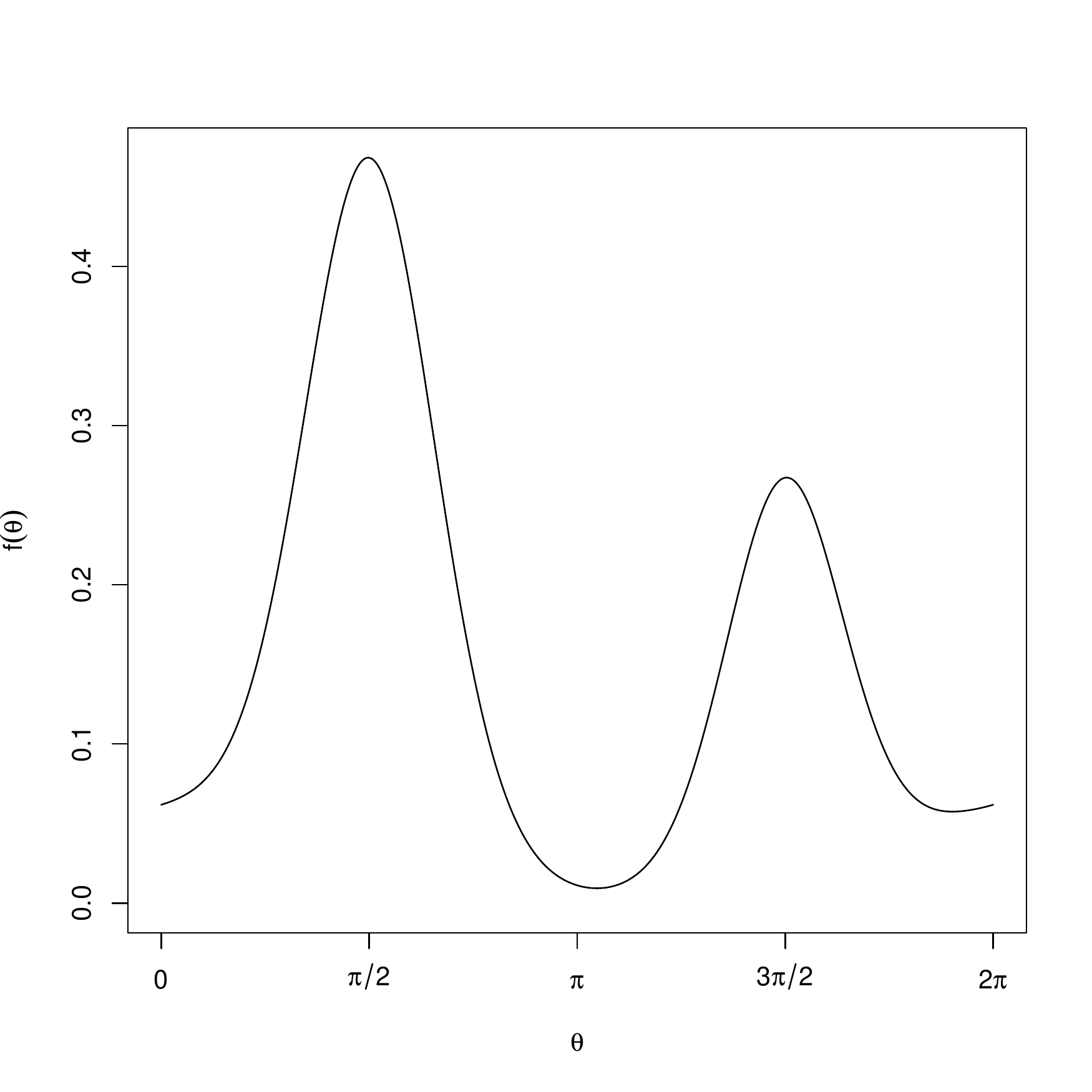}}& \multirow{4}{*}{\includegraphics[width=15mm]{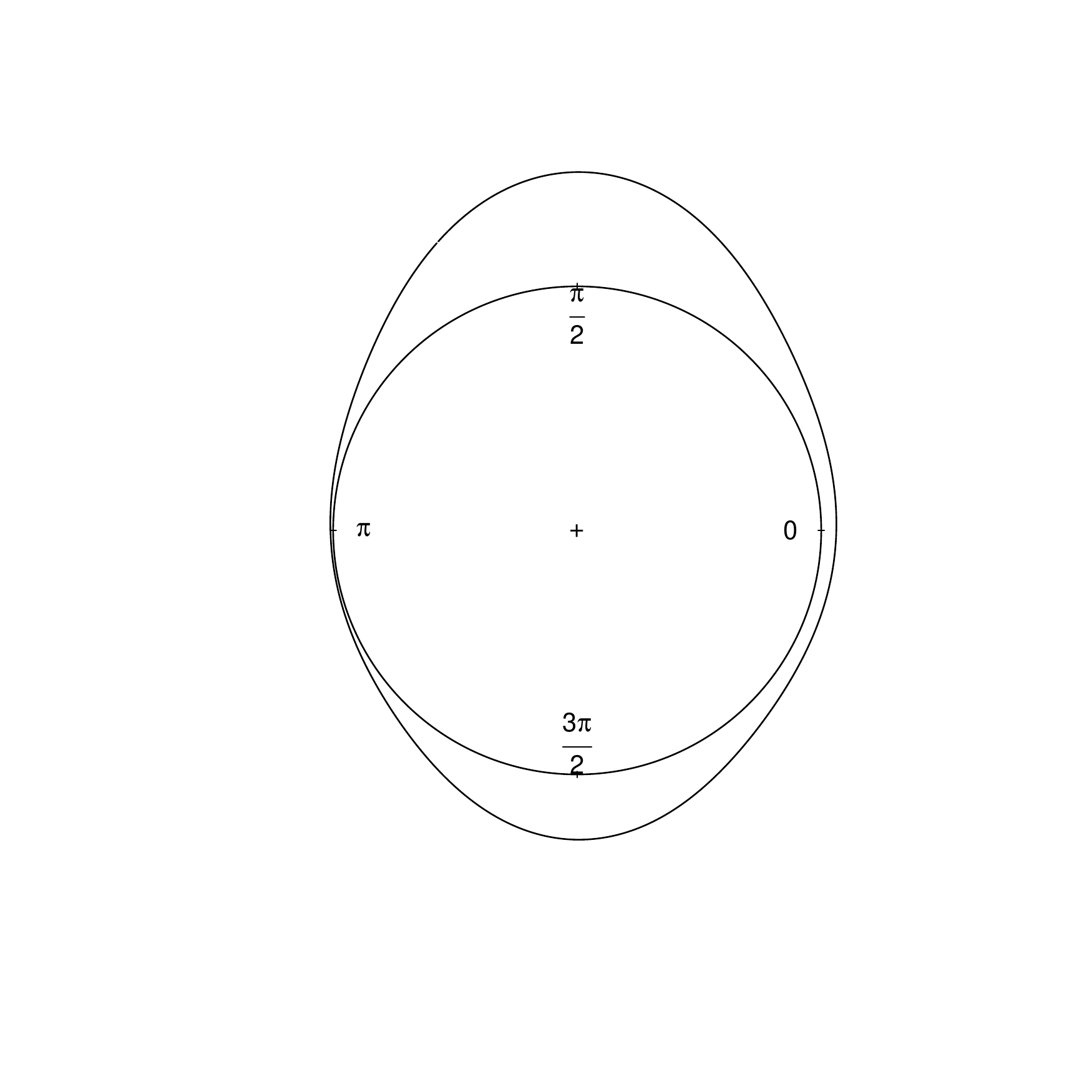}}& M18 &\multicolumn{3}{c|}{$U^2$ of Watson} &\multicolumn{3}{c|}{Excess mass}  \\ \cline{3-9}
 &  & $n=50$ & 0.002(0.004) & 0.036(0.016) & 0.088(0.025) & 0.010(0.009) & 0.036(0.016) & 0.072(0.023) \\ 
   &  & $n=200$ & 0.008(0.008) & 0.046(0.018) & 0.126(0.029) & 0.008(0.008) & 0.044(0.018) & 0.074(0.023) \\ 
   &  & $n=1000$ & 0.012(0.010) & 0.070(0.022) & 0.122(0.029) & 0.014(0.010) & 0.042(0.017) & 0.090(0.025) \\  
\hline
\multirow{4}{*}{\includegraphics[width=15mm]{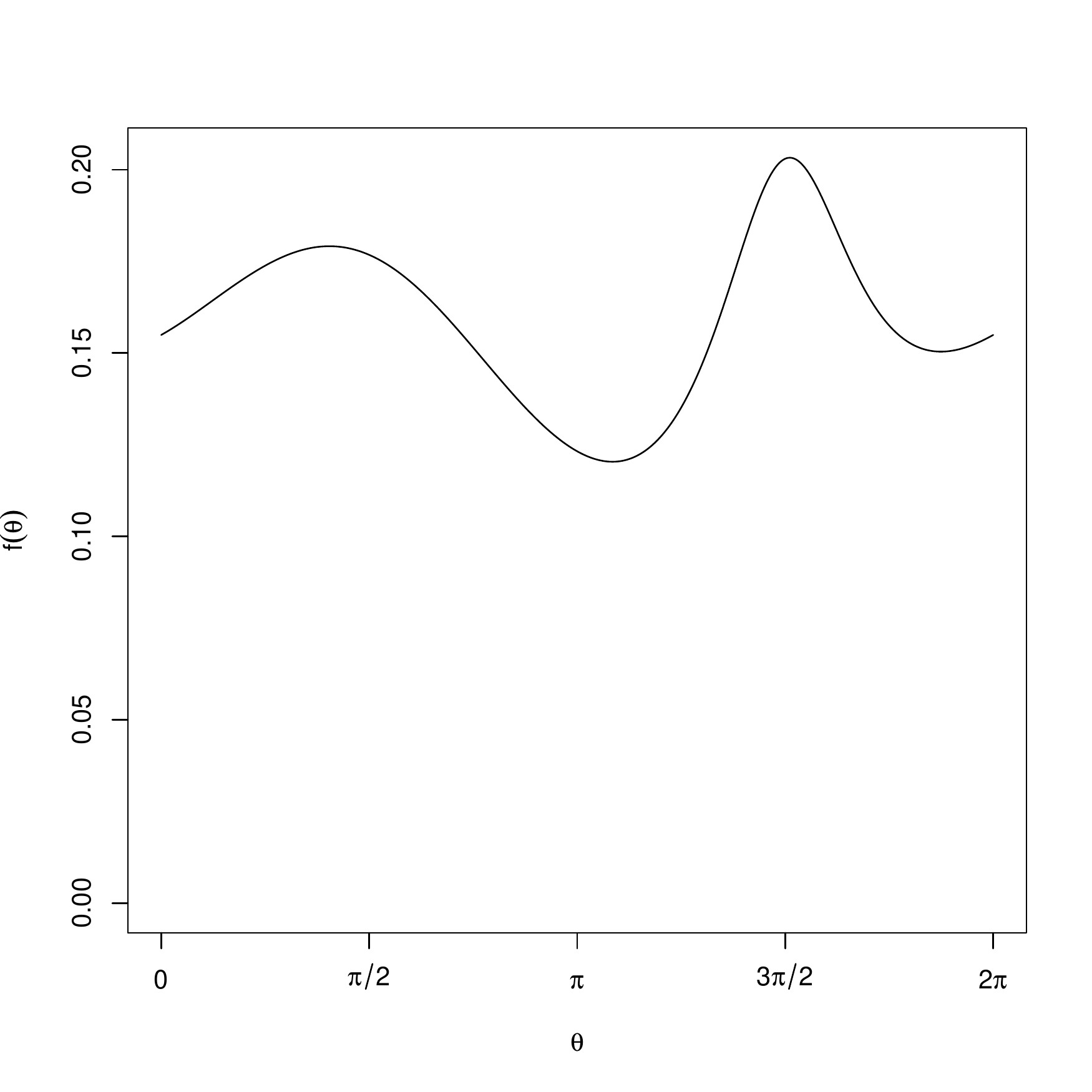}}& \multirow{4}{*}{\includegraphics[width=15mm]{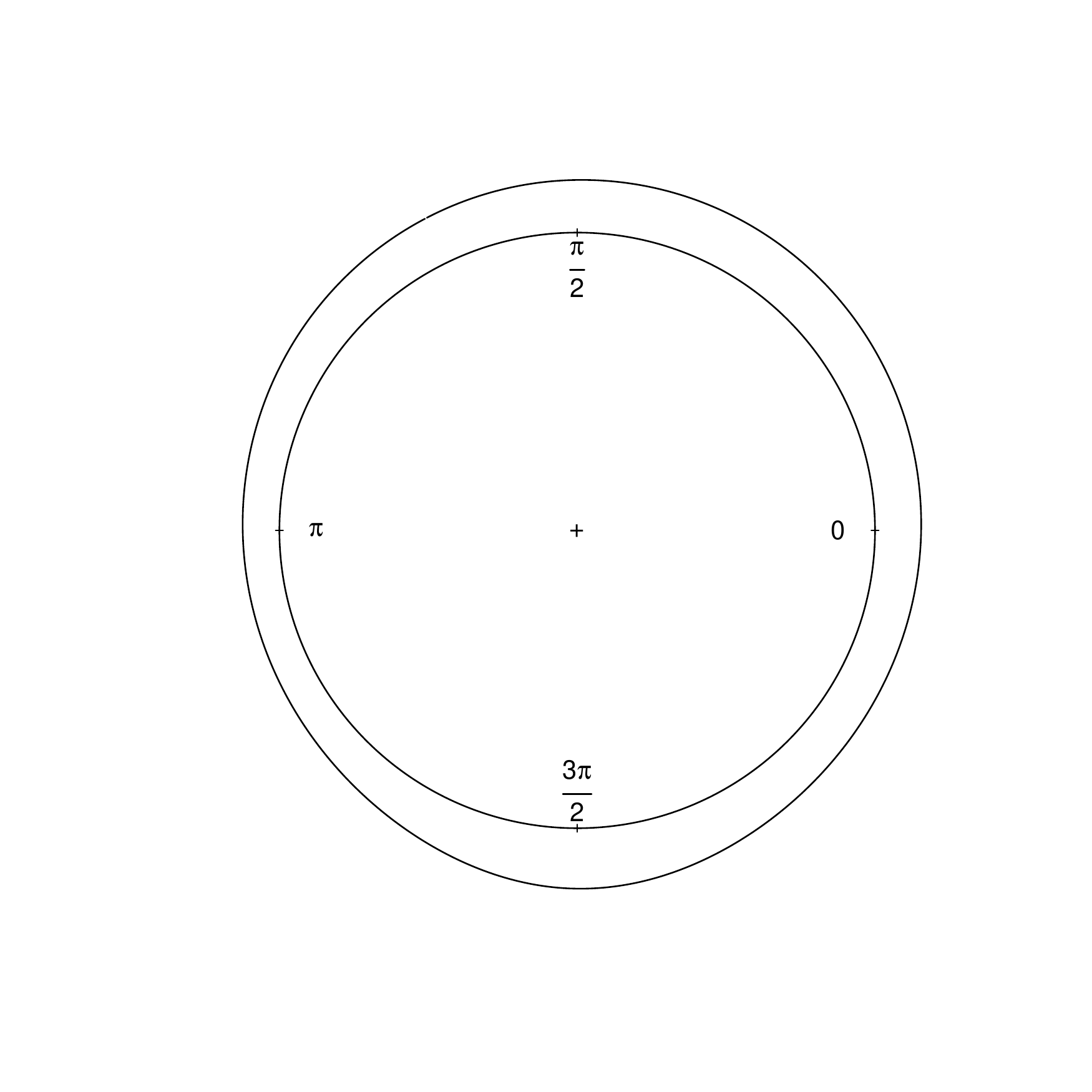}}& M19 &\multicolumn{3}{c|}{$U^2$ of Watson} &\multicolumn{3}{c|}{Excess mass}  \\ \cline{3-9}
 &  & $n=50$ & 0.010(0.009) & 0.044(0.018) & 0.098(0.026) & 0.006(0.007) & 0.046(0.018) & 0.096(0.026) \\ 
   &  & $n=200$ & 0.016(0.011) & 0.052(0.019) & 0.100(0.026) & 0.008(0.008) & 0.062(0.021) & 0.118(0.028) \\ 
   &  & $n=1000$ & 0.016(0.011) & 0.068(0.022) & 0.122(0.029) & 0.012(0.010) & 0.058(0.020) & 0.110(0.027) \\   
\hline
\multirow{4}{*}{\includegraphics[width=15mm]{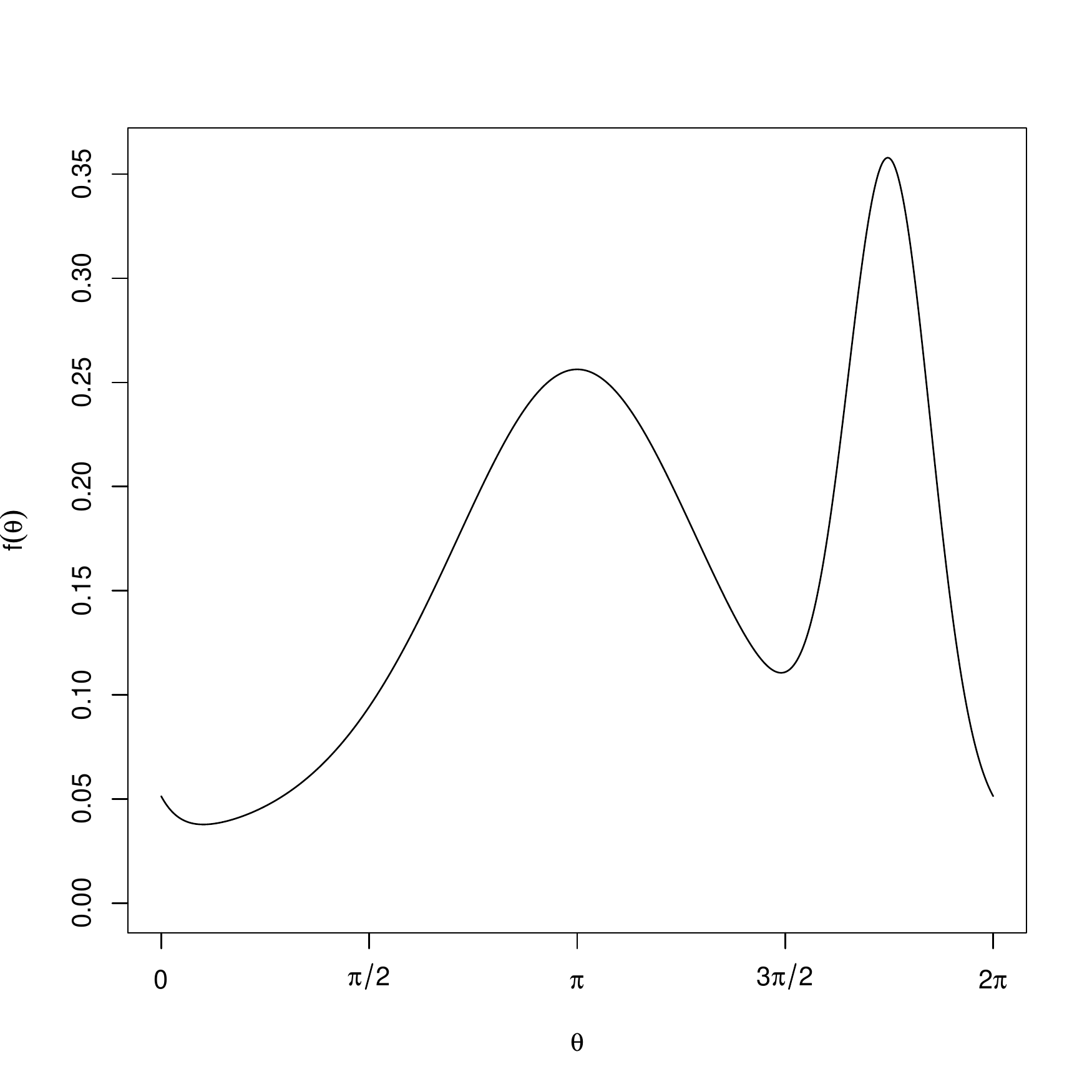}}& \multirow{4}{*}{\includegraphics[width=15mm]{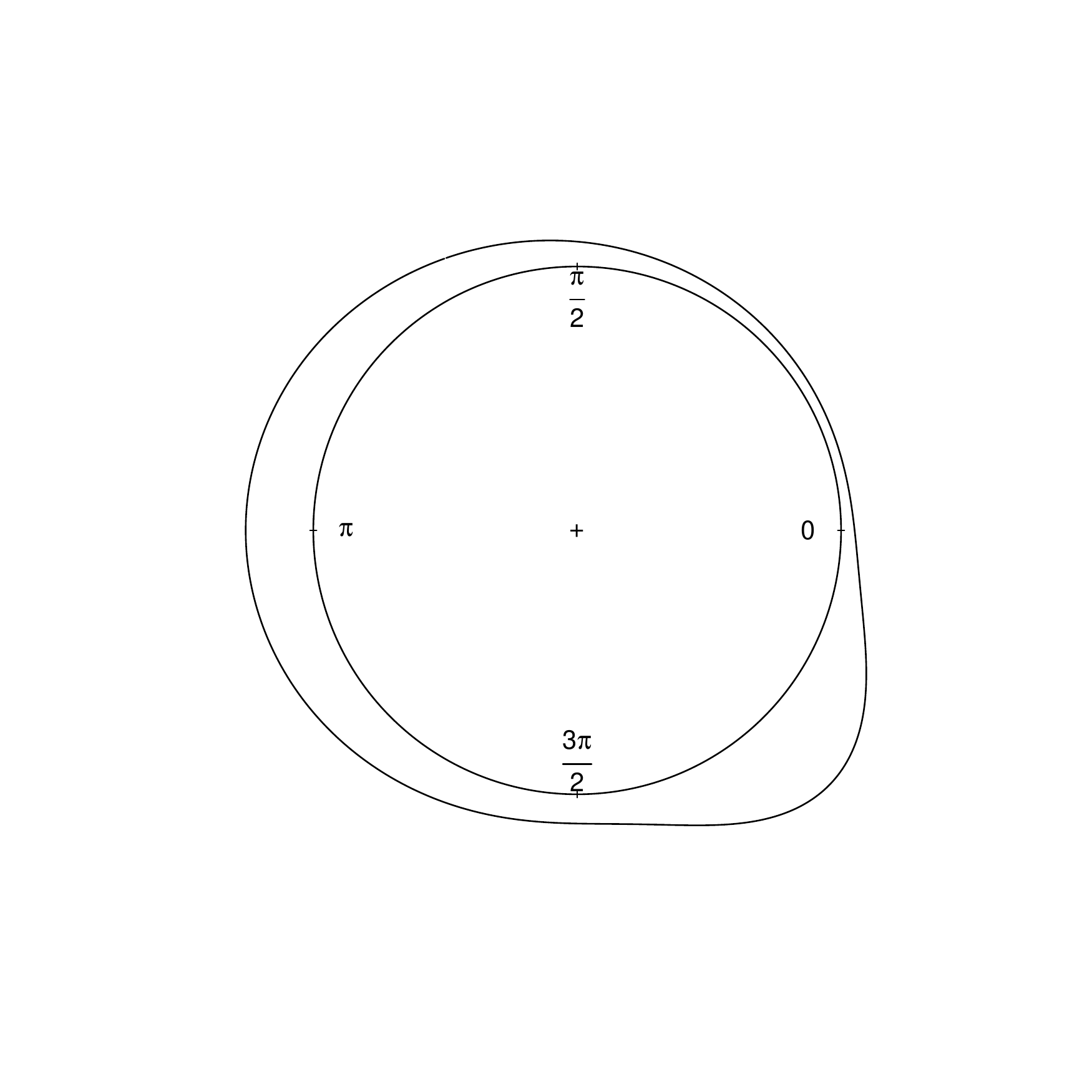}}& M20 &\multicolumn{3}{c|}{$U^2$ of Watson} &\multicolumn{3}{c|}{Excess mass}  \\ \cline{3-9}
 &  & $n=50$ & 0.032(0.015) & 0.144(0.031) & 0.214(0.036) & 0.018(0.012) & 0.078(0.024) & 0.140(0.030) \\ 
   &  & $n=200$ & 0.054(0.020) & 0.144(0.031) & 0.230(0.037) & 0.016(0.011) & 0.064(0.021) & 0.126(0.029) \\ 
   &  & $n=1000$ & 0.024(0.013) & 0.102(0.027) & 0.192(0.035) & 0.006(0.007) & 0.040(0.017) & 0.092(0.025) \\ 
\hline \hline
(b) & Trimodal &  $\alpha$ & 0.01 & 0.05& 0.10 & 0.01 & 0.05& 0.10 \\ \hline
\multirow{4}{*}{\includegraphics[width=15mm]{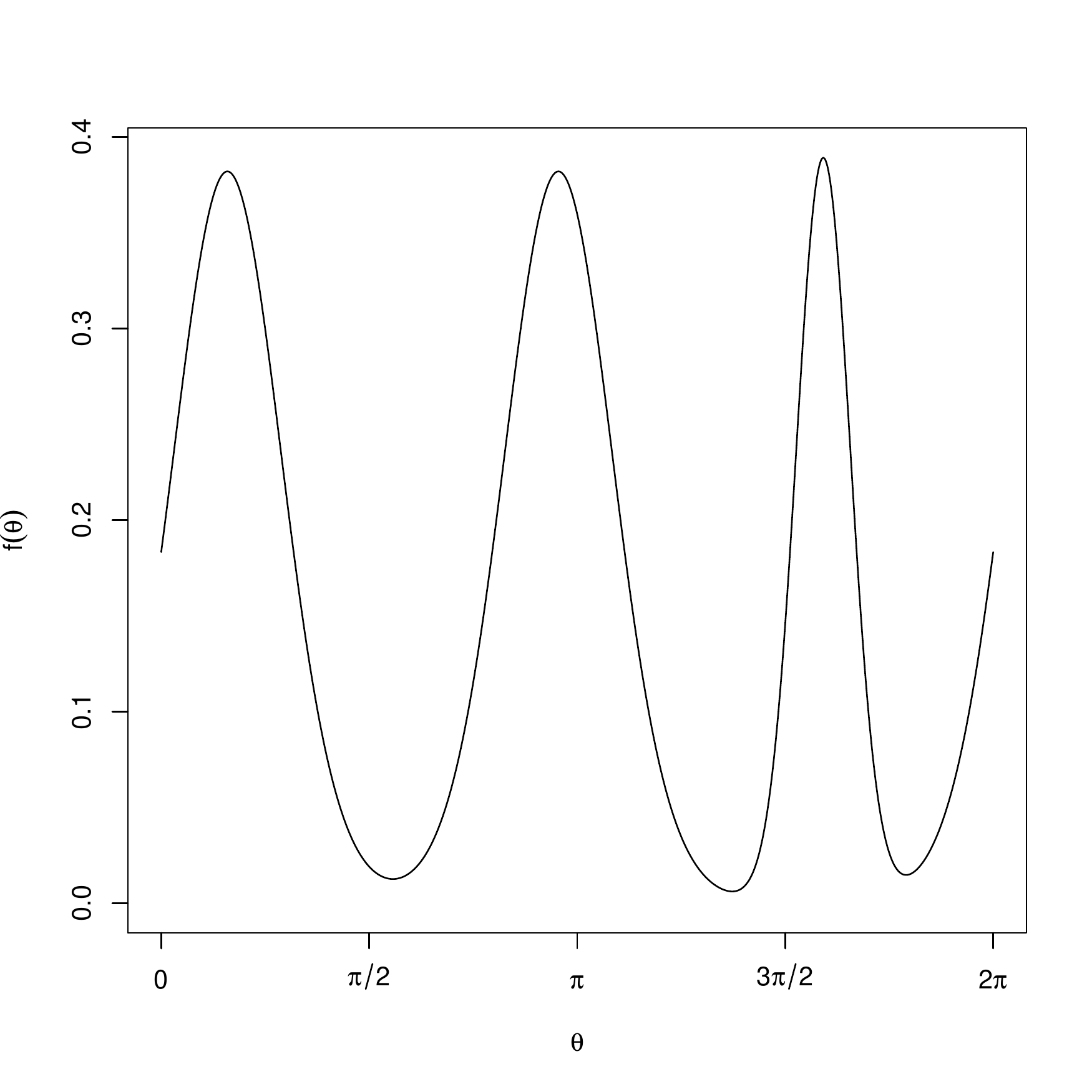}}& \multirow{4}{*}{\includegraphics[width=15mm]{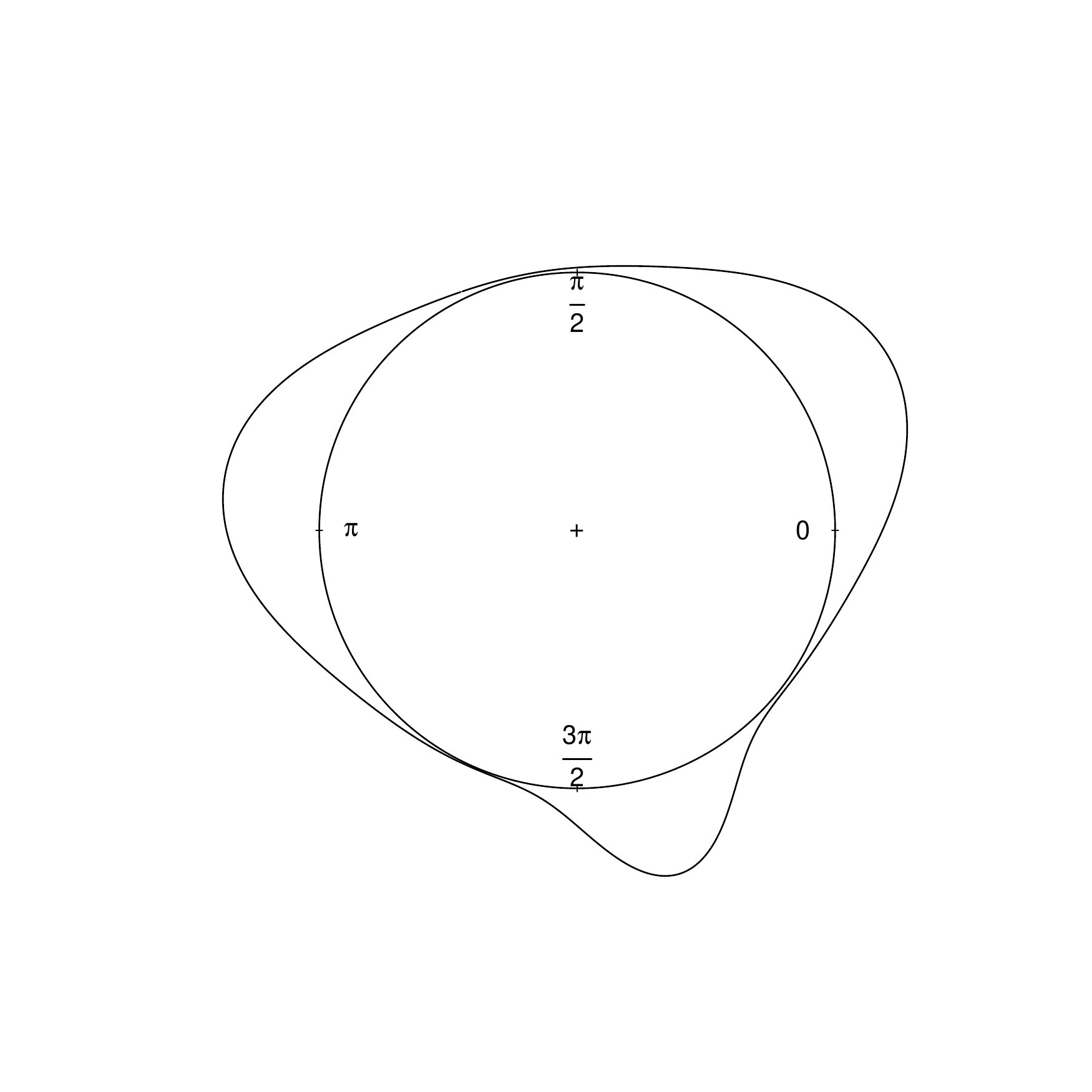}}& M21 &\multicolumn{3}{c|}{$U^2$ of Watson} &\multicolumn{3}{c|}{Excess mass}  \\ \cline{3-9}
  &  & $n=50$ & 0.870(0.029) & 0.964(0.016) & 0.978(0.013) & 0.600(0.043) & 0.782(0.036) & 0.842(0.032) \\ 
    &  & $n=100$ & 0.998(0.004) & 1(0) & 1(0) & 0.942(0.020) & 0.978(0.013) & 0.992(0.008) \\ 
    &  & $n=200$ & 1(0) & 1(0) & 1(0) & 1(0) & 1(0) & 1(0) \\ 
\hline
\multirow{4}{*}{\includegraphics[width=15mm]{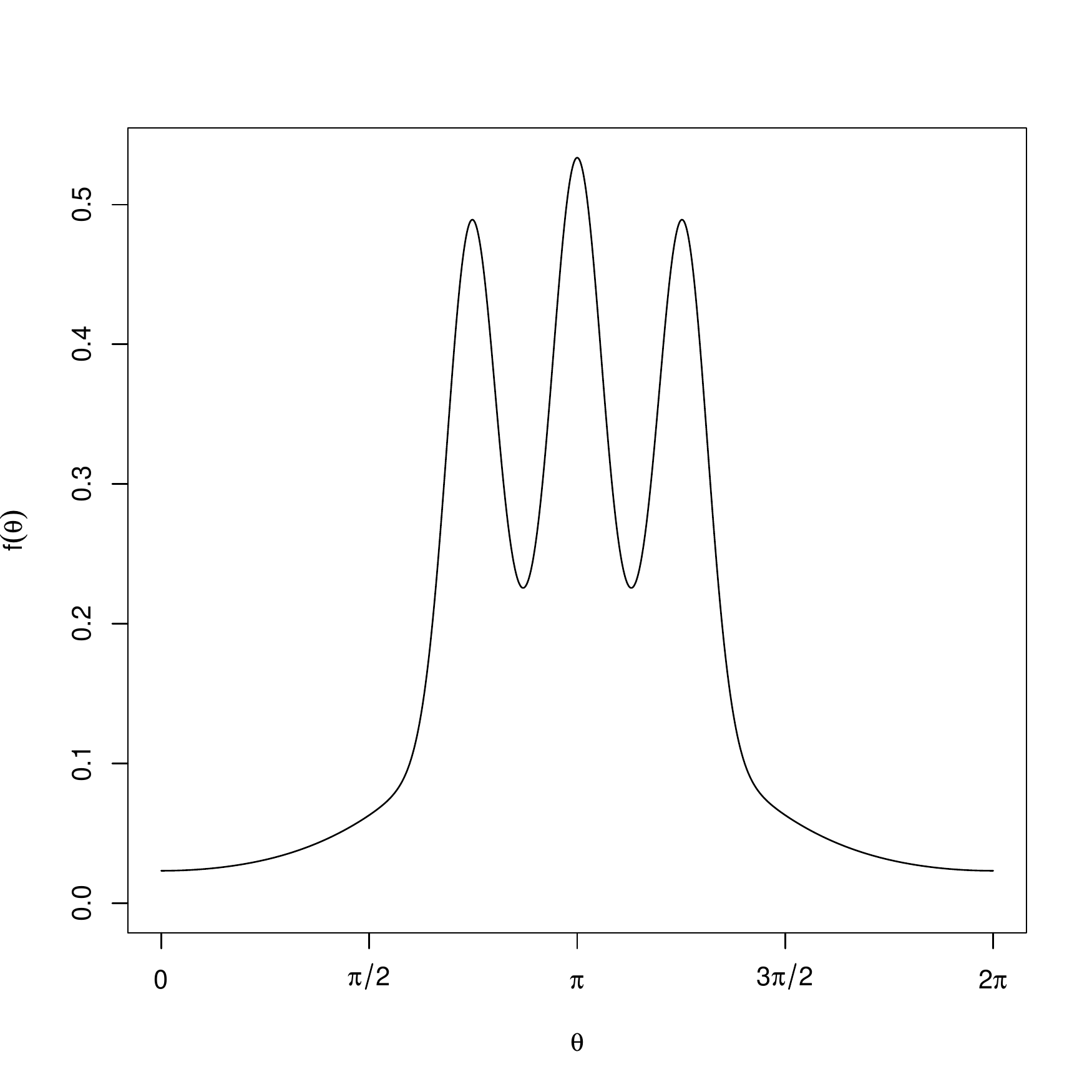}}& \multirow{4}{*}{\includegraphics[width=15mm]{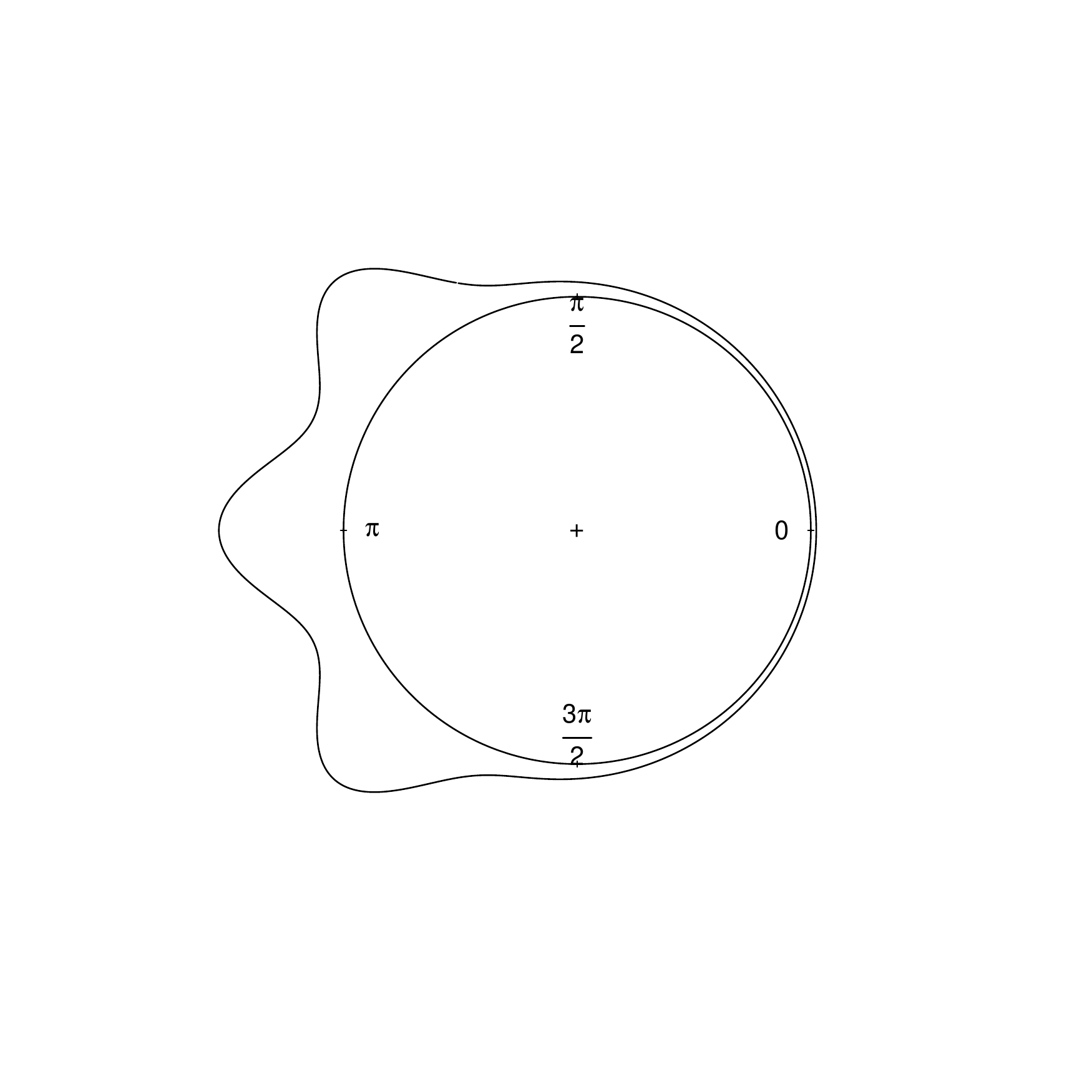}}& M22 &\multicolumn{3}{c|}{$U^2$ of Watson} &\multicolumn{3}{c|}{Excess mass}  \\ \cline{3-9}
  &  & $n=50$ & 0.050(0.019) & 0.180(0.034) & 0.288(0.040) & 0.062(0.021) & 0.204(0.035) & 0.282(0.039) \\ 
    &  & $n=100$ & 0.130(0.029) & 0.392(0.043) & 0.556(0.044) & 0.184(0.034) & 0.380(0.043) & 0.506(0.044) \\ 
    &  & $n=200$ & 0.344(0.042) & 0.700(0.040) & 0.852(0.031) & 0.436(0.043) & 0.692(0.040) & 0.780(0.036) \\ 
\hline
\multirow{4}{*}{\includegraphics[width=15mm]{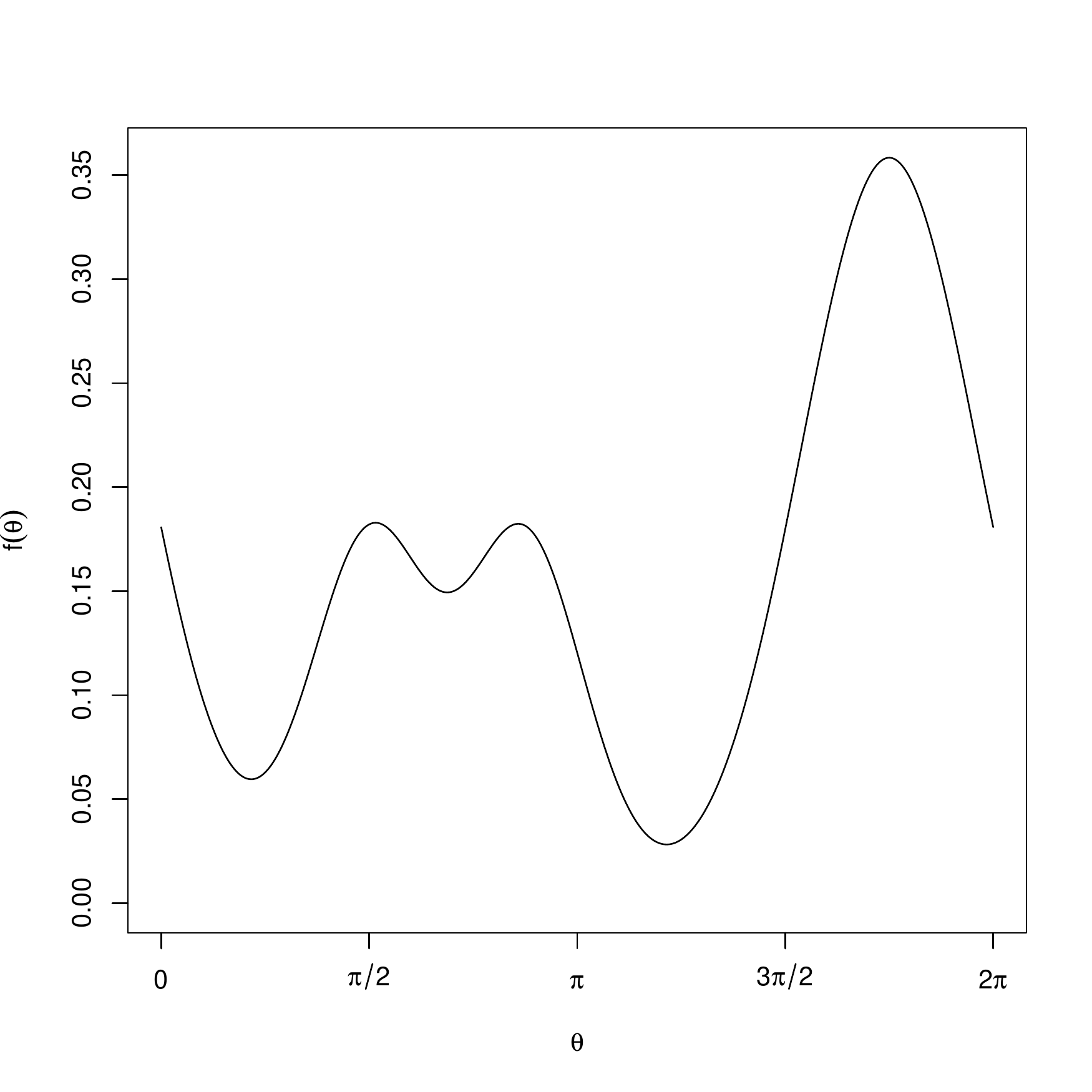}}& \multirow{4}{*}{\includegraphics[width=15mm]{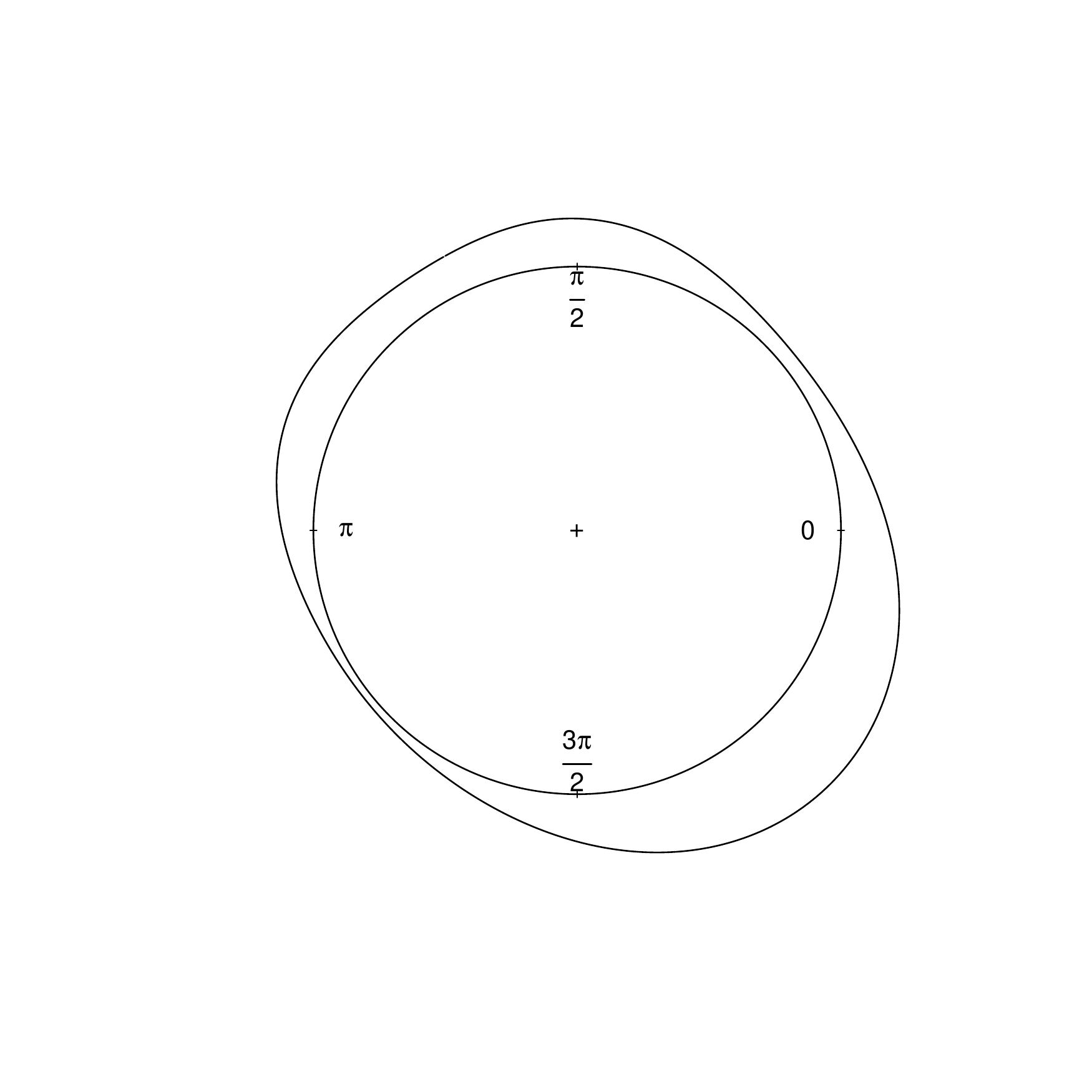}}& M23 &\multicolumn{3}{c|}{$U^2$ of Watson} &\multicolumn{3}{c|}{Excess mass}  \\ \cline{3-9}
  &  & $n=50$ & 0.032(0.015) & 0.102(0.027) & 0.164(0.032) & 0.008(0.008) & 0.054(0.020) & 0.102(0.027) \\ 
    &  & $n=100$ & 0.058(0.020) & 0.160(0.032) & 0.218(0.036) & 0.018(0.012) & 0.066(0.022) & 0.124(0.029) \\ 
    &  & $n=200$ & 0.060(0.021) & 0.188(0.034) & 0.262(0.039) & 0.036(0.016) & 0.088(0.025) & 0.178(0.034) \\ 
\hline
\multirow{4}{*}{\includegraphics[width=15mm]{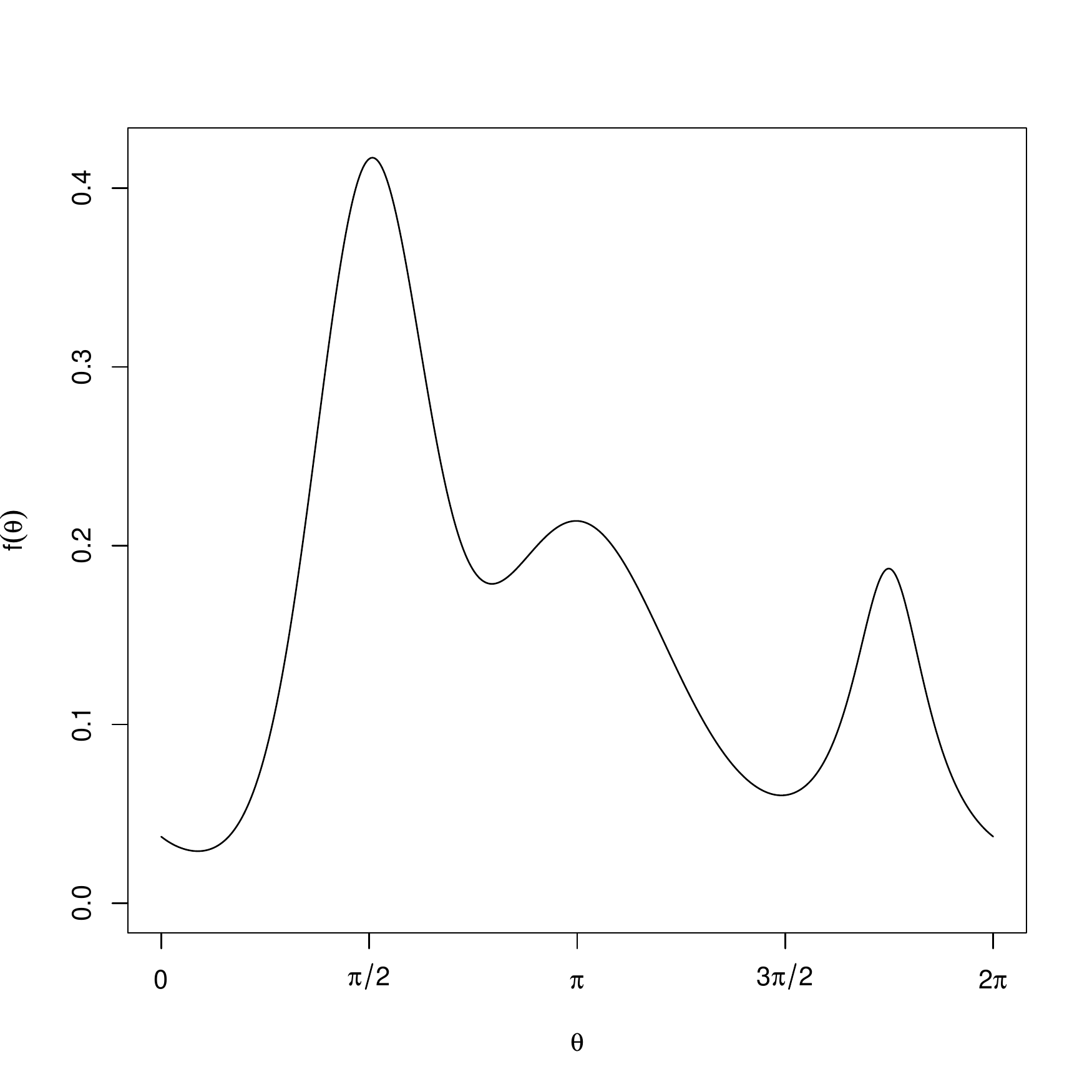}}& \multirow{4}{*}{\includegraphics[width=15mm]{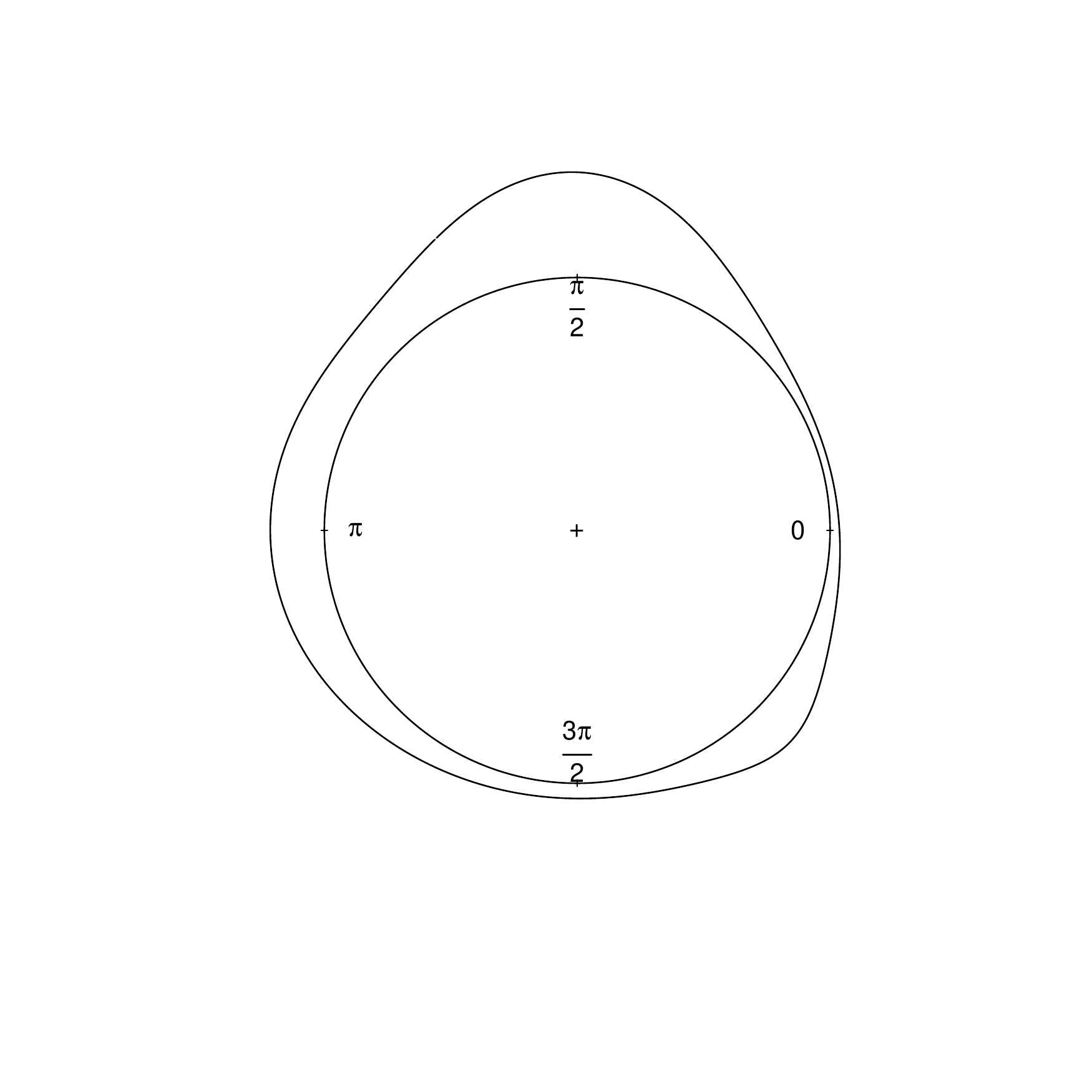}}& M24 &\multicolumn{3}{c|}{$U^2$ of Watson} &\multicolumn{3}{c|}{Excess mass}  \\ \cline{3-9}
	  &  & $n=50$ & 0.880(0.028) & 0.962(0.017) & 0.978(0.013) & 0.002(0.004) & 0.058(0.020) & 0.108(0.027) \\ 
    &  & $n=100$ & 1(0) & 1(0) & 1(0) & 0.022(0.013) & 0.078(0.024) & 0.154(0.032) \\ 
    &  & $n=200$ & 1(0) & 1(0) & 1(0) & 0.036(0.016) & 0.126(0.029) & 0.200(0.035) \\ 
\hline
\multirow{4}{*}{\includegraphics[width=15mm]{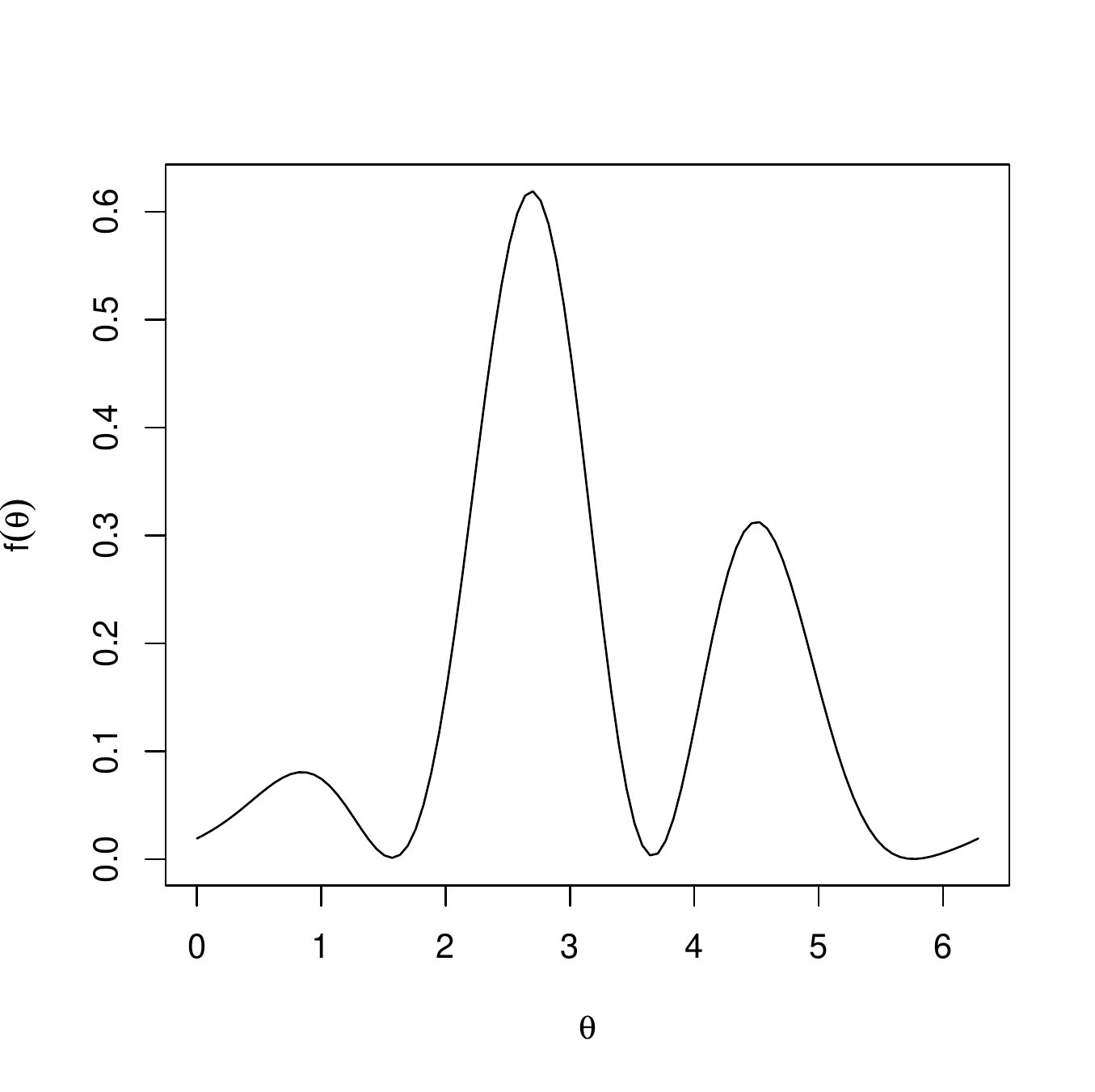}}& \multirow{4}{*}{\includegraphics[width=15mm]{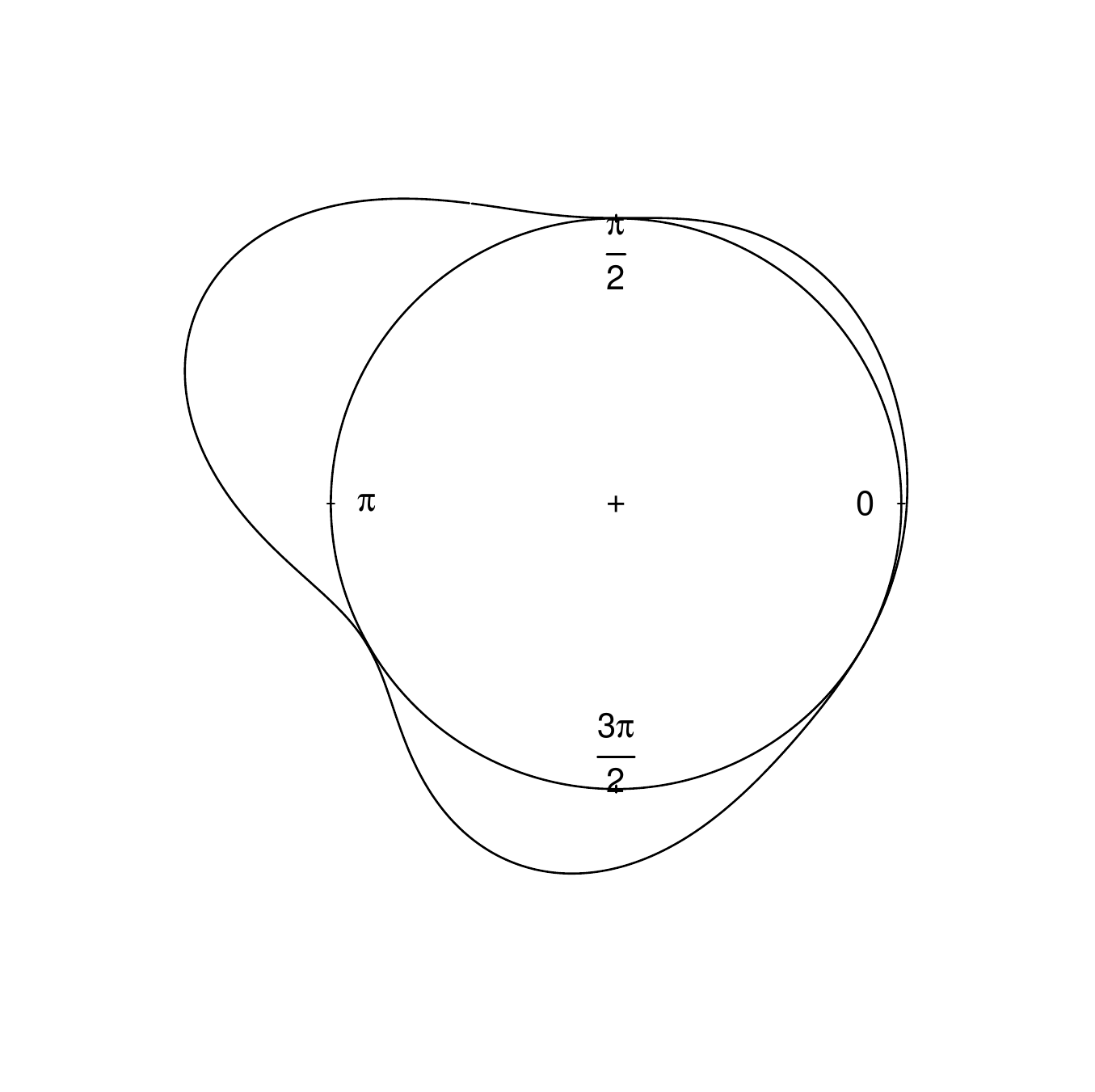}}& M25 &\multicolumn{3}{c|}{$U^2$ of Watson} &\multicolumn{3}{c|}{Excess mass}  \\ \cline{3-9}
&  & $n=50$ & 0.782(0.036) & 0.918(0.024) & 0.950(0.019) & 0.030(0.015) & 0.160(0.032) & 0.244(0.038) \\ 
   &  & $n=100$ & 0.980(0.012) & 1(0) & 1(0) & 0.154(0.032) & 0.340(0.042) & 0.448(0.044) \\ 
   &  & $n=200$ & 0.998(0.004) & 1(0) & 1(0) & 0.456(0.044) & 0.660(0.042) & 0.746(0.038) \\ 
\hline
\end{tabular}
}
\caption{Percentage of rejections for testing $H_0:j=2$ vs. $H_a:j>2$, with 500 simulations (1.96 times their estimated standard deviation in parenthesis) and $B=500$ bootstrap samples. For models under the null (a): M11--M20, and under the alternative hypothesis (b): M21--M25. First and second column: linear and circular representations.}
\label{estsimcirc2}
\end{table}

\section{Data analysis: detection of fire seasonality}\label{datafires}

As explained in Section~\ref{introfires}, determining the number of fire seasons has a special importance when the relationships between the climate and the land management are studied. Although a general $k$ is allowed in the testing problem, reflecting the possible appearance of more than one peak of climatological fires, it should be noted that only one season of fires, related with climatological reasons, is expected in the Russia--Kazakhstan border. Then, using the data described in Section~\ref{datadesc}, the goal is to assess if there are one or more fire seasons in the study area, which is divided in $0.5\degree$ cells. If just one cell is considered, this problem can be tackled employing the new procedure introduced in Section~\ref{method} for testing $H_0:j=1$, as the simulation study in Section~\ref{simstud} supports, in the finite--sample case, that the proposal presents a correct behavior in terms of calibration and power. However, since the goal is to analyze the number of fire seasons in the entire Russia--Kazakhstan border and this area is divided in a cell--grid, then, the proposed procedure can be applied systematically in each cell. As mentioned in the Introduction, a FDR procedure is required in order to control the incorrect rejections of the null hypothesis. For performing such correction, the spatial correlation between the test p--values computed at different cells must be considered. These two issues are solved in Section~\ref{sfdr} and the obtained results are provided in Section~\ref{datares}.

\subsection{Fire data}\label{datadesc}
The studied dataset contains the location and date of all active fires detected by the \textit{MODerate resolution Imaging Spectroradiometer} (MODIS), launched into Earth orbit by the \textit{National Aeronautics and Space Administration} (NASA) on board the Terra (\textit{EOS AM}) and the Aqua (\textit{EOS PM}) satellites, from 10 July 2002 to 9 July 2012. The MODIS algorithm \citep[see][for further details]{Giglioetal03} identifies pixels where one or more fires are actively burning at the time of satellite overpass, based on the contrasting responses of the middle-infrared and longwave infrared bands in areas containing hot targets. Cloud and water pixels are previously excluded from analysis using multiple numerical thresholds on visible and near--infrared reflectance, and thermal infrared temperature values. The size of the smallest flaming fire having at least a $50\%$ chance of being detected by the MODIS algorithm, under both ideal daytime and nighttime conditions is approximately 100 $\mbox{m}^2$. 

MODIS data are provided in a discretized scale, so in order to apply the testing procedure, it is necessary to recover the continuous underlying structure. For that purpose, denote by $(X_1,\ldots,X_n)$ the days of the year when the $n$ recorded fires occurred, with $X_i \in \{1,\ldots ,366\}$. The dataset used for the study of the number of fire seasons is the following

\begin{equation*}
\Theta_i=2\pi(X_i+\mathcal{E}_i)/366; \mbox{ with } i=1,\ldots n,
\end{equation*}
being $\mathcal{E}_i$ generated from the uniform distribution $U(-1,0)$.  This means that it is assumed that fires occurred at any time of the day. Provided that data are not repeated, as this issue can considerably alter the test statistic, other ways of modifying the data can be considered, but, in general, this perturbation does not show relevant impacts in the results. 

\begin{figure}
\centering
    \includegraphics[width=1\textwidth]{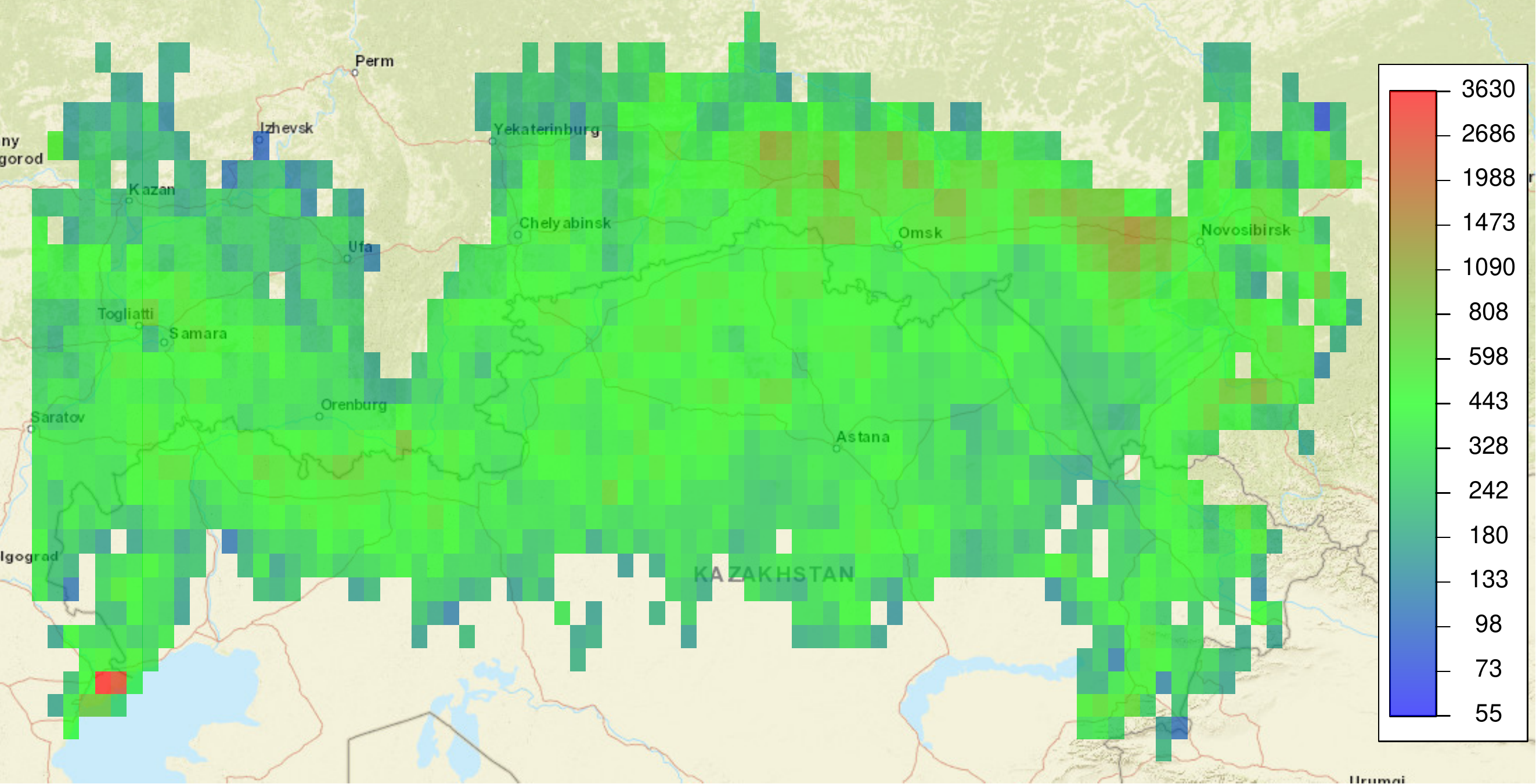}\\
 \caption{Number of fires detected by the MODIS in the different cells of the studied region from from 10 July 2002 to 9 July 2012 (in translucent color, as reflected in the right legend). Without superimposed translucent color, cells with low fire incidence or outside the studied area.}
 \label{fignfires}
\end{figure}

Once this modification is done, the analyzed area is divided in grid cells of size $0.5\degree$. Then, from the resulting cells, the ones with low fire incidence, i.e., cells having fewer than ten fires in more than seven out of ten years, will not be considered in the study. This leaves 1500 grid cells in the area, each one having between 55 and 3630 fires. A map including the studied area and a summary of total fire counts in the study period is shown in Figure \ref{fignfires}.

\subsection{Spatial False Discovered Rate}\label{sfdr}
The fires sample $\{\Theta_i;i=1,\ldots, 919654\}$ is organized in $1500$ cells inside the studied area. Hence, the proposed testing procedure (see Section~\ref{method}) will be applied systematically over these groups. It is clear that a FDR correction must be applied. In addition, the provided separation in grid cells is not necessarily designed for producing ``independent'' areas in the sense that occurrence of fires in a cell may not be independent of occurrence in the neighbouring ones. So, the spatial dependence must be taken into account for correcting the FDR procedure. This will be done following the ideas by \citet[Procedure 3]{BenjaminiHeller07}, with some modifications. Their method allows to control the FDR accounting for the spatial dependence of the data, employing prior information about the aggregation of different locations (where the size and shape of the different groups do not need to be equal). This is the case in our study, as it is expected that the behaviour of the fires will be similar in regions where the land is used for the same purpose. Once the aggregation is done, \cite{BenjaminiHeller07} propose testing first on those large units (\textit{patch}), allowing to have a single p--value for the large area. Then, if the null hypothesis is rejected, it controls the dependence of the aggregate and the cells p--values, in order to correct them in the multiple testing problem and properly detect the rejected locations.

The method employed here can be outlined as follows: in an initial step (Step 1), the testing procedure (introduced in Section~\ref{method}) is applied, locally, to each cell, obtaining the corresponding p--value. Secondly (Step 2), \textit{fire patches} by land cover are defined in order to create groups of cells that are related. Finally (Step 3), a hierarchical testing procedure \citep[similar to that one of][]{BenjaminiHeller07} is applied. This final step consists in, first, deciding in which of the previous patches the null hypothesis is rejected (patch testing; Step 3a) and, secondly, within each rejected patch, determining in which of its cells $H_0$ is rejected (trimming procedure; Step 3b). Further details in the specific problem and solution are provided below.

\textbf{Step 1. Local application of the test.} In each of the 1500 grid cells, the method proposed in Section~\ref{method} will be used (with $B=5000$ bootstrap replicates) for obtaining the corresponding p--values when it is tested if there is one season of fires or more.   

\begin{figure}
\centering
    \includegraphics[width=1\textwidth]{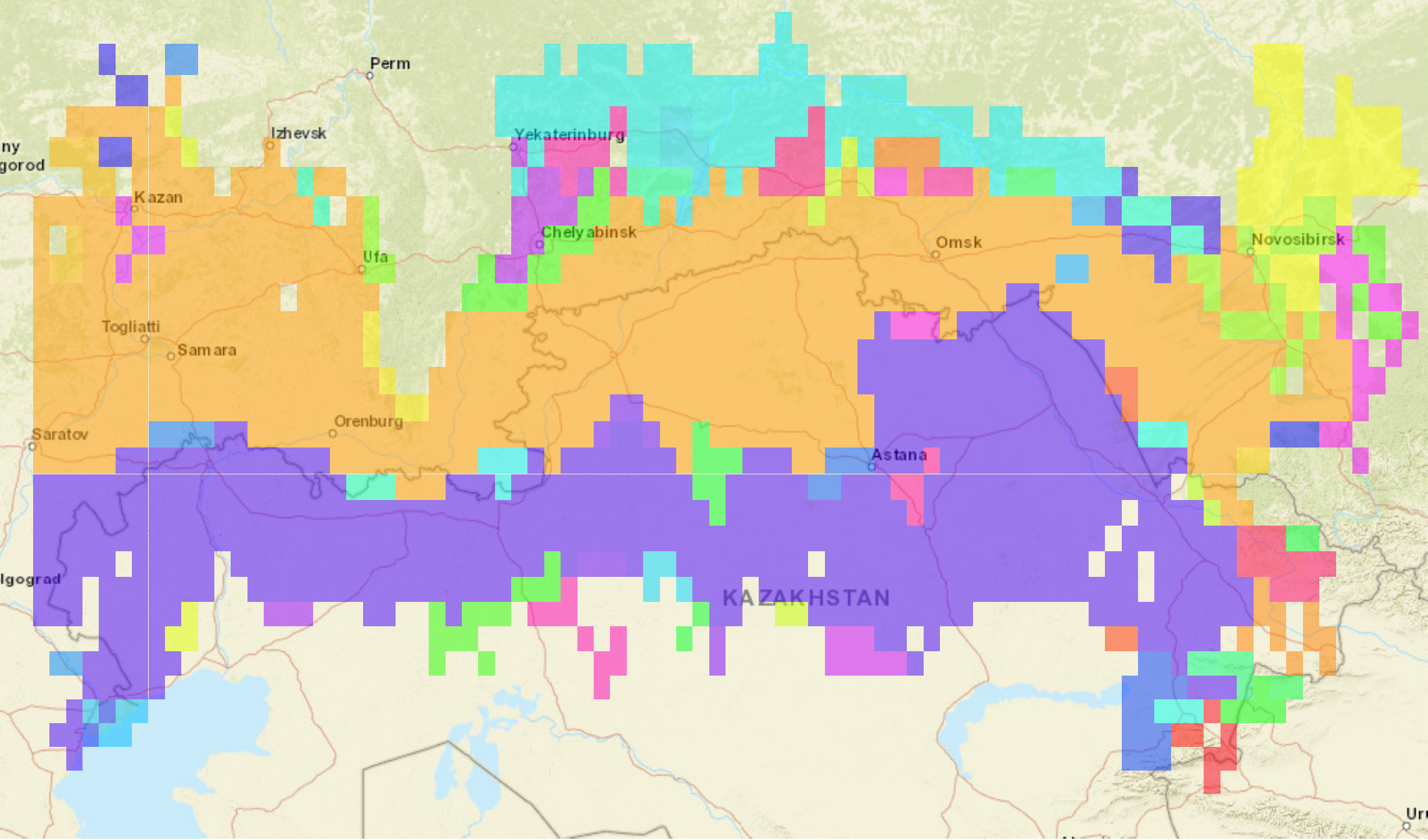}
 \caption{Each (translucent) color represents one homogeneous land cover patch in the studied area. Without superimposed translucent color, cells with low fire incidence or outside the studied area.}
 \label{figant}
\end{figure}

\textbf{Step 2. Fire patches.} To define the fire patches, a rule for determining the cells expected to display ``similar'' fire season modality patterns must be established. The fire patches were constructed using the information of the land cover data provided by the European Space Agency Climate Change Initiative project (\textit{Land Cover version 1.6.1}, data from 2008 to 2012, available at \url{http://www.esa-landcover-cci.org}), which describes the physical material at the surface of the earth, including various types of vegetation, bare rock and soil, water, snow and ice, and artificial surfaces. The construction of the fire patches is detailed in Appendix~\ref{firepatch} and the different patches in the studied area are represented in Figure \ref{figant}.

\textbf{Step 3. Hierarchical testing procedure.} Some notation is required for this part. Let $\mathfrak{j}=1,\ldots, J$ be the different patches created in Step 2, $\mathfrak{l}=1,\ldots,L_\mathfrak{j}$ be the cells within the patch $\mathfrak{j}$ and $\tilde{p}_{\mathfrak{l}\mathfrak{j}}$ the p--value obtained in Step 1 for the cell $\mathfrak{l}$ within the patch $\mathfrak{j}$. Then, $z_{\mathfrak{l}\mathfrak{j}}=\Phi^{-1} (1-\tilde{p}_{\mathfrak{l}\mathfrak{j}})$ will be the corresponding z--score for the cell $\mathfrak{l}$ in patch $\mathfrak{j}$, where $\Phi$ is the cumulative distribution of a standard normal distribution. Note that since a bootstrap procedure is used to approximate the p--values $\tilde{p}_{\mathfrak{l}\mathfrak{j}}$, they can be equal to 0 or 1 and, in that case, the z--score is non--finite. In order to correct that, if $\tilde{p}_{\mathfrak{l}\mathfrak{j}}$ is equal to 0, following the ideas of \cite{Jeffreys46} prior, if $B$ is the number of bootstrap resamples, this p--value is replaced by a random value from the distribution Beta$(1/2,B+1/2)$ and if it is equal to one then a random value from Beta$(B+1/2,1/2)$ is taken. Once the z--scores for each cell in the different patches are calculated, a hierarchical method will be employed. The testing procedure is divided in two steps: first it tests, at significance level $\alpha_c$, in which fire patches the null hypothesis can be rejected (Step 3a) and then tests $H_0$, at level $\alpha_r$, in the cells within the rejected patches (Step 3b).

\textbf{Step 3a. Patch testing.} In this stage, fire patches where the null hypothesis is rejected are identified. This step consists in computing a global p--value for each patch and then, since each patch has a different number of cells, the weighted FDR procedure of \cite{BenjaminiHochberg97}, at level $\alpha_c$, will be applied in order to correct for multiple testing. The global p--value of the patch $\mathfrak{j}$ is calculated as $\breve{p}_\mathfrak{j}=\tilde{\Phi}(\bar{Z}_\mathfrak{j}/\hat{\sigma}_{\bar{Z}_\mathfrak{j}})$, that is, the right tail probability of the standard normal distribution calculated in the standardized z--score average of the cells in a fire patch. More precisely:
\begin{enumerate}
\item In each patch $\mathfrak{j}$, with $\mathfrak{j}\in \{1,\ldots,J\}$, calculate the z--score average: $\bar{Z}_\mathfrak{j}=(1/L_\mathfrak{j})\sum_{\mathfrak{l}=1}^{L_\mathfrak{j}} z_{\mathfrak{l}\mathfrak{j}}$, where $z_{\mathfrak{l}\mathfrak{j}}$ are the z--scores defined above.
\item Compute its standard error: $\hat{\sigma}_{\bar{Z}_\mathfrak{j}}=(\hat{\sigma}_\mathfrak{j}/L_\mathfrak{j})\sqrt{L_\mathfrak{j}+2\sum_{\mathfrak{l}=1}^{L_\mathfrak{j}}\sum_{\mathfrak{m}=1}^{\mathfrak{l}-1} \hat{\rho}^\mathfrak{j}_{\mathfrak{l},\mathfrak{m}}}$, where $\hat{\rho}^\mathfrak{j}_{\mathfrak{l},\mathfrak{m}}=1-\hat{\gamma}(s_{\mathfrak{l}\mathfrak{j}}-s_{\mathfrak{m}\mathfrak{j}})/\hat{\sigma}_\mathfrak{j}$ is the estimated correlation between cells $\mathfrak{l}$ and $\mathfrak{m}$ within the patch $\mathfrak{j}$, $\hat{\gamma}(s_{\mathfrak{l}\mathfrak{j}}-s_{\mathfrak{m}\mathfrak{j}})$ is an estimation of the semivariogram evaluated at the distance (of the centroids in the map) between cells $\mathfrak{l}$ and $\mathfrak{m}$ and $\hat{\sigma}_\mathfrak{j}$ the estimated variance of the cells in patch $\mathfrak{j}$. Differently from \cite{BenjaminiHeller07}, the semivariogram estimator is obtained by (weighted) least squares on an exponential family, in order to ensure that such an estimator is indeed a valid semivariogram (something that may be not satisfied by nonparametric estimators) and to control the parameters of interest. Specifically, in an exponential family, two parameters drive the behaviour of the spatial covariance: the point variance and the range. Hence, the estimated variance is obtained from this parametric procedure. It should be noted that least squares procedures for variogram estimation require the use of a nonpametric pilot estimator. In this case, the robust version of the empirical variogram was used \citep[see][Ch.2]{Cressie93}.
\item The weighted FDR procedure, at level $\alpha_c$, is applied on the p--values $\breve{p}_\mathfrak{j}$, being the weight proportional to the patches size. Given the ordered p--values $\breve{p}_{(1)}\leq \ldots \leq \breve{p}_{(J)}$, unimodality is rejected in the $\mathfrak{k}$ patches with the smallest p--values, being $\mathfrak{k}=\max \{\upsilon: \breve{p}_{(\upsilon)} \leq (\sum_{\mathfrak{j}=1}^{\upsilon} L_{(\mathfrak{j})}/\sum_{\mathfrak{j}=1}^{J} L_{(\mathfrak{j})})\alpha_c\}$ and $L_{(\mathfrak{j})}$ the number of cells in the fire patch associated with $\breve{p}_{(\mathfrak{j})}$.
\end{enumerate}

\textbf{Step 3b. Trimming procedure.} Once a decision about which patches are candidates for rejecting the null hypothesis is made (and hence exhibiting a multimodal fire pattern), specific cells where this rejection holds are identified. It should be noted that the cell test statistic is correlated with the test statistic at the patch level. This means that a FDR correction cannot be directly applied over all the cells belonging to the same patch and a correction is proposed by \cite{BenjaminiHochberg97}. First, calculate the conditional p--value of a cell within a patch that was rejected, $\hat{p}_{\mathfrak{l}\mathfrak{j}}$. Then, over these p--values, apply the two--stage procedure introduced by \cite{Benjaminietal06}, at level $\alpha_r$, to enhance the power. This last method, in its first stage, consists in estimating the sum of weights of null cells, using for that purpose the classical FDR procedure, at level $\alpha_r$, and then using this quantity, in a second stage, to determine the number of rejected cells within the patch. To be more precise and summarize Step 3b, the following steps are detailed:

\begin{enumerate}
\setcounter{enumi}{3}
\item Calculate the conditional p--value of each cell $\mathfrak{l}$ within the patch that was rejected $\mathfrak{j}$, denoted as $\hat{p}_{\mathfrak{l}\mathfrak{j}}$. These p--values can be estimated as follows:
\begin{align*}
\hat{p}_{\mathfrak{l}\mathfrak{j}}=&\bigintss_{z_{\mathfrak{l}\mathfrak{j}}}^{\infty}{\left(\frac{\hat{J_0}}{J}\tilde{\Phi}\left( \frac{\tilde{\Phi}^{-1}(u_1) - \hat{\rho}_{\mathfrak{l}\mathfrak{j}} u}{ \sqrt{1-{\hat{\rho}_{\mathfrak{l}\mathfrak{j}}}^2}} \right) + \right.}\\
& \left. +\left(1-\frac{\hat{J_0}}{J}\right) \tilde{\Phi}\left( \frac{\tilde{\Phi}^{-1}(u_1) - \hat{\rho}_{\mathfrak{l}\mathfrak{j}} u-\hat{\mu}_\mathfrak{j}}{ \sqrt{1-{\hat{\rho}_{\mathfrak{l}\mathfrak{j}}}^2}}  \right) \right) \phi(u) du \times\\ 
& \times \left( \frac{\hat{J_0}}{J}u_1 +\left(1- \frac{\hat{J_0}}{J}\right) \tilde{\Phi}\left(\tilde{\Phi}^{-1}\left(u_1\right)-\hat{\mu}_\mathfrak{j}  \right) \right)^{-1},\\
\end{align*}
being $\phi$ a standard normal density and noting that
\begin{enumerate}
\item $u_1=(\sum_{\mathfrak{j}=1}^{\mathfrak{k}} L_{(\mathfrak{j})}/\sum_{\mathfrak{j}=1}^{J} L_{(\mathfrak{j})})\alpha_c$ is the cutoff point of the largest p--value rejected in Step 3a.
\item $\hat{J_0}=(J-\mathfrak{k})/(1-\alpha_c)$ is the estimated sum of weights of null patches.
\item $\hat{\mu}_\mathfrak{j}=((\sum_{\mathfrak{j}=1}^{J}\sum_{\mathfrak{l}=1}^{L_\mathfrak{j}} z_{\mathfrak{l}\mathfrak{j}})/(\sum_{\mathfrak{j}=1}^{J} L_\mathfrak{j}))/\hat{\sigma}_{\bar{Z}_\mathfrak{j}}$ is the estimation of the standardized expectation of the patch test statistic under the alternative.
\item $\hat{\rho}_{\mathfrak{l}\mathfrak{j}}=(1+ \sum_{\mathfrak{m}=1,\mathfrak{m}\neq \mathfrak{l}}^{L_\mathfrak{j}} \hat{\rho}_{\mathfrak{l},\mathfrak{m}}^\mathfrak{j})\hat{\sigma}_\mathfrak{j}/(L_\mathfrak{j} \hat{\sigma}_{\bar{Z}_\mathfrak{j}})$ is the estimated correlation between the z-score in a given cell and the average z-score of the patch.
\end{enumerate}
\item Given these $L_\mathfrak{j}$ p--values in the patch $\mathfrak{j}$, apply a two--stage procedure at level $\alpha_r$:
\begin{enumerate}
\item Apply the classic FDR procedure, at level $\alpha_r'=\alpha_r/(1+\alpha_r)$. Given the ordered p--values $\hat{p}_{(1) \mathfrak{j}}\leq \ldots \leq \hat{p}_{(L_\mathfrak{j}) \mathfrak{j}}$, let $\mathfrak{k}_{1\mathfrak{j}}=\max \{\upsilon: \hat{p}_{(\upsilon)\mathfrak{j}} \leq (\upsilon/L_\mathfrak{j})\alpha_r'\}$.
\item Apply again the classic FDR procedure at level $\alpha_r'$, being in this case the sum of weights of null cells: $\hat{J}_{0\mathfrak{j}}=L_\mathfrak{j}-\mathfrak{k}_{1\mathfrak{j}}$. Reject the unimodality in the $\mathfrak{k}_{2\mathfrak{j}}$ cells with the smallest p--values, being $\mathfrak{k}_{2\mathfrak{j}}=\max \{\upsilon: \hat{p}_{(\upsilon)\mathfrak{j}} \leq (\upsilon/\hat{J}_{0\mathfrak{j}})\alpha_r'\}$.
\end{enumerate}
\end{enumerate}

\subsection{Results}\label{datares}
\begin{figure}
\centering
    \includegraphics[width=1\textwidth]{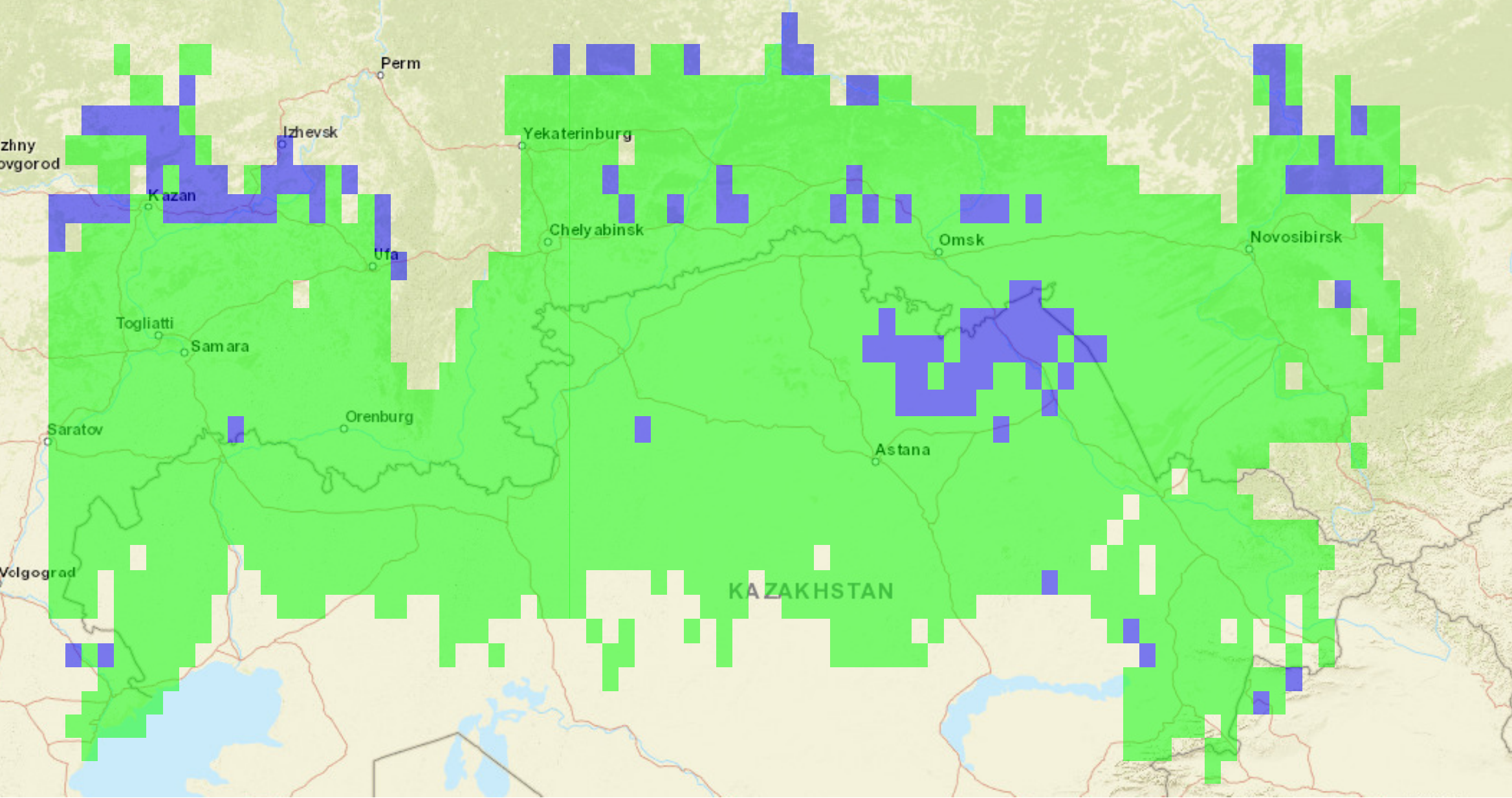}
 \caption{Results after applying the procedure described in Section~\ref{sfdr}, with $\alpha_c=0.01$ and $\alpha_r=0.01$, in the studied area divided in grids of size $0.5\degree$. In green: cells where $H_0$ is rejected (multimodal). In blue: cells where there is no evidence to reject $H_0$ (unimodal). Without superimposed translucent color, cells with low fire incidence or outside the studied area.}
 \label{worldresults}
\end{figure}

In what follows, the application of the new testing proposal, jointly with the FDR correction, to the wildfire dataset is presented. As a first step, the p--values, applying the new procedure provided in Section~\ref{method} (with $B=5000$ bootstrap replicates), were computed in all the cells of the study area. In a second step, the different fires patches were created using the land cover database. Finally, the hierarchical testing procedure was applied. First, to determine in which of the previously created patches the null hypothesis is rejected at significance level $\alpha_c=0.01$. Second, within the rejected patches, it was determined which cells can be definitely rejected at the trimming significance level $\alpha_r=0.01$. The rejected cells are shown in green color in Figure~\ref{worldresults}.

In view of the modality map (Figure~\ref{worldresults}), it can be concluded that a multimodal pattern prevails, these results support previous findings \citep[see, e.g.,][]{Magi12} in the use of fire as a land management tool in Russia--Kazakhstan area. This general multimodal fire regime is a consequence of the fact that human activity, in this area, is produced under meteorological conditions that are marginally suitable for fire, months prior to the start of the hotter and drier part of the year. Recognition that multimodal fire regimes are more widespread than previously thought, and typically associated with anthropogenic burning, is important for various reasons. It will contribute to improve the parametrization of dynamic global vegetation models used to predict environmental impacts of changes in land use and climate, to refine estimates of the seasonality of greenhouse gas emissions from vegetation fires, and to support fire management activities that aim to reduce the exposure of human populations and the losses of valuable resources and assets.

\section{Conclusions}\label{conclusions}

A new and effective nonparametric method for testing circular multimodality is presented, with the objective of assessing the number of fire seasons and their mismatch with the expected ones, driven by climatological conditions. The reported results support previous results of the anthropogenic character of fire seasonality in the Russia--Kazakhstan area.

For a better understanding of vegetation fires in the Earth system, future research may include the use of multimodality test as a preliminary tool in different regions of the world for exploring when the peaks of fires are produced and their associated mass. This would allow to review different works in the environmental science literature with nonparametric techniques. For instance, one could determine when the principal peaks of fires are produced in each $0.5\degree$ cell \citep{LePageetal10}, the delay of the agricultural fires with respect to the climatological ones \citep{Magi12} or the mass associated to each peak for better understanding the importance of the different human activities \citep{Korontzi06}. 

Related with the proposed multimodality test, due the growing interest in the last few years in more flexible models in circular data \citep[see][Ch. 2]{Ley17}, the new proposal can be used as a preliminary tool for determining if a multimodal model is needed. Also, this test could be employed for determining the minimum number of components in a mixture of parametric unimodal distributions when the objective is modeling the wildfires data. 

When the FDR, accounting for the spatial dependence of the data, needs to be considered, the presented method provides a useful algorithm for any context where prior information about the neighboring locations is known.

\newpage

\appendix

\section{Simulated models}\label{app_sim_models}

The specific formulas of the models considered in the simulation studies carried out in Section~\ref{simstud} are given here. The notation in the mixture models is $\sum_{i=1}^m p_i \cdot \psi_i$, where each $\psi_i$ represents a circular density of the mixture component and $p_i$ are the weights of these different components, $0\leq p_i \leq 1$, with $i=1,\ldots, m$; satisfying $\sum_{i=1}^m p_i =1$. 

\begin{figure}
\begin{multicols}{4}
\centering
\includegraphics[width=1\linewidth]{mc1}\\
M1
\includegraphics[width=1\linewidth]{mc5}\\
M5
\includegraphics[width=1\linewidth]{mca1}\\
M9
\includegraphics[width=1\linewidth]{mc2}\\
M2
\includegraphics[width=1\linewidth]{mc6}\\
M6
\includegraphics[width=1\linewidth]{mca2}\\
M10
\includegraphics[width=1\linewidth]{mc3}\\
M3
\includegraphics[width=1\linewidth]{mc7}\\
M7
\includegraphics[width=1\linewidth]{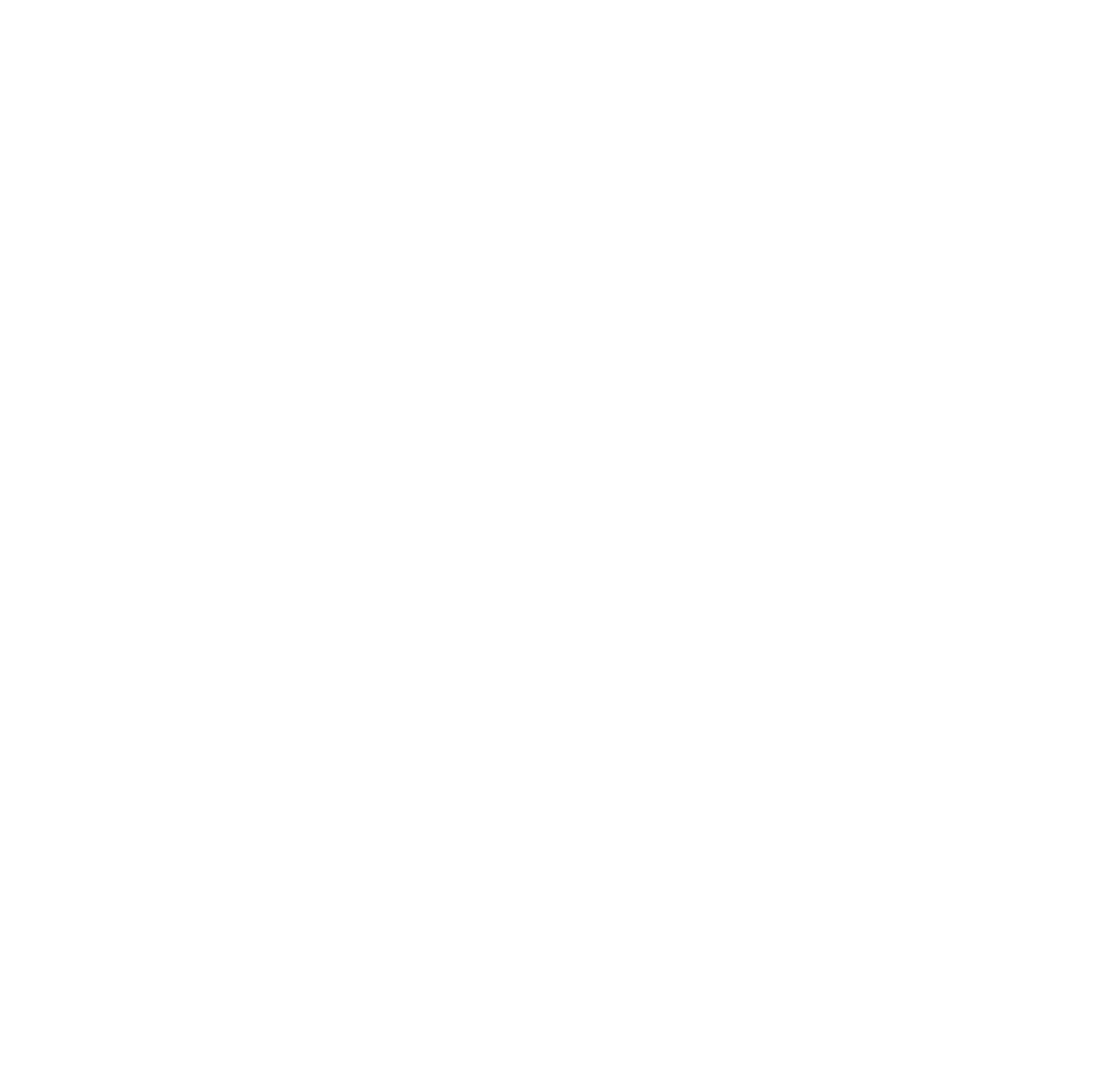}\\
\phantom{M27}
\includegraphics[width=1\linewidth]{mc4}\\
M4
\includegraphics[width=1\linewidth]{mc8}\\
M8
\includegraphics[width=1\linewidth]{white_image}\\
\phantom{M27}
\end{multicols}
\caption{Linear representation of the unimodal circular density functions: M1--M10.}
\label{figs3}
\end{figure}

\begin{figure}
\begin{multicols}{4}
\centering
\includegraphics[width=1\linewidth]{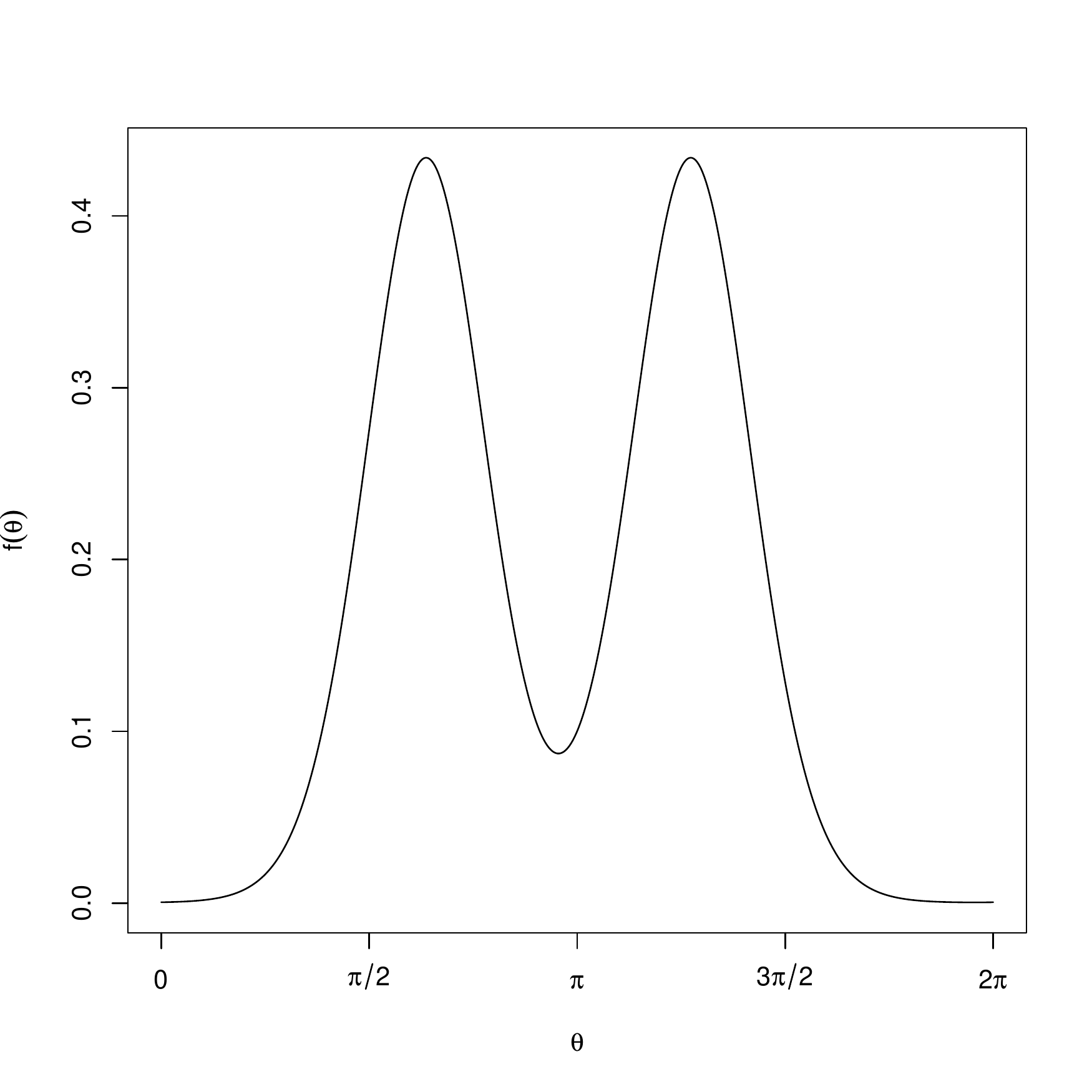}\\
M11
\includegraphics[width=1\linewidth]{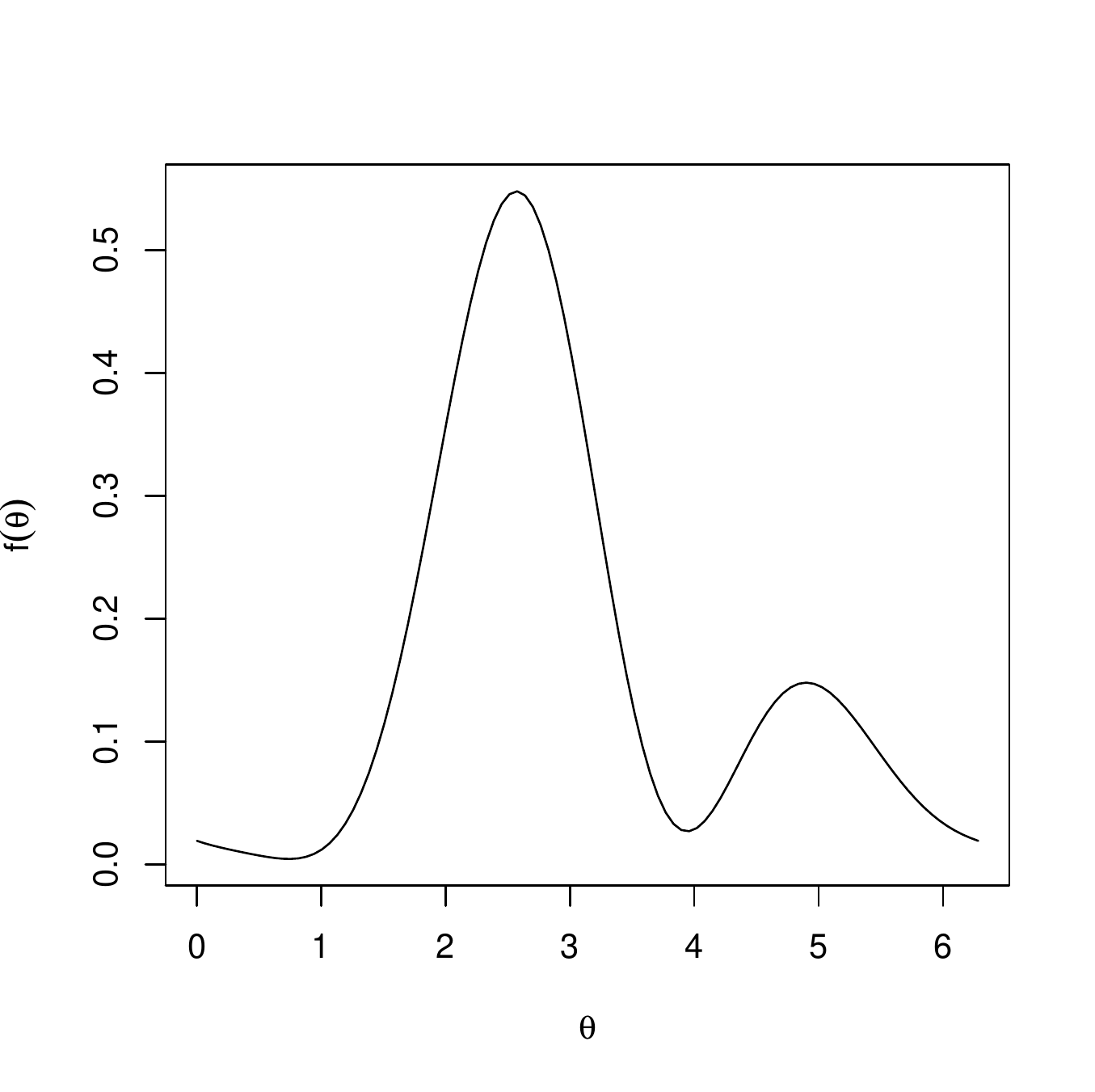}\\
M15
\includegraphics[width=1\linewidth]{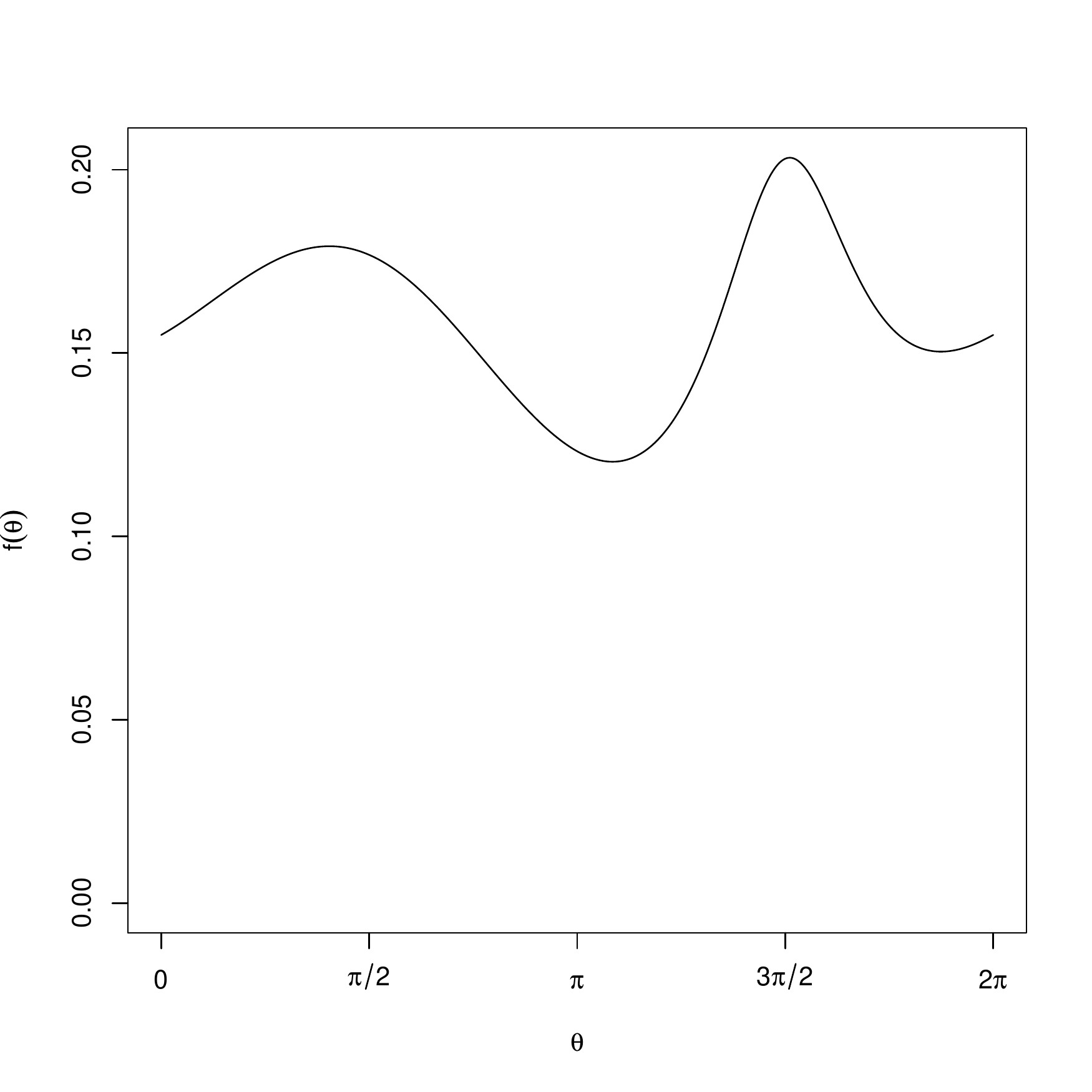}\\
M19
\includegraphics[width=1\linewidth]{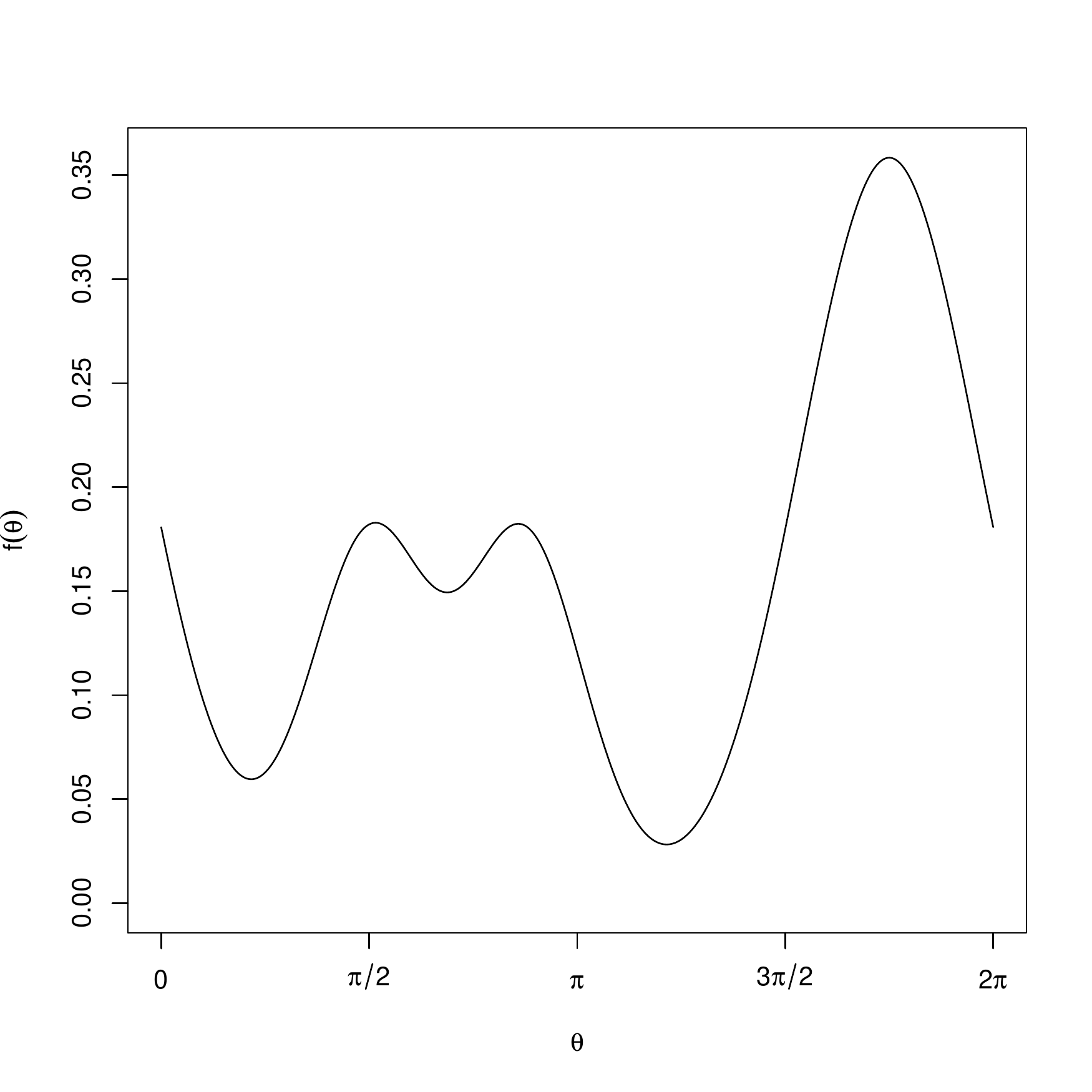}\\
M23
\includegraphics[width=1\linewidth]{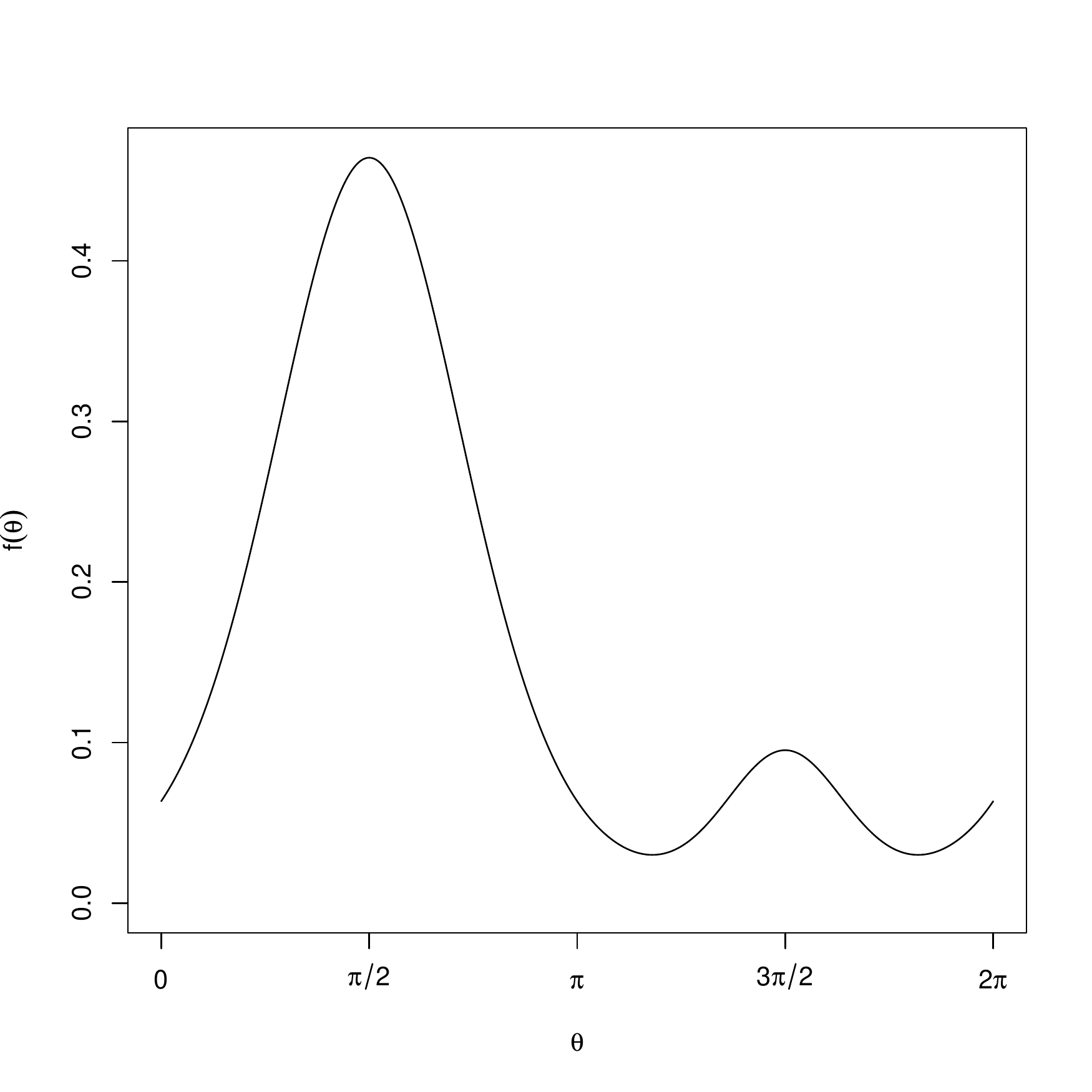}\\
M12
\includegraphics[width=1\linewidth]{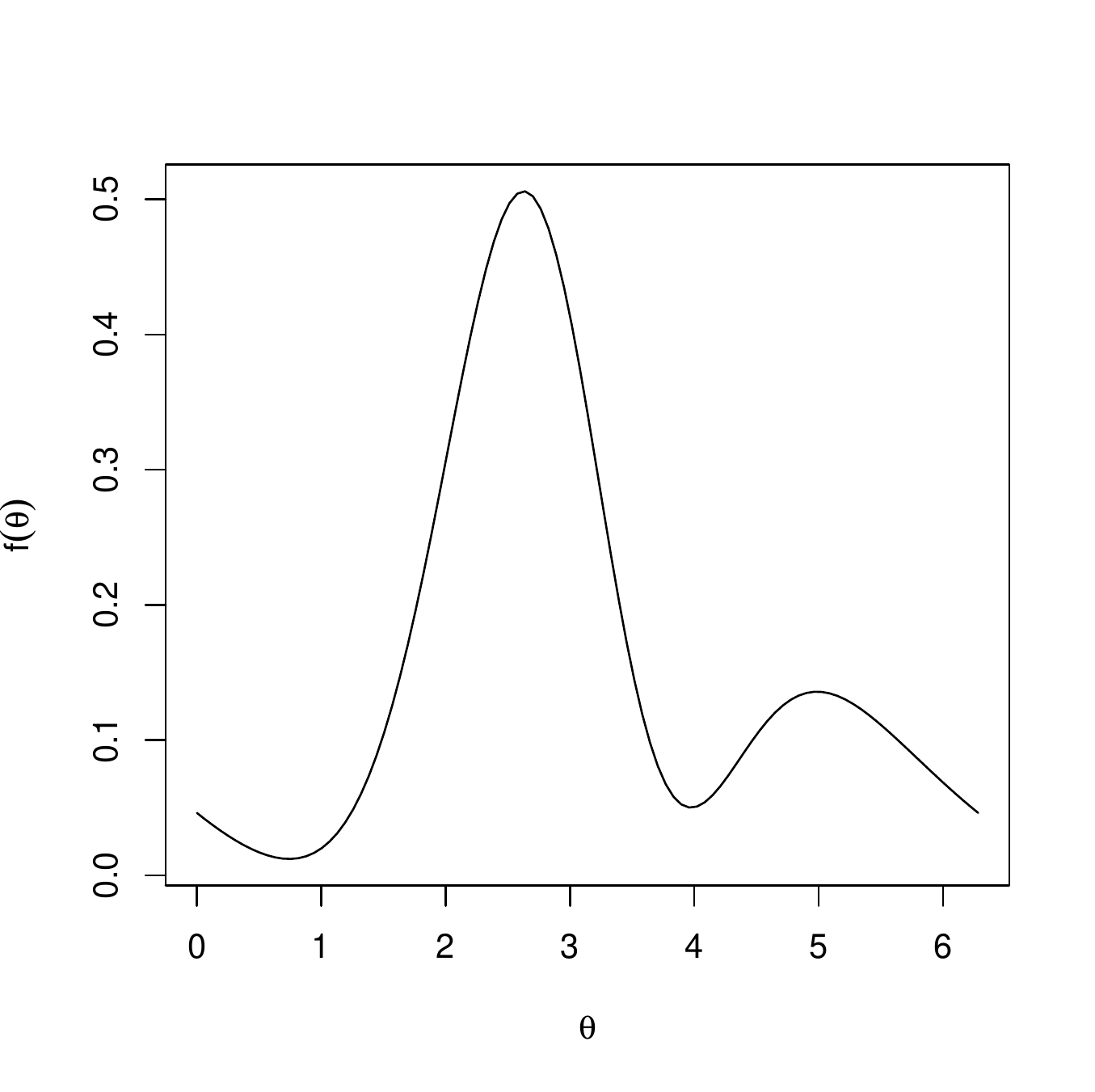}\\
M16
\includegraphics[width=1\linewidth]{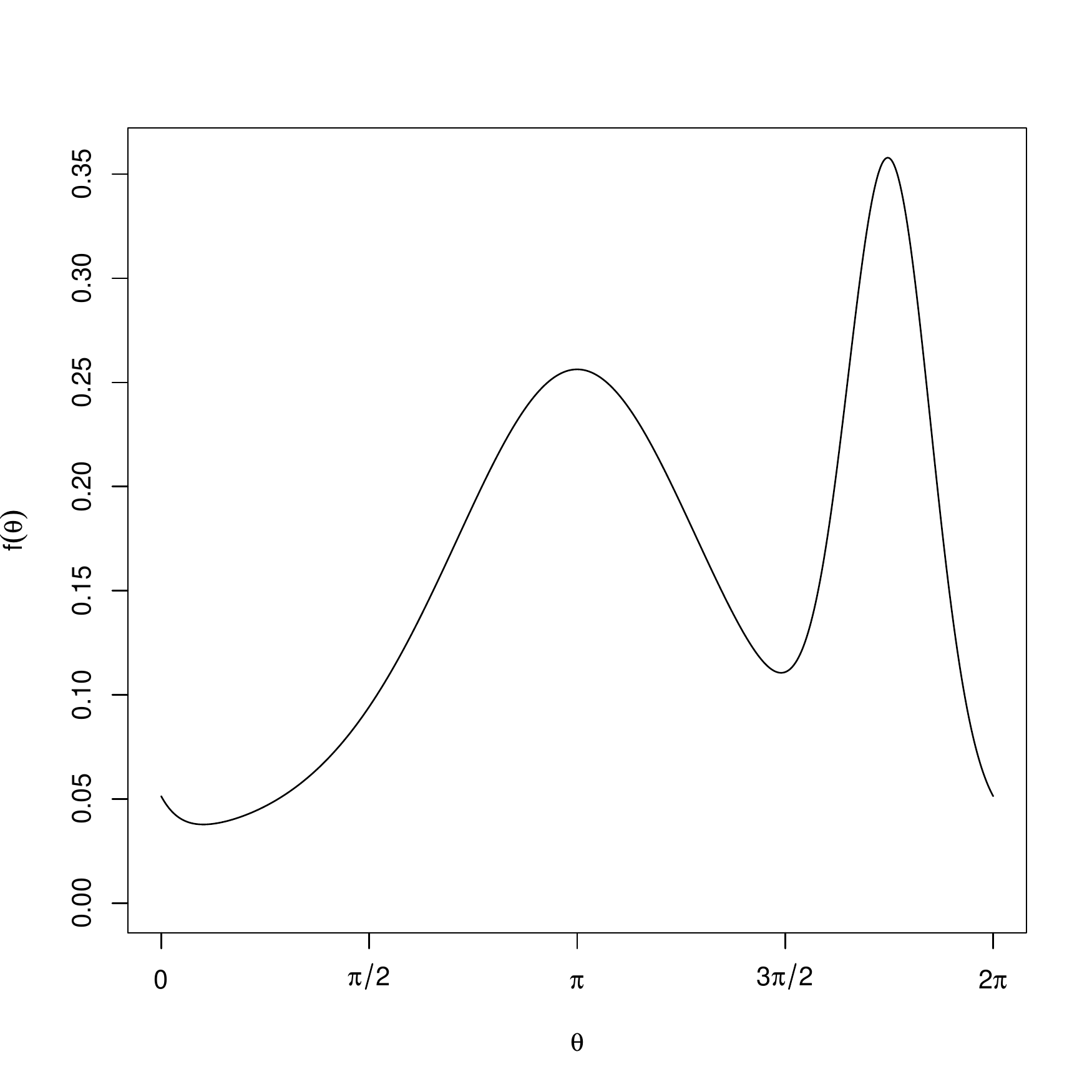}\\
M20
\includegraphics[width=1\linewidth]{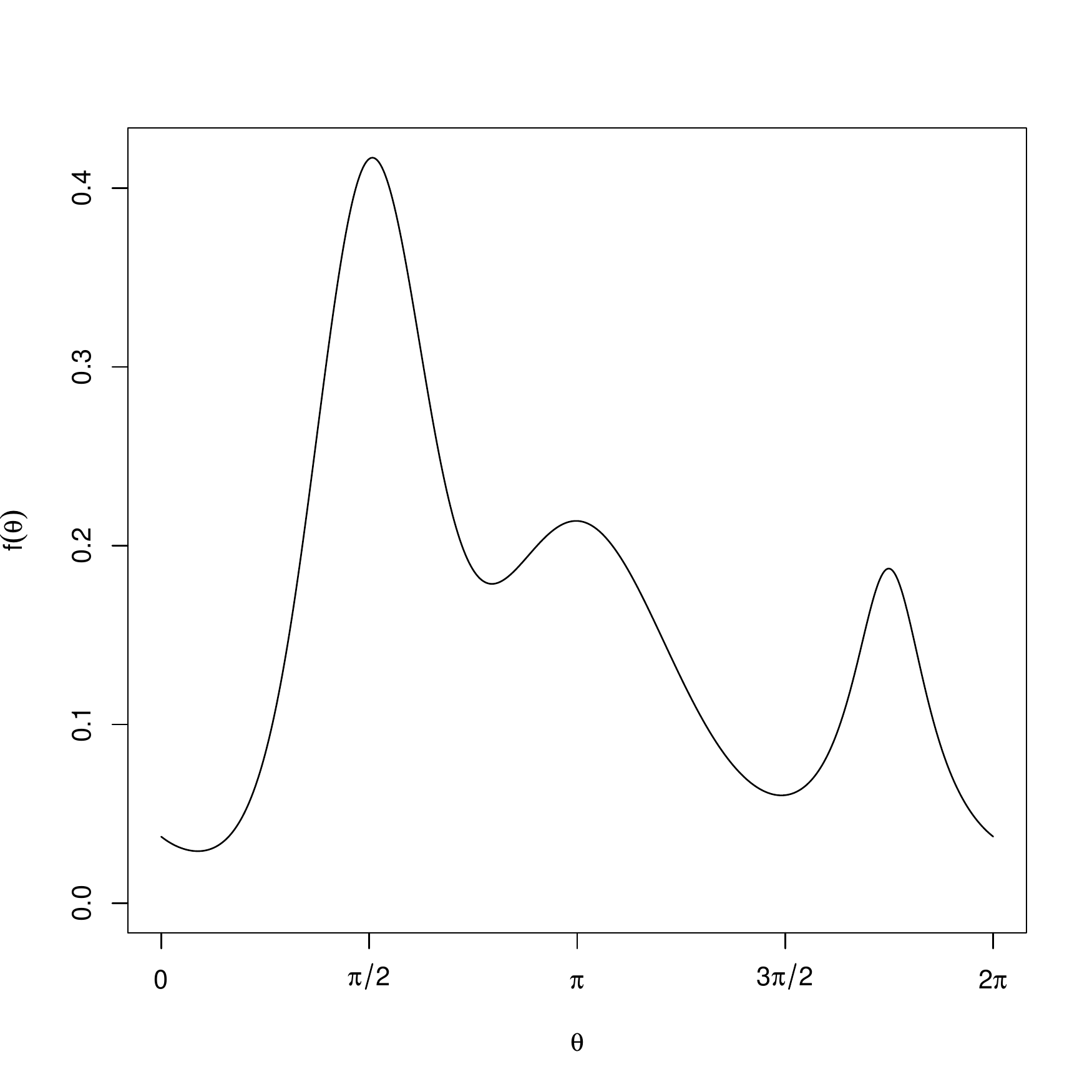}\\
M24
\includegraphics[width=1\linewidth]{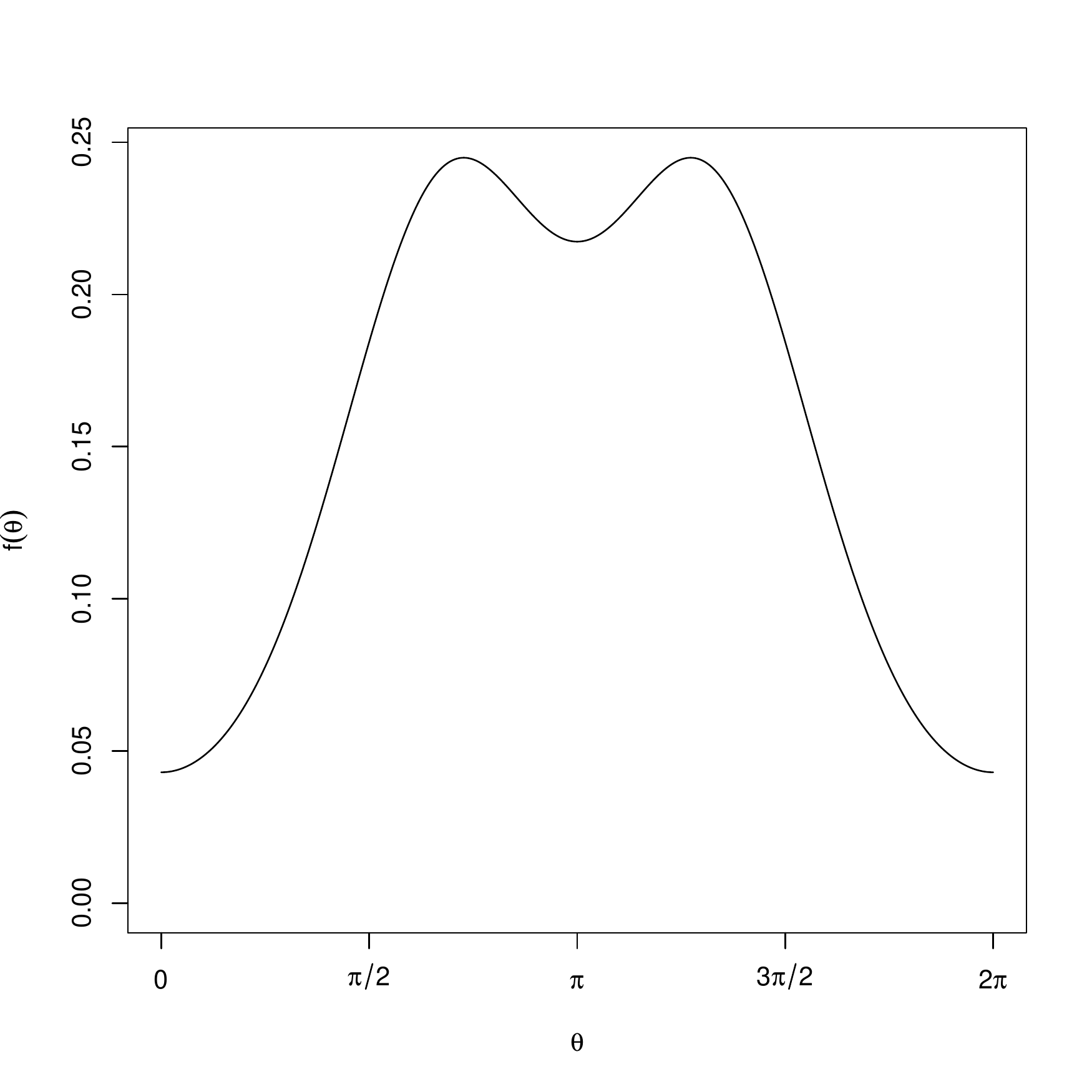}\\
M13
\includegraphics[width=1\linewidth]{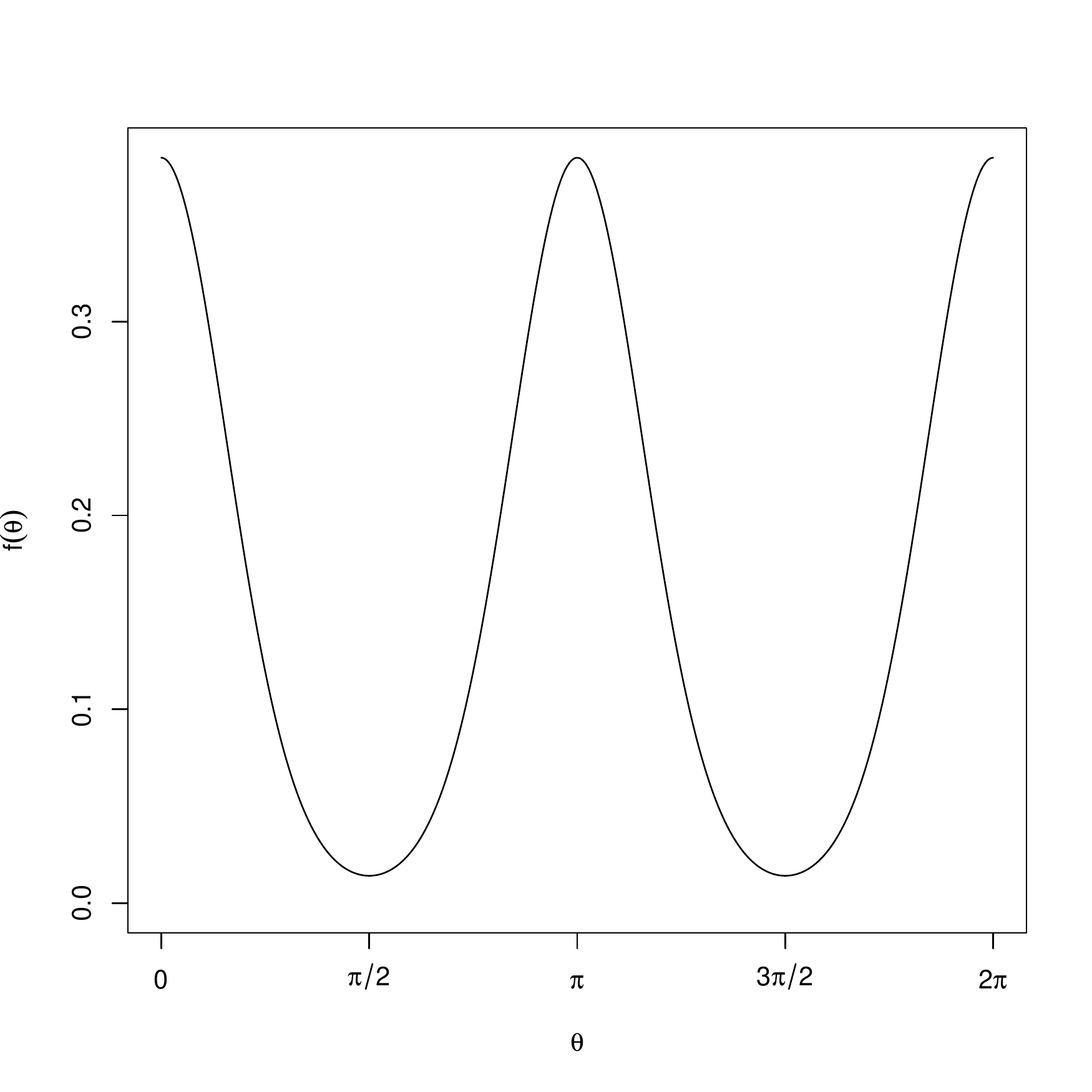}\\
M17
\includegraphics[width=1\linewidth]{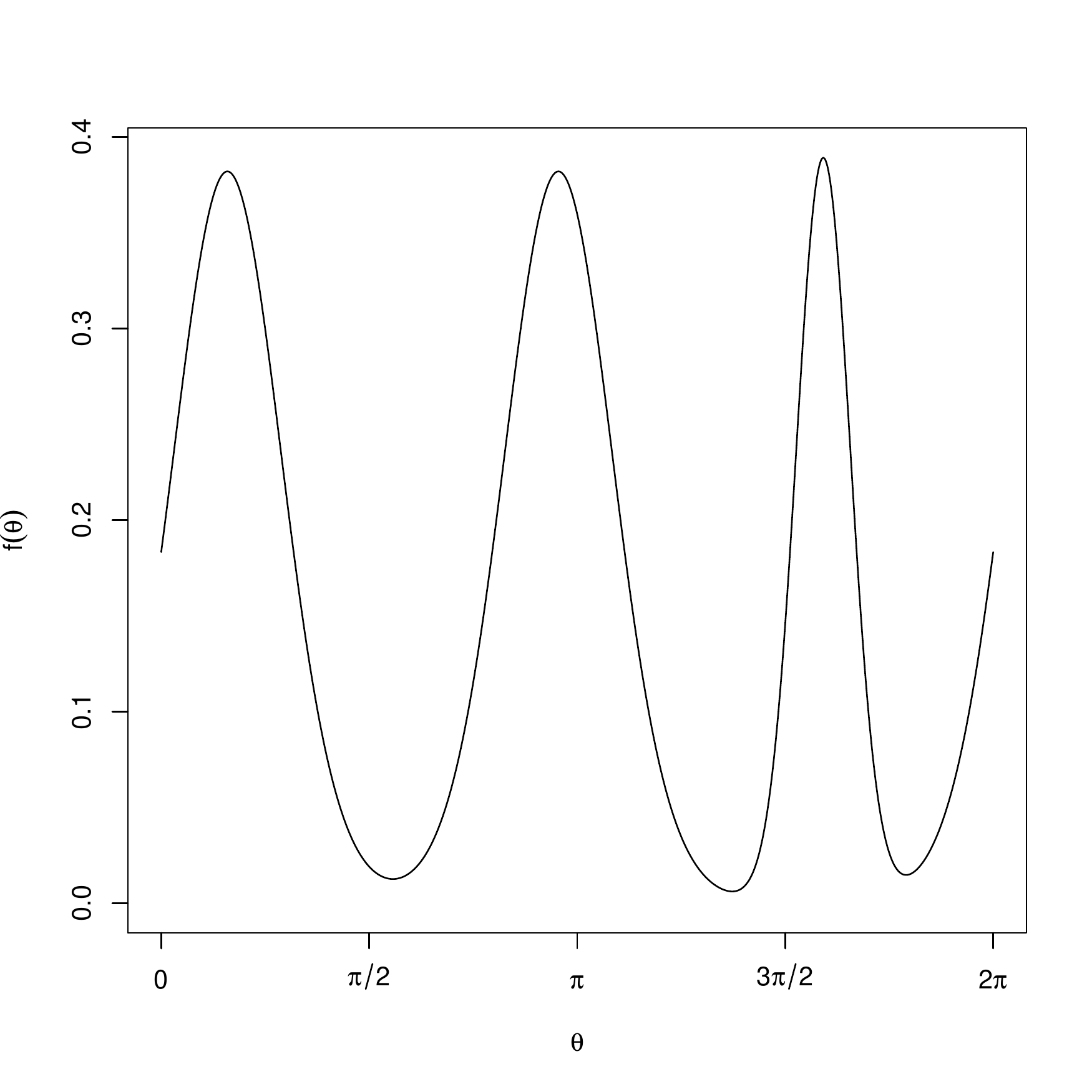}\\
M21
\includegraphics[width=1\linewidth]{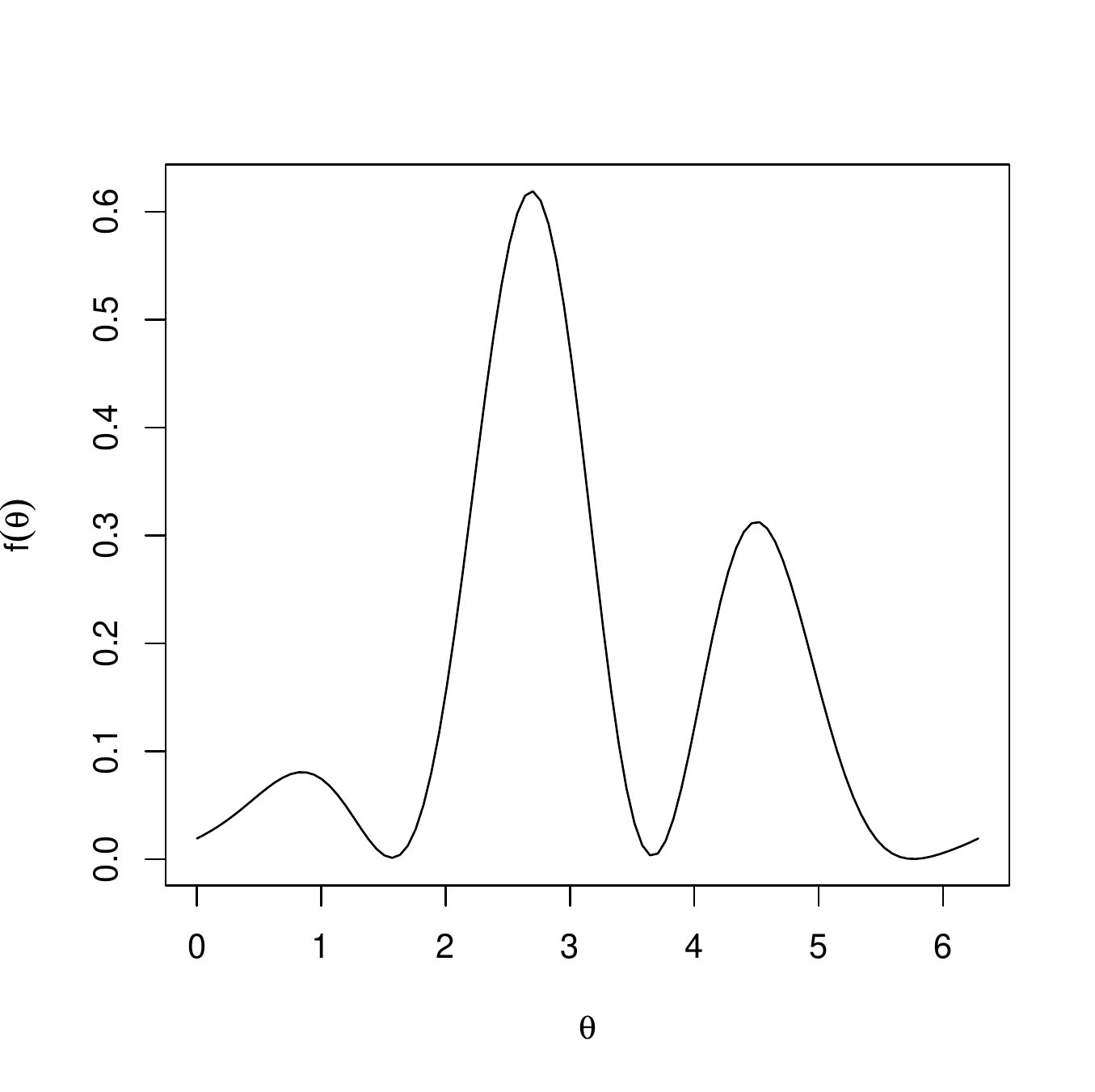}\\
M25
\includegraphics[width=1\linewidth]{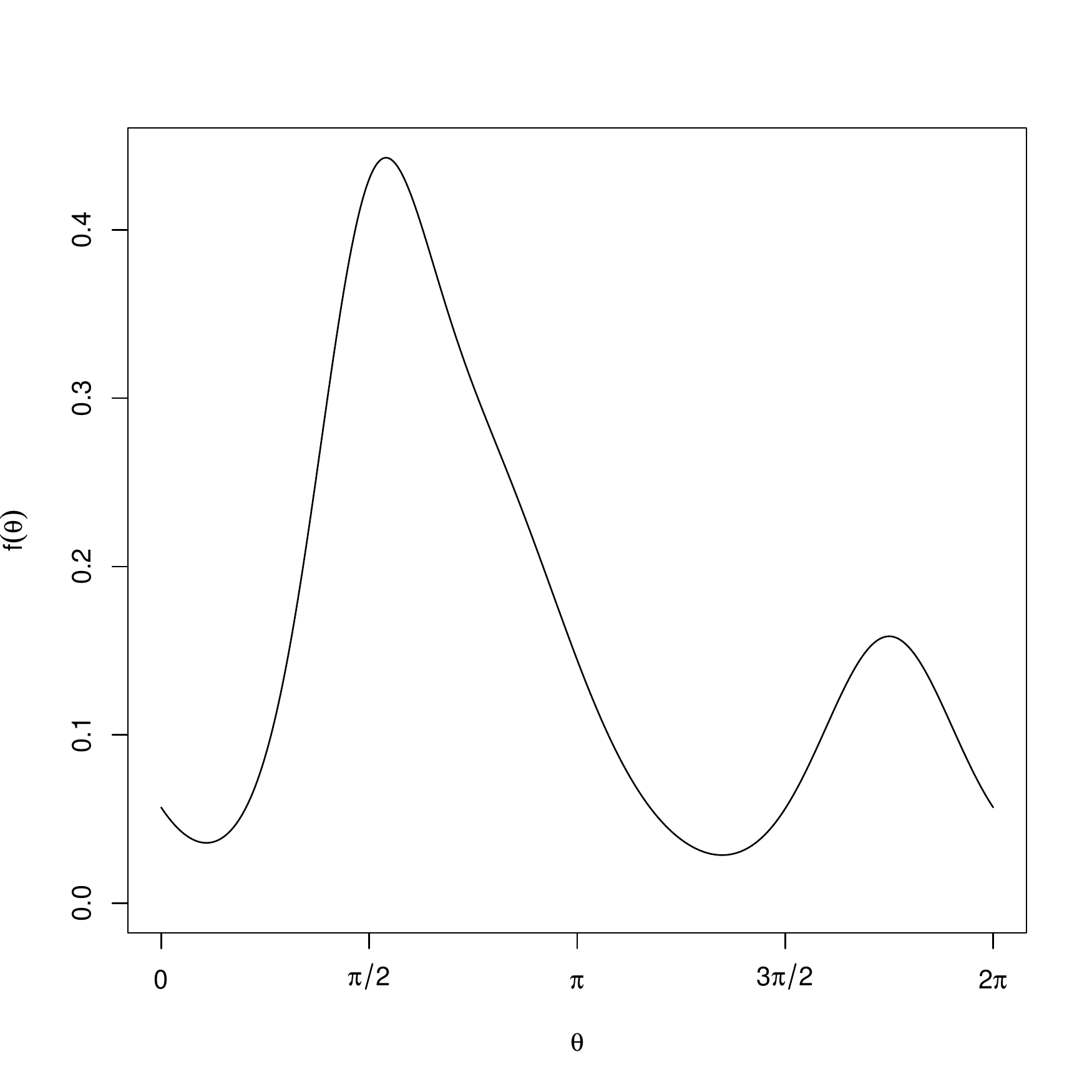}\\
M14
\includegraphics[width=1\linewidth]{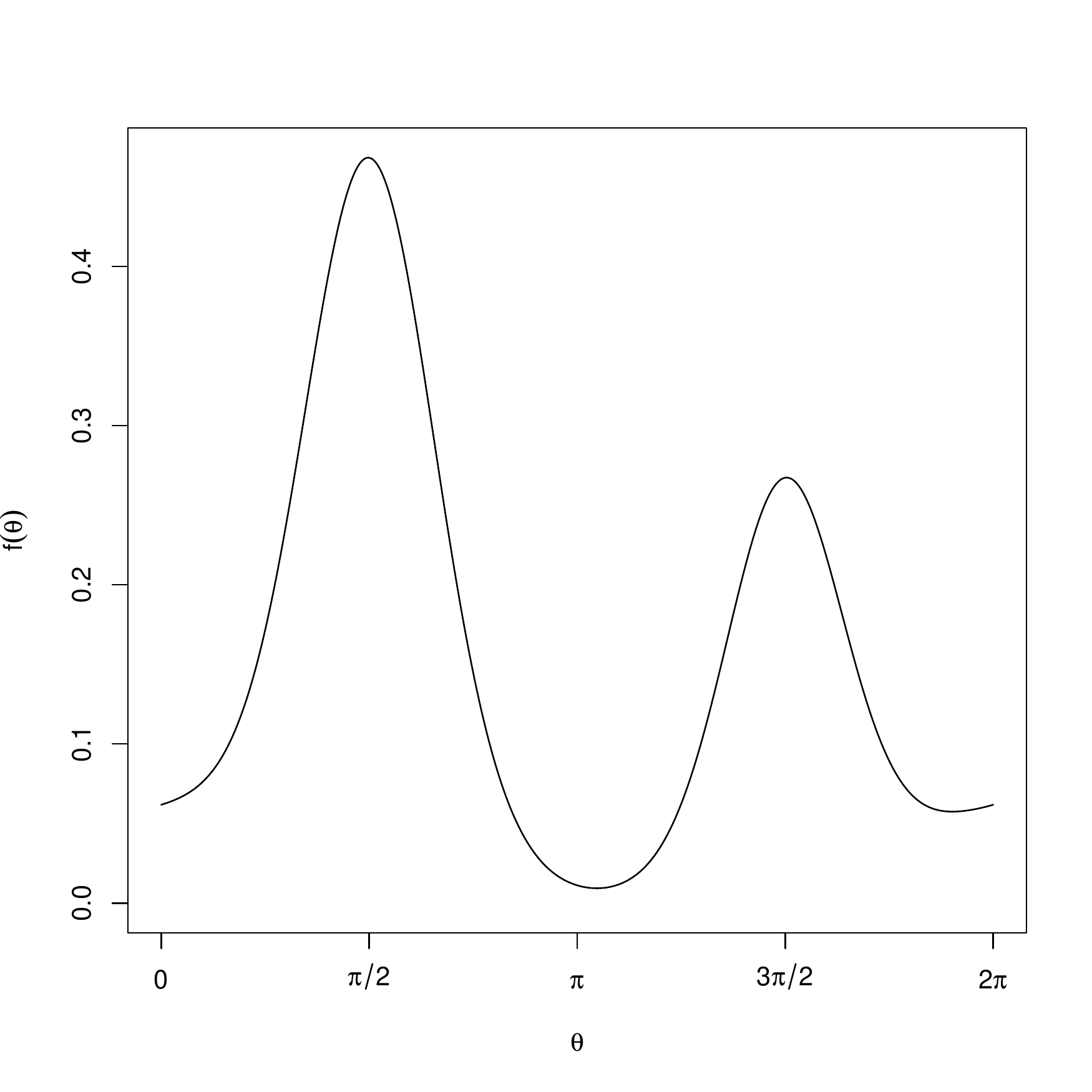}\\
M18
\includegraphics[width=1\linewidth]{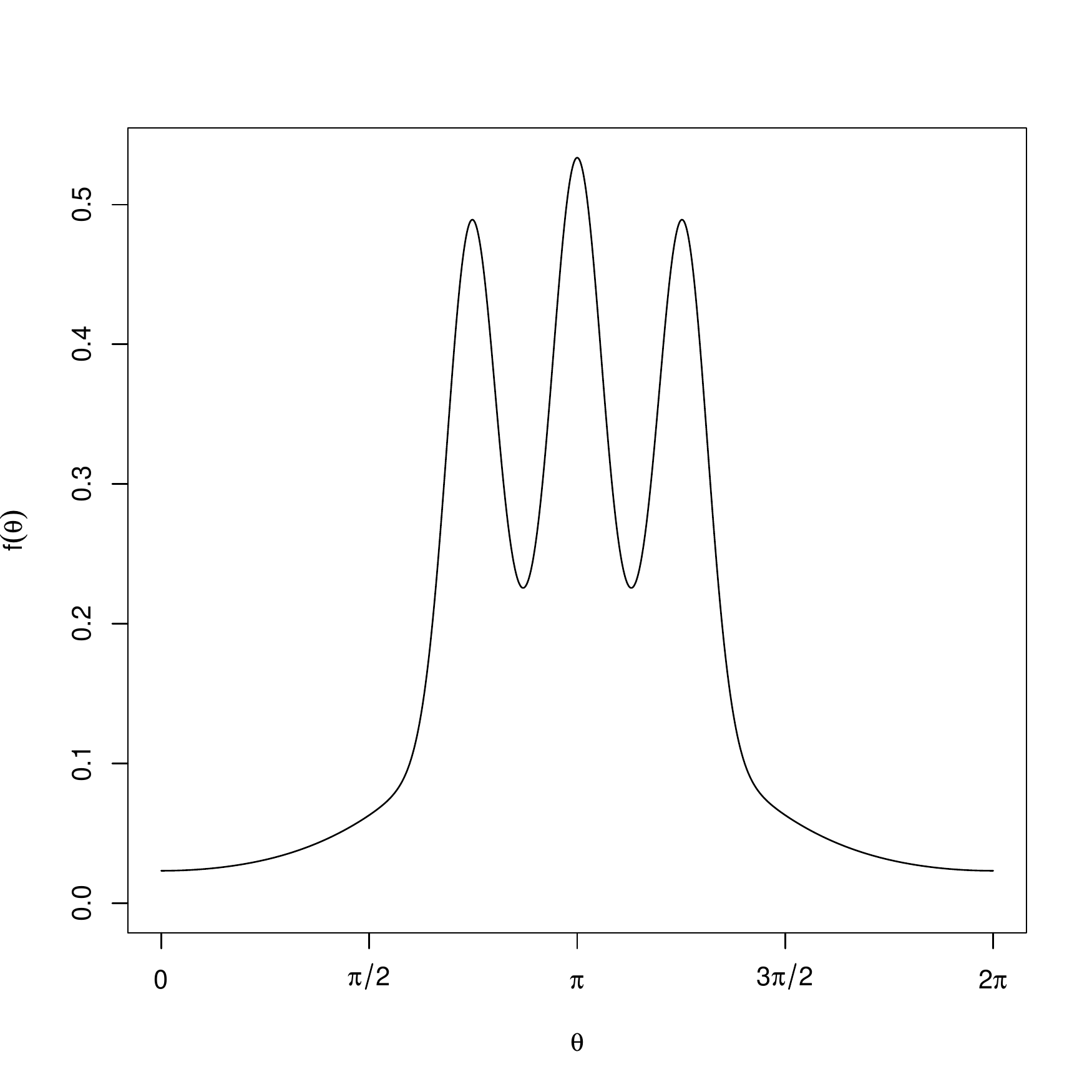}\\
M22
\includegraphics[width=1\linewidth]{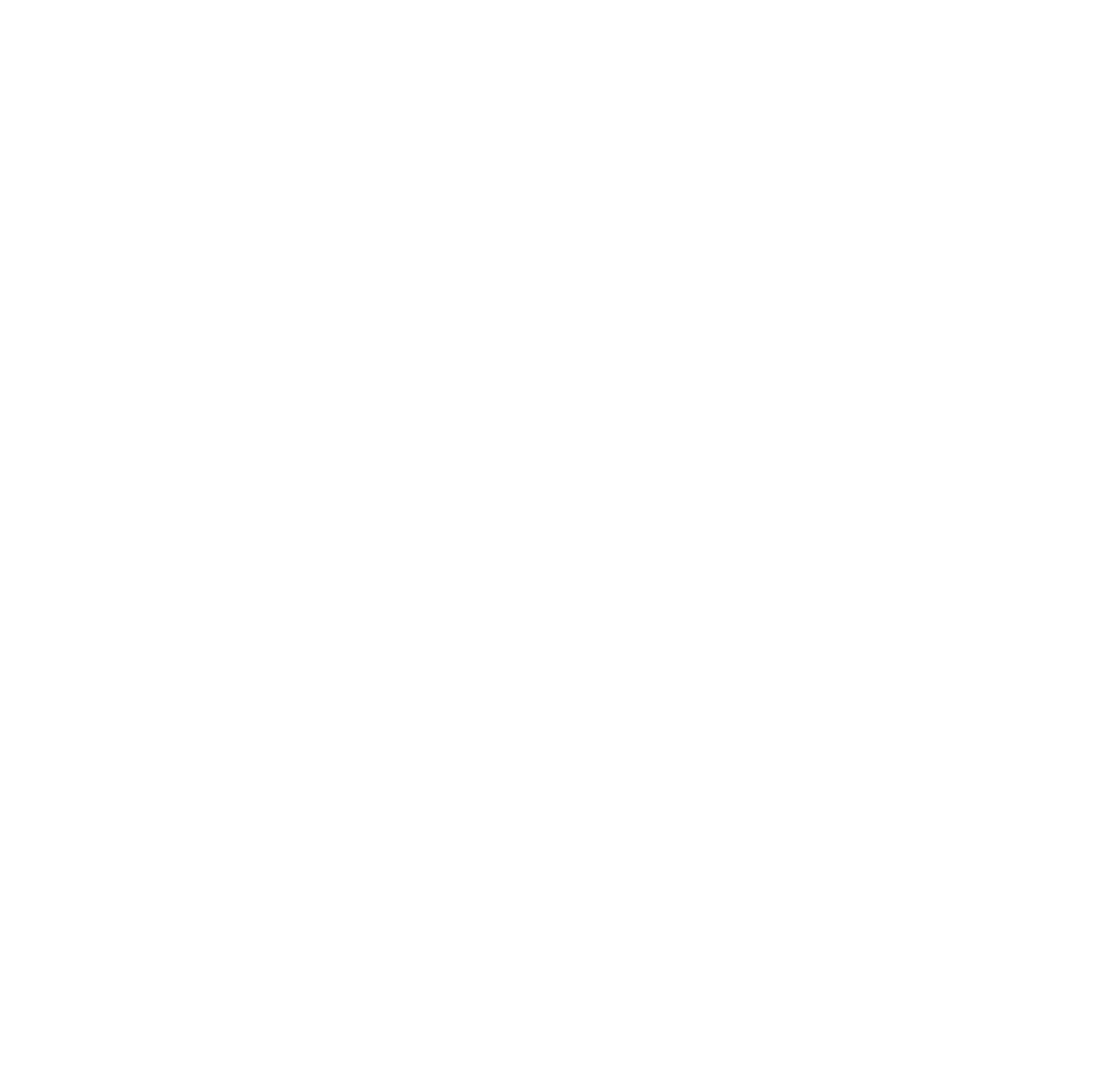}\\
\phantom{M28}
\end{multicols}
\caption{Linear representation of different circular density functions. M11--M20: bimodal models. M21--M25: trimodal models.}
\label{figs4}
\end{figure}

\begin{figure}
\begin{multicols}{4}
\centering
\includegraphics[width=1\linewidth]{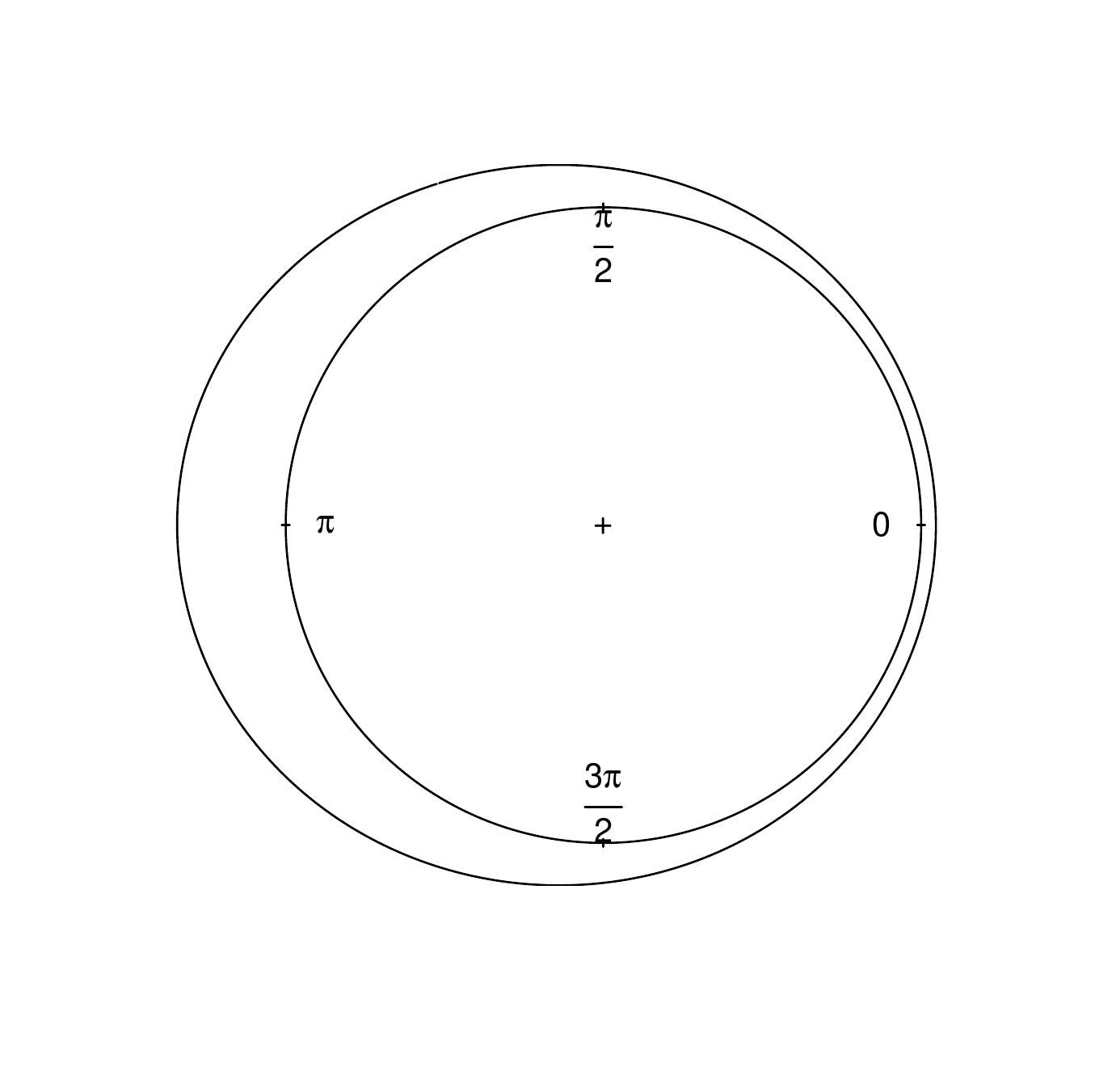}\\
M1
\includegraphics[width=1\linewidth]{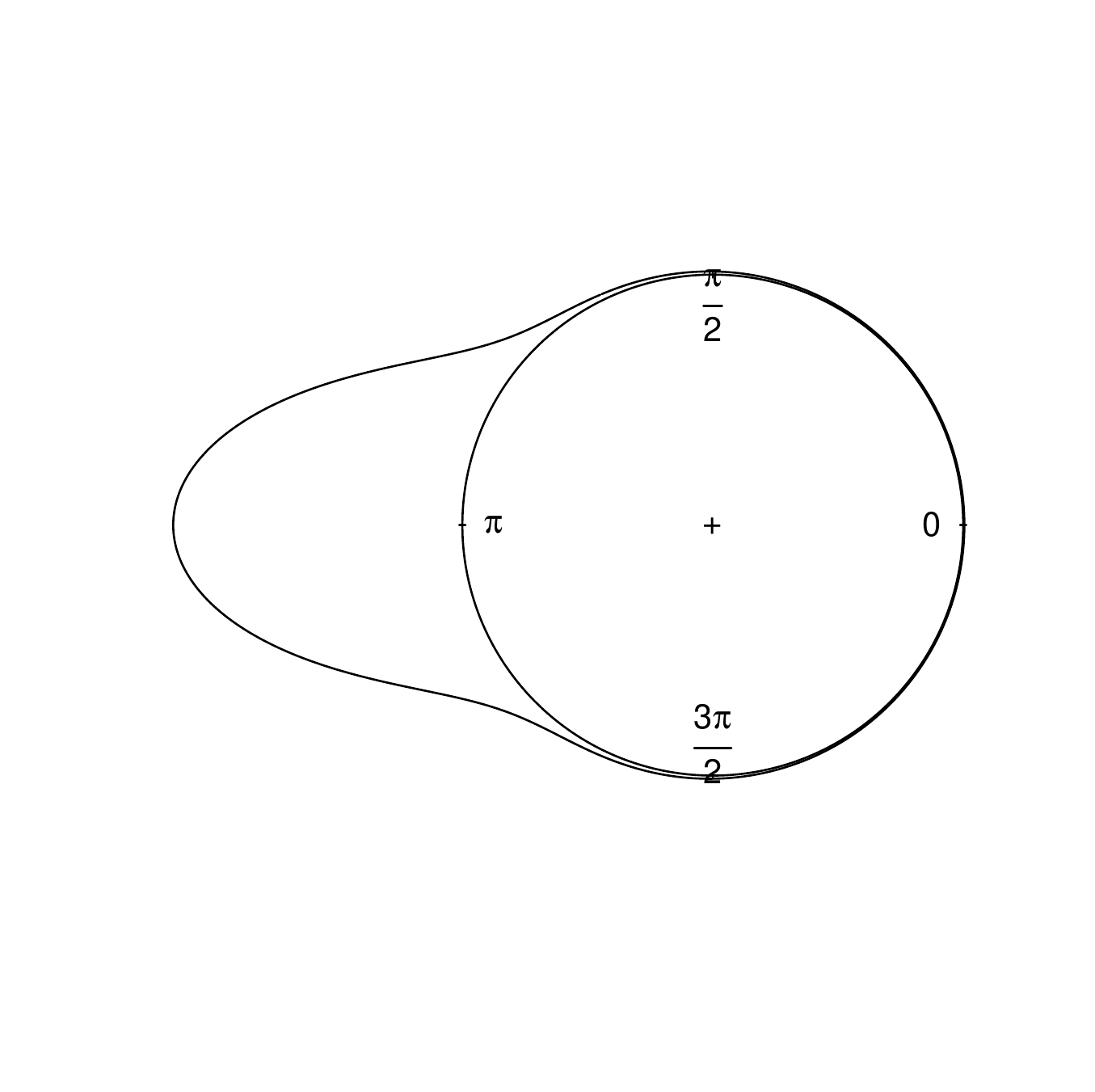}\\
M5
\includegraphics[width=1\linewidth]{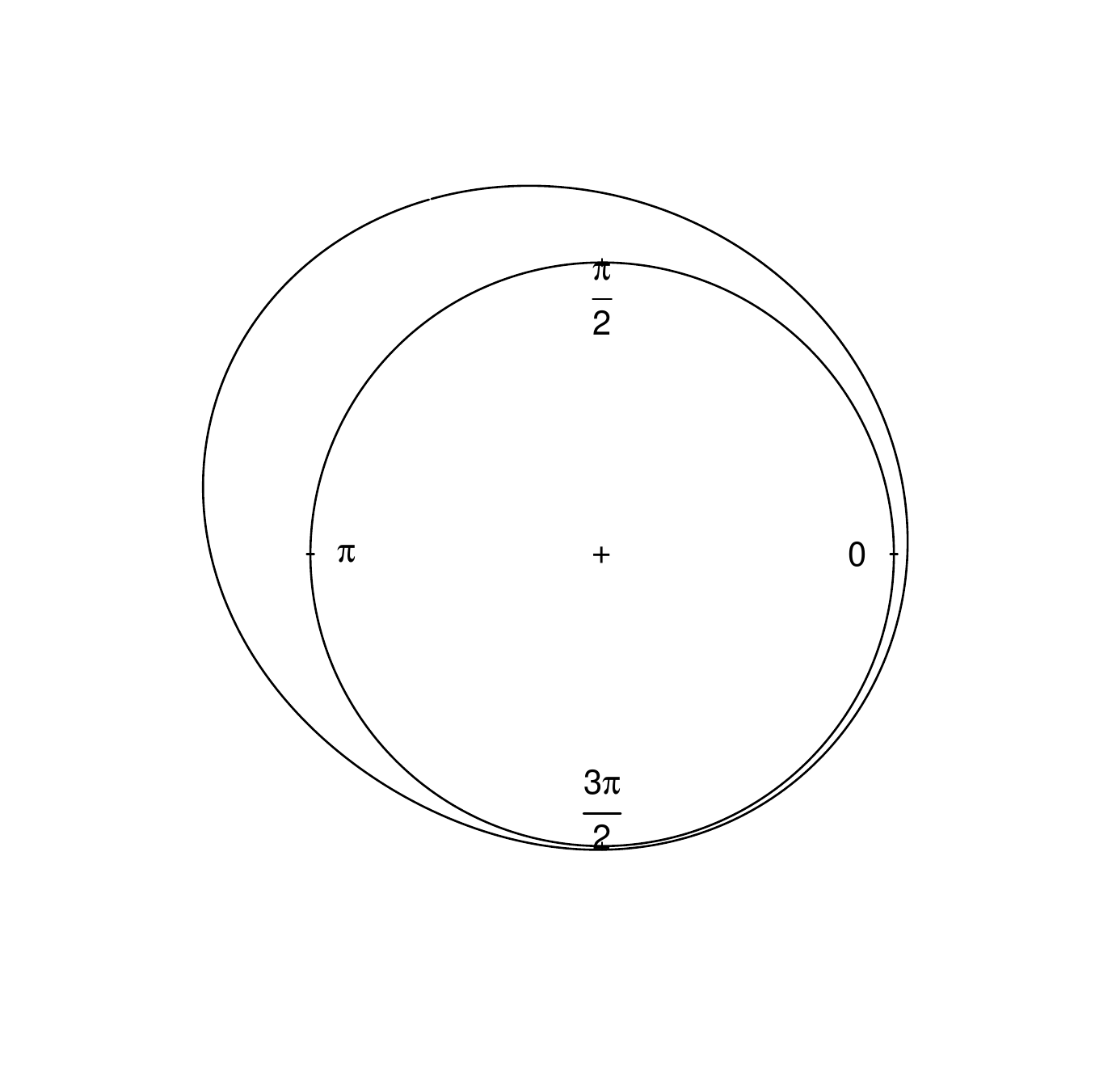}\\
M9
\includegraphics[width=1\linewidth]{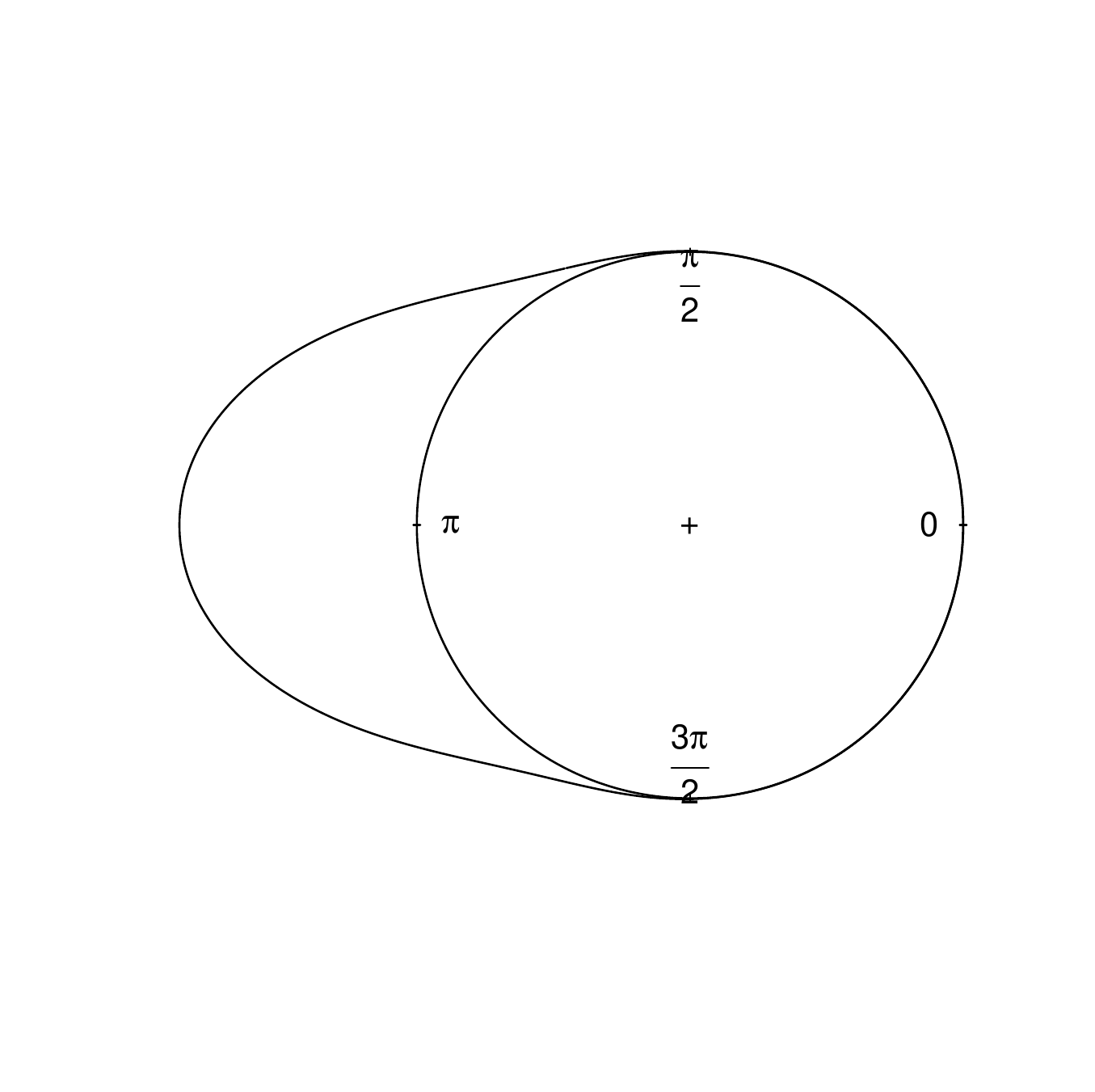}\\
M2
\includegraphics[width=1\linewidth]{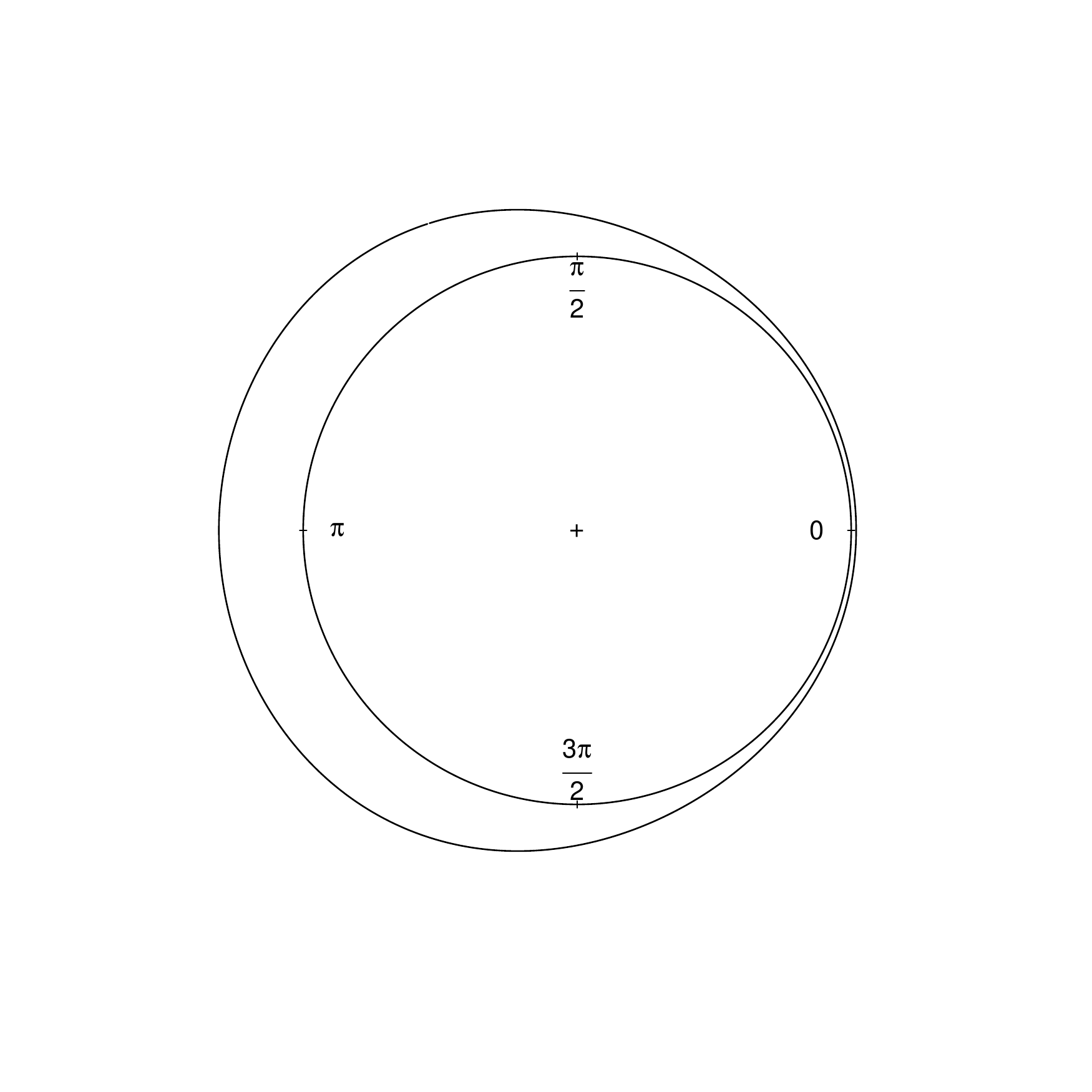}\\
M6
\includegraphics[width=1\linewidth]{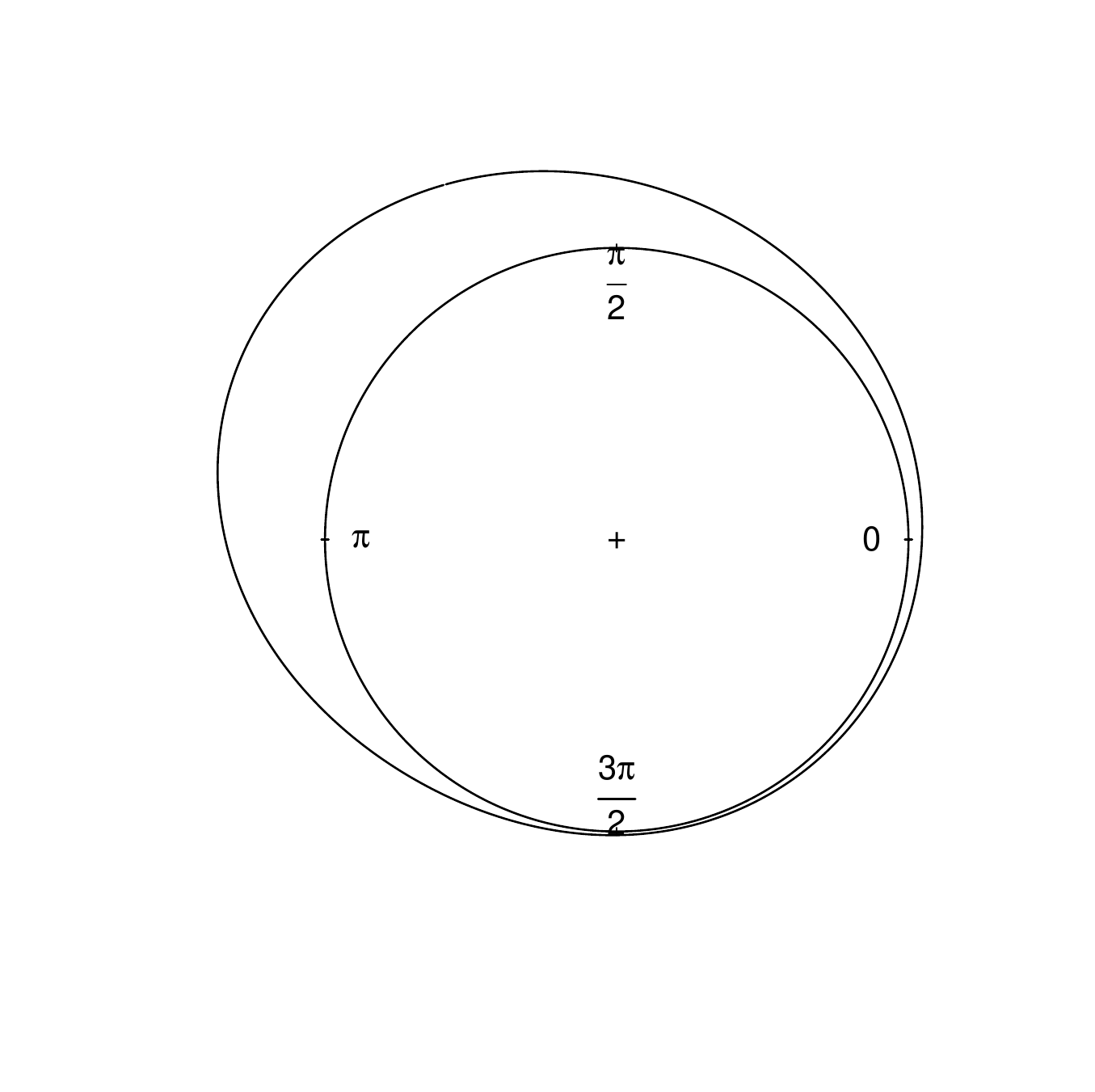}\\
M10
\includegraphics[width=1\linewidth]{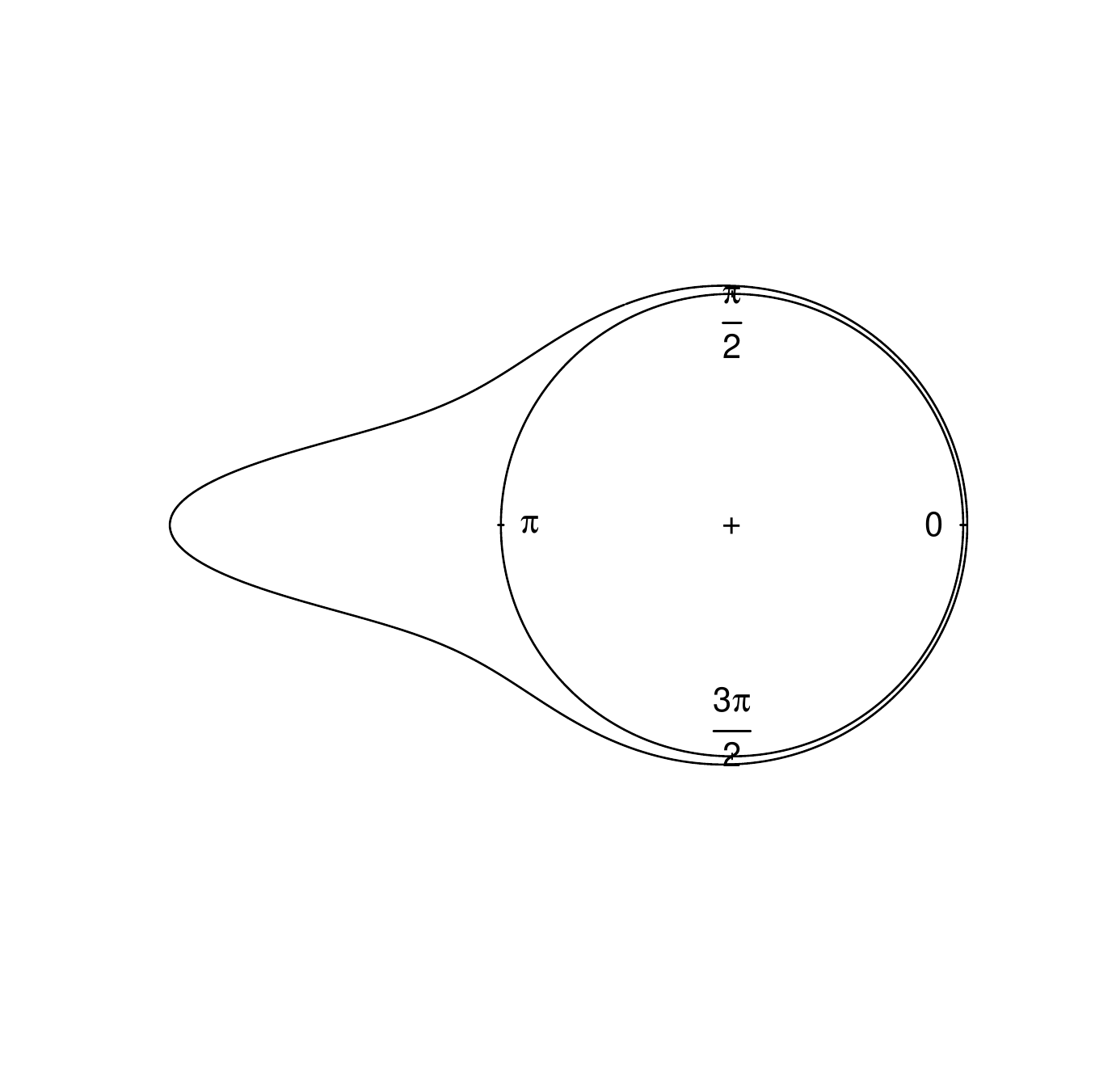}\\
M3
\includegraphics[width=1\linewidth]{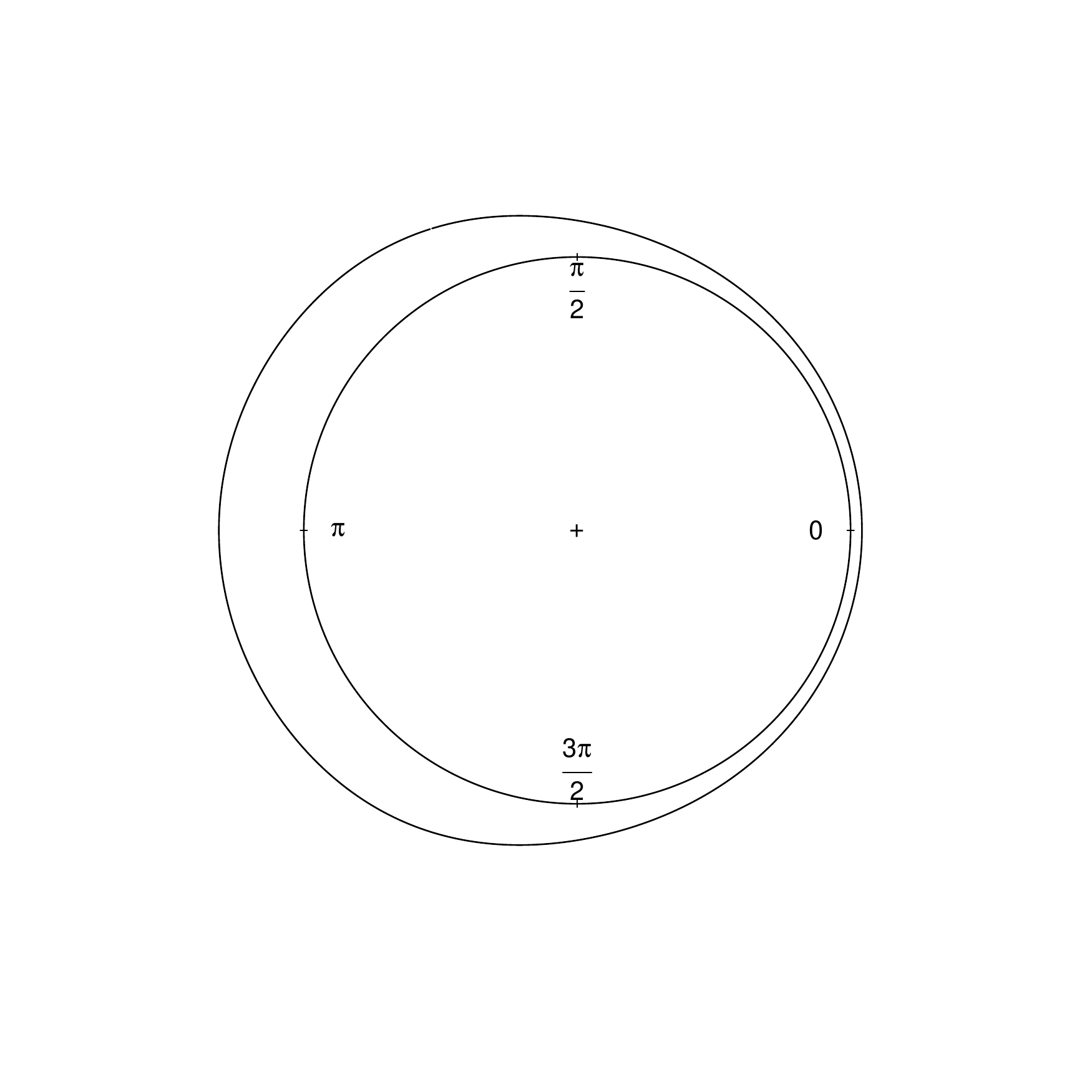}\\
M7
\includegraphics[width=1\linewidth]{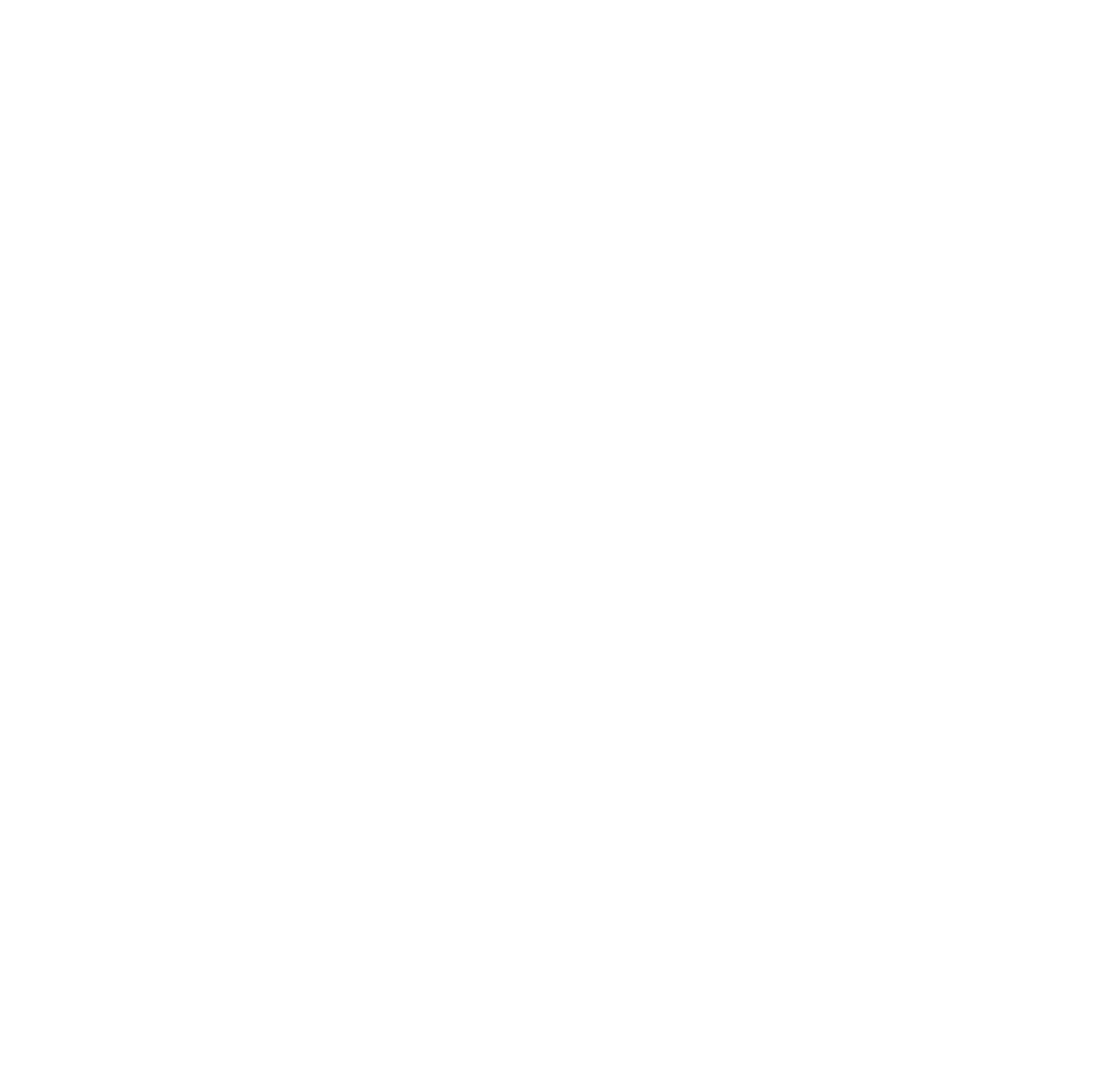}\\
\phantom{M27}
\includegraphics[width=1\linewidth]{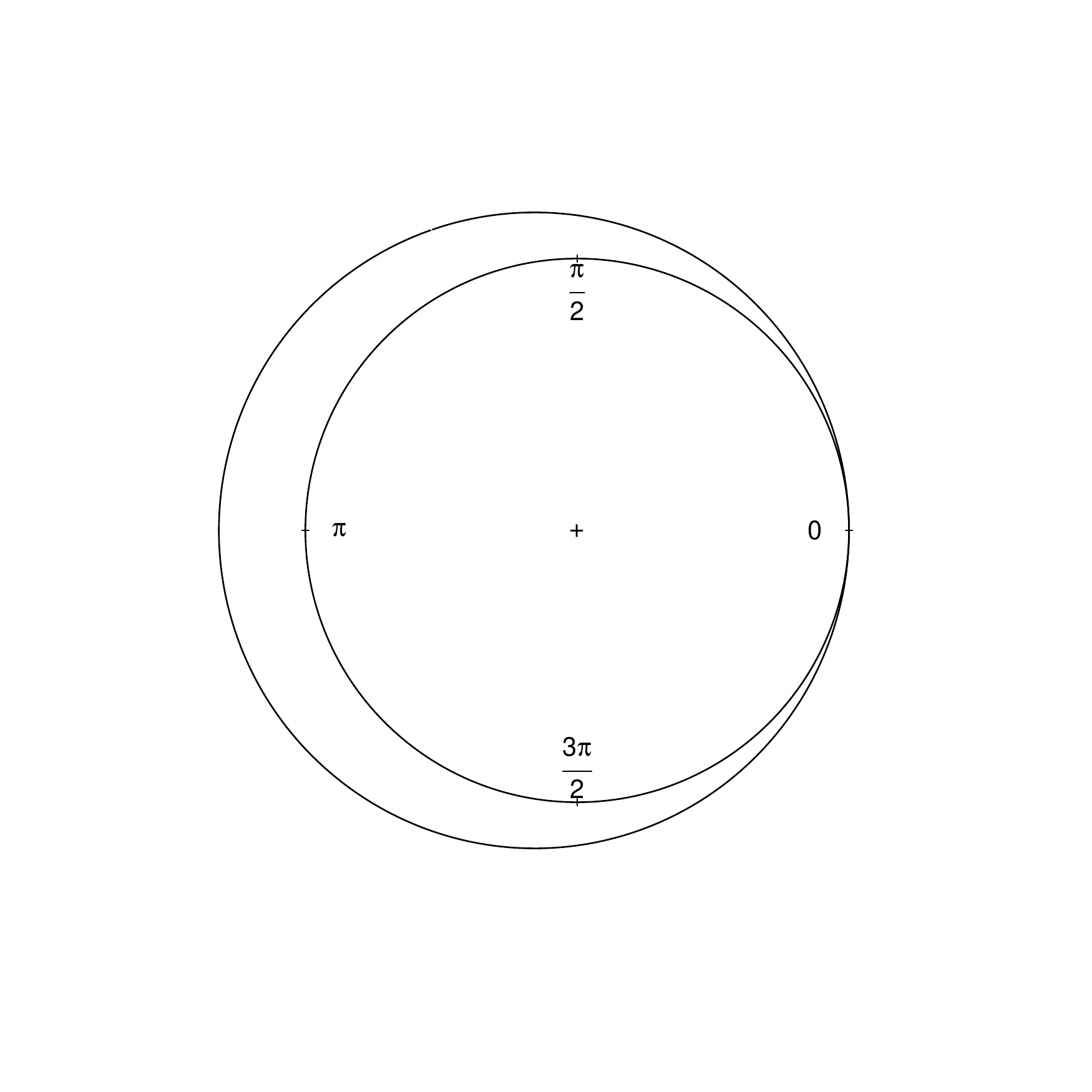}\\
M4
\includegraphics[width=1\linewidth]{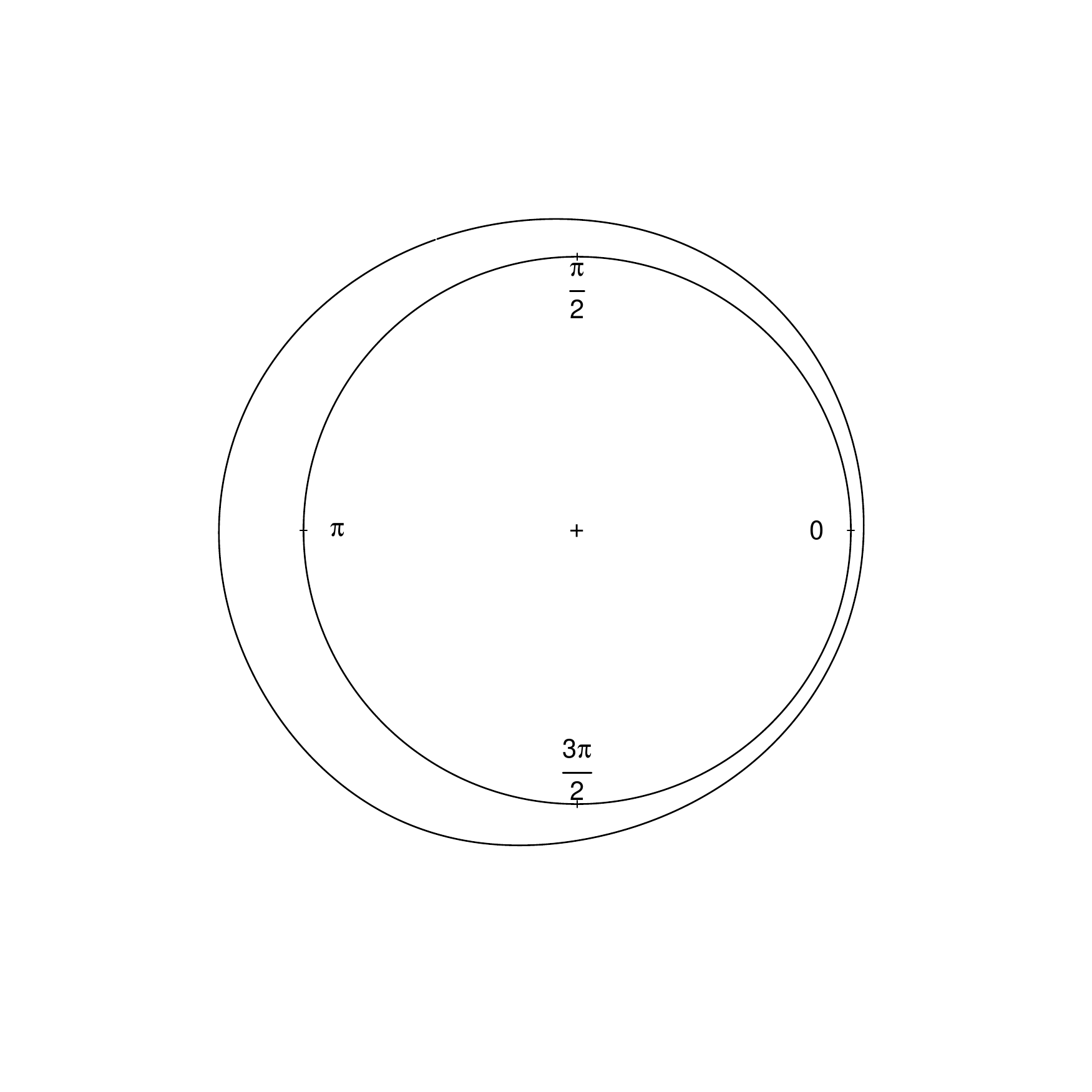}\\
M8
\includegraphics[width=1\linewidth]{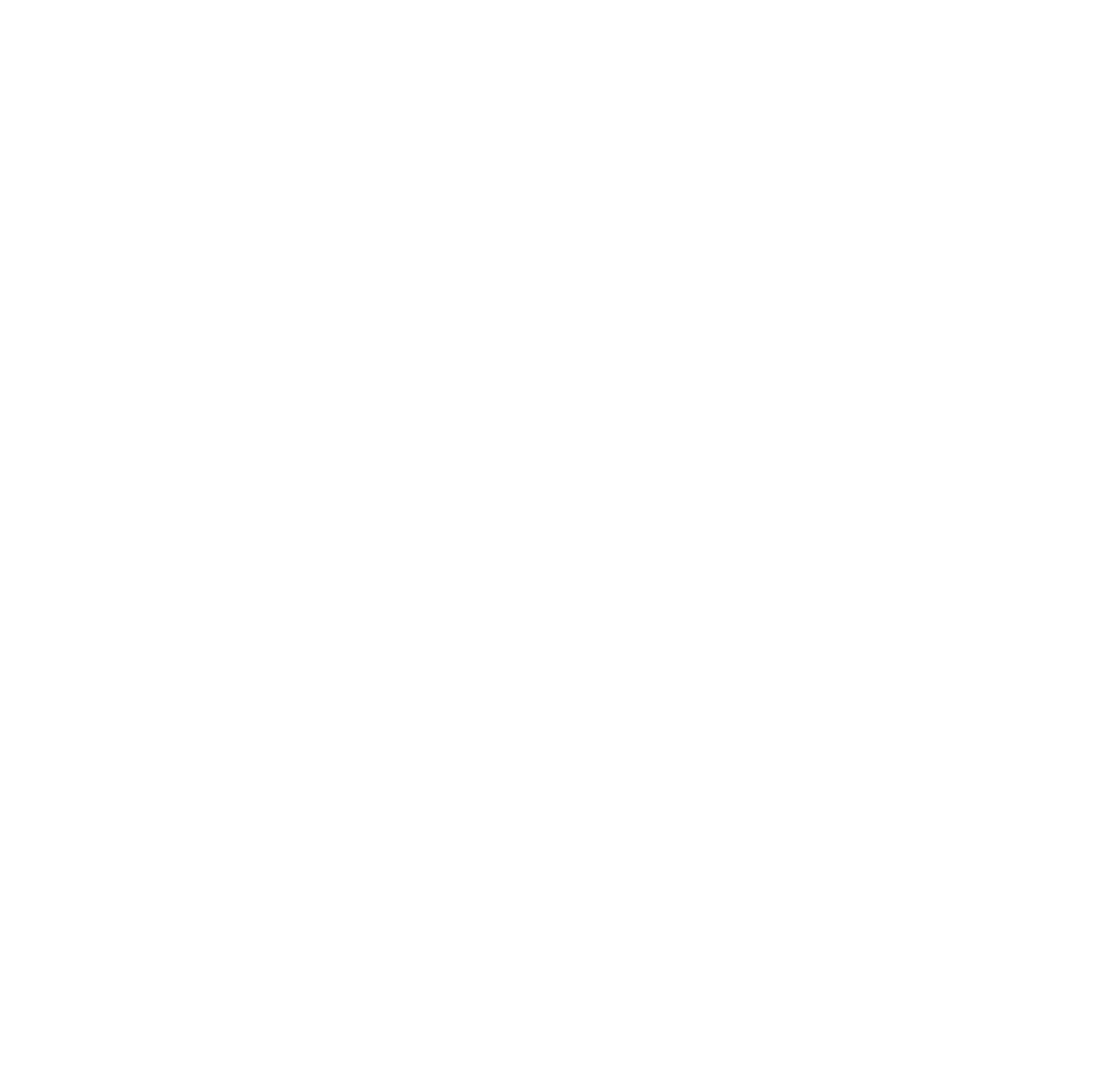}\\
\phantom{M27}
\end{multicols}
\caption{Circular representation of the unimodal density functions: M1--M10.}
\label{figs5}
\end{figure}

\begin{figure}
\begin{multicols}{4}
\centering
\includegraphics[width=1\linewidth]{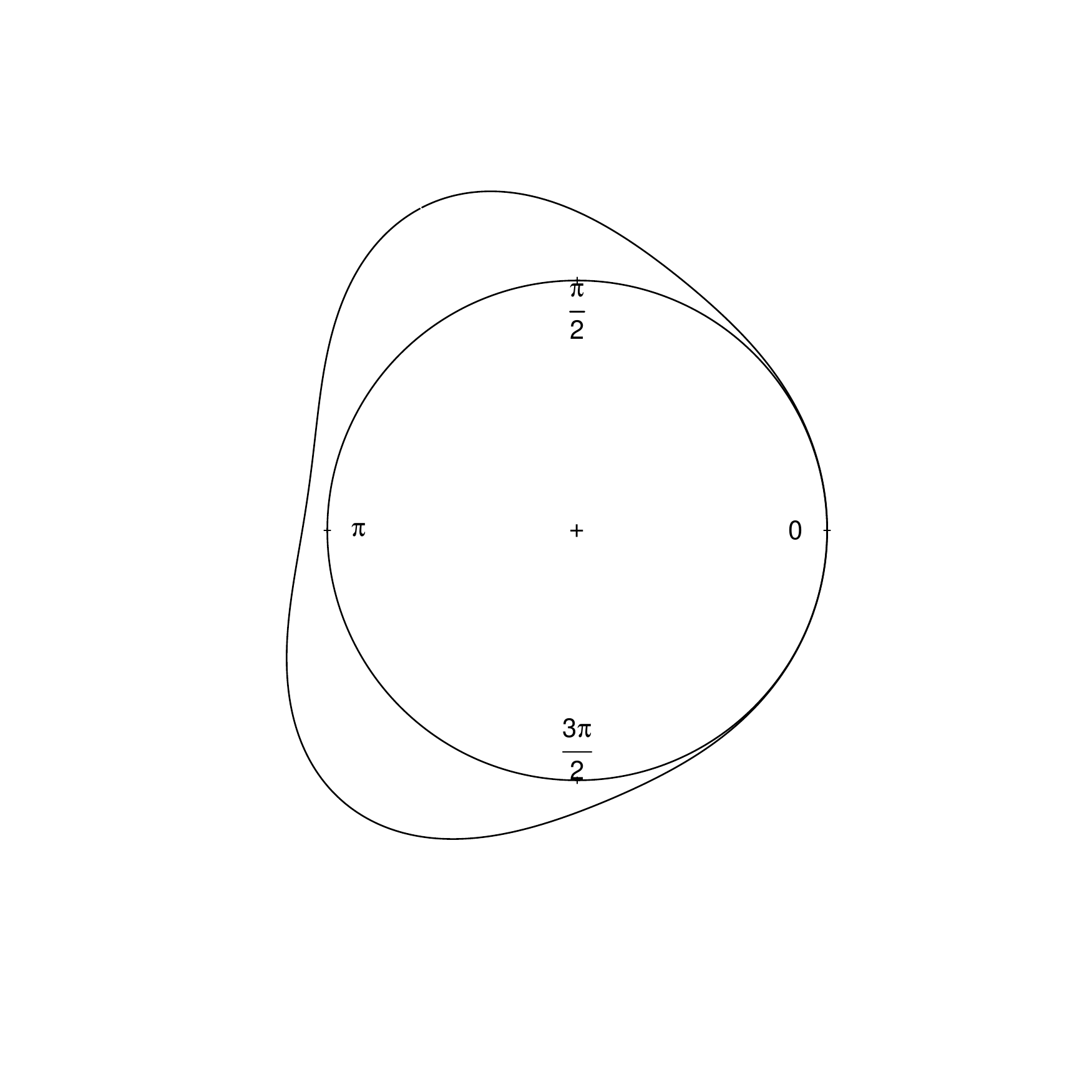}\\
M11
\includegraphics[width=1\linewidth]{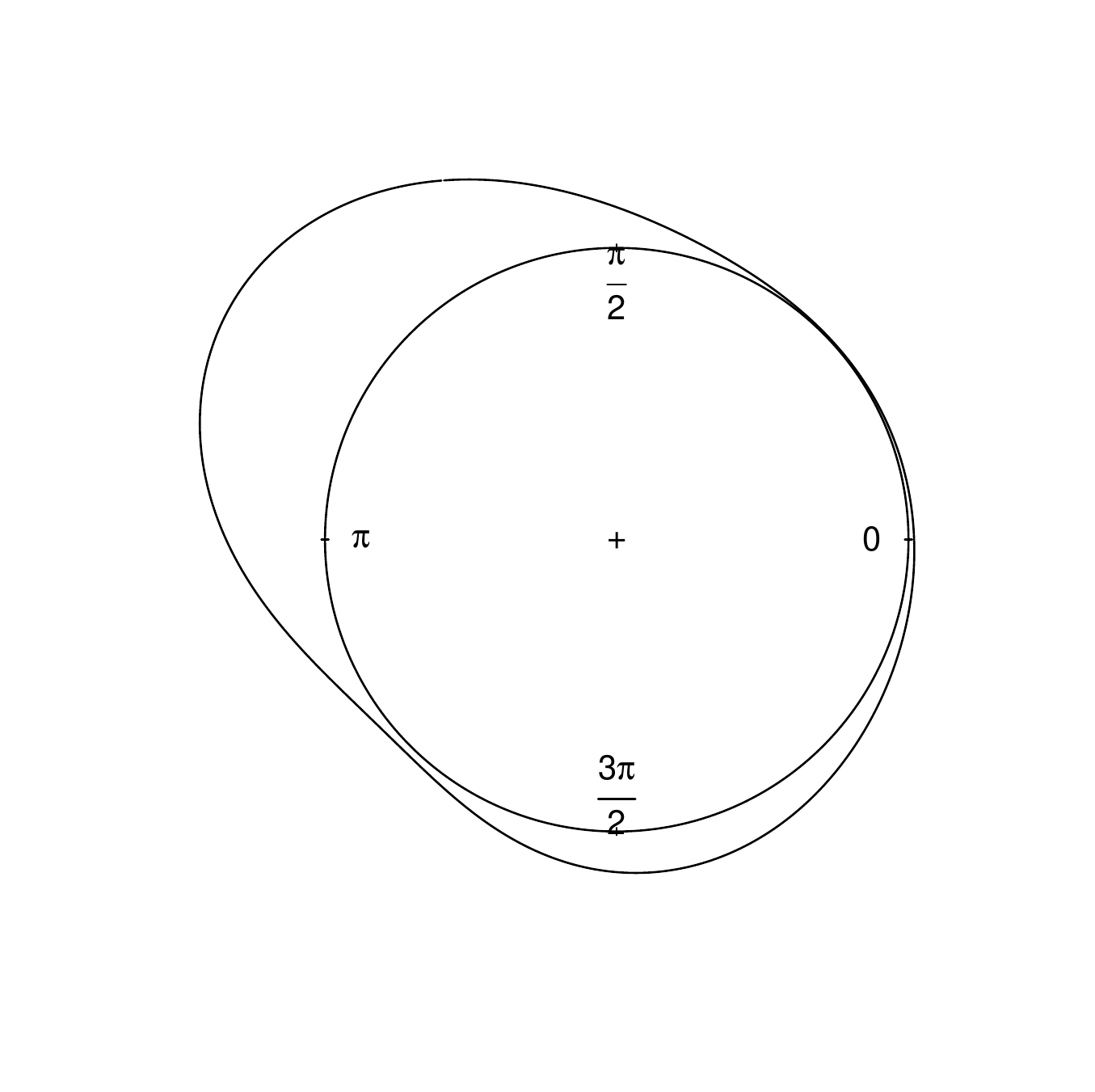}\\
M15
\includegraphics[width=1\linewidth]{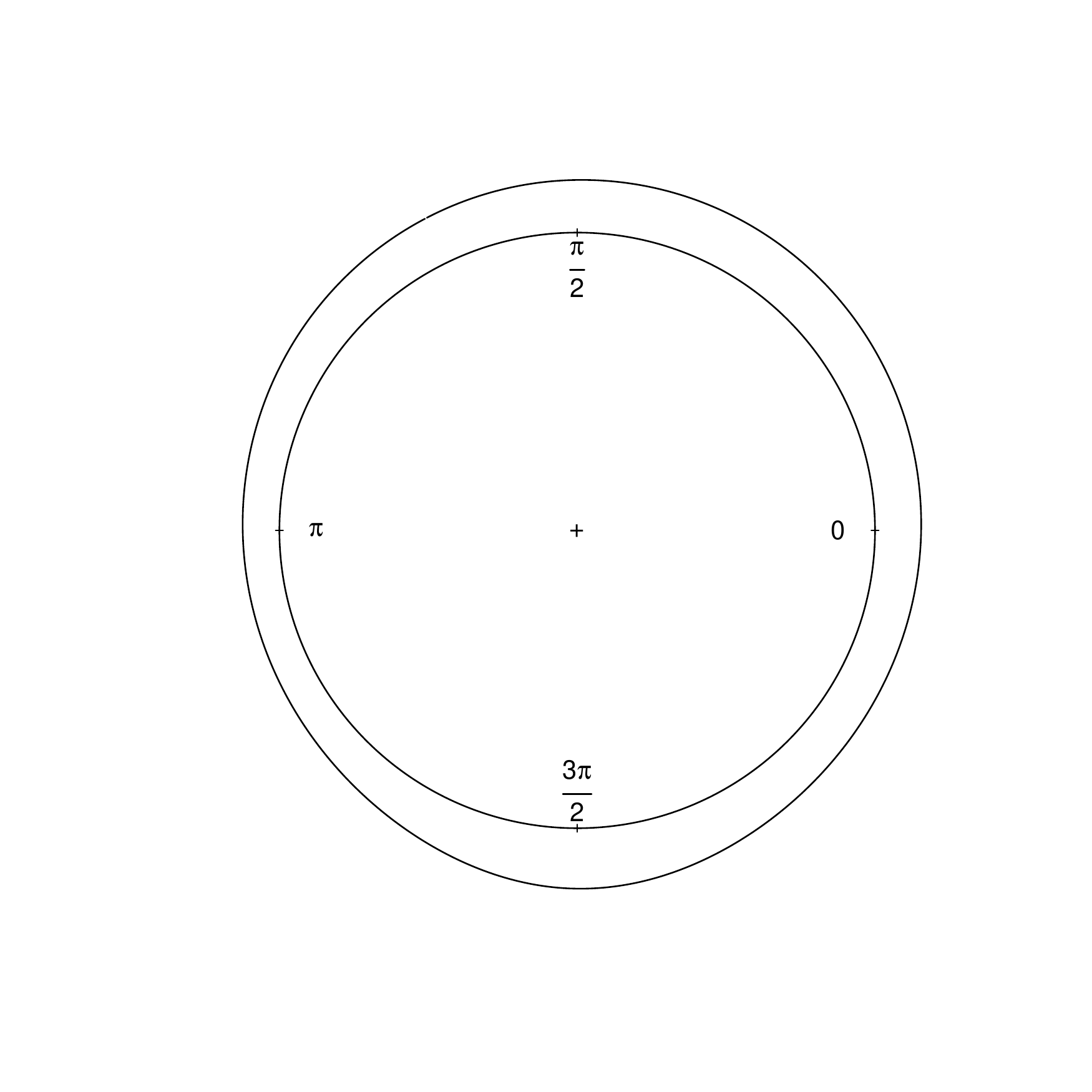}\\
M19
\includegraphics[width=1\linewidth]{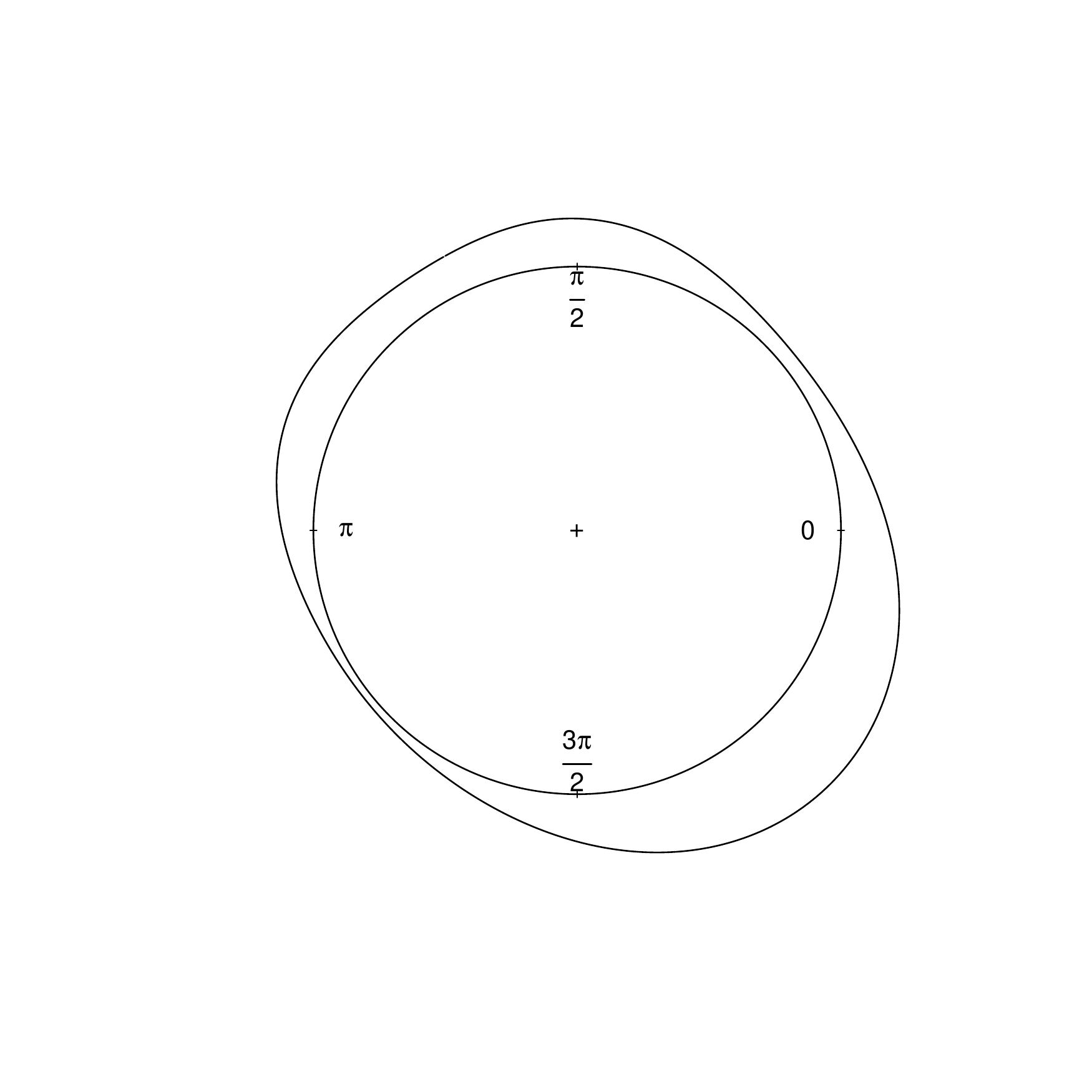}\\
M23
\includegraphics[width=1\linewidth]{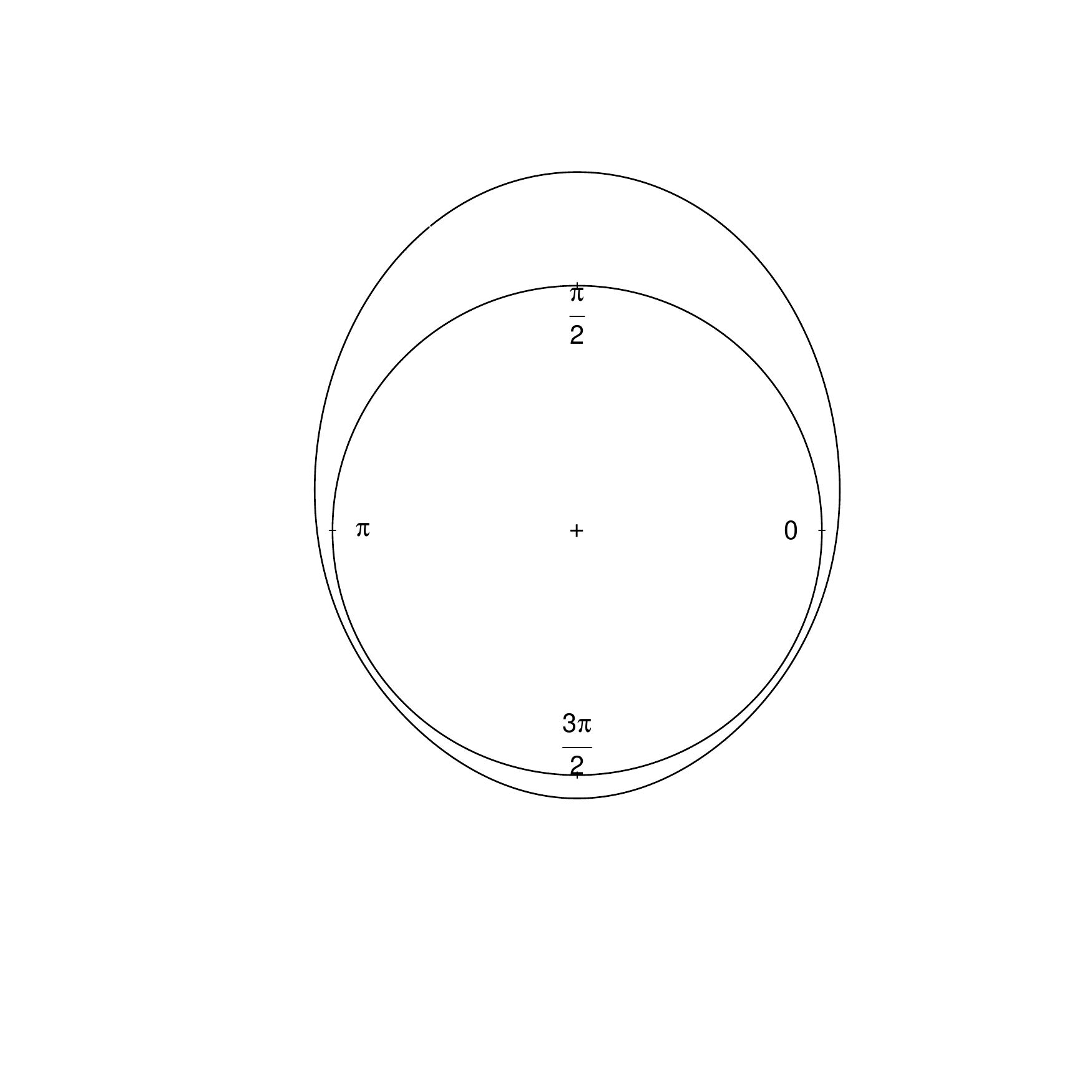}\\
M12
\includegraphics[width=1\linewidth]{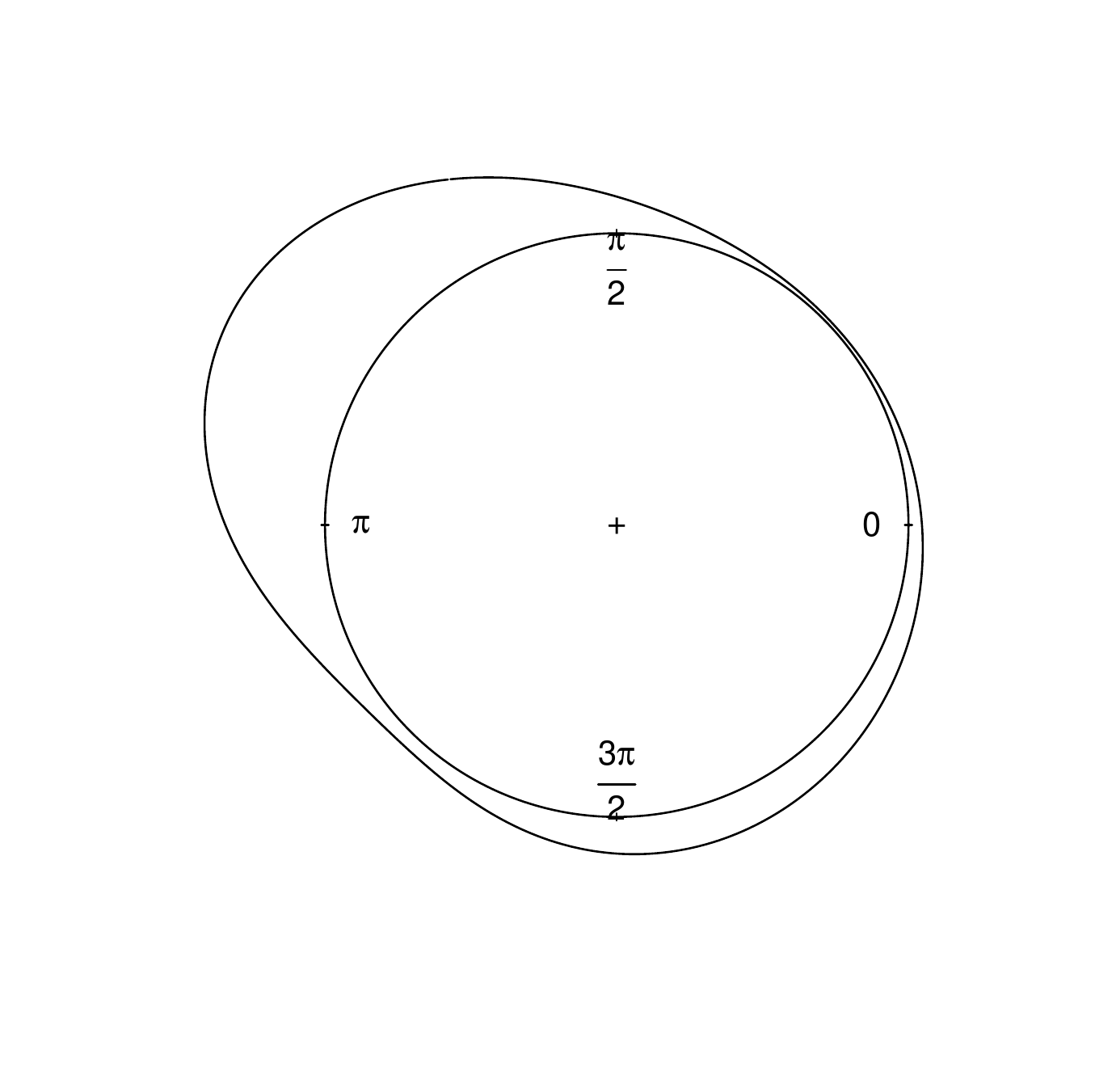}\\
M16
\includegraphics[width=1\linewidth]{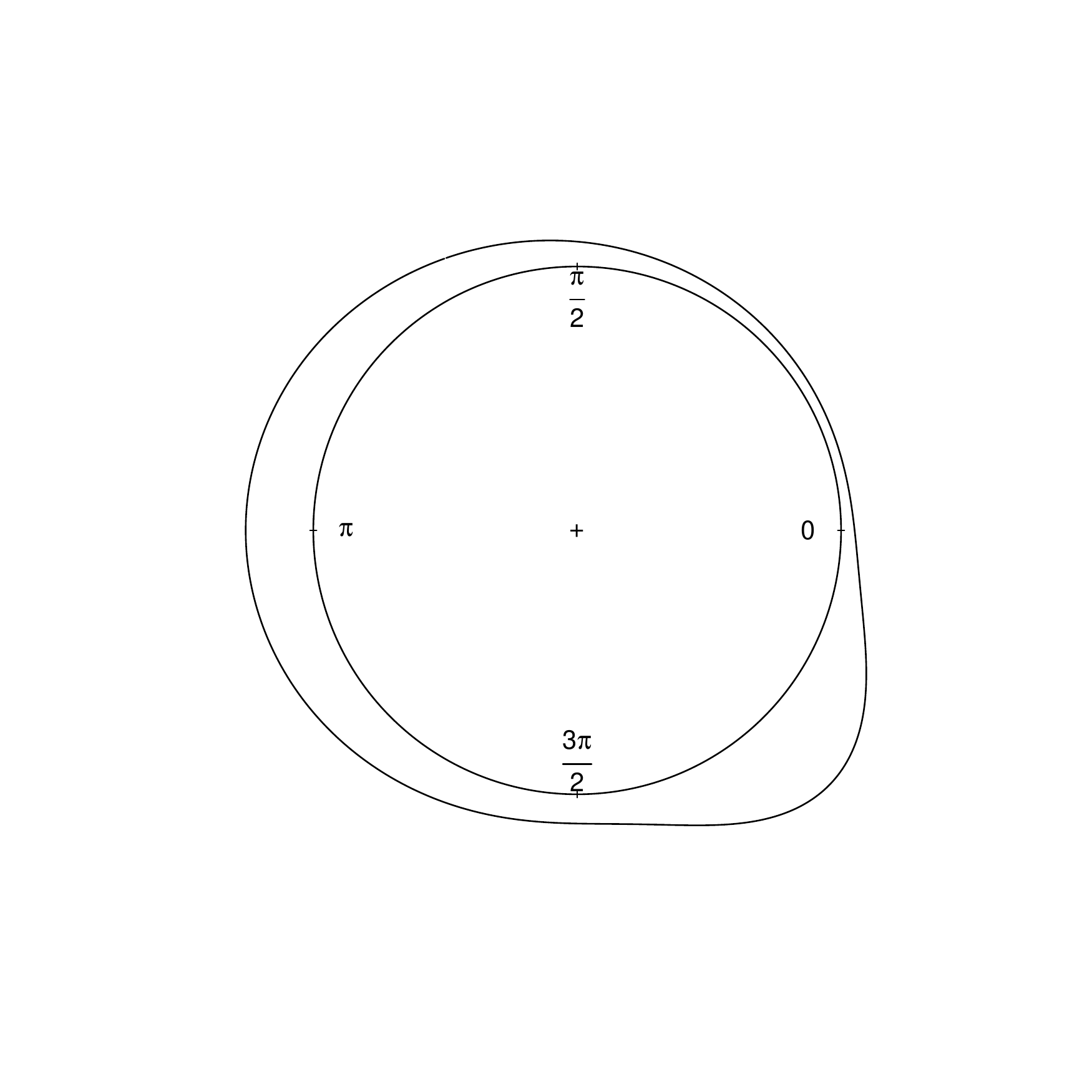}\\
M20
\includegraphics[width=1\linewidth]{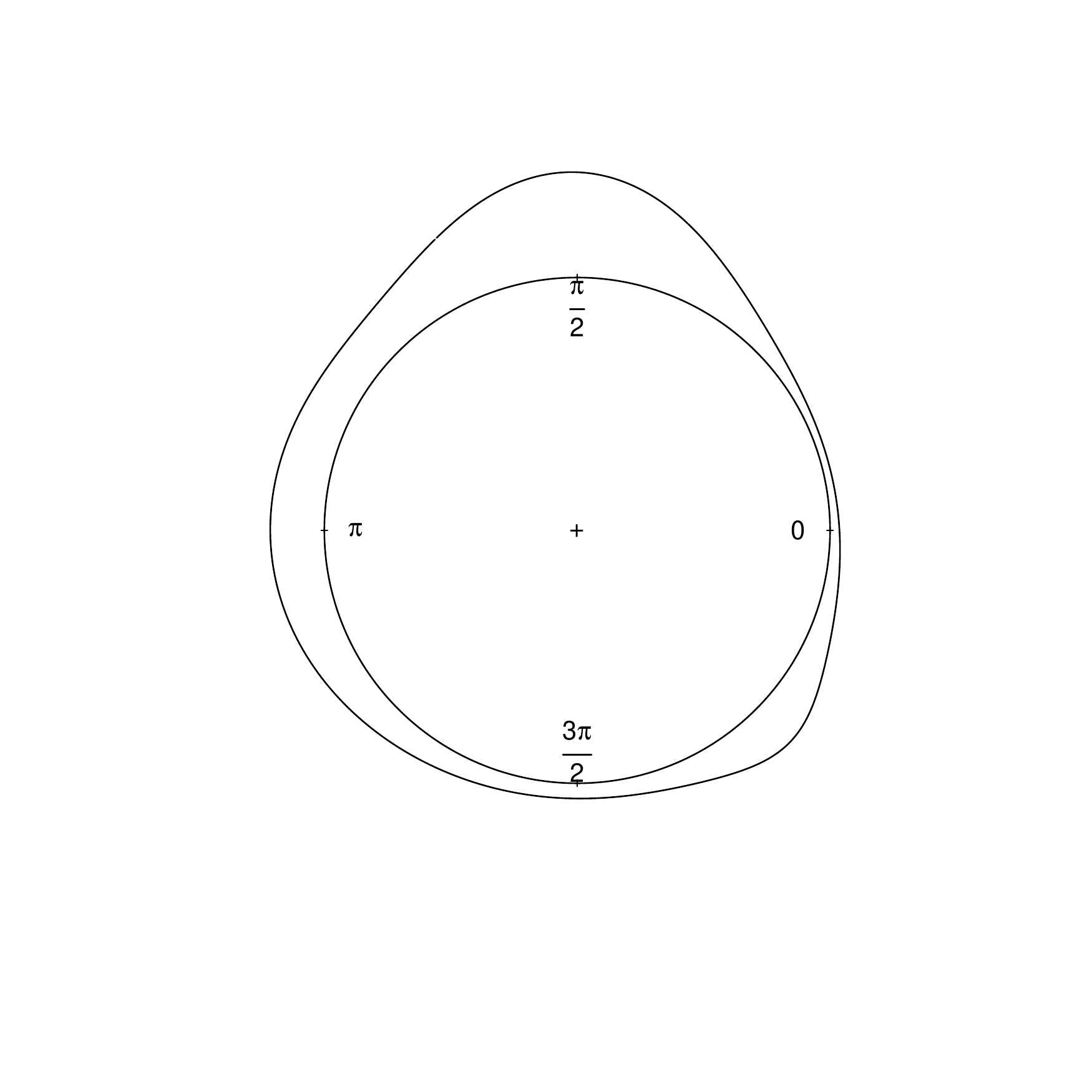}\\
M24
\includegraphics[width=1\linewidth]{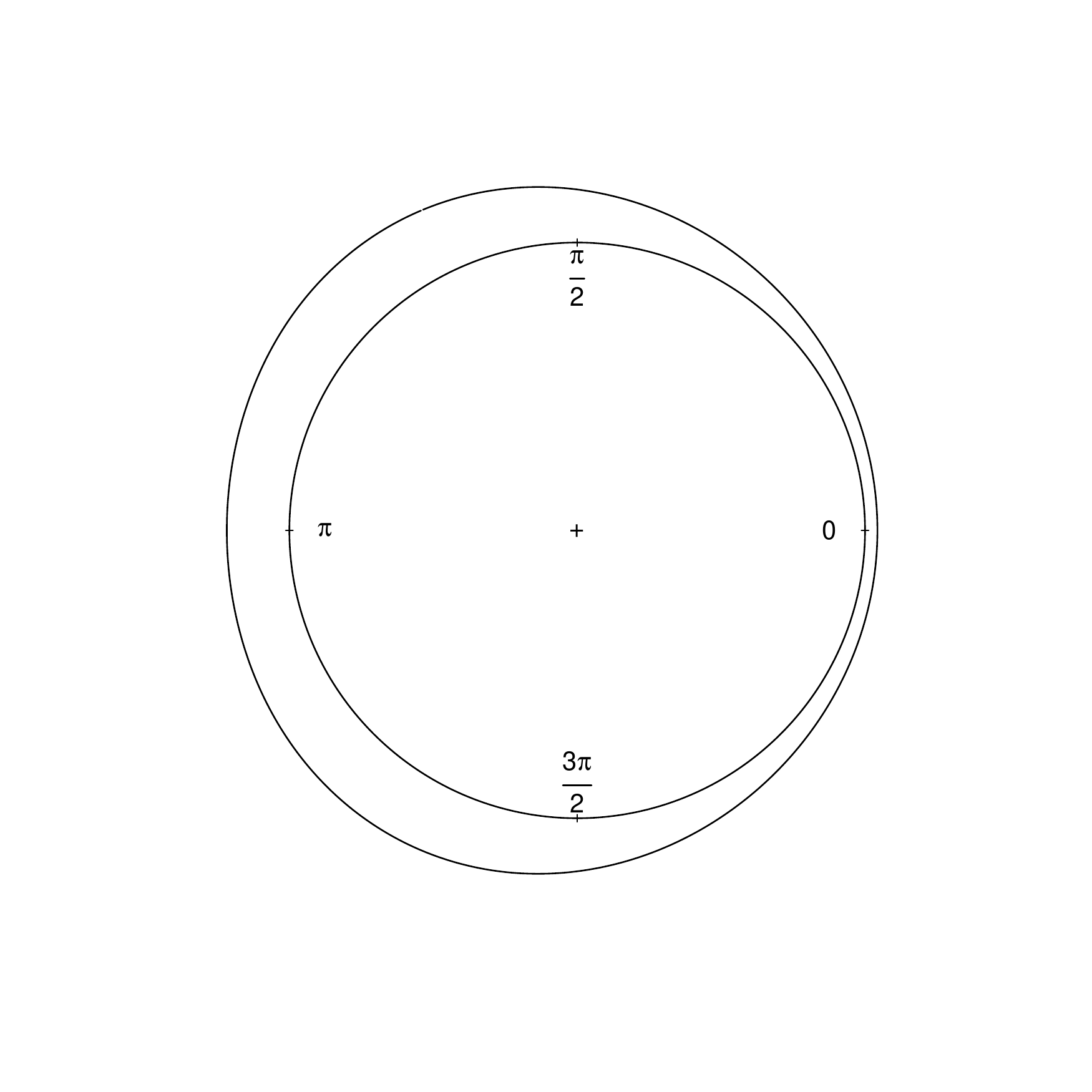}\\
M13
\includegraphics[width=1\linewidth]{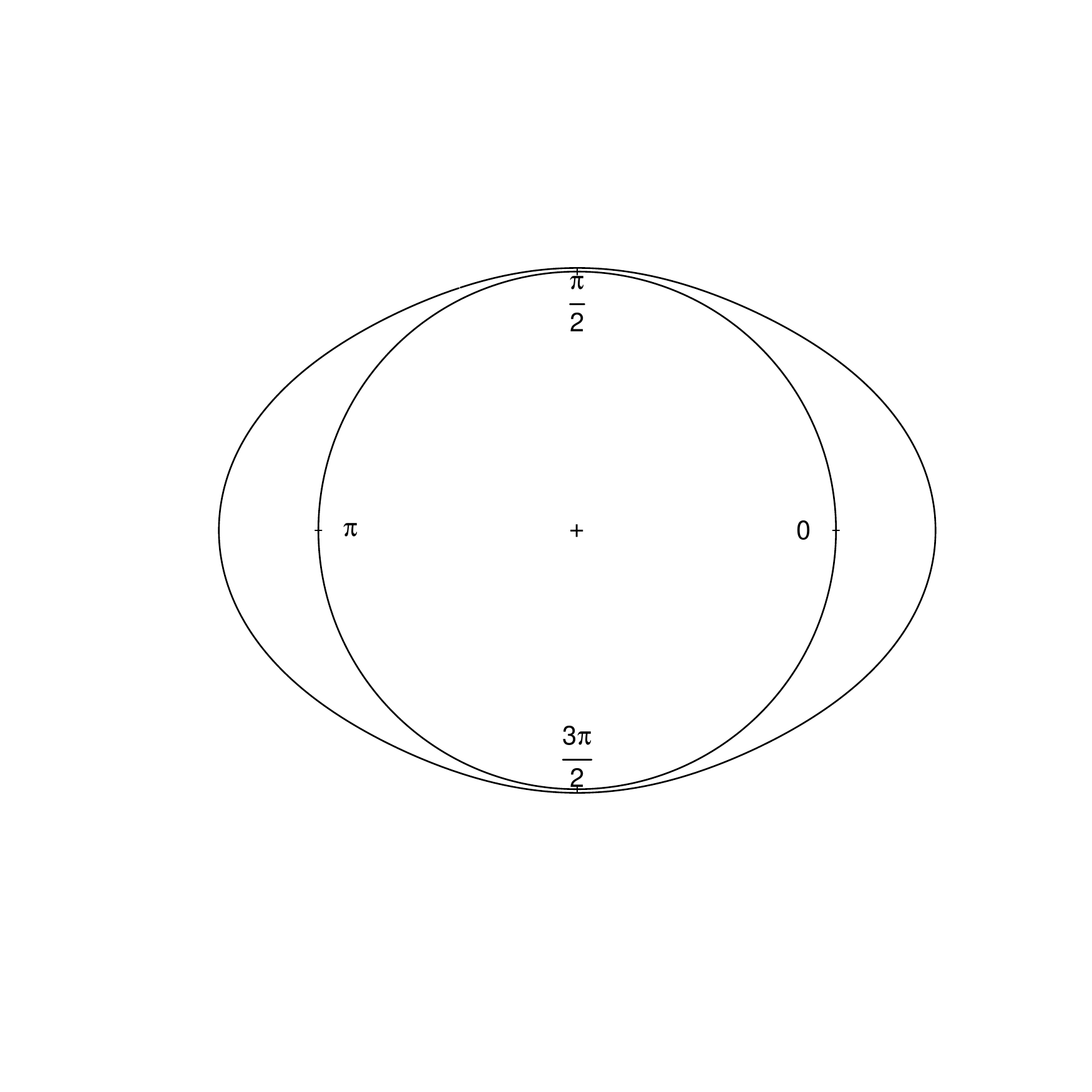}\\
M17
\includegraphics[width=1\linewidth]{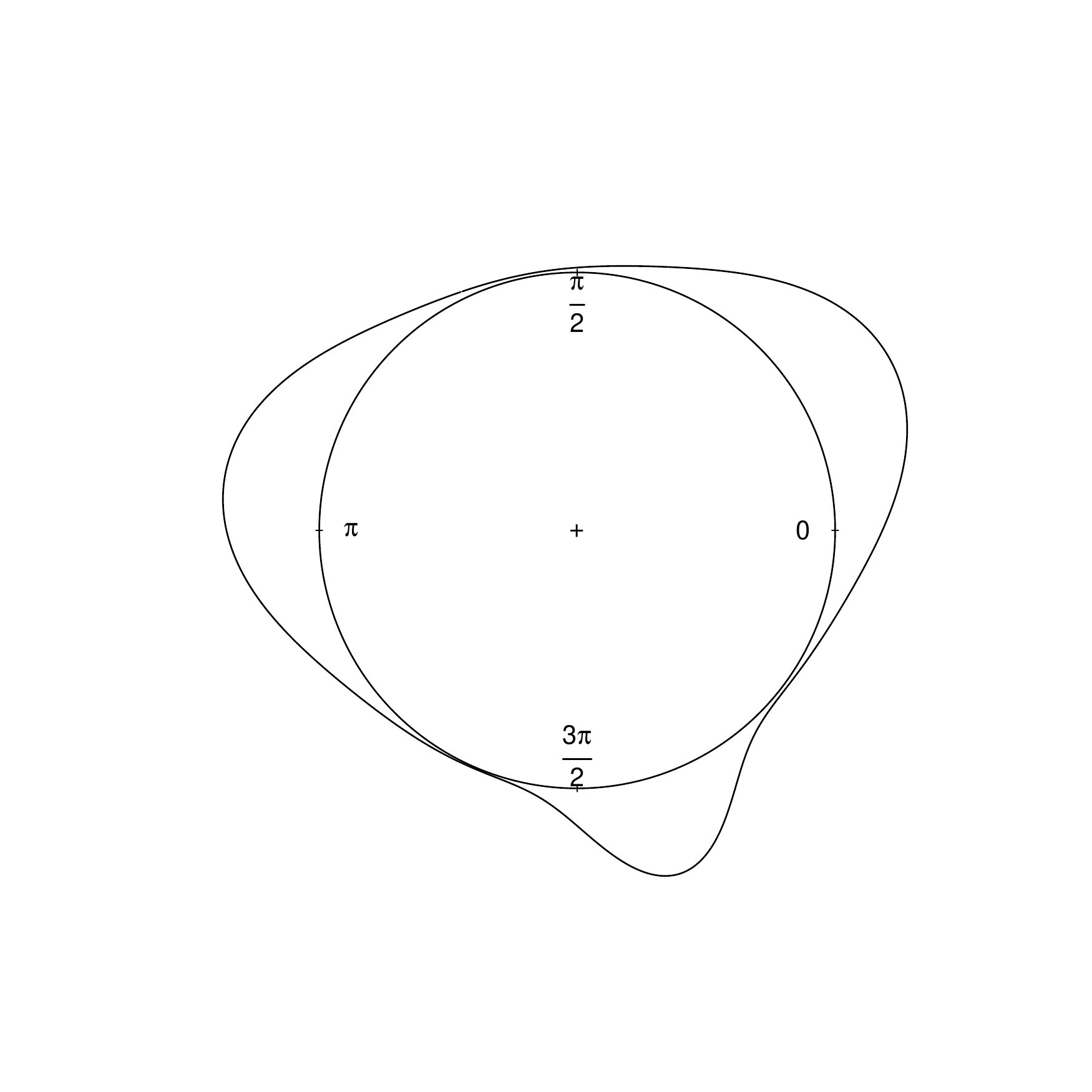}\\
M21
\includegraphics[width=1\linewidth]{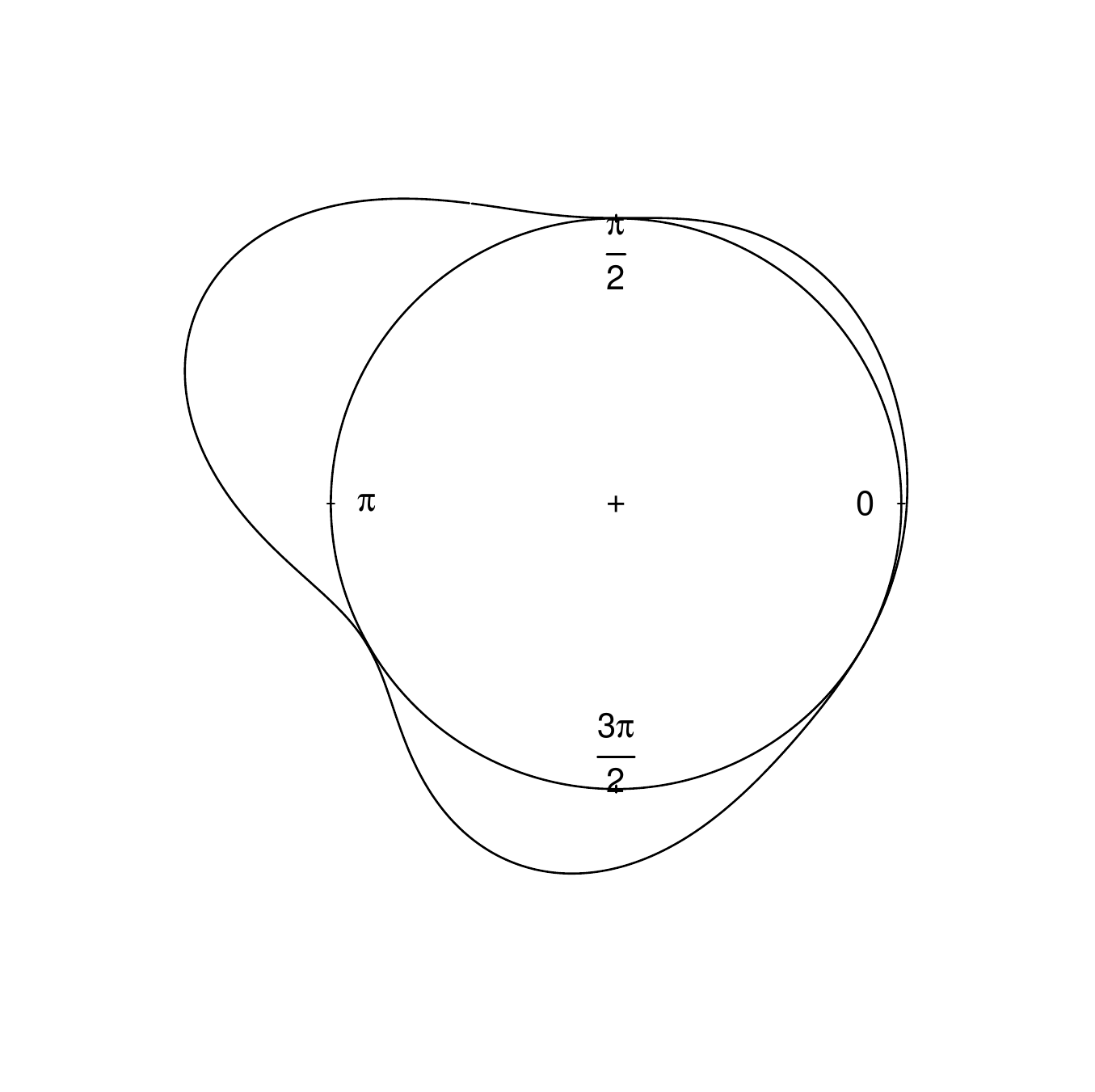}\\
M25
\includegraphics[width=1\linewidth]{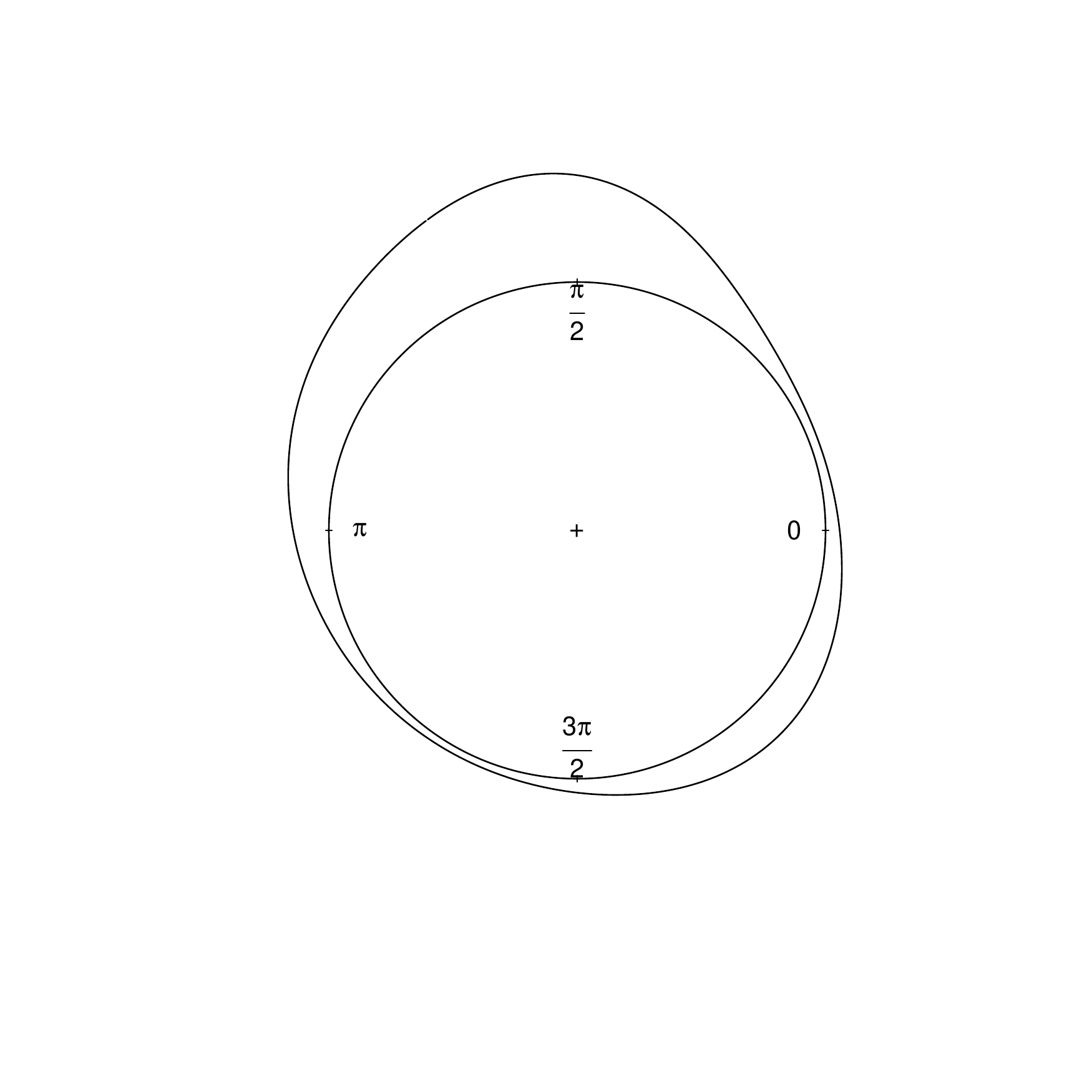}\\
M14
\includegraphics[width=1\linewidth]{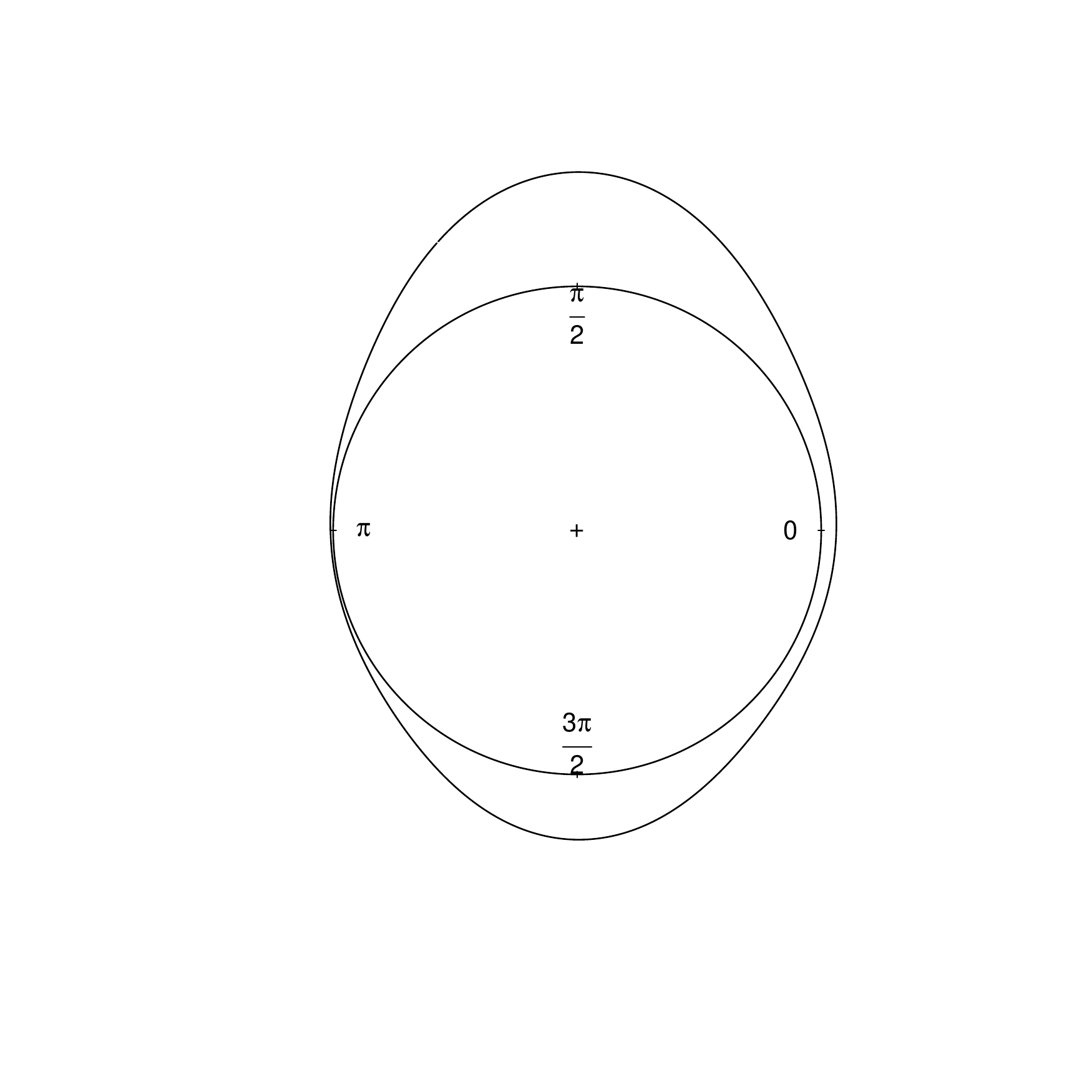}\\
M18
\includegraphics[width=1\linewidth]{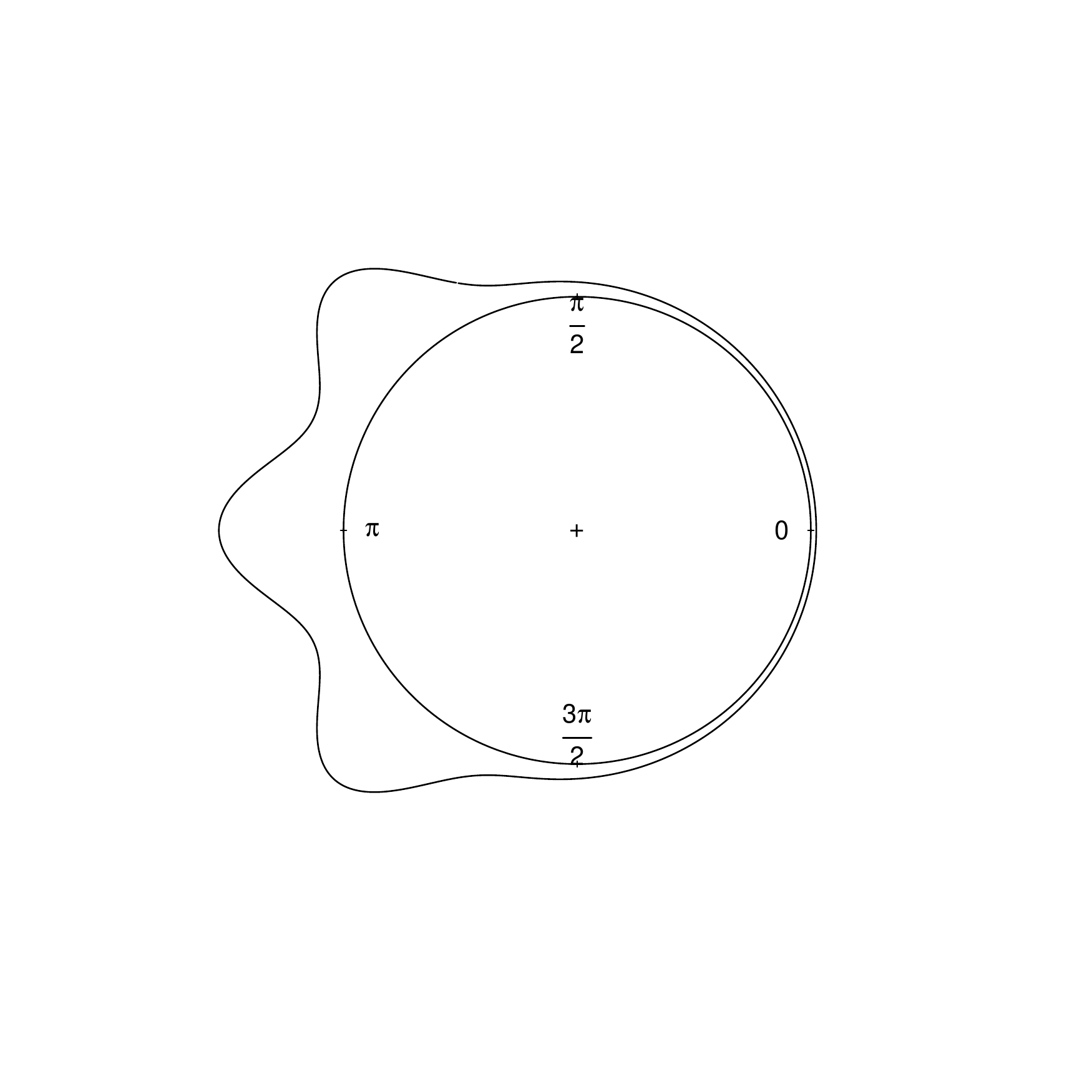}\\
M22
\includegraphics[width=1\linewidth]{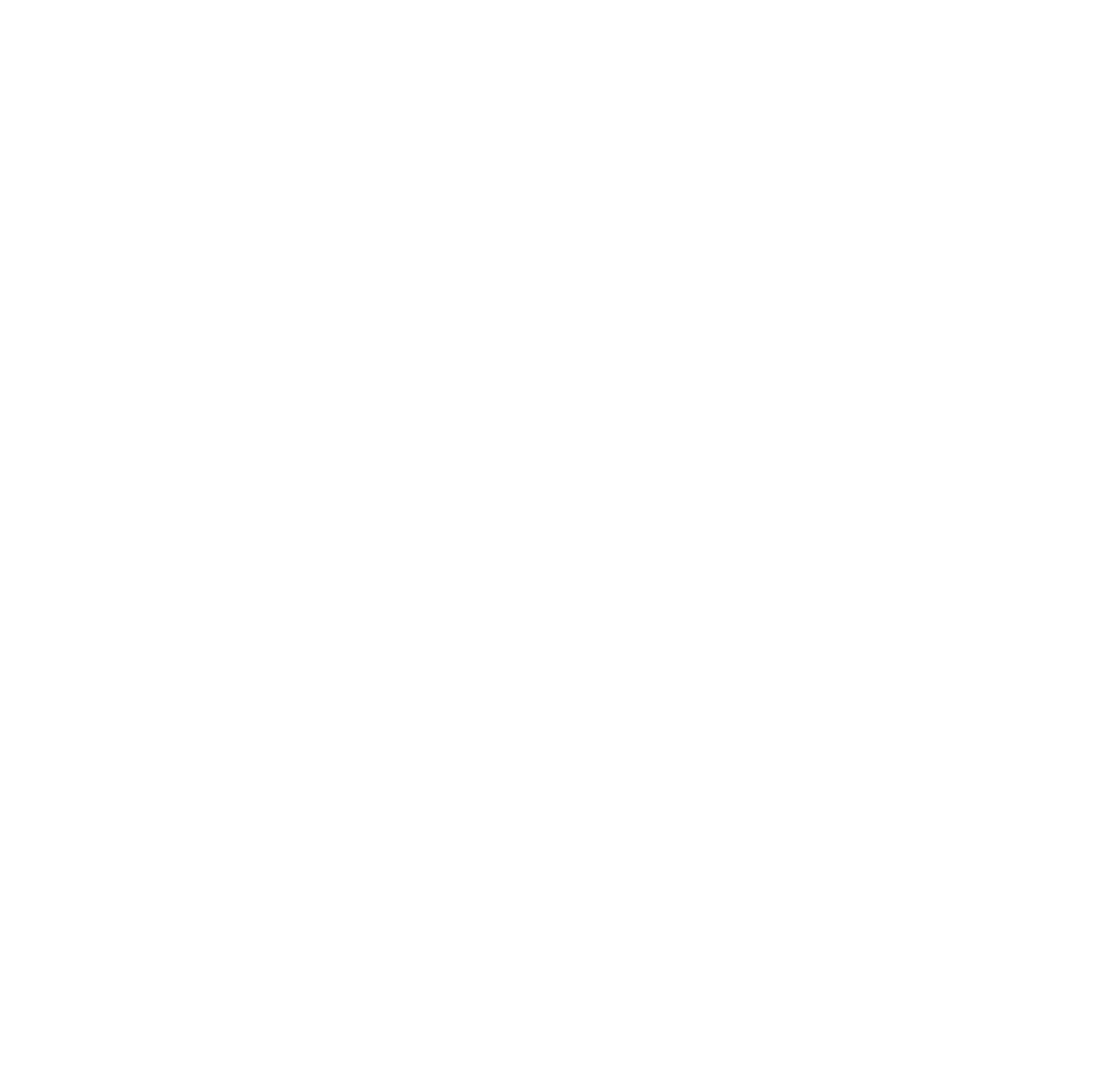}\\
\phantom{M28}
\end{multicols}
\caption{Circular representation of different density functions. M11--M20: bimodal models. M21--M25: trimodal models.}
\label{figs6}
\end{figure}

The unimodal circular density functions employed as $\psi_i$, (reflectively) symmetric with respect to $\mu_i\in[0,2\pi)$, are the following models, as defined in \cite{Jammalamadaka01}: von Mises $vM(\mu_i,\kappa_i)$, with $\kappa_i \geq 0$; wrapped Normal $WN(\mu_i,\rho_i)$, with $\rho_i \in (0,1)$; wrapped Cauchy $WC(\mu_i,\rho_i)$, with $\rho_i \in (0,1)$; and cardioid $C(\mu_i,\rho_i)$, with $\rho_i \in [-0.5,0.5]$. Also, to see the behavior of the test in asymmetric models, it was employed the family of $\mathtt{k}$--sine--skewed distributions \citep[see][]{AbePewsey11}. Given a circular unimodal and symmetric (around $\mu$) density $f$, this family is of the form $f(\theta)(1+\lambda\sin(\mathtt{k}(\theta-\mu)))$, with $\lambda\in (-1,1)$ and $\mathtt{k}\in\{1,2,\ldots\}$. The models employed were the $\mathtt{k}$--sine--skewed wrapped Normal $\mathtt{k}ssWN(\mu,\rho,\lambda,\mathtt{k})$ and the $\mathtt{k}$--sine--skewed von Mises $\mathtt{k}ssvM(\mu,\kappa,\lambda,\mathtt{k})$.  The linear representation of the unimodal circular probability density functions are showed in Figure~\ref{figs3}, the bimodal and trimodal models appear in Figure~\ref{figs4}. The circular representation of the unimodal models can be found in Figure~\ref{figs5}, the bimodal and trimodal models appear in Figure~\ref{figs6}.

\textit{Unimodal models:}

\begin{itemize}
\item M1: $vM(\pi,1)$
\item M2: $WN(\pi,0.9)$
\item M3: $WC(\pi,0.8)$
\item M4: $C(\pi,0.5)$
\item M5: $0.9 \cdot vM(\pi,10) + 0.1\cdot vM(\pi,1)$.
\item M6: $0.2 \cdot vM(2\pi/3,3) + 0.6\cdot vM(\pi,1.4)+ 0.2\cdot vM(4\pi/3,3)$.
\item M7: $0.05 \cdot vM(2\pi/3,7) + 0.9\cdot vM(\pi,1)+ 0.05\cdot vM(4\pi/3,7)$.
\item M8: $0.05 \cdot vM(2\pi/3,4) + 0.9\cdot vM(\pi,1)+ 0.05\cdot vM(4\pi/3,7)$.
\item M9: $\mathtt{k}ssWN(\pi,0.4,0.99,1)$.
\item M10: $\mathtt{k}ssvM(\pi,1,0.9,1)$.
\end{itemize}

\textit{Bimodal models:}

\begin{itemize}
\item M11: $0.5 \cdot vM(2,5) + 0.5\cdot vM(4,5)$.
\item M12: $0.9 \cdot vM(\pi/2,2) + 0.1\cdot vM(3\pi/2,5)$.
\item M13: $0.5 \cdot vM(\pi-1,1.5) + 0.5\cdot vM(\pi+1,1.5)$.
\item M14: $0.3 \cdot vM(\pi/2,6) + 0.5\cdot vM(3\pi/4,2)+ 0.2\cdot vM(7\pi/4,4)$.
\item M15: $\mathtt{k}ssWN(\pi,0.5,0.9,2)$.
\item M16: $\mathtt{k}ssvM(\pi,1,0.8,2)$.
\item M17: $0.5 \cdot vM(0,4) + 0.5\cdot vM(\pi,4)$.
\item M18: $0.1 \cdot vM(0,2) + 0.6\cdot vM(\pi/2,4)+ 0.3\cdot vM(3\pi/2,5)$.
\item M19: $0.5 \cdot vM(0,0.2) + 0.25\cdot WN(\pi/2,0.5) + 0.25\cdot WC(3\pi/2,0.5)$.
\item M20: $0.75 \cdot vM(\pi,1) + 0.25\cdot vM(7\pi/4,10)$.
\end{itemize}

\textit{Trimodal models:}

\begin{itemize}
\item M21: $0.4 \cdot vM(0.5,6) + 0.4\cdot vM(3,6)+ 0.2\cdot vM(5,24)$.
\item M22: $(1/6) \cdot vM(\pi-0.8,30) + 0.5\cdot vM(\pi,1)+(1/6)  \cdot vM(\pi,30) + (1/6) \cdot vM(\pi+0.8,30)$.
\item M23: $0.2 \cdot vM(\pi/2,5) + 0.2\cdot vM(7\pi/8,5)+ 0.6\cdot WN(7\pi/4,0.8)$.
\item M24: $0.2 \cdot vM(\pi/2,6) + 0.2\cdot vM(7\pi/8,2)+ 0.6\cdot WC(7\pi/4,0.7)$.
\item M25: $\mathtt{k}ssWN(\pi,0.5,0.99,3)$.
\end{itemize}

\section{The calibration function in detail}\label{calibrationfun}

In this section, details about the design of the calibration function employed to generate the resamples in the bootstrap procedure are given. As mentioned in Section~\ref{method}, in order to provide the asymptotic behavior of the excess mass test statistic~(\ref{emstat}) some regularity conditions over $f$ are needed. First, assuming that $f$ has $k$ modes, then the regularity conditions are:  

\begin{enumerate}[label={C.\arabic*}] 
\item\label{regcond1} $f$ is bounded and it has a continuous derivative,
\item\label{regcond2} when $\theta\in [0,2\pi)$, $f'(\theta)=0$ holds only for $f(\theta)=0$ or in the modes and antimodes, namely $\theta_i$, with $i=1,\ldots,2k$;
\item\label{regcond3} $f''$ exists and is H\"older continuous within a neighborhood of $\theta_i$.
\end{enumerate}

Under the null hypothesis of having $k$ modes ($H_0:j=k$), when these assumptions are satisfied, following \cite{ChengHall98}, $\Delta_{n,k+1}$ should be independent from $f$, except for the values in the modes and antimodes of

\begin{equation*}
d_i=\frac{|f''(\theta_i)|}{f(\theta_i)^3} \mbox{, with } i=1,\ldots,2k.
\end{equation*}  

Under the previous premises (\ref{regcond1}, \ref{regcond2} and \ref{regcond3}), from the results of \cite{ChengHall98}, the distribution of $\Delta_{n,k+1}$ can be approximated by $\Delta_{n,k+1}^*$, being $\Delta_{n,k+1}^*$ the excess mass statistic calculated from the resamples obtained from a `calibration distribution' $g$. To guarantee this property, the resampling function $g$ must satisfy the same regularity conditions as $f$ and also it should verify that values $\hat{d_i}=|g''(\hat{\theta_i})|/g(\hat{\theta_i})^3$ converge in probability to $d_i$, as $n \rightarrow \infty$, for $i=1,\ldots,2k$.

As it was pointed out in Section~\ref{method}, in order to have a function preserving the structure of the data under the assumption $f$ has $k$ modes, the kernel density estimation with the critical concentration can be used. This function satisfy the regularity conditions with the exception of (\ref{regcond2}) as saddle points may appear. In addition, a modification must to be done in a neighborhood of the modes and antimodes of $\hat{f}_{\nu_k}$, $\hat{\theta_i}\mbox{, with } i=1,\ldots,2k$; to obtain that $\hat{d_i}=|g''(\hat{\theta_i})|/g(\hat{\theta_i})^3$ is a proper estimator of $d_i$. The modifications over $\hat{f}_{\nu_k}$ should be done in a way that conditions (\ref{regcond1}) and (\ref{regcond3}) hold.

The aforementioned modifications in the kernel density estimation are obtained with the $J$ (for the turning points) and $L$ (for the saddle points) modifications introduced in the $g$ function defined in (\ref{gfunc}). But, before giving a complete characterization of these two functions, the link function $l$ needs to be introduced,

{\footnotesize
\begin{align}\label{lfunc}
l(\theta;u,v,a_0,a_1,b_0,b_1) = \frac{a_0-a_1}{2} \left( 1+2 \left( \frac{\theta-u}{v-u} \right)^3 - 3 \left( \frac{\theta-u}{v-u} \right)^2 \right) \exp \left({\frac{2 (\theta-u)b_0}{a_0-a_1}}\right) +& \nonumber\\
+  \frac{a_0-a_1}{2}\left( 2 \left( \frac{\theta-u}{v-u} \right)^3 - 3 \left( \frac{\theta-u}{v-u} \right)^2 \right) \exp \left({\frac{2 (v-\theta)b_1}{a_0-a_1}}\right) + \frac{a_0+a_1}{2},& \nonumber\\
\end{align}} 
\hspace*{-1.8mm}where $a_0 \neq a_1$, $v>u$ and $l$ is defined in $\theta \in [u,v]$; $a_0, a_1$ will be the values of the linked functions in $u$ and $v$; and $b_0$, $b_1$ the values of their derivatives on these points (and they are going to have the same sign). The function $l$ defined in (\ref{lfunc}) will ensure that $g$ is going to be of class one and also that its derivative will always have the same sign in $(u,v)$.

The construction of $J$ will involve two steps. First, provide a function $\mathcal{K}$ in an open neighborhood of $\widehat{\theta_i}$ satisfying: (i) condition (\ref{regcond3}), (ii) $\theta_i$ is the unique turning point in this neighborhood and (iii) $|J''(\hat{\theta_i})|/J(\hat{\theta_i})^3=\hat{d_i}$, where $\hat{d_i}$ is the quantity estimated in (\ref{dihat}). Second, the new function $\mathcal{K}$ will be connected with $\hat{f}_{\nu_k}$ using the link function $l$ defined in (\ref{lfunc}), in order to ensure condition (\ref{regcond1}). The aforementioned $\mathcal{K}$ is defined as follows

\begin{equation*}
\mathcal{K}(\theta; \widehat{\theta_i},\widehat{f}_{\nu_k}(\widehat{\theta_i}),\widehat{f}_{\nu_{\tiny{\mbox{PI}}}}''(\widehat{\theta_i}),\eta_i)=\widehat{f}_{\nu_k}\left(\widehat{\theta_i}\right)\left( 1+ \delta_i \left( \frac{\theta-\widehat{\theta_i}}{\eta_i} \right)^2 \right)^{\eta_i^2 \frac{\delta_i \cdot \hat{f}_{\nu_{\tiny{\mbox{PI}}}}''\left(\widehat{\theta_i}\right)}{2 \hat{f}_{\nu_k}\left(\widehat{\theta_i}\right)}}, 
\end{equation*}
being $\delta_i= (-1)$ if $\widehat{\theta_i}$ is a mode and $\delta_i= 1$ if it is an antimode. The parameter $\eta_i$, defined in (\ref{variabJ}), will directly depend on $\varsigma_i$ and this last parameter controls at which density height of $\hat{f}_{\nu_k}$ the modification of the $J$ function is done. 

From functions $\mathcal{K}$ and $l$, if $\boldsymbol{\rho}_i$ denotes $(\widehat{\theta_i},\widehat{f}_{\nu_k}(\widehat{\theta_i}),\widehat{f}_{\nu_{\tiny{\mbox{PI}}}}''(\widehat{\theta_i}))$, the function $J$ is constructed in the following way:

{\scriptsize
\begin{equation} \label{jfunc}
J(\theta;\widehat{\theta_i},\nu_k,\nu_{\tiny{\mbox{PI}}},\varsigma_i)= 
  \begin{cases}
     l\left(\theta;\mathfrak{r}_i,\mathfrak{v}_i, \widehat{f}_{\nu_k}(\mathfrak{r}_i),\mathcal{K}(\mathfrak{v}_i;  \boldsymbol{\rho}_i,\eta_i), \widehat{f}'_{\nu_k}(\mathfrak{r}_i), \mathcal{K}'(\mathfrak{v}_i;  \boldsymbol{\rho}_i,\eta_i) \right) &\mbox{ if } \theta\in(\mathfrak{r}_i,\mathfrak{v}_i), \\
		\mathcal{K}(\theta; \boldsymbol{\rho}_i,\eta_i)  &\mbox{ if } \theta\in[\mathfrak{v}_i, \mathfrak{w}_i], \\
		 l\left(\theta;\mathfrak{w}_i,\mathfrak{s}_i,\mathcal{K}(\mathfrak{w}_i;  \boldsymbol{\rho}_i,\eta_i), \widehat{f}_{\nu_k}(\mathfrak{s}_i),\mathcal{K}'(\mathfrak{w}_i;  \boldsymbol{\rho}_i,\eta_i), \widehat{f}'_{\nu_k}(\mathfrak{s}_i)\right) &\mbox{ if } \theta\in(\mathfrak{w}_i,\mathfrak{s}_i), \\
  \end{cases}
\end{equation}}
\hspace*{-1.8mm}being $\mathfrak{v}_i=\widehat{\theta_i}- \eta_i/2$ and $\mathfrak{w}_i=\widehat{\theta_i}+ \eta_i/2$. The function $J$ described in (\ref{jfunc}) depends on the constants $\varsigma_i \in (0,1/2)$. As mentioned before, in each mode or antimode, $\widehat{\theta_i}$, the value $\varsigma_i$ is employed to define at which density height the modification of the kernel density estimation is made. Values of $\varsigma_i$ close to 0 imply a modification in an ``small'' neighbourhood around $\widehat{\theta_i}$. To provide the remaining constants in (\ref{jfunc}), let first order (as it was a real sequence) the modes and antimodes $(0\leq \widehat{\theta_1}<\ldots<\widehat{\theta}_{2k}<2\pi)$ and denote as $\widehat{\theta_0}=\widehat{\theta}_{2k}-2\pi$ and $\widehat{\theta}_{2k+1}=\widehat{\theta_{1}}+2\pi$. Considering the periodicity of $\widehat{f}_{\nu_k}$ and the greater and lower inequality as defined in the real line, then

{\footnotesize
\begin{eqnarray}\label{variabJ}
 \vartheta_i&=& \widehat{f}_{\nu_k}(\widehat{\theta_i}) + \delta_i \cdot \varsigma_i \cdot \min\left(|\widehat{f}_{\nu_k}(\widehat{\theta_i}) - \widehat{f}_{\nu_k}(\widehat{\theta}_{i-1})|,|\widehat{f}_{\nu_k}(\widehat{\theta_i}) - \widehat{f}_{\nu_k}(\widehat{\theta}_{i+1})|\right), \nonumber\\
\mathfrak{r}_i&=&\inf\{\theta: \theta>\widehat{\theta}_{i-1}, \delta_i \cdot \widehat{f}_{\nu_k}(\theta)\leq  \delta_i \cdot \vartheta_i  \mbox{ and } \widehat{f}'_{\nu_k}(\theta)\neq 0\}, \nonumber\\
\mathfrak{s}_i&=&\sup\{\theta: \theta<\widehat{\theta}_{i+1}, \delta_i \cdot \widehat{f}_{\nu_k}(\theta)\leq \delta_i \cdot \vartheta_i \mbox{ and } \widehat{f}'_{\nu_k}(\theta)\neq 0 \} , \nonumber\\
\eta_i&=& \sup \{\gamma: \gamma \in (0,\min(\widehat{\theta_{i}}-\mathfrak{r}_i,\mathfrak{s}_i-\widehat{\theta_{i}})), \delta_i \mathcal{K}(\widehat{\theta_i}+\gamma/2; \boldsymbol{\rho}_i,\gamma) \leq \delta_i (\widehat{f}_{\nu_k}(\widehat{\theta_i}) +\vartheta_i)/2  \nonumber\\
&&\quad \quad \mbox{and }  \widehat{f}'_{\nu_k}(\widehat{\theta_i}\pm\gamma/2)\neq 0\}. \nonumber\\
\end{eqnarray}
}

\begin{figure}
 \centering
\begin{tabular}{cc}
\subfloat{
    \includegraphics[width=0.49\textwidth]{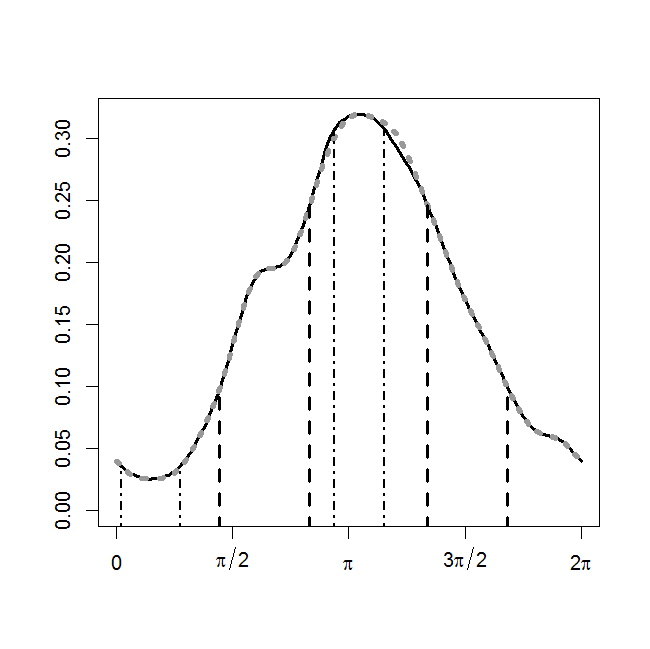}}
\subfloat{
    \includegraphics[width=0.49\textwidth]{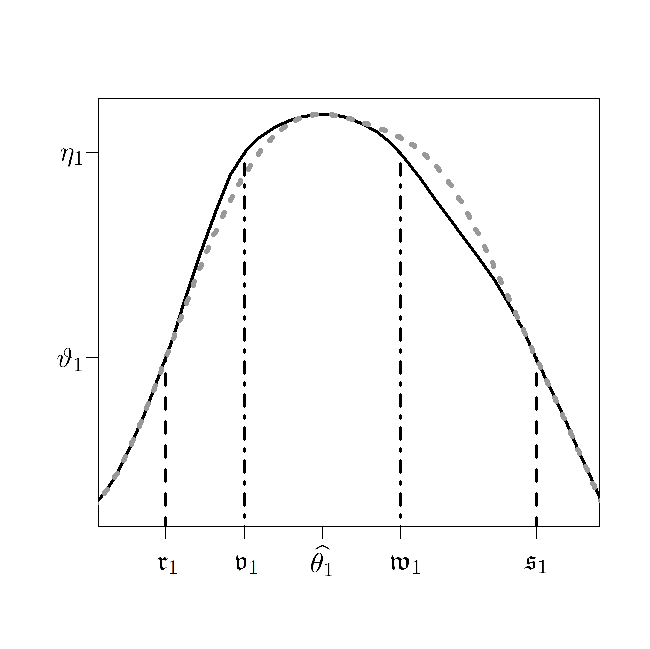}} 
\end{tabular}
 \caption{Sample of $n=200$ observations obtained from model M7 (described in Appendix~\ref{app_sim_models}). Dotted line (in grey): kernel density estimation with critical concentration parameter for one mode, $\hat{f}_{\nu_1}$. Solid line: calibration function $g$. Dashed line: neighborhood where the functions $J(\cdot;\widehat{\theta_i},\nu_1,\nu_{\tiny{\mbox{PI}}},0.25)$ are defined, with $i=1,2$. Dot--dashed line: neighborhood where the $\mathcal{K}$ functions are defined. Left: in the support $[0,2\pi)$. Right: in a neighborhood of the mode $\widehat{\theta_1}$.}
 \label{fig1}
\end{figure}

A representation of the modification achieved with the $J$ function is provided in Figure \ref{fig1}. For convenience, in practice, the values of $\varsigma_i$ are taken close enough to 0 to avoid an impact on the integral value of the calibration function.

As it was pointed out, the objective of the second modification, achieved with $L$, is to remove the remaining saddle points of the calibration function (those ones outside the neighborhoods of the modes and in $\widehat{f}_{\nu_k}$). Denoting the ordered saddle points as $\zeta_i$ $(0\leq \zeta_1< \ldots<\zeta_t<2\pi)$, $\xi$ is defined as $\xi=\min \{d(\dot{\theta},\ddot{\theta}): \dot{\theta},\ddot{\theta} \in (\zeta_1,\ldots,\zeta_t) \cup  (\mathfrak{r}_1,\mathfrak{s}_1,\ldots, \mathfrak{r}_{2k-1},\mathfrak{s}_{2k-1}) \}$, where $d$ is the circular (geodesic) distance. Then, the neighborhood used to remove the saddle points are delimited by $z_{(2p-1)}=\zeta_p-\varpi \cdot \xi$ and $z_{(2p)}=\zeta_p+\varpi \cdot \xi$, with $\varpi \in (0,1/4)$ and $p=1,\ldots,t$. In the simulation study, the value of $\varpi$ was also taken close enough to 0 to avoid an impact in the value of the integral associated to $g$. Once these points are calculated, the stationary and turning points can be removed from $g$ with the function $L$ constructed from the link function in the following way

{\footnotesize
\begin{equation}\label{Lgranfunc}
L(\theta;z_{(2p-1)},z_{(2p)},\nu_k)=l(\theta;z_{(2p-1)},z_{(2p)},\widehat{f}_{\nu_k}(z_{(2p-1)}),\widehat{f}_{\nu_k}(z_{(2p)}),\widehat{f}'_{\nu_k}(z_{(2p-1)}),\widehat{f}'_{\nu_k}(z_{(2p)})).
\end{equation} 
}

Finally, using the $J$ and $L$ functions defined, respectively, in (\ref{jfunc}) and (\ref{Lgranfunc}), the complete characterization of the calibration function (\ref{gfunc}), given the vector $\boldsymbol{\varsigma}=(\varsigma_1,\ldots,\varsigma_{2k})$, is obtained as follows

{\small
\begin{equation*}
g(\theta;\nu_k,\nu_{\tiny{\mbox{PI}}},\boldsymbol{\varsigma})=
  \begin{cases}
J(\theta;\widehat{\theta_i},\nu_k,\nu_{\tiny{\mbox{PI}}},\varsigma_i) & \mbox{if } \theta \in [\widehat{\theta_{1}},\widehat{\theta}_{2k}] \mbox{ and } \theta \in(\mathfrak{r}_i,\mathfrak{s}_i)\\
&   \mbox{for some } i \in \{1,\ldots,2k\}, \\
J(\theta;\widehat{\theta_1},\nu_k,\nu_{\tiny{\mbox{PI}}},\varsigma_1) & \mbox{if } \mathfrak{r}_1\geq 0 \mbox{ and } \theta \in(\mathfrak{r}_1,\widehat{\theta_{1}}) \mbox{ or }\\
&\mbox{if } \mathfrak{r}_1< 0 \mbox{ and } \theta \in[0,\widehat{\theta_{1}}), \\
J(\theta-2\pi;\widehat{\theta_1},\nu_k,\nu_{\tiny{\mbox{PI}}},\varsigma_1) & \mbox{if } \mathfrak{r}_1< 0 \mbox{ and } \theta \in(\mathfrak{r}_1,2\pi), \\
J(\theta;\widehat{\theta_{2k}},\nu_k,\nu_{\tiny{\mbox{PI}}},\varsigma_{2k}) & \mbox{if } \mathfrak{s}_{2k}<2\pi \mbox{ and } \theta \in(\widehat{\theta}_{2k},\mathfrak{s}_{2k})\mbox{ or }\\
&\mbox{if } \mathfrak{s}_{2k} \geq 2\pi \mbox{ and } \theta \in(\widehat{\theta}_{2k},2\pi), \\
J(\theta+2\pi;\widehat{\theta_{2k}},\nu_k,\nu_{\tiny{\mbox{PI}}},\varsigma_{2k}) & \mbox{if } \mathfrak{s}_{2k} \geq 2\pi \mbox{ and } \theta \in [0,\mathfrak{s}_{2k}-2\pi),\\
L\left(\theta;z_{(2p-1)},z_{(2p)},\nu_k\right) & \mbox{if } \theta \in [\zeta_1,\zeta_t] \mbox{ and } \theta \in(z_{(2p-1)},z_{(2p)}) \\
&\mbox{for some } p \in \{1,\ldots,t\}, \\
L\left(\theta;z_{(1)},z_{(2)},\nu_k\right) & \mbox{if } z_1\geq 0 \mbox{ and } \theta \in(z_1,\zeta_1) \mbox{ or }\\
&\mbox{if } z_1< 0 \mbox{ and } \theta \in[0,\zeta_1), \\
L\left(\theta-2\pi;z_{(1)},z_{(2)},\nu_k\right)& \mbox{if } z_1< 0 \mbox{ and } \theta \in(z_1,2\pi), \\
L\left(\theta;z_{(2t-1)},z_{(2t)},\nu_k\right) & \mbox{if } z_{2t}<2\pi \mbox{ and } \theta \in(\zeta_{t},z_{2t})\mbox{ or }\\
&\mbox{if } z_{2t} \geq 2\pi \mbox{ and } \theta \in(\zeta_{2k},2\pi), \\
L\left(\theta+2\pi;z_{(2t-1)},z_{(2t)},\nu_k\right) & \mbox{if } z_{2t} \geq 2\pi \mbox{ and } \theta \in [0,z_{2t}-2\pi),\\
\hat{f}_{\nu_k}(\theta) & \mbox{in other case.}
	\end{cases}
\end{equation*}
}

Note that this $g$ function is nothing else than the function defined in~(\ref{gfunc}) taken into account the circular structure of the support. This last function $g$ can be used as a calibration function for testing $H_0:j=k$ since it satisfies the following conditions
\begin{enumerate}
\item $g$ is bounded and it has a continuous derivative;
\item $g'(\theta)=0$ holds only in the turning points $\theta\in[0,2\pi)$ such as $\theta=\widehat{\theta_i}$, where $\widehat{\theta_i}$, with $i=1,\ldots,2k$, are the modes and antimodes of $\hat{f}_{\nu_k}$;
\item $g''$ exists and is H\"older continuous within a neighbourhood of $\widehat{\theta_i}$, with $g''(\widehat{\theta_i})\neq 0$;
\item $|g''(\widehat{\theta_i})|/g(\widehat{\theta_i})^3=\hat{d_i}$, where $\hat{d_i}$ is the the estimator defined in (\ref{dihat}).
\end{enumerate}

\section{The construction of the fire patches}\label{firepatch}

As pointed out in Section~\ref{datafires}, those areas where similar fire pattern behaviour is expected must be identified. The objective of this section is to show how the fires patches were created with the following two steps. First, since the original land cover database is divided in pixels of size 300 meters is necessary to define the principal land cover of the $0.5\degree$ cell. Second, the land cover cells are grouped into homogeneous patches, where a similar fire season modality patterns is expected.

\begin{figure}
\centering
\begin{tabular}{cc}
		\subfloat{
    \includegraphics[height=0.225\textheight]{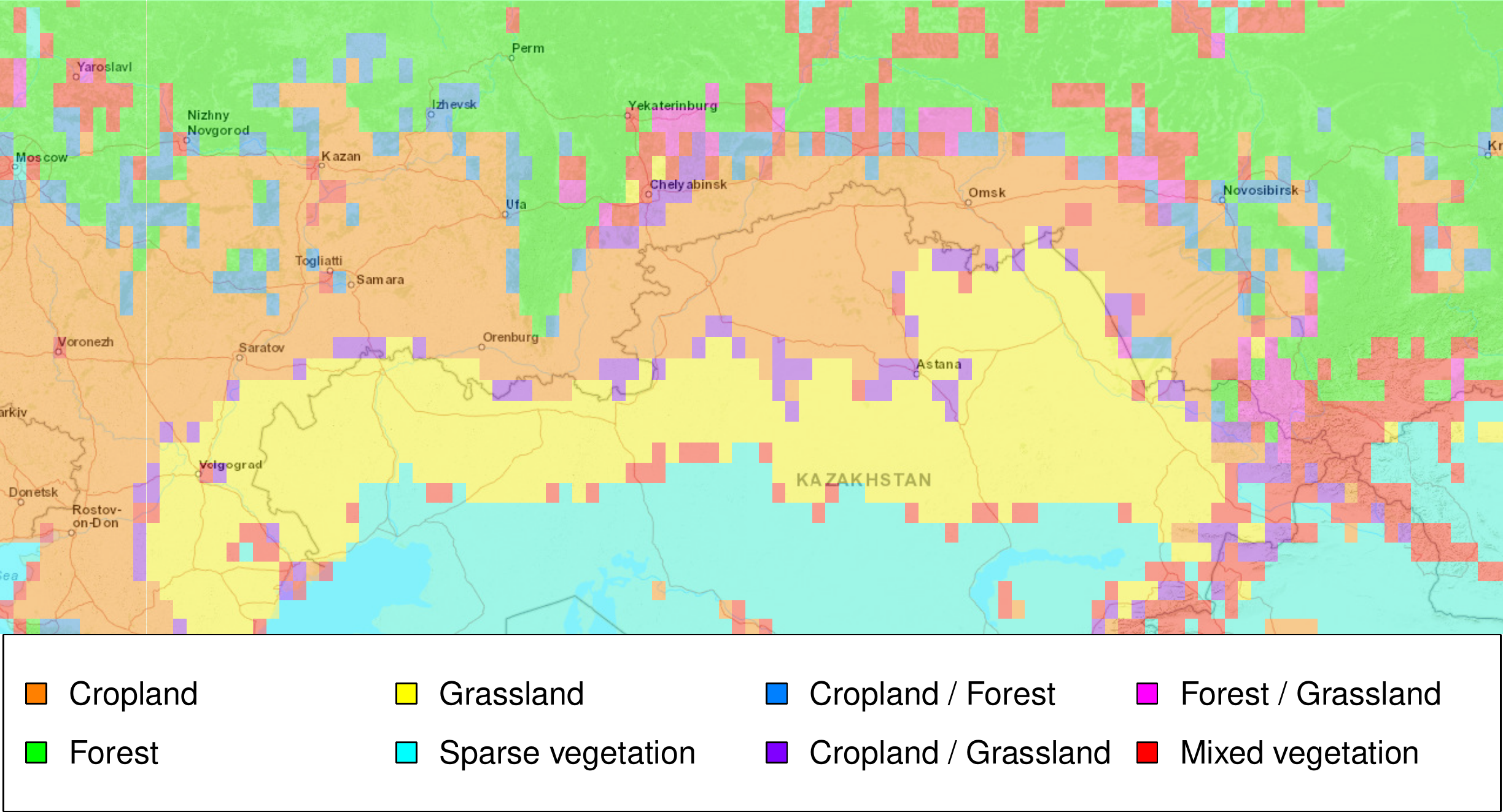}}
		\subfloat{
    \includegraphics[height=0.225\textheight]{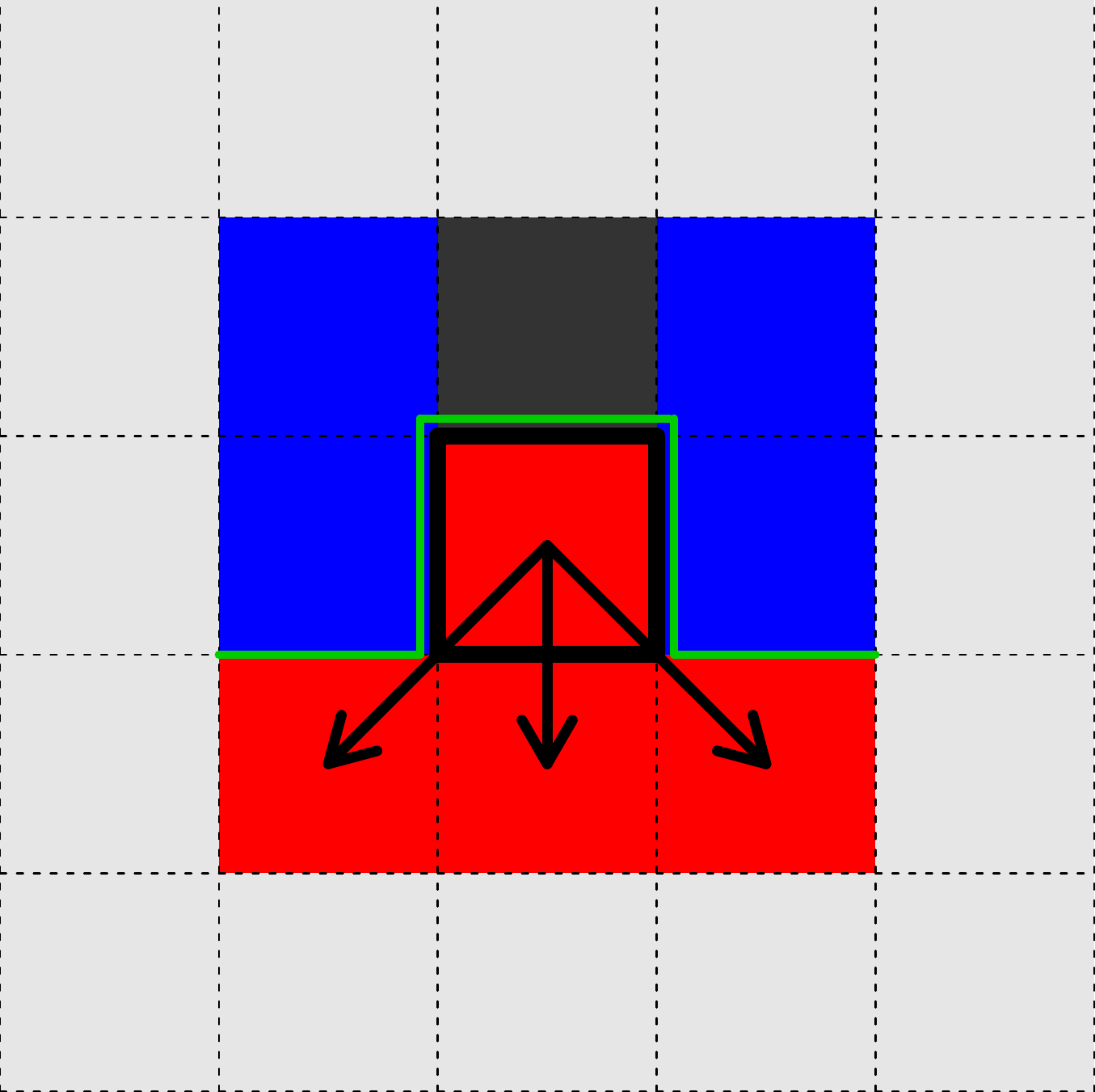}}
\end{tabular}
 \caption{Left: land cover, as specified in the legend, in different cells of an area between South--western Russia and Northern Kazakhstan. Right: construction of the fire patch; the cells are separated by dashed lines; in the middle (black square): focus cell; red and blue: different land uses in the cells; light gray: not considered cells; dark gray: region outside the studied area; the arrows indicate cells to be aggregated into a homogeneous land cover patch; green lines: part of the border of the patch.}
 \label{figant2}
\end{figure}

Then, the first objective is to decide which is the principal land cover in a given cell. For that purpose the original labels of the 300m--pixel are divided in six classes: \textit{cropland} (rainfed; irrigated or post--flooding), \textit{forest} (tree cover; broadleaved, needleleaved or mixed leaf type; evergreen or deciduous; closed or open), \textit{shrubland}, \textit{grassland} (herbaceous cover, grassland), \textit{lichens and mosses} and \textit{low vegetation} (sparse vegetation; tree cover, flooded; urban areas; bare areas; water bodies; permanent snow and ice). Also, some of the 300m--pixels have a mixed label, e. g., ``Mosaic tree and shrub $(>50\%)$ / herbaceous cover $(<50\%)$'', in that case the $75\%$ of the pixel is cataloged as the principal use and the $25\%$ of the pixel is cataloged as the secondary use. If the principal or the secondary use belong to a mixed label (in the last example the principal use is a mix of forest and shrubland), then its percentage is divided according to the surrounding $(180\times180)$ 300m--pixels in the $0.5\degree$--cell. In the last example if $20,000$ (out of $180\times180$) surrounding pixels are forest and $10,000$ shrubland, then the $75\cdot 20,000/30,000=50\%$ of the pixel is cataloged as forest and the $25\%$ as shrubland. Once the $(180\times180)$ 300m--pixels inside a given $0.5\degree$--cell are cataloged, if at least the $60\%$ of the land cover is of one type, then the cell is labeled like most pixels. If both the principal and secondary land cover are greater than $30\%$ and lower than $60\%$ (and not the principal nor the secondary land covers are low vegetation), then the cell is cataloged with a mixed label. Otherwise, the cell is labeled as \textit{mixed vegetation}. This classification leaves, in the studied region, the eight land covers represented in Figure \ref{figant2} (left).  

Now, in order to create the homogeneous land cover patches, given the grid cells surrounding a selected cell, if at least one of them has the same label as the selected one, then both cells will be considered in the same patch. Fire patches are created as follows: fix a cell and consider the eight surrounding cells; check which surrounding cells belong to the same class as the focal cell. All of them will be part of the same patch. This is shown in Figure~\ref{figant2} (right). Taking as a reference the central cell, this pixel will form part of the same fire patch as the red cells in the bottom. The remaining cells are going to be outside the patch: the blue ones have a different classification and the dark gray cell is outside of the studied area. Then, part of the border of this patch (green line) is determined by these non--red cells.

\newpage

\end{document}